\documentclass[
  ,draft            
  ,numberedheadings 
  ]
  {aippro}

\layoutstyle{6x9}

\usepackage{epsfig}       
\usepackage[dvips]{color}
\usepackage{mathrsfs}
\usepackage{amssymb}  

\definecolor{Red}{rgb}{1.00, 0.00, 0.00}

\definecolor{Green}{rgb}{0.00, 1.00, 0.00}

\definecolor{Blue}{rgb}{0.00, 0.00, 1.00}

\definecolor{Cyan}{rgb}{0.00, 1.00, 1.00}

\definecolor{Magenta}{rgb}{1.00, 0.00, 1.00}

\definecolor{Yellow}{rgb}{1.00, 1.00, 0.00}

\definecolor{DarkYellow}{rgb}{0.9, 0.8, 0.0}
\definecolor{Gold}{rgb}{1.0, 0.84, 0.00} 

\definecolor{DarkRed}{rgb}{0.75, 0.0, 0.0}

\definecolor{DarkGreen}{rgb}{0.20, 0.55, 0.20}

\definecolor{DarkBlue}{rgb}{0.00, 0.00, 0.70}
  
\definecolor{PaleBlue}{rgb}{0.40, 0.75, 1.00}
 
\definecolor{PaleGreen}{rgb}{0.50, 1.00, 0.50}

\definecolor{PaleRed}{rgb}{1.00, 0.30, 0.30}

\definecolor{Orange}{rgb}{1.00, 0.5, 0.00}
\definecolor{DarkOrange}{rgb}{0.8, 0.4, 0.00} 
\definecolor{DeepPink}{rgb}{1.00, 0.1, 0.6}
\definecolor{Violet}{rgb}{0.5, 0.00, 0.6}

\definecolor{DarkViolet}{rgb}{0.5, 0.00, 0.6}
\definecolor{Brown}{rgb}{0.54, 0.27, 0.07}

\def\db{\color{DarkBlue}}
\def\rr{\color{DarkRed}}

\renewcommand{\leadsto}{\to}

\newcommand{\ep}{\varepsilon}
\newcommand{\eps}{\epsilon}

\newcommand{\om}{\omega}

\newcommand{\ul}{\underline}
\newcommand{\txt}{\textstyle}

\newcommand{\beq}{\begin{equation}}
\newcommand{\eeq}{\end{equation}}
\newcommand{\ba}{\begin{array}}
\newcommand{\bea}{\begin{eqnarray}}
\newcommand{\ea}{\end{array}}
\newcommand{\eea}{\end{eqnarray}}
\newcommand{\bi}{\begin{itemize}\setlength{\itemsep}{0\parsep}}
\newcommand{\ei}{\end{itemize}}
\newcommand{\ben}{\begin{enumerate}\setlength{\itemsep}{0\parsep}}
\newcommand{\een}{\end{enumerate}}
\newcommand{\bc}{\begin{center}}
\newcommand{\ec}{\end{center}}
\newcommand{\bfl}{\begin{flushleft}}
\newcommand{\efl}{\end{flushleft}}
\newcommand{\bfr}{\begin{flushright}}
\newcommand{\efr}{\end{flushright}}

\renewcommand{\slash}{\!\!\!\!/\,}

\newcommand{\ol}{\overline}

\newcommand\comment[1]{ \hbox{[{\it Comment suppressed here.}\/]} }
\newcommand\hide[1]{}

\newcommand{\Or}{{\mathcal O}}
\newcommand{\C}{{\mathcal C}}
\newcommand{\tr}{\hbox{tr}}
\newcommand{\Tr}{\hbox{Tr}}

\renewcommand{\Re}{{\rm Re}\,}
\newcommand{\sign}{{\rm sign}}

\newcommand{\skipover}[1]{}
\newcommand{\nn}{\nonumber \\}
\newcommand{\nnn}{\nonumber}

\newcommand{\half} {{\txt \frac{1}{2}}}

\newcommand{\pa}{\parallel}
\newcommand{\pe}{\perp}

\newcommand{\bk}{{\mathbf k}}
\newcommand{\bp}{{\mathbf p}}
\newcommand{\bq}{{\mathbf q}}

\newcommand{\bv}{{\mathbf v}}
\newcommand{\bs}{{\mathbf s}}
\newcommand{\bx}{{\mathbf x}}
\newcommand{\by}{{\mathbf y}}
\newcommand{\bz}{{\mathbf z}}
\newcommand{\rmd}{{\mathrm d}}
\newcommand{\cl}{{\rm cl}}

\begin{document}

\title{Introduction to Nonequilibrium Quantum Field Theory}

\author{J{\"u}rgen Berges}{
  address={Institut f{\"u}r Theoretische Physik, Universit{\"a}t Heidelberg\\
           Philosophenweg 16, 69120 Heidelberg, Germany\\
           \url{http://www.thphys.uni-heidelberg.de/~berges}}
}

\begin{abstract}
There has been substantial progress in recent years in the
quantitative understanding of the nonequilibrium time evolution of 
quantum fields. Important topical applications, in particular 
in high energy particle physics and cosmology, involve
dynamics of quantum fields far away from the ground state
or thermal equilibrium. In these cases, standard 
approaches based on small deviations from equilibrium,
or on a sufficient homogeneity in time underlying kinetic
descriptions, are not applicable. A particular
challenge is to connect the far-from-equilibrium dynamics
at early times with the approach to thermal equilibrium
at late times. Understanding the ``link'' between the
early- and the late-time behavior of quantum fields 
is crucial for a wide range of phenomena. For the first
time questions such as the explosive particle production 
at the end of the inflationary universe, including the subsequent process of 
thermalization, can be addressed in quantum field theory from first
principles. The progress in this field is based
on efficient functional integral techniques,
so-called $n$-particle irreducible effective actions,
for which powerful nonperturbative approximation
schemes are available. 
Here we give an introduction to these techniques and
show how they can be applied in practice. Though 
we focus on particle physics and
cosmology applications, we emphasize that these techniques 
can be equally applied to other nonequilibrium phenomena
in complex many body systems.\\[0.2cm] 

\noindent
Based on summer school lectures presented at 
HADRON-RANP 2004, March 28 -- April 3, 2004, 
Rio de Janeiro, Brazil and
at QUANTUM FIELDS IN AND OUT OF EQUILIBRIUM,
September 23 -- 27, 2003, Bielefeld, Germany.
\end{abstract}

\maketitle

\newpage

\tableofcontents

\section{Motivations and overview}

Cosmology is time evolution. Solid theoretical descriptions exist
for the temporal and spatial local thermal equilibrium related
to the time evolution of a homogeneous and isotropic
metric and for small perturbations of this. Furthermore,
there are powerful numerical techniques for $N$-particle
simulations. In contrast, research concerning the interplay 
between fluctuations and the time evolution of fields is still 
scarce. Their interaction is of crucial importance for the generation
of density fluctuations during the inflationary phase of the
early universe and of entropy at the end of inflation. 
The corresponding temperature fluctuations in the
cosmic microwave background radiation have led to spectacular
high-precision measurements of cosmological parameters.
It is the dynamics of fluctuations which decides the question whether the
baryon asymmetry was created during a cosmological phase 
transition. Back reactions of large density fluctuations
on the evolution of the cosmic scale factor are also possible.

A frequently employed strategy is to concentrate on classical 
statistical field theory, which can be simulated numerically.
It gives important insights when the number of field quanta per mode 
is sufficiently large such that {\rr\em quantum fluctuations} 
are suppressed compared to statistical fluctuations.
However, classical Rayleigh-Jeans divergences and the lack of genuine quantum 
effects --- such as the approach to quantum thermal equilibrium
characterized by Bose-Einstein or Fermi-Dirac statistics --- limit
their use.
A coherent understanding of the {\rr\em time evolution in quantum field theory}
is required --- a program which has made substantial progress in recent years 
with the development of powerful theoretical techniques.
For the first time questions such as the explosive particle production 
at the end of the inflationary universe, including the subsequent process of 
{\rr\em thermalization}, can be addressed in quantum field theory from first
principles. Thermalization leads to the loss of a 
substantial part of the information about the conditions in the early universe.
The precise understanding of phenomena out of equilibrium play therefore 
a crucial role for our knowledge about the primordial universe. Important 
examples are the density fluctuations, nucleosynthesis or baryogenesis --- the
latter being responsible for our own existence.  

The abundance of experimental data on matter in extreme conditions from
relativistic heavy-ion collision experiments, 
as well as applications in astrophysics and cosmology urge a quantitative 
understanding of {\rr\em far-from-equilibrium} quantum field theory.
The initial stages of a heavy-ion collision require  
considering extreme nonequilibrium dynamics.
Connecting this far-from-equilibrium dynamics at early
times with the approach to thermal equilibrium at late
times is a challenge for theory. The experiments seem to indicate 
early thermalization whereas the present theoretical understanding of 
QCD suggests a much longer thermal equilibration time. 
To resolve these questions, it is important to understand  
to what ``degree'' thermalization is required to explain the 
observations. Different quantities effectively thermalize 
on different time scales and a complete thermalization of all 
quantities may not be necessary. For instance,
an approximately time-independent 
equation of state $p=p(\epsilon)$, characterized by an almost 
fixed relation between pressure $p$ and energy density $\epsilon$, 
may form very early --- even though the system is still 
far from equilibrium! Such prethermalized quantities approximately
take on their final thermal values already at a time
when the occupation numbers of individual momentum 
modes still show strong deviations from the late-time
Bose-Einstein or Fermi-Dirac distribution.
{\rr\em Prethermalization} is a universal 
far-from-equilibrium phenomenon which occurs on time scales 
dramatically shorter than the thermal equilibration time. 
In order to establish such a behavior it is crucial to be able 
to compare between the time scales of prethermalization
and thermal equilibration. Approaches based on small deviations 
from equilibrium, or on a sufficient homogeneity in time underlying 
kinetic descriptions, are not applicable in this case to describe 
the required ``link'' between the early and the late-time behavior.

A successful description of the dynamics of quantum fields away
from equilibrium is tightly related to the basic problem of how 
macroscopic irreversible behavior arises from time-reversal invariant
dynamics of quantum fields. This is a fundamental question with most
diverse applications. The basic field theoretical techniques, which 
are required to understand the physics of heavy-ion collision experiments or
dynamics in the early universe, are equally relevant for instance
for the description of the dynamics of Bose--Einstein condensates 
in the laboratory.

\subsection{How to describe nonequilibrium quantum fields 
            from first principles?}
\label{sec:how}

There are very few ingredients. {\rr\em Nonequilibrium dynamics 
requires the specification of an initial state} at some given time $t_0$. 
This may include a density matrix  
$\rho_D(t_0)$ in a mixed ($\Tr \rho_D^2(t_0) < 1$) or pure state
($\Tr \rho_D^2(t_0) = 1$). Nonequilibrium means that the
initial density matrix does not correspond to a thermal
equilibrium density matrix:
$\rho_D(t_0) \not =  \rho_D^{\rm (eq)}$ 
with for instance $\rho_D^{\rm (eq)} \sim e^{-\beta H}$ for the
case of a canonical thermal ensemble with inverse temperature $\beta$. 
In contrast to close-to-equilibrium field theory,
the initial density matrix $\rho_D(t_0)$ may deviate 
substantially from thermal 
equilibrium. This preempts the use of (non-)linear response theory,
which is based on sufficiently small deviations from equilibrium, 
or assumptions about the validity of a fluctuation-dissipation relation. 
Since time-translation invariance is explicitly 
broken at initial times, there will be no assumption about a sufficient 
homogeneity in time underlying effective kinetic descriptions.
In their range of applicability these properties should come out of the 
calculation. 

Completely equivalent to the specification of the initial density 
matrix $\rho_D(t_0)$ is the knowledge of all initial correlation functions: 
the initial one-point function
$\Tr\{\rho_D(t_0) \Phi(t_0,\bx)\}$, two-point function
$\Tr\{\rho_D(t_0) \Phi(t_0,\bx)\Phi(t_0,\by)\}$,
three-point function etc., where
$\Phi(t,\bx)$ denotes a Heisenberg field operator.
Typically, the ``experimental setup'' 
requires only knowledge about a few lowest correlation functions
at time $t_0$, whereas complicated higher correlation functions
often build up at later times. The question 
that nonequilibrium quantum field theory addresses concerns the 
behavior of these correlation functions for $t > t_0$ from which
one can extract the time evolution of other quantities such
as occupation numbers. This is depicted schematically in the figure below.
Of course, one could equally evolve the density matrix in time and compute
observable quantities such as correlations from it at some later
time. However, this is in general much less efficient
than directly expressing the dynamics in terms of correlations,
which we will do here (cf.~Sec.~\ref{sec:nPI2}).  
\begin{figure}[b]
\centerline{
\epsfig{file=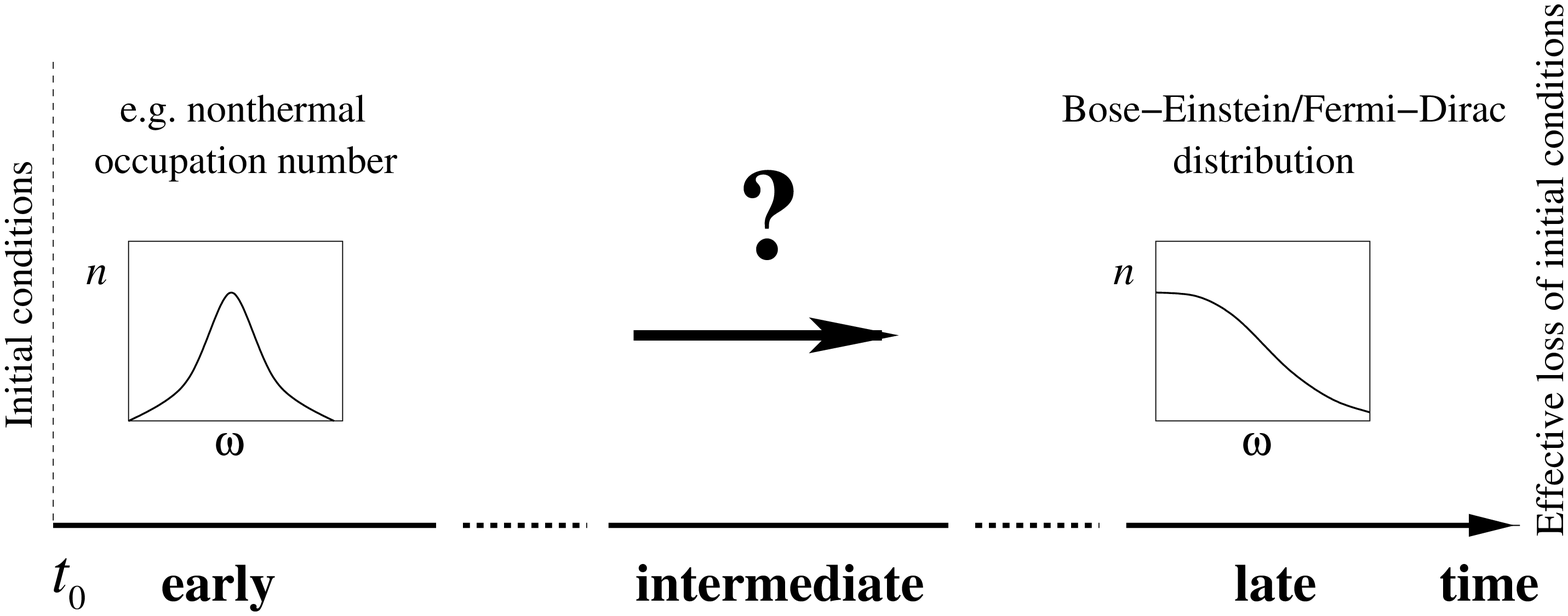,width=14.cm}}
\end{figure}

Once the nonequilibrium initial state is 
specified, the time-evolution is completely determined by  
the Hamiltonian or, equivalently, 
the dynamics can be described in terms of a functional path integral 
with the classical action $S$. From the latter one obtains 
the {\rr\em effective action $\Gamma$}, which is the generating
functional for all correlation functions of the quantum theory
(cf.~Sec.~\ref{sec:genfunc1}). 
{\rr\em There are no further ingredients involved concerning the dynamics 
than what is known from standard vacuum quantum field theory.}
Here we consider closed systems without coupling to an
external heat bath or external fields. There is no
course graining or averaging involved in the dynamics.
In this respect, the analysis is very different from 
irreversible phenomenological approaches. 
The fact that the dynamics is obtained from an effective action
automatically guarantees a number of crucial properties for the time
evolution. The most important one is conservation of energy.
The analogue in classical mechanics is well known: if the 
equations of motion can be derived from a given action then they
will not admit any friction term without further approximations.  

It should be stressed that during the nonequilibrium time evolution 
there is no loss of information in any strict sense.
The important process of thermalization is a nontrivial question in
a calculation from first principles. Thermal equilibrium keeps no memory about 
the time history except for the values of a few conserved charges.
Equilibrium is time-translation invariant and cannot be reached from
a nonequilibrium evolution on a fundamental level. It is striking 
that we will observe below that the evolution can go very closely 
towards thermal equilibrium results without
ever deviating from them for accessible times.
The observed {\rr\em effective loss of details about
the initial conditions} can mimic very accurately
the irreversible dynamics obtained from phenomenological
descriptions in their range of applicability.

\subsubsection{Standard approximation methods 
fail out of equilibrium} 

For out-of-equilibrium calculations there are additional 
complications which do not appear in vacuum or thermal 
equilibrium.\footnote{This does not concern restrictions to
sufficiently small deviations from thermal equilibrium, such as described
in terms of (non-)linear response theory, which only involve
thermal equilibrium correlators in real time.} 
The first new aspect concerns {\rr\em secularity:}
The perturbative time evolution suffers from
the presence of spurious, so-called secular terms, which grow with 
time and invalidate the expansion even in the presence of a weak 
coupling. Here it is important  
to note that the very same problem appears as well for nonperturbative
approximation schemes such as standard $1/N$ expansions, where
$N$ denotes the number of field components.\footnote{Note 
that restrictions to mean-field type approximations such as
leading-order large-$N$ are insufficient. 
They suffer from the presence of an infinite number of spurious 
conserved quantities, and are known to fail to describe
thermalization. Secularity enters the required next-to-leading
order corrections and beyond. (Cf.~Sec.~\ref{sec:neqevolution}.)} 
Typically, secularity is a not a very difficult problem and for a given
approximation there can be various ways to resolve it. 
There is a requirement, however, which poses very strong restrictions 
on the possible approximations: {\rr\em Universality,} i.e.~the 
insensitivity of the late-time behavior to the details of the 
initial conditions. If thermal equilibrium is approached then the 
late-time result is universal in the sense that it becomes uniquely 
determined by the energy density and further conserved charges.
To implement the necessary nonlinear dynamics is demanding.
Both requirements of a non-secular and universal behavior can indeed
be fulfilled using efficient functional integral techniques:
so-called {\rr\em $n$-particle irreducible effective actions},
for which powerful nonperturbative approximation
schemes are available. They provide a practical means to 
describe far-from-equilibrium dynamics as well
as thermalization from first principles.

\paragraph{An illustrative example from classical mechanics}

It is instructive to consider for a moment the simple example of the time
evolution of a classical anharmonic oscillator. This exercise will
explain some general problems, which one encounters using perturbative
techniques for initial-value problems, and indicates how to resolve them. 
Below we will consider an illustrative ``translation'' of the
outcome to the case of nonequilibrium quantum fields. 
One of the benefits will be that important concepts such as 
$n$-particle irreducible effective actions appear here 
in a very intuitive way, before they will be thoroughly
discussed in later sections.    

Our damped oscillator with time-dependent amplitude $y(t)$ 
is characterized by an infinite number of anharmonic terms:
\bea\db 
\ddot{y} + y  \,=\,  - \eps \dot{y} - (\eps y)^3 - (\eps y)^5 - (\eps y)^7 - 
\ldots \quad , \quad \eps \ll 1 
\label{eq:anharmosc}
\eea
Here each dot denotes a derivative with respect to time $t$. 
The presence of anharmonic terms to arbitrarily high order is
reminiscent of the situation in quantum field theory (QFT), where the
presence of quantum fluctuations can induce self-interactions 
to high powers in the field. Consider the
ideal case for perturbative estimates, i.e.~the presence of a 
small parameter $\eps$ which suppresses the contributions
from higher powers in $y$. The example is chosen such that it 
can be easily solved without further approximations numerically
by summing the geometric series with  
$\sum_{n=3}^{\infty} (\eps y)^n = (\eps y)^3/(1- \eps^2 y^2)$.
For the above second-order differential
equation the intial-value problem is defined by 
giving the amplitude and its 
first derivative at initial time $t=0$. We employ e.g.~$y(0)=1$, 
$\dot{y}(0)= -\eps/2$ and consider the time evolution for $t \ge 0$.
One finds the expected damped oscillator behavior
displayed below for $\eps = 0.1$:

\centerline{
\epsfig{file=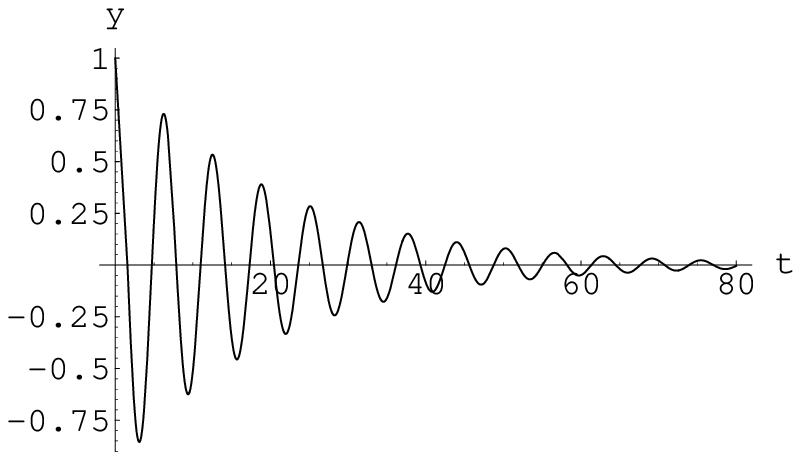,width=8.cm}
}

The question is whether an accurate approximate description of the 
full $y(t)$ is possible if contributions 
from higher powers of $\eps$ are neglected?
In view of the presence of the small expansion parameter 
this is certainly possible, however, standard
perturbation theory fails dramatically to provide
a good description. A perturbative expansion of $y$ in $\eps$, 
\bea \db
y_{\rm pert}(t) = y_0(t) + \eps y_1(t) 
+ \frac{\eps^2 }{2} y_2(t)
+ \Or(\eps^3) \, ,
\label{eq:pertexp}
\eea
yields the standard hierarchy of equations:
\bea \db
\quad \ddot{y}_0(t) + y_0(t) = 0 \quad , \quad
\ddot{y}_1(t) + y_1(t) =\,  {\rr - \dot{y}_0}  \quad , \quad
\ldots 
\label{eq:perthier}
\eea
These can be iteratively solved as
\bea
{\db y_0(t) = \frac{1}{2} e^{it} + \mbox{c.c.} } \quad , \quad
{\db y_1(t) = - \frac{1}{4} e^{it} {\rr t} + \mbox{c.c.} } \quad , \quad
\ldots  
\eea
Doing this to second order in $\eps$ one finds: 
\bea \db
y_{\rm pert}(t) = \frac{1}{2} e^{it} \left( 1 - \frac{\eps}{2} 
{\rr t} + \frac{\eps^2}{8} \left[{\rr t^2} - i {\rr t}\right]  
\right) + \mbox{c.c.} 
\label{eq:pertsol}
\eea
Beyond the lowest order, the solution contains {\rr\em secular terms} 
which grow with powers of the time $t$ and one arrives at the 
important conclusion:
\bc
\framebox{\rr The perturbative expansion is only valid for $\eps\, t \ll 1$ 
}
\ec

\vspace*{0.2cm}

\noindent
We now consider an {\rr\em alternative (``$n$PI type'') expansion:} 
For this we classify the terms of the equation of motion (\ref{eq:anharmosc}) 
according to powers of the small parameter $\eps$. The lowest order
takes into account all terms of the equation to order $\eps^0$,
which to this order gives the same as in perturbation theory:
\bea \db
  \ddot{y}_{n{\rm PI}}^{(0)} + y_{n{\rm PI}}^{(0)} = 0 \quad
\Rightarrow \quad 
y_{n{\rm PI}}^{(0)}(t) = \frac{1}{2} e^{i t} + \mbox{c.c.} 
\eea 
The next order takes into account all terms of (\ref{eq:anharmosc}) 
to order $\eps$:
\bea \db
\quad\qquad \ddot{y}_{n{\rm PI}}^{(2)} + y_{n{\rm PI}}^{(2)} 
= - \eps \dot{y}_{n{\rm PI}}^{(2)} \quad
\Rightarrow \quad
y_{n{\rm PI}}^{(2)}(t) 
= \frac{1}{2} e^{i t \sqrt{1-\eps^2/4} -\eps t/2} + \mbox{c.c.}
\label{eq:nPItype} 
\eea
Since there are no $\eps^2$ terms appearing in (\ref{eq:anharmosc}), 
this also agrees with the second order of the expansion.
The next order would take into account all terms of the equation to 
order $\eps^3$ and so on. This provides a systematic approximation
procedure in terms of powers of the small parameter $\eps$.
However, in contrast to the perturbative expansion the ``$n$PI type''
expansion turns out to be non-secular in time. One explicitly
observes that both the lowest order approximation and 
$y_{n{\rm PI}}^{(2)}(t)$ do not exceed $\Or(\eps^0)$ for all times.
Therefore, the result (\ref{eq:nPItype}) provides an approximation
to the equation (\ref{eq:anharmosc}) up to quantitative corrections 
of order $\eps^3$ at all times. The already very good agreement with the
exact result can be checked explicitly.  

It is instructive to compare the ``$n$PI type'' solution 
to the perturbative one by expanding (\ref{eq:nPItype})
in powers of $\eps$:
\bea \db
y_{n{\rm PI}}^{(2)}(t) 
&\db =&\db \frac{1}{2} e^{i t \sqrt{1-\eps^2/4} -\eps t/2} + \mbox{c.c.} 
\nn
&\db =&\db \frac{1}{2} e^{it} \left( 1 - \frac{\eps}{2} {\rr t} 
+ \frac{\eps^2}{8} \left[{\rr t^2} - i {\rr t}\right]  
\right) + \Or(\eps^3) + \mbox{c.c.} \nn
&\db \stackrel{(!)}{=}& 
\db y_{\rm pert}(t) + \Or(\eps^3) + \mbox{c.c.} \nonumber
\eea
We conclude that
\bi
\item {\rr the ``$n$PI'' result corresponds to the 
perturbative one up to the order of approximation (here up to 
$\Or(\eps^3)$ corrections).}
\item {\rr the ``$n$PI'' result resums all secular terms to 
infinite perturbative order in $\eps$.}
\ei
Stated differently: infinite summation of perturbative orders 
is required to obtain a {\rr\em uniform approximation to the exact solution,} 
i.e.~that the error stays of a given order at all times.

What was the reason for the ``success'' of this alternative expansion?
It is based on a standard procedure for differential equations  
in order to enlarge the convergence radius of an expansion.
In the differential equation one neglects 
contributions from higher powers of the expansion parameter.
However, in contrast to perturbation theory one does not expand
in addition the variable $y$. This procedure is sometimes
called ``self-consistent'' since at each order in the approximation
the dynamics is solely expressed in terms of the dynamical degree
of freedom itself. For instance, in Eq.~(\ref{eq:nPItype})
both on the left and on the right hand side of the differential
equation appear derivatives of the same variable $y_{n{\rm PI}}^{(2)}$.
In contrast, in the perturbative hierarchy (\ref{eq:perthier})
the presence of ``external'' oscillating ``source terms'' such 
as $y_0$ driving $y_1$ leads to secular behavior. The procedure
of the ``$n$PI type'' expansion is simple and turns out to be very 
efficient to achieve non-secular behavior:
{\rr\em For given dynamical degrees of freedom truncate
the dynamics according to powers of a small expansion parameter.}
Of course, the choice of the degrees of freedom is always based on 
physics. The price to be paid for this expansion scheme is that
at some order one necessarily has to solve nonlinear equations 
without further approximations.
However, as we will see next, it is precisely the {\rr\em nonlinearity}
which is required to be able to obtain {\rr\em universality}
in the sense mentioned above.

In order to illustrate properly the aspect of universality the single 
oscillator example (\ref{eq:anharmosc}) is too simple. In order
to include more degrees of freedom we add a ``three-momentum'' label
to the variables and consider:   
\beq\db
\dot{y}_\bp  \,\sim\,   \int_{\bq\bk} \left[
(1+y_\bp ) (1+y_\bq ) y_{\bk} y_{\bp -\bq -\bk} \,
-\, y_{\bp} y_{\bq} (1+y_\bk ) (1+y_{\bp -\bq -\bk} ) \right]  \, . 
\label{eq:uniex}
\eeq
Here the integrals on the r.h.s.~involve momenta $\bq$ and $\bk$
above some suitable value, and one observes that the time derivative 
of $y_\bp$ is proportional to a ``gain'' and a ``loss'' term.
This is like in a Boltzmann equation, which describes the
rate of scattering of particles into the state with 
momentum ${\bf p}$ minus the rate of scattering out of that 
momentum (cf.~Sec.~\ref{sec:detourboltzmann}).
If the nonlinear equation (\ref{eq:uniex}) is solved without 
further approximations then the late-time solution is given by
the well-known result
\beq\db
y_\bp \to \frac{1}{e^{\beta(|\bp| - \mu)} - 1} \, ,
\label{eq:be}
\eeq
with real parameters $\beta$ and $\mu$, which for the Boltzmann
equation are identified as temperature and chemical potential
respectively. The form of the solution (\ref{eq:be}) is 
universal, i.e.~completely independent of the details of the 
initial condition for the solution of (\ref{eq:uniex}). 
We emphasize that the Boltzmann type equation (\ref{eq:uniex})
is ``self-consistent'' in the sense mentioned above.
In contrast, a non--``self-consistent'' approximation 
will not show the desired universality in general. 
For instance, consider (\ref{eq:uniex}) in a linearized 
approximation
\beq\db 
\dot{y}_\bp \,=\, (1+y_\bp) {\rr\sigma^0_\bp} 
- y_{\bp} {\rr \overline{\sigma}^0_\bp} \, ,
\label{eq:lin}
\eeq
with time-independent $\sigma^0_\bp \sim \int_{\bq\bk}
(1+y_\bq(0)) y_{\bk}(0) y_{\bp -\bq -\bk}(0)$ and equivalently
for $\overline{\sigma}^0_\bp$. Eq.~(\ref{eq:lin}) has the solution
\beq\db
y_\bp \,=\, {\rr \frac{\sigma^0_\bp}{\gamma^0_\bp}} 
+ \left[ y_\bp(0) - \frac{\sigma^0_\bp}{\gamma^0_\bp} \right] 
e^{-\gamma^0_\bp t} \, ,
\eeq
with $\gamma^0_\bp = \sigma^0_\bp - \overline{\sigma}^0_\bp$.
For late times $\gamma^0_\bp t \gg 1$ this always depends on
the chosen initial $\sigma^0_\bp/\gamma^0_\bp$ and is, of course, only
useful if they are chosen to be already the equilibrium values.

\subsubsection{$n$-Particle irreducible expansions: universality and
non-secularity}
\label{sec:introuniv}

One should not misunderstand the following illustrative ``translation'' 
of the above mechanics examples to QFT. It is intended to give intuitive
insight. The questions of secularity and universality
are of course settled by actual nonequilibrium calculations 
in QFT, which is the topic of the main body
of this text. For a moment, let us transcribe the above results
to the language of QFT. The amplitude $y$ of the above oscillator
example plays the role of the one-point function, i.e.~the field expectation 
value or {\rr\em macroscopic field}. The amplitude squared plays the role of
the two-point function or {\rr\em propagator}, the cubic amplitude that of
the three-point function or proper {\rr\em three-vertex,} etc.:    
\bea
\db y(t)\,\, \qquad &\mbox{''}\leadsto\mbox{''}& \qquad  
\db \phi(x) = \langle \hat{\Phi}(x) \rangle 
\nonumber\\
&& \qquad \mbox{(macroscopic field)}
\nonumber\\[0.1cm]
\db y^2(t) \qquad &\mbox{''}\leadsto\mbox{''}& \qquad  
\db G_2(x,y) = \langle T \hat{\Phi}(x)\hat{\Phi}(y) \rangle 
- \phi(x) \phi(y) \equiv G(x,y)
\nonumber\\
&& \qquad \mbox{(propagator)}
\nonumber\\[0.1cm]
\db y^3(t) \qquad &\mbox{''}\leadsto\mbox{''}& \qquad  
\db G_3(x,y,z)\,\, \mbox{or, with}\,\, G_3 \sim GGG V_3:
\nonumber\\
&&{\db \qquad V_3(x,y,z)} \qquad\,\,\,\,\,\,\,\,\,\,\,
\mbox{(proper three-vertex)}
\nonumber\\[0.1cm]
\db y^4(t) \qquad &\mbox{''}\leadsto\mbox{''}& \qquad
{\db V_4(x,y,z,w)} \qquad
\mbox{(proper 4-vertex)}
\nonumber\\
\vdots \qquad && \nonumber
\eea
where the symbol $T$ denotes time-ordering.
The knowledge of all $n$-point functions provides a full description
of the quantum theory. In contrast to the classical example,
the information contained e.g.~in the two-point function cannot be 
recovered from the one-point function: $G \not \sim \phi^2$ etc.
In QFT all $n$-point functions $\phi$, $G$, $V_3$, $V_4$ \ldots 
constitute the set of dynamical ``degrees of freedom''. 

The equations of motions for all $n$-point functions are conveniently
encoded in a generating functional or effective action.
There are different functional representations of the effective
action. The so-called {\rr\em $n$-particle irreducible ($n$PI) effective
action} $\db \Gamma[\phi,G,V_3,\ldots,V_n]$ is expressed
in terms of $\phi$, $G$, $V_3$, \ldots, $V_n$ and is particularly
efficient for a description of suitable approximation schemes in
nonequilibrium quantum field theory. It
determines the equations of motion for $\phi$, $G$, $V_3$, \ldots,
$V_n$ by first-order functional derivatives or so-called
{\rr\em stationarity conditions:} 
\bea \db \qquad
\frac{\delta \Gamma[\phi,G,V_3,\ldots]}{\delta \phi} = 0 \,\, ,\,\,\,
\frac{\delta \Gamma[\phi,G,V_3,\ldots]}{\delta G} = 0 \,\, ,\,\,\,
\frac{\delta \Gamma[\phi,G,V_3,\ldots]}{\delta V_3} = 0 \,\, ,\,\,\ldots
\label{eq:eomnPI}
\eea
The above ``$n$PI'' approximation scheme for the oscillator example reads 
now:
\bi
\item {\rr Classify the contributions to 
$\Gamma[\phi,G,V_3,\ldots]$ according to powers of a 
suitable expansion parameter (e.g.~coupling/loops, or $1/N$).
In terms of the ``variables'' $\phi$, $G$, $V_3$ \ldots
there is a definite answer for the approximate effective
action to a given order of the expansion.}  
\item {\rr The equations of motion for $\phi$, $G$, $V_3$ \ldots 
are then determined from the approximate effective action 
according to (\ref{eq:eomnPI}) without any further
assumptions.}
\ei
In general this leads to nonlinear integro-differential equations,
which typically require numerical solution techniques if one
wants to provide the ``link'' between the early and the
late-time behavior. We emphasize that nonlinearity is a 
crucial ingredient for late-time universality. The reward is that one can
compute far-from-equilibrium dynamics as well as the subsequent
approach to thermal equilibrium in quantum field theory from
first principles (cf.~Sec.~\ref{sec:neqevolution}). In turn this can be used
to derive powerful effective descriptions and to determine their range 
of applicability. 

In order to be useful in practice it is crucial to observe
that it is not necessary to calculate the most general
$\Gamma[\phi,G,V_3,\ldots,V_n]$ for arbitrarily large $n$. 
There exists an equivalence hierarchy between $n$PI
effective actions. For instance, in the context of a
loop expansion one has:
\bea
\rr \Gamma^{\rm (1loop)}[\phi]
&\! \rr = \!& \rr \Gamma^{\rm (1loop)}[\phi,D] \,=\,\ldots\, ,
\nonumber\\
\db \Gamma^{\rm (2loop)}[\phi]
&\! \db \not = \!& \rr \Gamma^{\rm (2loop)}[\phi,D]
\,=\, \Gamma^{\rm (2loop)}[\phi,D,V_3] \,=\,\ldots \, ,
\label{eq:hierarchy}\nonumber\\
\db \Gamma^{\rm (3loop)}[\phi] 
&\! \db \not = \!& {\db \Gamma^{\rm (3loop)}[\phi,D]
\, \not = \,\,\, }{\rr \Gamma^{\rm (3loop)}[\phi,D,V_3] 
\, =\,  \Gamma^{\rm (3loop)}[\phi,D,V_3,V_4] = \ldots  \, ,}
\nonumber\\
\vdots \nonumber
\eea   
where $\Gamma^{{\rm (}n{\rm -loop)}}$ denotes the approximation of the 
respective effective action to $n$-th loop order in the
absence of external sources. E.g.~for a two-loop approximation
all $n$PI descriptions with $n \ge 2$ are equivalent and
the 2PI effective action captures already the 
complete answer for the self-consistent description
up to this order. In contrast, a self-consistently complete 
result to three-loop order requires at least the 3PI effective action 
in general, etc. 
There are, however, other simplifications which can decrease 
the hierarchy even further such that lower $n$PI effective 
actions are often sufficient in
practice.  For instance, for a vanishing macroscopic field one has  
$\Gamma^{\rm (3loop)}_{2{\rm PI}}[\phi=0,G] = 
\Gamma^{\rm (3loop)}_{4{\rm PI}}[\phi=0,G,V_3=0,V_4] = \ldots$ in the
absence of sources.
Typically the 2PI, 3PI or maybe the 4PI effective action  
captures already the complete answer for the self-consistent 
description to the desired/computationally feasible order of 
approximation.

\section{$n$-Particle irreducible generating functionals I} 
\label{sec:genfunc1}

In this section we discuss the construction of the one-particle
irreducible (1PI) and the {\rr\em two-particle irreducible (2PI) 
effective action}. The latter
provides a powerful starting point for
systematic approximations in nonequilibrium quantum field theory. 
To be specific, we consider first a quantum field theory for a real, 
$N$--component scalar field 
$\varphi_a$ ($a=1,\ldots,N$) with 
{\rr\em {classical action $S[\varphi]$}:}\footnote{We 
always work in units where $h\!\!\bar{}\,\,=c=1$ such that 
the mass of a particle is equal to its rest energy ($m c^2$)
and also to its inverse Compton wavelength ($m c/h\!\!\bar{}\,\,$).}  
\beq\db
S[\varphi]=\int_x
	\left(\frac{1}{2}\partial^\mu\varphi_a(x)\partial_\mu\varphi_a(x)  
	-\frac{m^2}{2}\varphi_a\!(x)\varphi_a\!(x) 
	-\frac{\lambda}{4!N}\left(\varphi_a\!(x)\varphi_a\!(x)\right)^2
 \right)  \, .
\label{eq:classical}
\eeq
Here summation over repeated indices is implied and
we use the shorthand notation 
$\int_x \equiv \int_{\C} \rmd x^0 \int {\rmd}^d x$ with $x \equiv (x^0,\bx)$.
The time path $\C$ will be specified later.

The {\rr\em {generating functional $W[J,R]$ for connected Green's functions}} 
in the presence of {\em two} source terms $\sim J_a(x)$ and 
$\sim R_{ab}(x,y)$ is given by
\bea \db
Z[J,R] &\db = & \db \exp\left(i W[J,R] \right) \nonumber\\
	&\db = & \db \int\! \mathscr{D}
\varphi\, \exp\left(i \left[S[\varphi]+ \int_{x}\!
J_a(x)\varphi_a(x)+\frac{1}{2}
\int_{x y}\!\! R_{ab}(x,y) \varphi_a(x)\varphi_b(y)
	\right] \right) \, . 
\label{modZ}
\eea
We define the {\rr\em macroscopic field $\phi_a$} and the 
{\rr\em connected two-point function $G_{ab}$} by variation of $W$ 
in the presence of the source terms:
\bea \db
\frac{\delta W[J,R]}{\delta J_a(x)}
=
{\rr \phi_a(x)}\quad , \quad
\frac{\delta W[J,R]}{\delta R_{ab}(x,y)}
=
\frac{1}{2}\Big(\phi_a(x)\phi_b(y)+ {\rr G_{ab}(x,y)} 
\Big) \, .
\label{deltaWdeltaR}
\eea
Before constructing the 2PI effective action, we 
consider first the {\rr\em 1PI effective action}. It is
obtained by a Legendre transform
with respect to the source term which is linear in the field, 
\bea \db
\Gamma^{R}[\phi] \,=\, W[J,R]-\int_x \frac{\delta W[J,R]}{\delta J_a(x)} J_a(x)
\,=\, W[J,R]-\int_x \phi_a(x) J_a(x) \, .
\label{GammaR}
\eea
We note that:
\begin{enumerate}
\item $\Gamma^{R\equiv0}[\phi]$ corresponds to the standard 
{\rr\em 1PI effective action}. 
\item For $R \not = 0$ it can be formally viewed as the 1PI effective 
action for a theory governed by the modified classical action, 
$S^{R}[\varphi]$, in the
presence of a non-constant ``mass term'' $\sim R_{ab}(x,y)$
quadratic in the fields:
\bea \db
S^{R}[\varphi] = S[\varphi] + \frac{1}{2} \int_{xy} {\rr R_{ab}(x,y)} 
\varphi_a(x) \varphi_b(y) \, .   
\eea
\end{enumerate}
This can be directly observed from (\ref{modZ}) and the fact that
the standard 1PI effective action is obtained from the same defining
functional integral in the presence of a linear source term only.
As a consequence, it is straightforward to recover for $\Gamma^{R}[\phi]$
all ``textbook'' relations for the 1PI effective action taking into account
$R$. For instance, $\Gamma^R[\phi]$ to one-loop order is given by 
\bea\db 
\Gamma^{R ({\rm 1loop})}[\phi] \,=\, 
S^R[\phi] + \frac{i}{2} \Tr\ln [G_0^{-1}(\phi) {\rr \, -\, i R}] \, ,
\label{G1Roneloop}
\eea
which is the familiar result for the 1PI effective action with
$S[\phi] \to S^R[\phi]$ and 
$G_0^{-1}(\phi) \to G_0^{-1}(\phi) - i R$.
Similarly, one finds that the exact inverse propagator is obtained by
second functional field differentiation, 
\bea\db
\frac{\delta^2 \Gamma^{R}[\phi]}{\delta \phi_a(x)\delta \phi_b(y)}
&\db =&\db i G_{ab}^{-1}(x,y) \nn
&\db =&\db i \left[ G_{0,ab}^{-1}(x,y) {\rr\, -\, i R_{ab}(x,y)} 
- {\Sigma_{ab}^{R}(x,y)} 
\right] \, .
\label{exactpropagatorG1R} 
\eea
Here the classical inverse propagator $i G_{0,ab}^{-1}(x,y;\phi)=
\delta^2 S[\phi]/\delta \phi_a(x) \delta \phi_b(y)$ reads
\bea \db
i G^{-1}_{0,ab}(x,y;\phi) &\db =&\db - \left( \square_x + m^2 
+ \frac{\lambda}{6 N}\, \phi_c(x)\phi_c(x) \right) \delta_{ab}
\delta(x-y) \nonumber\\ 
&&\db - \frac{\lambda}{3 N}\, \phi_a(x) \phi_b(x) \delta(x-y) \, , 
\label{classprop}
\eea
and $\Sigma_{ab}^{R}(x,y)$ is the {\rr\em proper self-energy, to which only 
one-particle irreducible Feynman diagrams contribute}, i.e.~diagrams
which cannot be separated by opening one line.
These relations will be used below.

We now perform a further Legendre transform of $\Gamma^{R}[\phi]$ with
respect to the source~$R$ in order to arrive at the {\rr\em 2PI effective
action}:
\bea \db
\Gamma[\phi,G] & \db =& \db \Gamma^{R}[\phi] 
- \int_{xy} 
\underbrace{\frac{\delta \Gamma^{R}[\phi]}{\delta R_{ab}(x,y)}} R_{ba}(y,x)
\nonumber\\
&& \rr \qquad\quad \qquad \frac{\delta W[J,R]}{\delta R_{ab}(x,y)}
=\frac{1}{2}\Big(\phi_a(x)\phi_b(y)+ G_{ab}(x,y)\Big) \nn
&\db =& \db \Gamma^{R}[\phi] - \frac{1}{2} \int_{xy}
R_{ab}(x,y) \phi_a(x)\phi_b(y) - \frac{1}{2} \Tr\, R\, G  \, .
\label{twostepLT}
\eea
Here we have used that the relation between $\phi$ and $J$ is $R$-dependent,
i.e.\ inverting {\rr $\phi = \delta W[J,R]/\delta J$} yields
$J= J^{R}(\phi)$. From the above definition of $\Gamma^{R}[\phi]$ one
therefore finds 
\bea\db
\frac{\delta \Gamma^{R}[\phi]}{\delta R_{ab}(x,y)}&\db =&\db
\frac{\delta W[J,R]}{\delta R_{ab}(x,y)}
+ \int_z {\rr \frac{\delta W[J,R]}{\delta J_c(z)} } 
\frac{\delta J_c(z)}{\delta R_{ab}(x,y)}
-\int_z {\rr \phi_c(z)}\frac{\delta J_c(z)}{\delta R_{ab}(x,y)} \nonumber\\
&\db =&\db
\frac{\delta W[J,R]}{\delta R_{ab}(x,y)} \, .
\eea
Of course, the two subsequent Legendre transforms which have
been used to arrive at
(\ref{twostepLT}) agree with a simultaneous Legendre 
transform of $W[J,R]$ with respect to both source terms:
\bea \db
\Gamma[\phi,G] &\db = & \db W[J,R]-\int_x \frac{\delta W[J,R]}{\delta J_a(x)} 
J_a(x) - \int_{xy} \frac{\delta W[J,R]}{\delta R_{ab}(x,y)} R_{ab}(x,y)
\nonumber\\
&\db = & \db  W[J,R]-\int_x \phi_a(x) J_a(x) - \frac{1}{2} \int_{xy}
R_{ab}(x,y) \phi_a(x) \phi_b(y) - \frac{1}{2} \Tr\, G\, R 
\label{2PIGdirect}
\eea
From this one directly observes the {\rr\em stationarity conditions:}
\bea \db
\frac{\delta \Gamma[\phi,G]}
{\delta \phi_a(x)}& \db = &\db - J_a(x) - \int_{y} R_{ab}(x,y) \phi_b(y)  \, ,
\label{stationphi} \\
\db \frac{\delta \Gamma[\phi,G]}{\delta G_{ab}(x,y)}
&\db =& \db -\frac{1}{2} R_{ab}(x,y) \, , 
\label{eq:station}
\eea 
which give the {\rr\em equations of motion for $\phi$ and $G$} in
the absence of the sources, i.e.~$J=0$ and $R=0$. 

To get familiar with Eq.\ (\ref{twostepLT}) or (\ref{2PIGdirect}),
we may directly calculate 
$\Gamma[\phi,G]$ to one-loop order using the above results 
for $\Gamma^{R}[\phi]$. Plugging (\ref{G1Roneloop}) into (\ref{twostepLT})
one finds to this order
\beq \db
\Gamma[\phi,G] \,\simeq\,  
S[\phi] + \frac{i}{2} \Tr\ln \left[ G_0^{-1}(\phi) - i R \right]
- \frac{1}{2} \Tr\, R\, G \, .
\label{eq:onelopp2PIR}
\eeq
If we set $\rr G^{-1} = G_0^{-1}(\phi) - i R$ then we can write
\beq \db
\Gamma[\phi,G] \,\simeq\, 
S[\phi] + \frac{i}{2} \Tr\ln G^{-1} 
+ \frac{i}{2} \Tr\, G_0^{-1} G +
{\rm const}\, ,  
\label{2PIGoneloop}
\eeq
with $\Tr\, G^{-1} G \sim \Tr\, {\rm \bf 1} = {\rm const}$.
To verify this one can check from the stationarity condition
(\ref{eq:station}) that indeed to this order
\bea
\db \frac{\delta \Gamma[\phi,G]}{\delta G}
\simeq -\frac{i}{2} G^{-1} + \frac{i}{2} G_0^{-1}(\phi) =  -\frac{1}{2} R
\quad \Rightarrow \quad {\rr G^{-1} = G_0^{-1}(\phi) - i R} \,\, .
\nonumber
\eea
To go beyond this order it is convenient to write the exact 
$\Gamma[\phi,G]$ as the one-loop type expression (\ref{2PIGoneloop})
and a {\rr\em `rest'}: 
\bea
\mbox{\framebox{$\displaystyle 
{\db \,\, \Gamma[\phi,G] = S[\phi] + \frac{i}{2} \Tr\ln G^{-1} 
          + \frac{i}{2} \Tr\, G_0^{-1}\! (\phi)\, G } 
{\rr          \, + \, \Gamma_2[\phi,G]}\db + {\rm const} \,\, $}}
\label{2PIaction}
\eea
Here we have added an irrelevant constant which can be adjusted 
for normalization. 
To get an understanding of the {\rr\em `rest' term $\Gamma_2[\phi,G]$} 
we vary this expression with respect to $G$, which yields  
\bea \db
G_{ab}^{-1}(x,y) &\db =& \db G_{0,ab}^{-1}(x,y;\phi)  
- i R_{ab}(x,y) 
- {\rr \Sigma_{ab}(x,y;\phi,G)} \, ,
\label{SchwingerDysonR}
\eea
where we have written 
\bea
\mbox{\framebox{\rr
$\displaystyle \,\, \Sigma_{ab}(x,y;\phi,G) \equiv   
2 i\, \frac{\delta \Gamma_2[\phi,G]}{\delta G_{ab}(x,y)}\,\,$}}
\label{exactsigma}
\eea
Comparing with the exact expression (\ref{exactpropagatorG1R}) one observes 
that 
\beq \db
\Sigma_{ab}(x,y;\phi,G) = \Sigma^R_{ab}(x,y;\phi)\, ,
\label{sigmarelation}
\eeq
which relates the functional $G$-derivative of $\Gamma_2[\phi,G]$ 
to the proper self-energy. As mentioned above, 
{\rr\em to the proper self-energy 
only 1PI diagrams contribute} with propagator lines associated
to the effective classical propagator $(G_0^{-1} - i R)^{-1}$. The mapping  
between $\Sigma_{ab}(x,y;\phi,G)$ and $\Sigma^R_{ab}(x,y;\phi)$ is provided 
by (\ref{SchwingerDysonR}), which can be used 
to express the full propagator $G$ as an infinite series in terms
of the classical propagator $G_0$ and $\Sigma$: 
\bea\db
G &\db =&\db (G_0^{-1} - i R)^{-1} + (G_0^{-1} - i R)^{-1}\, \Sigma\,
(G_0^{-1} - i R)^{-1} \nonumber\\
&&\db + (G_0^{-1} - i R)^{-1}\, \Sigma\, (G_0^{-1} - i R)^{-1}
\, \Sigma\, (G_0^{-1} - i R)^{-1} + \ldots \, ,
\label{expandG}
\eea
where we employ an obvious matrix notation.
As a consequence there is a direct correspondence between a 
1PI diagram with propagator lines associated to $G$
and an infinite set of 1PI diagrams with propagator 
lines associated to the classical propagator $G_0$.
Below we will consider some explicit examples.
Most importantly, from the fact that to $\Sigma(\phi,G)$
only 1PI diagrams contribute one can conclude that 
\bi
\item {\rr\em$\Gamma_2[\phi,G]$ only contains contributions 
from two--particle irreducible (2PI) diagrams}.
\ei
A diagram is said to be two-particle
irreducible if it does not become disconnected by opening two lines. 
Suppose $\Gamma_2[\phi,G]$ 
had a {\rr\em two--particle reducible} contribution. The latter
could be written as $\tilde{\Gamma} G G \tilde{\Gamma}'$, where $GG$ denotes 
in a matrix notation two propagator lines connecting two parts 
$\tilde{\Gamma}$ and $\tilde{\Gamma}'$ of a diagram.
Then $\Sigma(\phi,G)$ would contain a contribution of the form 
$\tilde{\Gamma} G \tilde{\Gamma}'$ since it is given by a derivative
of $\Gamma_2$ with respect to $G$. Such a structure is {\rr\em 
one-particle reducible} and cannot occur for the proper self-energy. Therefore 
two-particle reducible contributions to $\Gamma_2[\phi,G]$ are absent.

Diagrammatically, the graphs contributing to $\Sigma(\phi,G)$
are obtained by opening one propagator line in graphs 
contributing to $\Gamma_2[\phi,G]$. This is exemplified
for a two- and a three-loop graph for $\Gamma_2$ and the
corresponding self-energy graphs:

\vspace*{0.3cm}

$\qquad$ \parbox{1.cm}{$\db \Gamma_2$:}
\parbox{2cm}
{\centerline{\epsfig{file=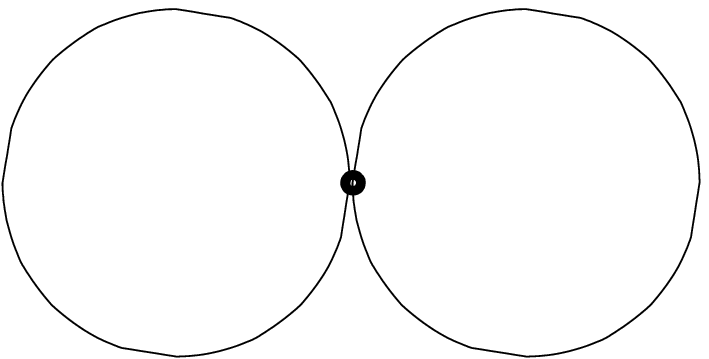,width=2.25cm}}

\vspace*{0.1cm}

\centerline{\epsfig{file=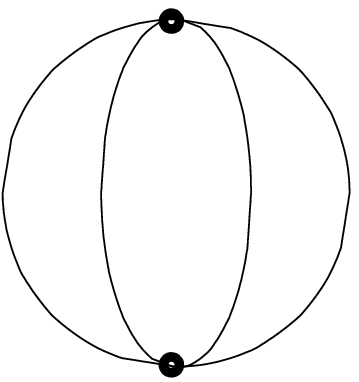,width=1.2cm}}
} 
\parbox{2.3cm}{

\vspace*{0.4cm}

$\qquad\longrightarrow\qquad\quad$}
\parbox{2.8cm}{$\db \Sigma \sim \delta \Gamma_2/\delta G$:}
\parbox{2cm}
{\centerline{\epsfig{file=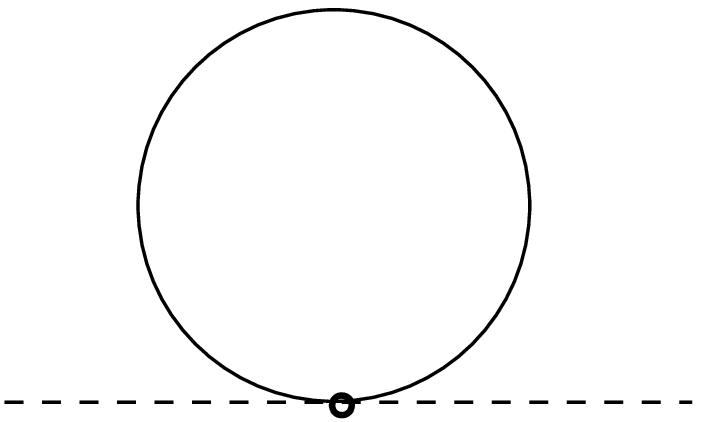,width=2.cm}}
\vspace*{0.1cm}

\centerline{\epsfig{file=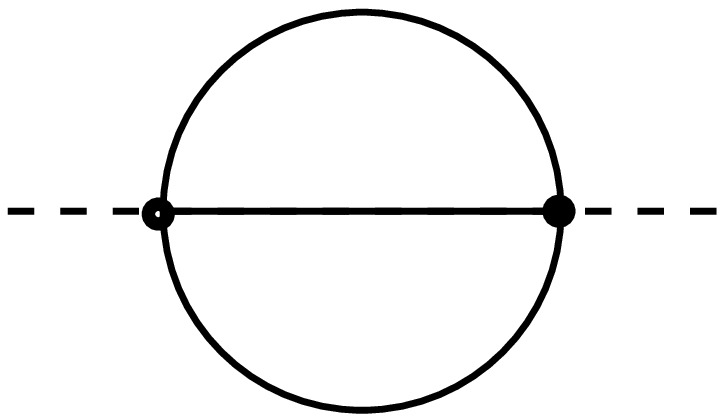,width=2.1cm}}
}

\vspace*{0.3cm}

\noindent
Because of (\ref{expandG}) each 2PI diagram corresponds to infinite 
series of 1PI diagrams, and the above 2PI two- and three-loop diagrams
contain e.g.~the full series of so-called ``daisies'' and
``ladder'' resummations: 
\begin{figure}[h]
\centerline{\epsfig{file=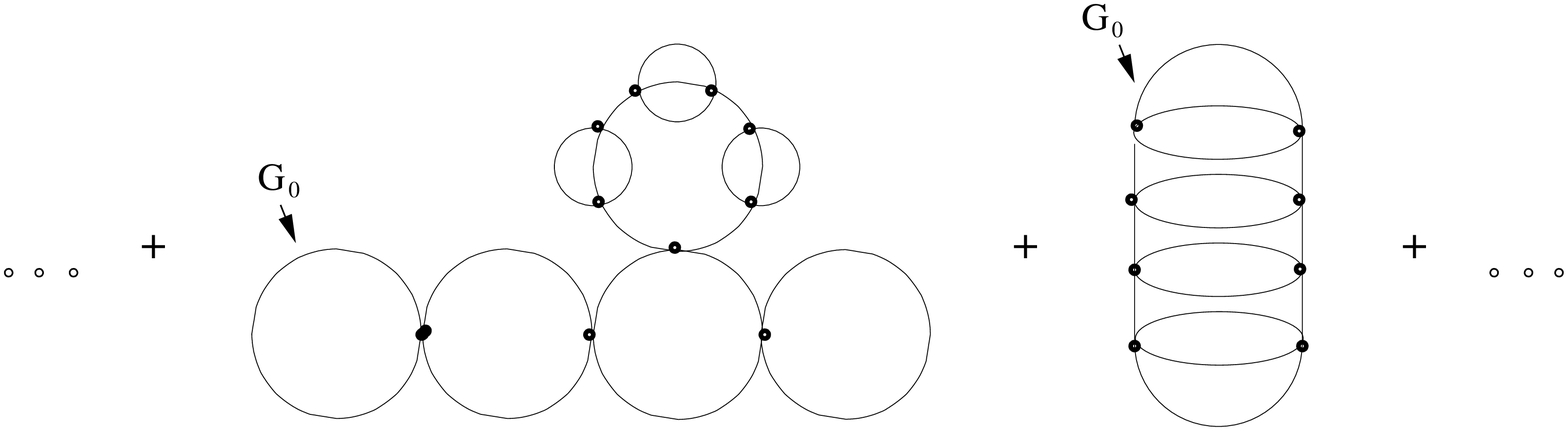,width=11.cm}}
\end{figure}

\subsection{Loop or coupling expansion of the 2PI effective action} 
\label{sec:loopexp}

Loop or coupling expansions of the 2PI effective action proceed 
along the same lines as the corresponding expansions for the standard 
1PI effective action, with the only difference that 
\bi
\item {\rr the full propagator $G$ is associated to
propagator lines of a diagram} 
\item {\rr and only 2PI graphs are kept.}
\ei
To be specific, for the considered $N$-component scalar field theory 
the diagrams are constructed in the standard way 
from the effective interaction 
\beq \db
i S_{\rm int}[\phi,\varphi] =  
{\rr 
- \int_x i \frac{\lambda}{6 N} \phi_a(x)\varphi_a(x)\varphi_b(x)\varphi_b(x)}
- \int_x i \frac{\lambda}{4! N} \Big(\varphi_a(x)\varphi_a(x)\Big)^2 \, ,
\label{interactionS}
\eeq
which is obtained from the classical action (\ref{eq:classical})
by shifting $\varphi_a(x) \to \phi_a(x) + \varphi_a(x)$
and collecting all terms cubic and quartic in the fluctuating field
$\varphi_a(x)$. As for the 1PI effective action, in 
addition to the quartic interaction there is
an {\rr\em effective cubic interaction for non-vanishing field expectation
value, i.e.~$\phi_a(x) \not = 0$}. 

\begin{figure}[t]
\centerline{\epsfig{file=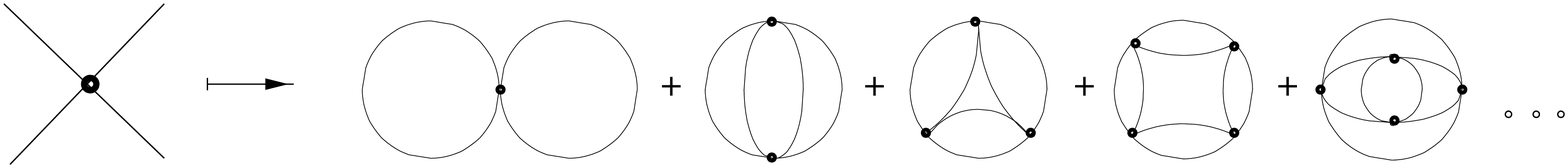,width=14.5cm}}
\caption{Topologically distinct diagrams in the 2PI loop expansion
up to five-loop order for $\phi = 0$. The suppressed prefactors are given in 
Eq.\ (\ref{eq:2PIGfiveloop}).}
\label{fig:fiveloop}
\end{figure}

Since $\Gamma[\phi,G]$ is a functional, which associates a number
to the fields $\phi$ and $G$, only closed loop diagrams can appear. 
We consider for a moment the real scalar theory for $N=1$. 
To lowest order one has $\Gamma_2[\phi,G] = 0$ and we recover 
the one-loop result given in Eq.\ (\ref{2PIGoneloop}). At two-loop 
order there are two contributions 
\bea \db
\Gamma_2^{\rm (2loop)}[\phi,G] &\db =&\db 
-i\, 3\, \Big(-i \frac{\lambda}{4!}\Big) \int_x G^2(x,x) 
\nonumber\\
&& {\rr -i\, 6\, \frac{1}{2}\, 
\int_{xy} \Big(-i \frac{\lambda}{6}\phi(x) \Big) 
\Big(-i \frac{\lambda}{6}\phi(y) \Big)
G^3(x,y)}\, ,  
\eea
where we have made explicit the 
different factors coming from the overall $-i$
in the defining functional integral for $\Gamma[\phi,G]$ 
(cf.\ Eq.\ (\ref{modZ})), the combinatorics and the
vertices. The 2PI loop expansion exhibits of course
much less topologically distinct diagrams than the
respective 1PI expansion. For instance, in the symmetric
phase ($\phi = 0$) one finds that up to four 
loops only a single diagram 
contributes at each order. 
At fifth order there are two distinct diagrams which
are shown along with the lower-loop graphs in Fig.\ \ref{fig:fiveloop}.
For later discussion, we give here
also the results up to five loops for general~$N$. For $\phi = 0$ by 
virtue of $O(N)$ rotations the propagator can be taken to be diagonal:
\beq\db
G_{ab}(x,y) \,=\, G(x,y) \delta_{ab} \, . 
\label{eq:symmetricG}
\eeq
One finds to five-loop order:
\bea\db
\lefteqn{
\Gamma_2^{\rm (5loop)}[G]_{|G_{ab}=G\delta_{ab}} 
\,\,=\,\, \sum_{l=2}^{5} \Gamma_2^{(l)}\, , \label{2PIGfiveloop}} \\[0.2cm] 
\db\Gamma_2^{(2)} &\db=&\db 
-\frac{\lambda}{8}\frac{(N+2)}{3}\int_x G^2(x,x)\quad ,\quad 
\Gamma_2^{(3)} \,\,=\,\, \frac{i\lambda^2}{48}\frac{(N+2)}{3N}\int_{xy} 
G^4(x,y)\,,\nonumber\\
\db\Gamma_2^{(4)} &\db=&\db 
\frac{\lambda^3}{48}\frac{(N+2)(N+8)}{27N^2}\int_{xyz} 
G^2(x,y)G^2(x,z)G^2(z,y) \, ,\nonumber\\
\db\Gamma_2^{(5)} &\db=&\db 
-\frac{i\lambda^4}{128}\frac{(N+2)(N^2+6N+20)}{81N^3} \int_{xyzw} \!\!\!
G^2(x,y)G^2(y,z)G^2(z,w)G^2(w,x) \nonumber\\
&&\db 
-\frac{i\lambda^4}{32}\frac{(N+2)(5N+22)}{81N^3} \int_{xyzw} \!\!\!
G^2(x,y)G(x,z)G(x,w)G^2(z,w)G(y,z)G(y,w) \, .\nonumber
\label{eq:2PIGfiveloop}
\eea

\subsection{Renormalization}
\label{sec:renormalize}

The 2PI resummed effective action $\Gamma[\phi,G(\phi)]$
is defined in a standard way by suitable
regularization, as e.g.~lattice regularization or 
dimensional regularization, and renormalization conditions
which specify the field theory to be considered. 
We employ renormalization conditions for the  
two-point function, $\Gamma^{(2)}$, and four-point function,
$\Gamma^{(4)}$, given by 
\bea\db
\Gamma^{(2)}(x,y) &\db \equiv&\db 
\frac{\delta^2 \Gamma[\phi,G(\phi)]}{\delta \phi(x) \delta \phi(y)}
\Big|_{\phi=0} \, , 
\\\db
\Gamma^{(4)}(x,y,z,w) &\db\equiv&\db 
\frac{\delta^4 \Gamma[\phi,G(\phi)]}{\delta \phi(x) 
\delta \phi(y) \delta \phi(z) \delta \phi(w)} 
\Big|_{\phi=0}\, ,
\eea
where we consider a one-component field for notational simplicity.
Without loss of generality we use renormalization conditions
for $\phi = 0$ which in Fourier space 
read: 
\bea\rr
Z\, \Gamma^{(2)}(p^2)|_{p = 0} &\rr =&\rr - m_R^2 \, ,
\label{eq:r1}\\\rr
Z\, \frac{d}{d p^2} \Gamma^{(2)}(p^2)|_{p = 0} &\rr =&\rr - 1 \, ,
\label{eq:r2}\\\rr
Z^2\, \Gamma^{(4)}(p_1,p_2,p_3)|_{p_1 = p_2 = p_3 = 0} &\rr =&\rr 
- \lambda_R \, ,
\label{eq:r3}
\eea
with the wave function renormalization $Z$.\footnote{For the 
Fourier transform of the $n$-th derivative one has
\bea
\Gamma^{(n)}(x_1,\ldots x_n) =  
\int \frac{{\rm d}^4 p_1}{(2\pi)^4} e^{-i p_1 x_1} \ldots
\int \frac{{\rm d}^4 p_n}{(2\pi)^4} e^{-i p_n x_n}
(2 \pi)^4 \delta^4(p_1 + \ldots p_n) 
\Gamma^{(n)}(p_1,\ldots p_n) \, .
\nonumber
\eea} Here the renormalized
mass parameter $m_R$ corresponds to the inverse correlation 
length. The physical four-vertex at zero momentum is given by $\lambda_R$.

\subsubsection{2PI renormalization scheme to order $\lambda_R^2$}

In order to impose the renormalization conditions 
(\ref{eq:r1})--(\ref{eq:r3}) one first has to calculate 
the solution for the two-point field $G(\phi)$ for $\phi = 0$, 
which encodes the resummation and which is obtained from the stationarity
condition for the 2PI effective action (\ref{eq:station}). 

The renormalized field is 
\beq\db
\phi_R = Z^{-1/2} \phi \, .
\eeq
It is convenient to introduce the counterterms relating the bare
and renormalized variables in a standard way with
\begin{eqnarray}\db
Z m^2 = m_R^2 + {\rr \delta m^2} 
\,\, , \qquad 
Z^2 \lambda = \lambda_R + {\rr \delta \lambda} 
\,\, , \qquad {\rr \delta Z} = Z - 1 \, ,
\label{eq:introcount}
\end{eqnarray}
and we write
\begin{equation}\db
G(\phi) = Z G_R(\phi_R)  \,\, . 
\end{equation}
In terms of the renormalized quantities the classical action 
(\ref{eq:classical}) reads
\begin{eqnarray}\db
S =
\int_x \left( \frac{1}{2} \partial_{\mu} \phi_R \partial^{\mu} \phi_R
- \frac{1}{2} m_R^2 \phi_R^2 - \frac{\lambda_R}{24} \phi_R^4
+ \frac{1}{2} {\rr \delta Z}\,     
\partial_{\mu} \phi_R \partial^{\mu} \phi_R
- \frac{1}{2} {\rr \delta m^2} \phi_R^2
- \frac{\rr \delta \lambda}{24} \phi_R^4 \right) \, .
\label{eq:counterS}
\end{eqnarray}
Similarly, one can write for the one-loop part 
$\Tr\ln G^{-1} = \Tr\ln G_R^{-1}$ up to an irrelevant constant 
and with $G = G(\phi)$:
\bea\db
\frac{i}{2} \Tr\, G_{0}^{-1}(\phi) G(\phi) &\db =&\db 
- \frac{1}{2} \int_x \left(\square_x + m_R^2 + {\rr \delta Z_1}\, \square_x
+ {\rr \delta m_1^2} \right)
G_R(x,y;\phi_R)|_{x=y}
\nonumber\\
&& \db - \frac{\lambda_R + {\rr \delta \lambda_1}}{4}  \int_x 
\phi_R^2(x) G_R(x,x;\phi_R)\, . 
\label{eq:counter1loop}
\eea
Here $\delta Z_1$, $\delta m_1^2$ and $\delta \lambda_1$
denote the same counterterms as introduced in (\ref{eq:introcount}),
however, approximated to the given order.
To express $\Gamma_2$ in terms of renormalized quantities it is
useful to note the identity
\begin{equation} \db
\Gamma_2[\phi,G(\phi)]|_{\lambda} 
= \Gamma_2[\phi_R,G_R(\phi_R)]|_{\lambda_R + {\rr \delta \lambda}}
\, , 
\label{eq:counter2PI}
\end{equation}
which simply follows from the standard relation between 
the number of vertices, lines and fields by counting factors of $Z$.  
Therefore, one can replace in $\Gamma_2$ the bare field and 
propagator by the renormalized ones if one replaces bare by renormalized
vertices as well. We emphasize that mass and wavefunction renormalization
counterterms, $\delta Z$ and $\delta m^2$, do not appear explicitly 
in $\Gamma_2$. The counterterms in the classical action (\ref{eq:counterS}),
in the one-loop term (\ref{eq:counter1loop}) and beyond one-loop
contained in $\Gamma_2$ have to be calculated for a given approximation
of $\Gamma_2$. For an explicit example, 
we consider here the 2PI effective action to order $\lambda_R^2$ with    
\begin{eqnarray}\db
\Gamma_2[\phi_R,G_R(\phi_R)] &\db\!\! = \!\!&\db
- \frac{1}{8} \lambda_R \int_x G_R^2(x,x;\phi_R)
+  i \frac{1}{12} \lambda_R^2
\int_{x y} \phi_R(x) G_R^3(x,y;\phi_R) \phi_R(y) 
\nonumber\\
&&\db +  i \frac{1}{48} \lambda_R^2 \int_{x y}  G_R^4(x,y;\phi_R) 
- \frac{1}{8} {\rr \delta \lambda_2} 
\int_x G_R^2(x,x;\phi_R)
\, ,
\label{eq:G2orderg4}
\end{eqnarray}
where the last term contains the respective coupling 
counterterm at two-loop. There are no three-loop
counterterms since the divergences arising
from the three-loop contribution in (\ref{eq:G2orderg4}) are
taken into account by the lower counterterms.

One first has to calculate 
the solution $G_R(\phi_R)$ obtained from the stationarity
condition (\ref{eq:station}) for the 2PI effective action. 
For this one has to impose the 
same renormalization condition as for the propagator 
(\ref{eq:r1}) in Fourier space:
\begin{equation} \rr
i G_R^{-1}(p^2;\phi_R)|_{p = 0,\phi_R=0} = - m_R^2 \, ,
\label{eq:rencondD}
\end{equation}
for given finite renormalized ``four-point'' field
\begin{equation}\db
V_R(x,y;z,w) \equiv 
\frac{\delta^2 i G_R^{-1}(x,y;\phi_R)}{\delta \phi_R(z) 
\phi_R(w)}\Big|_{\phi_R=0} \, .
\label{eq:4field}
\end{equation}
For the above approximation we note the identity
\begin{equation}\db
\frac{\delta^2 \Gamma[\phi_R,G_R(\phi_R)]}
{\delta \phi_R(x) \phi_R(y)}\Big|_{\phi_R = 0}
\equiv i G_R^{-1}(x,y;\phi_R)|_{\phi_R = 0}
\label{eq:Dident}
\end{equation}
for 
\begin{equation} \rr
\delta Z = \delta Z_1 \quad ,\quad
\delta m^2 = \delta m_1^2 \quad , \quad 
\delta \lambda_1 = \delta \lambda_2 \, ,
\label{eq:renconst}
\end{equation}
such that (\ref{eq:rencondD}) for $G_R$ is trivially fulfilled
because of (\ref{eq:r1}). 
In contrast to the exact theory, for the 2PI effective action
to order $\lambda_R^2$ a similar identity does not connect the proper 
four-vertex with $V_R$.\footnote{We
emphasize that for more general approximations the equation  
(\ref{eq:Dident}) may only be valid up to higher order corrections as well.
This is a typical property of self-consistent resummations, and it does
not affect the renormalizability of the theory.  
In this case the proper renormalization procedure still 
involves, in particular, the conditions (\ref{eq:rencondD}) and 
(\ref{eq:Vren}).} Here the respective condition for the 
four-point field $V_R$ in Fourier space reads
\begin{equation}\rr
V_R(p_1,p_2,p_3)|_{p_1 = p_2 = p_3 = 0} = - \lambda_R \, .
\label{eq:Vren}
\end{equation}
Note that this has to be 
the same than for the four-vertex (\ref{eq:r3}).
For the universality class of the $\phi^4$ theory there are 
only two independent input parameters, which we take to
be $m_R$ and $\lambda_R$, and for the exact theory
$V_R$ and the four-vertex agree.  
The renormalization conditions (\ref{eq:r1})--(\ref{eq:r3}) for the propagator
and four-vertex, together with the scheme 
(\ref{eq:rencondD})--(\ref{eq:Vren})
provides an efficient fixing of all the above
counter terms. In particular, it can be very 
conveniently implemented numerically, which turns out to be crucial for 
calculations beyond order~$\lambda_R$. 

We emphasize that the approximation (\ref{eq:G2orderg4}) 
for the 2PI effective action can only be expected to be valid for sufficiently
small $\phi_R \ll m_R/g_R$.  If the latter is not fulfilled there are
additional ${\mathcal O} (\lambda_R^2)$ contributions at three-loop 
$\sim \lambda_R^3 \phi_R^2$ and $\sim \lambda_R^4 \phi_R^4$. 
This approximation should therefore not be
used to study the theory in the spontaneously broken phase or near the critical
temperature of the second-order phase transition. Quantitative
studies of the latter can be performed in practice using
$1/N$ expansion of the 2PI effective action to NLO.

\subsubsection{Renormalized equations for the two- and four-point functions}
\label{ssec:renequations}

From the 2PI effective action to order $\lambda_R^2$ we find with
$\delta \lambda_1 = \delta \lambda_2$ from (\ref{eq:renconst}) 
for the two-point function:
\begin{eqnarray}\db
i G_R^{-1}(x,y;\phi_R) &\db\!\!=\!\!&\db 
- \Big[ (1+ {\rr \delta Z_1}) \square_x + m_R^2 + {\rr \delta m_1^2} 
\nonumber\\
&&\db + \frac{1}{2} (\lambda_R + {\rr \delta \lambda_1}) 
\left( G_R(x,x;\phi_R)+\phi_R^2(x) \right) \Big] \delta(x-y)
\nonumber\\
&&\db
 + \frac{i}{2} \lambda_R^2 G_R^2(x,y;\phi_R)\phi_R(x)\phi_R(y)
 + \frac{i}{6} \lambda_R^2 G_R^3(x,y;\phi_R) 
\, .
\label{eq:invD}
\end{eqnarray}
According to (\ref{eq:Dident}) this expression coincides with the one for the
propagator 
$\delta^2 \Gamma[\phi_R,G_R(\phi_R)]/\delta \phi_R(x) \delta \phi_R(y)$ 
at $\phi_R = 0$. It is straightforward to verify this using 
\begin{equation}\db
\frac{\delta \Gamma[\phi_R,G_R(\phi_R)]}{\delta \phi_R(x)}
\equiv \frac{\delta \Gamma[\phi_R,G_R]}{\delta \phi_R(x)} \, ,
\end{equation}
which is valid since the variation of $G_R(\phi_R)$ with $\phi_R$ does not
contribute due to the stationarity condition (\ref{eq:station}). 
The four-point field (\ref{eq:4field}) in this approximation is 
given by 
\begin{eqnarray}\db
V_R(x,y;z,w) &\db =&\db - (\lambda_R + {\rr \delta \lambda_1}) \delta(x-y) 
\delta(x-z) \delta(x-w) 
\nonumber\\
&&\db - \frac{1}{2} (\lambda_R + {\rr \delta \lambda_1})
\frac{\delta^2 G_R(x,x;\phi_R)}{\delta \phi_R(z) \delta
\phi_R(w)}\Big|_{\phi_R=0} \delta(x-y) 
\nonumber\\
&&\db + \frac{i}{2} \lambda_R^2
\Bigg( \delta(x-w) \delta(y-z) + \delta(y-w)
\delta(x-z) 
\nonumber\\
&&\db + \frac{\delta^2 G_R(x,y;\phi_R)}{\delta \phi_R(z) 
\delta \phi_R(w)}\Big|_{\phi_R=0}\Bigg) G_R^2(x,y;\phi_R=0) \, .
\label{eq:fourfieldg4}
\end{eqnarray}
Inserting the chain rule formula
\begin{equation}\db
\frac{\delta^2 i G_R(x,y;\phi_R)}{\delta \phi_R(z) \delta
\phi_R(w)}\Big|_{\phi_R = 0} =
- \left.
\int_{u,v} G_R(x,u;\phi_R) V_R(u,v;z,w) G_R(v,y;\phi_R)\right|_{\phi_R = 0} \, 
\label{eq:fourfieldchain}
\end{equation} 
one observes that
Eqs.~(\ref{eq:fourfieldg4}) and 
(\ref{eq:invD}) form a closed set of equations for 
the determination of the counterterms $\delta Z_1$, $\delta m_1^2$ 
and $\delta \lambda_1$. Together with $(\ref{eq:renconst})$ one 
notes that $\delta \lambda$ would be undetermined from these equations
alone. This counterterm
is determined by taking into account the equation for the physical 
four-vertex, which is obtained from the above 2PI effective action as 
\begin{eqnarray}\db 
\frac{\delta^4 \Gamma[\phi_R,G_R(\phi_R)]}{\delta \phi_R(x) 
\delta \phi_R(y) \delta \phi_R(z) \delta \phi_R(w)} 
\Big|_{\phi_R=0} = - (\lambda_R + {\rr \delta \lambda})
\delta(x-y) \delta(x-z) \delta(x-w) 
\nonumber\\\db
+ V_R(x,y;z,w) + V_R(x,z;y,w) + V_R(x,w;y,z) 
\nonumber\\[0.2cm]
\db - 2 \left.\left( \frac{\delta i G_R^{-1}(x,y;\phi_R)}{\delta G_R(z,w)} 
+ \frac{\delta i G_R^{-1}(x,z;\phi_R)}{\delta G_R(y,w)}
+ \frac{\delta i G_R^{-1}(x,w;\phi_R)}{\delta G_R(y,z)}
\right)\right|_{\phi_R=0} \, ,
\label{eq:fourvertex} 
\end{eqnarray}
where from (\ref{eq:invD}) one uses the relation
\begin{eqnarray}\db
\left.\frac{\delta i G_R^{-1}(x,y;\phi_R)}{\delta G_R(z,w)}\right|_{\phi_R=0}
&\db =&\db - \frac{1}{2} (\lambda_R + {\rr \delta \lambda_1}) 
\delta(x-y) \delta(x-z) \delta(x-w)
\nonumber\\
&&\db + \frac{i}{2} \lambda_R^2 
G_R^2(x,y;\phi_R=0) \delta(x-z) \delta(y-w) \, .
\end{eqnarray}
We emphasize that
the counterterm $\delta \lambda$
plays a crucial role in the broken phase, since it is
always multiplied by the field~$\phi_0$ and hence it is essential
for the determination of the effective potential.
It is also required in the symmetric phase, in particular, 
when one calculates the momentum-dependent 
four-vertex using Eq.~(\ref{eq:fourvertex}).

It is instructive to consider for a moment the 2PI effective action
to two-loop order for which much can be discussed analytically.
In this case $Z=1$ and the renormalized vacuum
mass $m_R$ of Eq.~(\ref{eq:r1}) or,
equivalently, of Eq.~(\ref{eq:rencondD}) is
given to this order by 
\begin{eqnarray}\db
\frac{m_R^2}{\lambda} 
&\db =&\db \frac{1}{2}\, \mu^{\eps}\! \int \frac{{\rm d}^d k}{(2\pi)^d} 
\left( k^2 + m_R^2 \right)^{-1}
+ \frac{m^2}{\lambda} 
\nonumber\\ 
&\db =&\db - \frac{m_R^2}{16 \pi^2}
\left( \frac{1}{\eps} - \ln \frac{m_R}{\bar{\mu}} + \frac{1}{2} \right)
+ \frac{m^2}{\lambda} \, ,
\label{eq:massgap}
\end{eqnarray}
where $m^2 = m_R^2 + {\rr \delta m_1^2}$ and 
$\lambda = \lambda_R + {\rr \delta \lambda_1}$
and we have used (\ref{eq:renconst}).
Here we have employed dimensional regularization 
and evaluated the integral
in $d = 4 - \eps$ for Euclidean momenta $k$. The bare coupling in the
action (\ref{eq:classical}) has been rescaled
accordingly, $\lambda \to \mu^{\eps} \lambda$, and is dimensionless;
$\bar{\mu}^2 \equiv 4 \pi e^{-\gamma_E} \mu^2$ and
$\gamma_E$ denotes Euler's constant. Below we will consider
lattice regularizations for comparison and to go beyond two-loop order.

Similarly, the zero-temperature 
four-point function resulting from Eq.~(\ref{eq:Vren}) for the
2PI effective action to order $\lambda_R$ is given by
\begin{eqnarray}\db
\lambda_R &\db =&\db \lambda - \frac{\lambda \lambda_R}{2} \,
\mu^{\eps}\! \int \frac{{\rm d}^d k}{(2\pi)^d} 
\left( k^2 + m_R^2 \right)^{-2}
\nonumber\\
&\db =&\db \lambda - \frac{\lambda \lambda_R}{16 \pi^2} 
\left( \frac{1}{\eps} - \ln \frac{m_R}{\bar{\mu}} \right) \, 
\label{eq:rencoup}
\end{eqnarray}
with (\ref{eq:renconst}). We emphasize that the same zero-temperature
equation is obtained starting from
the renormalization condition for the proper four-vertex (\ref{eq:r3})
with $\rr \delta \lambda = 3 \delta \lambda_1$. One observes that all
counterterms are uniquely fixed by the renormalization procedure
discussed in the previous section. 

Though dimensional regularization is elegant for analytical
computations, it turns out that high momentum
cutoff regularizations are often more efficient for numerical
implementations. We will discuss below cutoff regularizations that
are obtained by formulating the theory on a discrete space-time 
lattice. In particular, {\rr\em it is often convenient to carry out the 
numerical calculations using unrenormalized equations}, or equations
where only the dominant (quadratically) divergent contributions
in the presence of scalars are subtracted. 
In order to obtain results that are insensitive to variations
of the cutoff, it is then typically sufficient to consider 
momentum cutoffs that are sufficiently large compared to the
characteristic energy-momentum scale of the process of interest.
It should be also stressed that a number of important
renormalized quantities such as
renormalized masses or damping rates can be directly inferred from
the oscillation frequency or the damping of the time-evolution for
the unrenormalized propagator $G$.

\subsection{2PI effective action for fermions}
\label{sec:2PIfermion}

The construction of the 2PI effective action for fermionic fields
proceeds along very similar lines than for bosons. However, 
one has to take into account the anti-commuting (Grassmann) behavior of
the fermion fields. The main differences compared to the bosonic case
can be already observed from the one-loop part ($\Gamma_2\equiv 0$) of the 
2PI effective action. For vanishing field expectation values
it involves the integrals:  
\bea
\!\!\!\!\! \mbox{Fermions:}\,\,\,\,\,
{\db -i \ln  \int \mathscr{D} \bar{\psi}\mathscr{D} \psi\,
e^{iS_0^{(f)}} = -i \ln \left( \det \Delta_0^{-1} \right)} 
&\db =& \db {\rr -i}\, \Tr \ln \Delta_0^{-1} \, ,\nn
\!\!\mbox{Bosons:} \qquad 
{\db -i \ln  \int \mathscr{D} \varphi\,
e^{iS_0} = -i \ln \left(\det G_0^{-1}\right)^{-\frac{1}{2}} } 
&\db =& \db {\rr \frac{i}{2}} \Tr \ln G_0^{-1}  \, ,
\eea
where $S_0^{(f)} = \int {\rm d}^4 x {\rm d}^4 y\, 
\bar{\psi}(x) i \Delta_{0}^{-1}(x,y)
\psi(y)$ denotes a fermion action that is bilinear in the
Grassmann fields. For Dirac fermions with mass $m^{(f)}$ the free inverse 
propagator reads
\bea \db
i \Delta_0^{-1} (x,y) 
= [i \partial\!\slash_{\!x}\, - m^{(f)} ]\, \delta (x-y)\, ,
\eea
where $\partial\!\slash \equiv \gamma^\mu \partial_\mu$ with Dirac matrices
$\gamma_\mu$ ($\mu =0,\ldots,3$), and $\bar{\psi} = \psi^\dagger \gamma^0$.
For the bosons $S_0$ is given by the quadratic part of (\ref{eq:classical}).
Comparing the two integrals one observes that the factor $\rr 1/2$
for the bosonic case is replaced by $\rr -1$ for the fermion fields
because of their anti-commuting property. With this difference,
following along the lines of Sec.~\ref{sec:genfunc1} one finds
that the 2PI effective action for fermions can be written in
complete analogy to (\ref{2PIaction}). Accordingly,
for the case of vanishing fermion field expectation
values, $\langle \Psi \rangle = \langle \bar{\Psi} \rangle = 0$, 
one has:
\beq
\mbox{\framebox{\db$\,\displaystyle \Gamma[\Delta] \, = \, {\rr -i}\, 
\Tr\ln \Delta^{-1} {\rr -i}\, \Tr\, \Delta_0^{-1} \Delta
+ {\rr \Gamma_2[\Delta]} + {\rm const}\,$}}
\label{eq:fermion2PI}
\eeq
Here {\rr\em $\Gamma_2[\Delta]$ contains all 2PI
diagrams} with lines associated to the time ordered 
propagator $\Delta(x,y) = \langle T \Psi(x) 
\bar{\Psi}(y) \rangle$. As for the 1PI effective action diagrams get  
an additional minus sign from each closed fermion loop.
The trace ``$\Tr$'' includes integration over time and
spatial coordinates, as well as summation over field indices.

As for the bosonic case of Eq.~(\ref{eq:station}), the 
equation of motion for $\Delta$ in absence of external sources is 
obtained by extremizing the effective action: 
\beq\db
\frac{\delta\Gamma[\Delta]}{\delta \Delta(x,y)} = 0 \, .
\label{eq:fermstat}
\eeq 
Using (\ref{eq:fermion2PI}) this stationarity condition
can be written as
\beq\db
\Delta^{-1}(x,y)\, =\, 
\Delta_{0}^{-1}(x,y) - {\rr \Sigma^{(f)}(x,y;\Delta)} \, ,
\label{eq:fermSD}
\eeq
with the proper fermion self-energy: 
\beq 
\mbox{\framebox{\rr$\,\displaystyle
\Sigma^{(f)}(x,y;\Delta) \equiv -i 
\frac{\delta \Gamma_2[\Delta]}{\delta \Delta(y,x)}\,$}}
\label{eq:fermselfen}
\eeq  

\subsubsection{Chiral quark-meson model}
\label{sec:chiralqmm}

As an example we consider a quantum field theory involving two fermion
flavors (``quarks'') coupled in a chirally invariant way to a scalar
$\sigma$--field and a triplet of pseudoscalar ``pions'' $\pi^a$
($a=1,2,3$). The classical action reads
\bea\db
S &\db =&\db \int {\rm d}^4 x \Big\{\bar{\psi} i \partial\!\slash \psi 
+\frac{1}{2}\left[\partial_\mu \sigma \partial^\mu \sigma
+ \partial_\mu \pi^a  \partial^\mu \pi^a \right] \nonumber\\
&&\db \qquad\quad\!\!\!  
+\, \frac{h}{N_f} \bar{\psi} \left[\sigma + i\gamma_5 \tau^a \pi^a \right] \psi
- V(\sigma^2 + \pi^2) \Big\} \, ,
\label{chiralfermact}
\eea
where $\pi^2\equiv \pi^a \pi^a$. 
Here $\tau^a$ denote the standard 
Pauli matrices and $h/N_f$ is the Yukawa coupling. 
The number of fermion flavors is $N_f \equiv 2$ 
and $N_f^2$ is the number of scalar components. 
The above action 
is invariant under chiral 
$SU_L(2)\times SU_R(2) \sim O(4)$ transformations.  
For a quartic scalar self-interaction, 
\beq\db
V(\sigma^2 + \pi^2) = \frac{1}{2} m^2 (\sigma^2 + \pi^2) 
+ \frac{\lambda}{4! N_f^2} \left(\sigma^2+ \pi^2\right)^2 \, ,
\label{eq:potterm}
\eeq
this model corresponds to the well known linear $\sigma$--model, incorporating
the chiral symmetry of massless two-flavor QCD.

Here we do not consider 
the possibility of a spontaneously broken symmetry, thus field
expectation values vanish. 
For this model the 2PI effective action is then a functional 
of fermion as well as scalar propagators. The scalar fields form
an $O(4)$ vector $\phi_a(x) \equiv (\sigma(x), \vec{\pi}(x)\,)$ and
we denote the full scalar propagator by $G_{ab}(x,y)$ 
with $a,b=0,\ldots,3$. In addition to its Dirac structure the 
fermion propagator carries flavor indices $i,j$ for $N_f=2$ flavors.
The 2PI effective action for this Yukawa model is given by
\beq\db
  \Gamma[G,{\rr \Delta}] =  \frac{i}{2} \Tr\ln G^{-1} 
 + \frac{i}{2} \Tr G_0^{-1} G\,
 {\rr\, -i\, \Tr\ln \Delta^{-1} -i\, \Tr\, \Delta_0^{-1} \Delta}
 + \Gamma_2[G,{\rr \Delta}] + {\rm const} \, ,
\label{totaleffact}
\eeq
with the free scalar and fermion inverse propagators 
\beq
\label{classscalar} \db
 i G_{0,ab}^{-1} (x,y) = -(\square_x + m^2 )
 \delta(x-y) \delta_{ab} \,\,\,\, , \,\,\,\, 
{\rr i \Delta_{0,ij}^{-1} (x,y) 
= (i \partial\!\slash_{\!x} - m^{(f)} ) \delta(x-y) \delta_{ij}} \, .
\eeq
The equation of motions for the scalar and fermion propagators
are obtained from the stationary conditions 
\beq\db
\frac{\delta\Gamma[G,{\rr \Delta}]}{\delta G_{ab}(x,y)}=0 \qquad ,\qquad
\frac{\delta\Gamma[G,{\rr \Delta}]}{\delta {\rr \Delta_{ij}(x,y)}} = 0 \, .
\eeq
The first non-zero order in a loop
expansion of $\Gamma_2[G,{\rr \Delta}]$ consists of 
a purely scalar contribution 
corresponding to the two-loop diagram of Fig.~(\ref{fig:fiveloop}),
as well as a fermion-scalar contribution depicted here graphically: 

\vspace*{0.2cm}

\centerline{\parbox{3cm}{
\centerline{\epsfig{file=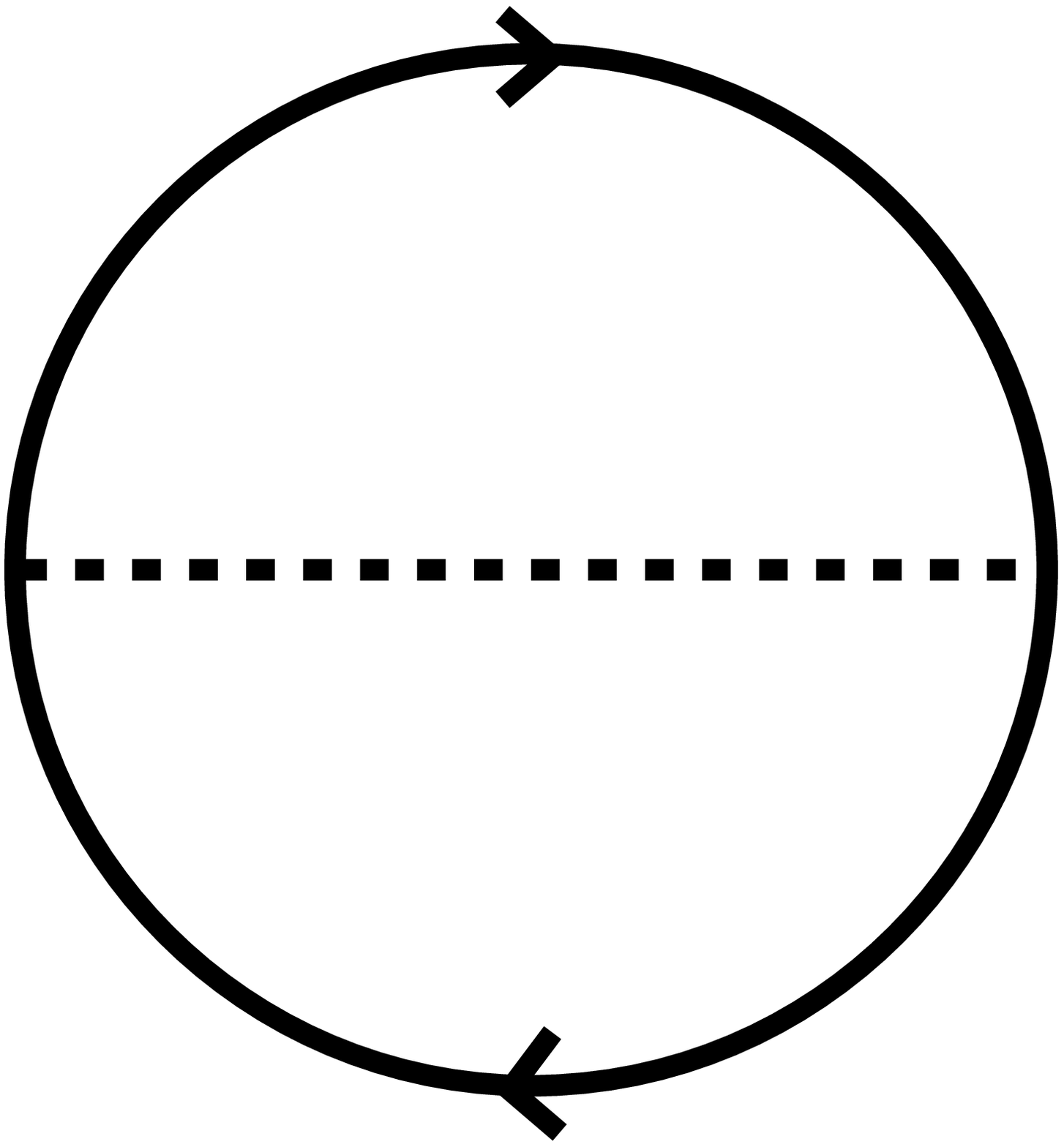,width=1.6cm}}}
\parbox{5cm}{\small
\mbox{solid line: fermion propagator $\Delta$} \newline
\mbox{dashed line: scalar propagator $G$}
}}

\vspace*{0.2cm}

\noindent
Without loss of generality in the absence of spontaneous
chiral symmetry breaking, the effective action and 
its functional derivatives can be evaluated
for $G_{ab}$ taken to be the unit matrix in $O(4)$--space. 
Similarly, the most general fermion two-point function 
can be taken to be proportional to unity in flavor space
and we can write: 
\beq\db
 G_{ab} (x,y) = G (x,y) \, \delta_{ab} 
\qquad , \qquad
 {\rr \Delta_{ij} (x,y) = \Delta(x,y) \,\delta_{ij}}\,\, .
\label{eq:propsym}
\eeq 
The two-loop approximation then reads   
\bea \db
 \Gamma_2^{\rm (2loop)} [G,{\rr \Delta}]|_{G_{ab}=G\delta_{ab},
 {\rr \Delta_{ij} = \Delta \delta_{ij}}}
 &\db=&\db - \frac{\lambda}{8} \frac{(N_f^2 + 2)}{3} \int_x G^2(x,x)
\nonumber\\ 
 &&\db - i h^2\,\frac{N_f}{2}\int_{x y} 
 \,{\rr \tr [\Delta(x,y)\,\Delta(y,x)]}\, G(x,y) \,\, ,
\label{eq:chiraltwoloop}
\eea
where the trace ``$\tr$'' acts only in Dirac space. According 
to (\ref{exactsigma}) and (\ref{eq:fermselfen}) the
self-energies can then be obtained by
\bea\db
\Sigma_{ab}(x,y) &\db \!=\!&\db 2 i\, 
\int_{x' y'} \frac{\delta \Gamma_2|_{G\delta_{ab},
 {\rr \Delta \delta_{ij}}}}{\delta G(x',y')}
\frac{\delta G(x',y')}{\delta G^{ab}(x,y)}
\,\,=\,\, 2 i\, \frac{\delta_{a b}}{N_f^2} 
\frac{\delta \Gamma_2|_{G\delta_{ab},
 {\rr \Delta \delta_{ij}}}}{\delta G(x,y)} \, , \\
\rr \Sigma^{(f)}_{ij}(x,y) &\db \!=\!&\db - i\, \frac{\delta_{a b}}{N_f}
\frac{\db \delta \Gamma_2|_{G\delta_{ab},
 {\rr \Delta \delta_{ij}}}}{\delta {\rr \Delta(y,x)}} \,\, , 
\eea
where we have used $G = G_{ab} \delta_{ab}/N_f^2$ with $\delta_{aa} = N_f^2$,
and equivalently for the fermion propagator with $\delta_{ii} = N_f$.

\subsection{Two-particle irreducible $1/N$ expansion} 
\label{sec:2PIN}

In this section we discuss a systematic nonperturbative 
approximation scheme for the 2PI effective action. It
classifies the contributions to the 2PI effective action
according to their scaling with powers of $1/N$, where
$N$ denotes the number of field components:
\bea\db
\Gamma_2[\phi,G] &\db\!=\!&\db  \Gamma_2^{\rm LO}[\phi,G] 
          \,+\, {\rr \Gamma_2^{\rm NLO}[\phi,G]} 
          \,+\, {\color{DarkViolet} \Gamma_2^{\rm NNLO}[\phi,G]}
          \,+\, \ldots \nonumber\\
 && \quad
{\db\sim N^1} \qquad\quad {\rr \sim N^0} \qquad\quad
{\color{DarkViolet}\sim N^{-1}} 
\nonumber
\eea
Each subsequent contribution $\Gamma_2^{\rm LO}$,
$\Gamma_2^{\rm NLO}$, $\Gamma_2^{\rm NNLO}$ etc.~is  
down by an additional factor of $1/N$. The importance of an expansion 
in powers of $1/N$ stems from the fact that it provides a controlled 
expansion parameter that is not based on weak couplings.
It can be applied to describe physics characterized by
nonperturbatively large fluctuations, such as encountered
near second-order phase transitions in thermal equilibrium,
or for extreme nonequilibrium phenomena such as 
parametric resonance. For the latter cases a 2PI coupling
or loop expansion is not applicable. The method
can be applied to bosonic or fermionic theories alike if
a suitable field number parameter is available,
and we exemplify it here for the case of the scalar
$O(N)$-symmetric theory with classical action (\ref{eq:classical}).
We comment on its application to the chiral quark-meson model
below. 

\subsubsection{Classification of diagrams} 
\label{sec:classofdiag}

We present a simple classification
scheme based on $O(N)$--invariants which parametrize the 2PI diagrams
contributing to $\Gamma[\phi,G]$. The interaction term of
the classical action in Eq.\ (\ref{eq:classical}) is written such that
$S[\phi]$ scales proportional to $N$. From the fields $\phi_a$ alone one
can construct only one independent invariant under $O(N)$ rotations, which
can be taken as $\tr\, \phi\phi \equiv \phi^2 = \phi_a \phi_a \sim
N$.  The minimum $\phi_0$ of the classical effective potential for this
theory is given by $\phi_0^2 = N (-6 m^2/\lambda)$ for negative
mass-squared $m^2$ and scales proportional to $N$.  Similarly, the trace
with respect to the field indices of the classical propagator $G_{0}$ is
of order~$N$.

The 2PI effective action is a singlet under $O(N)$ rotations and
parametrized by the two fields $\phi_a$ and $G_{ab}$.  To write down the
possible $O(N)$ invariants, which can be constructed from these fields, we
note that the number of $\phi$--fields has to be even in order to
construct an $O(N)$--singlet. For a compact notation we use 
$(\phi \phi )_{ab} = \phi_a \phi_b$. 
All functions of $\phi$ and $G$, which are singlets under $O(N)$, can be
built from the irreducible (i.e.\ nonfactorizable in field-index space)  
invariants
\beq\rr 
\phi^2, \quad\quad \tr (G^n) \quad\quad \mbox{and}  
\quad\quad \tr (\phi \phi G^n). 
\label{oninvariants}
\eeq
We note that for given $N$ only the invariants with $n \le N$ are
irreducible --- there cannot be more independent invariants than fields. 
We will see below that for lower orders in the $1/N$ expansion and
for sufficiently large $N$ one has $n < N$. In particular, for the
next-to-leading order approximation one finds that only invariants with $n
\le 2$ appear, which makes the expansion scheme appealing from a
practical point of view. 

Since each single graph contributing to $\Gamma[\phi,G]$ is an
$O(N)$--singlet, we can express them with the help of the set of
invariants in Eq.\ (\ref{oninvariants}). The factors of $N$ in a given
graph have two origins: 
\bi
\item {\rr each irreducible invariant is taken to scale
proportional to $N$ since it contains exactly one trace over the field
indices,} 
\item {\rr while each vertex provides a factor of $1/N$.} 
\ei
\begin{figure}[t]
\centerline{\epsfig{file=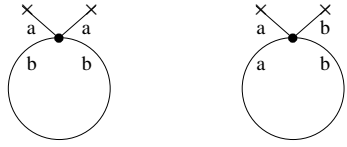,width=4.5cm}}
\caption{
Graphical representation of the $\phi$--dependent contributions for
$\Gamma_2 \equiv 0$. The crosses denote field insertions $\sim
\phi_a\phi_a$ for the left figure, which contributes at leading order, and
$\sim \phi_a\phi_b$ for the right figure contributing at next-to-leading
order.
}
\label{fig:oneloopfig}
\end{figure}
The expression (\ref{2PIaction}) for the 2PI effective
action contains, besides the classical action, the one-loop contribution
proportional to $\Tr\,\ln G^{-1} + \Tr\, G_0^{-1}(\phi) G$ and a
nonvanishing $\Gamma_2[\phi,G]$ if higher loops are taken into account.
The one-loop term contains both leading order (LO) and 
next-to-leading order (NLO) contributions.  The logarithmic
term corresponds, in absence of other terms, simply to the free field
effective action and scales proportional to the number of field
components $N$. To separate the LO and NLO contributions at the one-loop
level consider the second term $\Tr\, G_0^{-1}(\phi) G$. From the form of
the classical propagator (\ref{classprop}) one observes that it can be
decomposed into a term proportional to $\tr(G) \sim N$ and terms
$\sim (\lambda/6N) \left[ \tr(\phi\phi)\,\tr(G) + 2\,\tr(\phi\phi G) \right]$.
This can be seen as the sum of two ``2PI one-loop graphs'' with field
insertion $\sim \phi_a\phi_a$ and $\sim \phi_a\phi_b$, respectively,
as shown in Fig.~\ref{fig:oneloopfig}.
Counting the factors of $N$ coming from the traces and the prefactor, one
sees that only the first   
contributes at LO, while the second one is NLO.

According to the above rules one draws all topologically
distinct 2PI diagrams and counts the number of closed lines as well as 
the number of lines connecting two field insertions in a diagram
following the indices. For instance, the left diagram of
Fig.~\ref{fig:oneloopfig} admits one line connecting two field insertions
and one closed line. In contrast, the right figure admits only one line 
connecting two field insertions by following the indices. Therefore,
the right graph exhibits one factor of $N$ less and becomes subleading.
Similarly, for the two-loop graph below one finds:

\vspace*{0.2cm}

\centerline{\epsfig{file=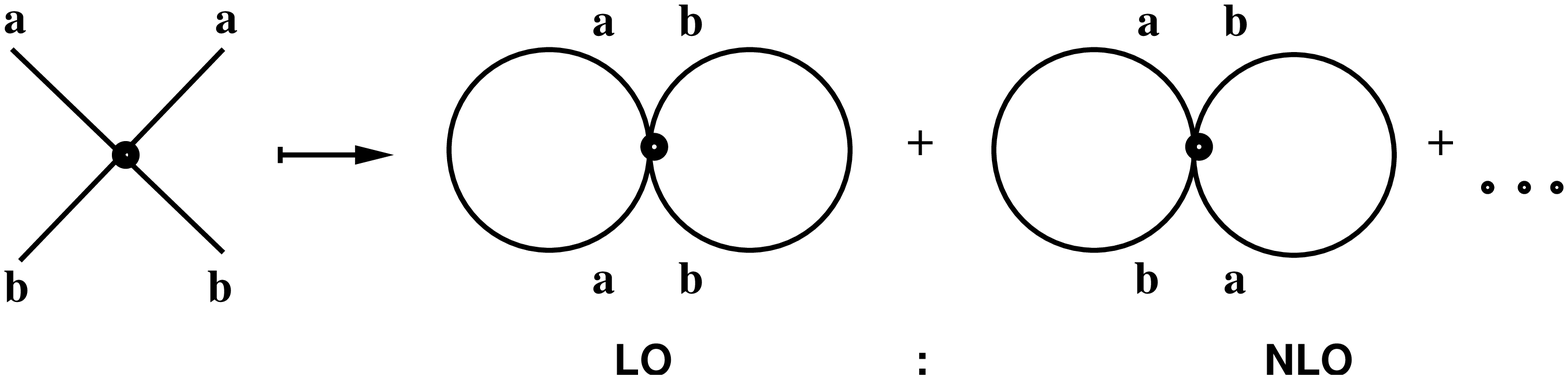,width=8.8cm}}
\hspace*{5.cm}{\db$(\tr G)^2/N \sim N$}  
\hspace*{0.7cm}{\rr$\tr G^2/N \sim N^{0}$}

\vspace*{0.2cm}

\noindent
The same can be applied to higher orders. We 
consider first the contributions to 
\mbox{$\Gamma_2[\phi=0,G] \equiv \Gamma_2[G]$},
i.e.~for a vanishing field expectation value and discuss $\phi \not = 0$
below. The LO contribution to $\Gamma_2[G]$ consists of only one 
two-loop graph, whereas to NLO there is an infinite series of
contributions which can be analytically summed:
\bea\db
\Gamma_2^{\rm LO}[G] &\db =&\db - \frac{\lambda}{4! N} 
  \int_x G_{aa}(x,x) G_{bb}(x,x) \, ,  
\label{LOcont} \\[0.2cm]
\rr \Gamma_2^{\rm NLO}[G] &\rr =&\rr  \frac{i}{2} \Tr  
\ln [\, B(G)\, ] \, , 
\label{eq:NLOcont} \\ 
\rr B(x,y;G) &\rr =&\rr \delta(x-y)
+ i \frac{\lambda}{6 N}\, G_{ab}(x,y)G_{ab}(x,y) \, .
\label{eq:Feq} 
\eea
In order to see that (\ref{eq:NLOcont}) with (\ref{eq:Feq}) 
sums the following infinite series of diagrams

\vspace*{0.2cm}

\centerline{\epsfig{file=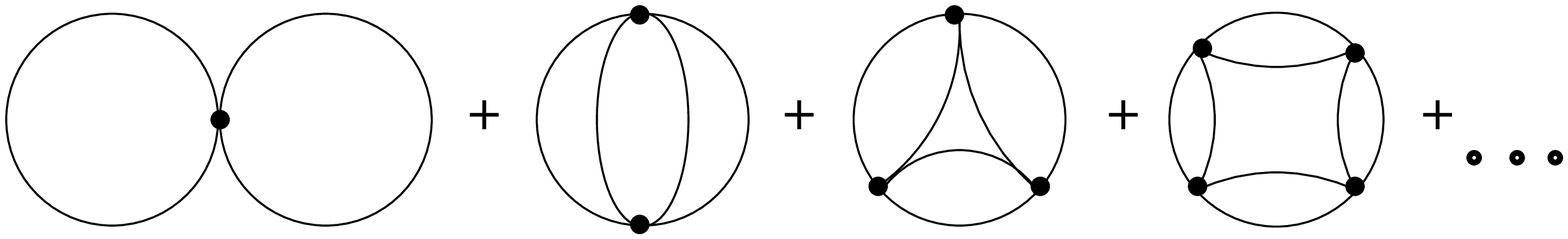,width=8.cm}}

\vspace*{0.2cm}

\noindent
one can expand:
\bea \rr
\Tr \ln [\, B(G)\, ] 
&\rr=&\rr  \int_{x} 
 \left( i \frac{\lambda}{6 N} G_{ab}(x,x)G_{ab}(x,x) \right)
\nonumber\\
&\rr -&\rr \frac{1}{2} \int_{xy}
\left( i \frac{\lambda}{6 N} G_{ab}(x,y)G_{ab}(x,y) \right)
\left( i \frac{\lambda}{6 N}\, G_{a'b'}(y,x)G_{a'b'}(y,x) \right)
\nonumber\\
&\rr +&\rr \ldots  \label{eq:logexpansion}
\eea
The first term on the r.h.s.~of (\ref{eq:logexpansion}) 
corresponds to the two-loop graph with the index structure 
exhibiting one trace such that the contribution scales as
$\tr G^2/N \sim N^0$. One observes that each additional
contribution scales as well proportional to 
$(\tr G^2/N)^n \sim N^0$ for all $n \ge 2$. Thus
all terms contribute at the same order. 

The terms appearing in the presence of a nonvanishing field expectation
value are obtained from the effectively cubic interaction term in 
(\ref{interactionS}) for $\phi \not = 0$. One first notes that 
there is no $\phi$-dependent graph contributing at LO. To NLO there
is again an infinite series of diagrams $\sim N^0$ which can be summed:
\bea\db
\Gamma_2^{\rm LO}[\phi,G] &\db \equiv&\db   
\Gamma_2^{\rm LO}[G] \, , 
\label{eq:LObrok} \\[0.3cm]
\rr \Gamma_2^{\rm NLO}[\phi,G] &\rr =& \rr   
\Gamma_2^{\rm NLO}[\phi\equiv 0,G] + 
\frac{i\lambda}{6N} \int_{xy} I (x,y;G)\,
\phi_a(x) \, G_{ab} (x,y) \, \phi_b (y) \, ,  \label{eq:NLObrok} \\
\rr I (x,y;G) &\rr =&\rr \frac{\lambda}{6 N}\, G_{ab}(x,y) G_{ab}(x,y)
- i\, \frac{\lambda}{6 N} \int_{z}\, I (x,z;G)
 G_{ab}(z,y) G_{ab}(z,y)  \, . \label{eq:Ifunc}
\eea
The series of terms contained in (\ref{eq:NLObrok}) with (\ref{eq:Ifunc})
corresponds to the diagrams:

\vspace*{0.2cm}
\centerline{\epsfig{file=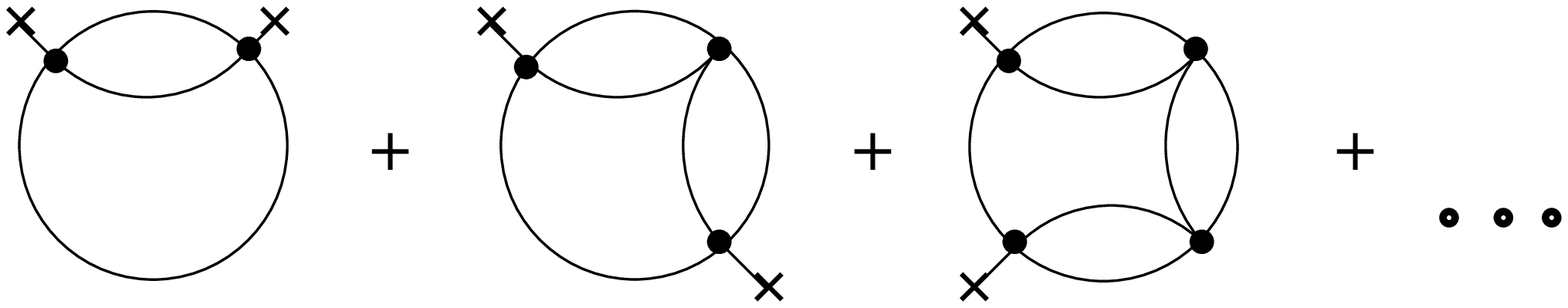,width=6.6cm}}

\vspace*{0.2cm}

\noindent
The functions $I (x,y;G)$ and the inverse 
of $B(x,y;G)$ are closely related by
\beq\db
B^{-1}(x,y;G) = \delta(x-y) - i I (x,y;G) \, ,
\label{Binverse}
\eeq
which follows from convoluting (\ref{eq:Feq}) with $B^{-1}$ and
using (\ref{eq:Ifunc}). We note that $B$ and $I$ do not depend
on $\phi$, and $\Gamma_2[\phi,G]$ is only quadratic in $\phi$ at NLO.

The 2PI can be straightforwardly applied to other theories as well.
For instance, the chiral quark-meson model of Sec.~\ref{sec:chiralqmm} 
admits a $1/N_f$ expansion. For the latter the LO contribution scales
$\sim N_f^2$ and is given by a purely scalar two-loop term as in
Eq.~(\ref{eq:chiraltwoloop}). Note that this two-loop contribution
contains a LO part $\sim N_f^2$ as well as a NNLO part $\sim N_f^0$. 
At NLO the first fermion 
contributions to $\Gamma_2$ appear, which scale $\sim N_f$
(cf.~Eq.~(\ref{eq:chiraltwoloop})). If no small coupling is
available the expansion parameter $N_f \equiv 2$ in this case
might not be suitable for a quantitative estimate at low orders. 
See however the precision tests of Sec.~\ref{sec:precision} for the scalar
$O(N)$ model, which exhibit a good convergence already 
for moderate values of $N \gtrsim 2$. 

\subsubsection{Symmetries and validity of Goldstone's theorem}

We emphasize that by construction each order in the 
2PI $1/N$--expansion respects $O(N)$ symmetry.  In particular, this 
is crucial for the validity of Ward identities. For the case of 
spontaneous symmetry breaking 
note that Goldstone's theorem 
is fulfilled at any order in the 2PI $1/N$ expansion. 
According to (\ref{oninvariants}) the 2PI effective action 
can be written as a function of the $O(N)$ invariants:
\beq\db
 \Gamma [\phi,G] \equiv 
 \Gamma \left[ \phi^2, \tr (G^n), \tr (\phi \phi G^p) \right].
\label{Gammainv}
\eeq 
In the case of spontaneous symmetry breaking one has a constant $\phi \not
= 0$ and the propagator can be parametrized as 
\beq\db
 G_{ab} (\phi) = G_L(\phi^2) P_{ab}^L + 
 G_T(\phi^2) P_{ab}^T \, ,
\label{Gconstphi}
\eeq   
where $P_{ab}^L = \phi_a \phi_b/\phi^2$ and $P_{ab}^T = \delta_{ab} -
P_{ab}^L$ are the longitudinal and transverse projectors with
respect to the field direction.
The 2PI resummed effective action $\Gamma[\phi,G(\phi)]$ is obtained by
evaluation at the stationary value
(\ref{eq:station}) for $G$. The mass matrix
${\cal M}_{ab}$ can then be obtained from
\beq\db  
{\cal M}_{ab} \sim \left. \frac{\delta^2 
\Gamma[\phi,G (\phi)]}{\delta
\phi_a\delta \phi_b}\right|_{\phi=\rm const}.
\eeq
If $\Gamma[\phi,G(\phi)]$ is calculated from (\ref{Gammainv}) and 
(\ref{Gconstphi}) one observes that indeed it  
depends only on one invariant, $\phi^2$:
$\Gamma[\phi,G(\phi)] = \Gamma[\phi^2]$. The form of the mass matrix 
${\cal M}_{ab}$ can now be inferred straightforwardly. 
To obtain the effective potential $U(\phi^2/2)$, we write
\beq\db
 \left. \Gamma [\phi^2] \right|_{\phi=\rm const} 
= \Omega_{d+1} U (\phi^2/2),
\eeq
where $\Omega_{d+1}$ is the $d+1$ dimensional Euclidean volume. The 
expectation value of the field is given by the solution of the 
stationarity equation (\ref{stationphi}) which becomes
\beq\db
\label{minphi}
 \frac{\partial U(\phi^2/2)}{\partial \phi_a}  = 
 \phi_a \, U^\prime = 0,
\eeq
where $U^\prime \equiv \partial U/ \partial(\phi^2/2)$ and similarly for 
higher derivatives. The mass matrix reads  
\beq\db
 {\cal M}^2_{ab} =
 \frac{\partial^2 U(\phi^2/2)}{\partial \phi_a \partial \phi_b} 
 =  \delta_{ab} U^\prime + 
 \phi_a \phi_b U^{\prime \prime} = 
 ( U^\prime + \phi^2 U^{\prime \prime} ) P_{ab}^L + U^\prime P_{ab}^T.
\eeq
In the symmetric phase ($\phi_a=0$) one finds that all modes have equal
mass squared ${\cal M}^2_{ab} = U^\prime\delta_{ab}$.  In the broken
phase, with $\phi_a \neq 0$, Eq.~(\ref{minphi}) implies that the mass of
the transverse modes $\sim U^\prime$ vanishes identically in agreement
with Goldstone's theorem. Truncations of the 2PI effective action 
may not show manifestly the presence of massless
transverse modes if one considers the solution of the 
stationarity equation (\ref{stationphi}) for the two-point field $G$ 
encoding the resummation. It is important to realize that 
the 2PI resummed effective action $\Gamma[\phi,G(\phi)]$ complies
fully with the symmetries.

\section{Nonequilibrium quantum field theory}
\label{sec:nPInoneq}

Out of equilibrium dynamics requires the specification of an initial
state. This may include a density matrix at a given time 
$\rho_D(t_0=0)$ in a mixed ($\Tr \rho_D^2(0) < 1$) or pure state
($\Tr \rho_D^2(0) = 1$). Nonequilibrium means that the
initial density matrix does not correspond to a thermal
equilibrium density matrix: $\rho_D(0) \not =  \rho_D^{\rm (eq)}$ 
with for instance $\rho_D^{\rm (eq)} \sim e^{-\beta H}$ for the
case of a canonical thermal ensemble.  
Once the initial state is specified, the dynamics 
can be described in terms of a functional path integral 
with the classical action $S$ as employed in the previous sections above.
The corresponding {\rr\em nonequilibrium effective action} is the
generating functional for all correlation functions with the
initial correlations determined by $\rho_D(0)$.

\subsection{Nonequilibrium generating functional}
\label{sec:nonequgenfunc}

All information about nonequilibrium quantum field
theory is contained in the {\rr\em nonequilibrium generating functional}
for correlation functions:
\bea\db 
Z[J,R;{\rr \rho_D}] &\db =&\db 
\Tr\left\{{\rr \rho_D(0)}\, T_{\C} e^{i \left( \int_{x}\!
J(x) \Phi(x)+\frac{1}{2}
\int_{x y}\! R(x,y)\Phi(x)\Phi(y)\right)}\right\}  
\label{eq:definingZneq} \, .
\eea
Here $T_{\C}$ denotes time-ordering along the time path
$\C$ appearing in the source term integrals with 
$\int_x \equiv \int_{\C} \rmd x^0 \int {\rmd}^d x$  
as specified below.
In complete analogy to Eq.~(\ref{modZ}) we have introduced the 
generating functional with two source terms, 
$J$ and $R$, in order to construct the corresponding  
nonequilibrium 2PI effective action below. We suppress
field indices in the notation, which can be directly recovered 
from (\ref{modZ}). The {\rr\em nonequilibrium correlation functions}, 
i.e.~expectation values of time-ordered
products of Heisenberg field operators $\Phi(x)$, are obtained
by functional differentiation. For instance the two-point
function reads~({\small $\Tr \rho_D = 1$}):
\bea\db
\Tr\{{\rr \rho_D(0)}\, T_{\C} \Phi(x)\Phi(y)\} &\db \equiv&\db
{\rr\langle} T_{\C} \Phi(x)\Phi(y) {\rr\rangle} 
\,\,=\,\, \frac{\delta^2 Z[J,R;{\rr\rho_D}]}
{i \delta J(x)\, i \delta J(y)}\Big|_{J=R=0} \, . 
\label{eq:twopointneq}
\eea
There is a simple functional integral representation of
$Z[J,R;\rho_D]$ similar to~(\ref{modZ}). One writes the trace as  
\bea \db
Z [J,R;{\rr \rho_D}] &\db =& {\rr
\int {\rmd} \varphi^{(1)}(\bx) {\rm d} \varphi^{(2)}(\bx)\, 
\langle \varphi^{(1)} | \rho_D(0) | \varphi^{(2)} \rangle } \nn
&&\db \langle \varphi^{(2)} | T e^{i \left( \int_{x}\!
J(x)\Phi(x)+\frac{1}{2}
\int_{x y}\! R(x,y)\Phi(x)\Phi(y)\right)} 
| \varphi^{(1)} \rangle  \, ,
\label{eq:Zfirststep}
\eea
where the matrix elements are taken with respect to 
eigenstates of the Heisenberg field operators at intial time, 
$\Phi(t=0,\bx) |\varphi^{(i)} \rangle = \varphi^{(i)}(\bx) 
|\varphi^{(i)} \rangle$, $i=1,2$. The source-dependent matrix element
can be expressed in terms of a path integral using standard
techniques if one considers a {\rr\em finite, closed real-time contour 
$\mathcal{C}$}:

\vspace*{0.2cm}

\centerline{\epsfig{file=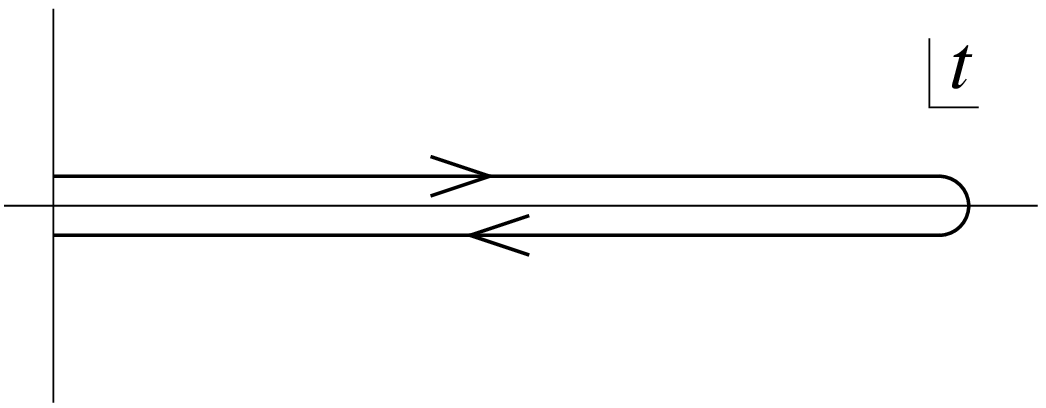,width=4.cm}}

\vspace*{0.2cm}

\noindent
Contour time ordering along this path corresponds to usual 
time ordering along the forward piece $\C^+$ and antitemporal 
ordering on the backward piece $\C^-$. Note that any time on 
$\C^-$ is considered later than any time on $\C^+$. One can then
use
\bea \db
\lefteqn{\langle \varphi^{(2)} | T e^{i \left( \int_{x}\!
J(x)\Phi(x)+\frac{1}{2}
\int_{x y}\! R(x,y)\Phi(x)\Phi(y)\right)} 
| \varphi^{(1)} \rangle } \nn
&\db =&\db \int\limits_{
\begin{array}{c} 
\scriptstyle \varphi^{(1)}(\bx) = \varphi(0^+,\bx)  
\end{array}}^{
\begin{array}{c}  
\scriptstyle \varphi^{(2)}( \bx) = \varphi(0^-,\bx) 
\end{array}}\!\!\!\!\!\!\!\!\!\!\!\!\!
\mathscr{D}'\! \varphi\, e^{i  \left(S[\varphi] 
+ \int_{x} J(x)\varphi(x)  + \frac{1}{2}
\int_{x y}\! R(x,y)\varphi(x)\varphi(y)\right)} \, ,
\label{eq:matrixelements}
\eea
which is the same relation as employed to obtain standard 
path integral expressions for vacuum or equilibrium matrix elements.
Note that the closed time contour starting from the initial time $t_0=0$
and ending at $t_0$ is required because the
density matrix element $\langle \varphi^{(1)} | \rho_D(0) 
| \varphi^{(2)} \rangle$ is taken on both sides with respect to 
states at time $t_0$. Using (\ref{eq:matrixelements}) in (\ref{eq:Zfirststep})
one finds that the expression for $Z[J,R;\rho_D]$ 
directly displays the ingredients entering nonequilibrium
quantum field theory --- the quantum fluctuations described by the
functional integral with action $S$, and the statistical fluctuations encoded 
in the weighted average with the initial-time elements:   
\beq \db
Z [J,R;{\rr \rho_D}] \,=\, {\rr\underbrace{
\int\limits_{_{_{_{_{_{}}}}}} {\rmd} \varphi^{(1)} {\rm d} \varphi^{(2)}  
\langle \varphi^{(1)} | \rho_D(0) | \varphi^{(2)} \rangle}} 
\underbrace{\int\limits_{{\rr \varphi^{(1)}}}^{{\rr \varphi^{(2)}}}
\mathscr{D}'\! \varphi e^{i  \left(S[\varphi] 
+ \int_{x} J(x)\varphi(x)  + \frac{1}{2}
\int_{x y}\! R(x,y)\varphi(x)\varphi(y)\right)}} 
\label{eq:neqgen}
\eeq
\hspace*{3.6cm}{\rr initial conditions}\hspace*{3.2cm}{\db quantum
dynamics} 

\vspace*{0.3cm}

\noindent
Of course, this generating functional can be equally applied to
standard vacuum physics. In this case the closed time contour just
ensures the normalization $Z_{|J=R=0} = 1$, which can simplify
calculations. In particular, the density matrix is time independent
for the vacuum as well as for the thermal equilibrium case.
In contrast to thermal equilibrium, the nonequilibrium
density matrix cannot be interpreted as an evolution operator 
in imaginary time, as is possible for instance for the
canonical $\rho^{\rm (eq)} \sim e^{-\beta H}$. 

In the literature formulations of closed time path generating
functionals often exhibit an infinite time interval $]- \infty,\infty[$. 
In this case
the number of field labels has to be doubled to distinguish the fields
on the underlying closed contour.
Causality implies that for any $n$-point function with finite time
arguments contributions of an infinite time path cancel for times 
exceeding the largest time argument of the $n$-point function. 
To avoid unnecessary
cancellations of infinite time path contributions we always
consider finite time paths. The largest time of the path  
is kept as a parameter and is evolved
in the time evolution equations as described
below. We stress that the initial time of the path has to
be finite --- a system that can thermalize will be already in equilibrium
at any finite time if the initial time is sent to the infinite past.
 
\subsection{Initial conditions}
\label{sec:initialconditions}

To understand in more detail how the initial density matrix 
enters calculations, we consider first the example of a 
Gaussian density matrix whose most general form can be written as 
\bea
&&\db \!\!\!\!\!\!  
\langle \varphi^{(1)}| \rho_D(0) | \varphi^{(2)} \rangle = \nn
&&\db \!\! \frac{1}{\sqrt{2 \pi {\rr \xi}^2}}
\exp \Big\{ i {\rr \dot{\phi}} (\varphi^{(1)} - \varphi^{(2)})
\!-\! \frac{{\rr \sigma}^2+1}{8 {\rr \xi}^2} 
\left[(\varphi^{(1)} - {\rr \phi})^2 
\!+\! (\varphi^{(2)} - {\rr \phi})^2 \right] \nn
&&\db \!\! + i\, \frac{\rr \eta}{2 {\rr \xi}} 
\left[(\varphi^{(1)} - {\rr \phi})^2 
- (\varphi^{(2)} - {\rr \phi})^2 \right]
+ \frac{{\rr \sigma}^2-1}{4 {\rr \xi}^2} 
(\varphi^{(1)} - {\rr \phi}) 
(\varphi^{(2)} - {\rr \phi}) \Big\} \, , 
\label{eq:GaussianrhoD}
\eea
where we neglect the spatial dependencies for 
a moment. Note that for homogeneous field 
expectation values taking into account spatial
dependencies simply amounts to adding a 
momentum label in Fourier space (cf.~also Sec.~\ref{sec:noneqeveq}). 
In order to see that this is the most general
Gaussian density matrix, we first note that (\ref{eq:GaussianrhoD}) 
is equivalent to the following set of initial conditions for one- and 
two-point functions:
\bea 
{\rr \phi} &\db \!=\!&\db 
\Tr\left\{\rho_D(0) \Phi(t) \right\}_{|t=0} \quad , \quad
{\rr \dot{\phi}} \,=\, 
\Tr\left\{\rho_D(0) \partial_t\Phi(t) \right\}_{|t=0} \, ,
\label{eq:equirhoa}
\\[0.3cm]
{\rr \xi^2} &\db\!=\!&\db \Tr\left\{\rho_D(0)
\Phi(t)\Phi(t') \right\}_{|t=t'=0} - {\rr \phi \phi} 
\nonumber \, ,\\ 
{\rr \xi \eta} &\db\!=\!&\db 
\frac{1}{2} \Tr\left\{\rho_D(0)
\left( \partial_t\Phi(t)\Phi(t') + \Phi(t) \partial_{t'}\Phi(t') 
\right) \right\}_{|t=t'=0} - {\rr \dot{\phi} \phi}
\, ,\label{eq:equirhoc}\\
{\rr \eta^2 + \frac{\sigma^2}{4 \xi^2}}
&\db\!=\!&\db 
  \Tr\left\{\rho_D(0)
\partial_t\Phi(t)\partial_{t'}\Phi(t') \right\}_{|t=t'=0} 
- {\rr \dot{\phi} \dot{\phi}} \, . 
\nnn
\eea   
It is straightforward to check explicitly
the equivalence between the initial density 
matrix and the initial conditions for the correlators:
\bea\db
\Tr \rho_D(0) &\db =&\db \int_{-\infty}^{\infty} {\rm d} \varphi\, 
\langle \varphi| \rho_D(0) | \varphi \rangle \nn
&\db =&\db \frac{1}{\sqrt{2 \pi {\rr \xi}^2}} \int_{-\infty}^{\infty}
{\rm d} \varphi \exp \Big\{- \frac{1}{2 {\rr\xi}^2}(\varphi - {\rr \phi})^2
\Big\} = 1 \, , \label{eq:initialdis}\\
\db \Tr \left\{\rho_D(0) \Phi(0)\right\} &\db =&\db
\frac{1}{\sqrt{2 \pi {\rr \xi}^2}} \int_{-\infty}^{\infty}
{\rm d} \varphi\, \varphi 
\exp \Big\{- \frac{1}{2 {\rr \xi}^2}(\varphi - {\rr \phi})^2
\Big\} \stackrel{\varphi \to \varphi+{\rr \phi}}{=} {\rr \phi} \, ,
\eea
etc. Similarly, one finds since only Gaussian integrations 
appear that all
initial $n$-point functions with $n > 2$ are given 
in terms of the one- and two-point functions.
Note also that
the anti-symmetrized initial correlator involving the 
commutator of $\Phi$ and $\partial_t{\Phi}$ is not independent 
because of the field commutation relation.

The crucial observation is that {\rr\em higher initial time 
derivatives are not independent} as can be observed
from the exact field equation of motion, which  
reads for the real scalar $\Phi^4$-theory:
\bea \db
\langle \partial^2_t{\Phi} \rangle =  - m^2 \langle \Phi \rangle 
- \frac{\lambda}{6 N} \langle \Phi^3 \rangle \, .
\eea
Since $\langle \Phi^3 \rangle$ is given in terms
of one- and two-point functions for Gaussian 
$\rho_D(0)$ also second and higher time derivatives
are not independent. We conclude that the most general
Gaussian density matrix is indeed described by the five parameters
appearing in (\ref{eq:GaussianrhoD}).
In particular, all observable information contained in the density
matrix can be conveniently expressed in terms of correlation
functions and their derivatives (\ref{eq:equirhoa}) and (\ref{eq:equirhoc}). 

For further interpretation of the initial conditions 
we note that
\beq\db 
\Tr\, \rho_D^2(0) = \int_{-\infty}^{\infty} {\rm d} \varphi
\int_{-\infty}^{\infty} {\rm d} \varphi'
\langle \varphi| \rho_D(0) | \varphi' \rangle
\langle \varphi'| \rho_D(0) | \varphi \rangle
= \frac{1}{\rr \sigma} \, . 
\eeq
The latter shows that for $\rr \sigma > 1$ the density
matrix describes a mixed state. A pure state requires $\rr \sigma = 1$ or,
equivalently, using (\ref{eq:equirhoc}) this condition can be
expressed in terms of initial-time correlators as 
\bea \db
\left[ \Tr\left\{\rho_D(0)
\Phi(t)\Phi(t') \right\}_{|t=t'=0} - {\phi \phi} \right]
\left[\Tr\left\{\rho_D(0)
\partial_t\Phi(t)\partial_{t'}\Phi(t') \right\}_{|t=t'=0} 
- {\dot{\phi} \dot{\phi}}\right] && \\
\db - \left[\frac{1}{2} \Tr\left\{\rho_D(0)
\left( \partial_t\Phi(t)\Phi(t') + \Phi(t) \partial_{t'}\Phi(t') 
\right) \right\}_{|t=t'=0} - {\dot{\phi} \phi}
\right]^2 &\db\!=\!&\db \frac{1}{4} \,\, . \,\,\, \nonumber
\eea
In field theory such a Gaussian initial condition is
associated with a vanishing initial particle number, as will
be discussed in Sec.~\ref{sec:neqevolution}. For $\rr \sigma = 1$
the ``mixing term'' in (\ref{eq:GaussianrhoD}) is absent and one obtains
a pure-state density matrix of the product form:
\beq \db 
\rho_D(0) =
| \Omega \rangle \langle \Omega |
\eeq
with Schr{\"o}dinger wave function
\beq \db 
\langle \varphi | \Omega \rangle
= \frac{1}{(2 \pi \xi^2)^{1/4}}
\exp \Big\{ i {\rr \dot{\phi}} \varphi
-\Big(\frac{1}{4 {\rr \xi}^2} + i\, \frac{\rr \eta}{2 {\rr \xi}} \Big)
(\varphi - {\rr \phi})^2 \Big\} \, .
\eeq

In order to go beyond Gaussian initial density matrices
one can generalize the above example and parametrize the most 
general density matrix as 
\beq\db
\langle \varphi^{(1)} | \rho_D(0) | \varphi^{(2)} \rangle 
= \mathcal{N}\, e^{i {\db f_\C[\varphi]}} \, ,
\label{eq:denpara}
\eeq
with normalization factor $\mathcal{N}$ and 
$f_\C[\varphi]$ expanded in powers of the fields:  
\bea\db
f_\C[\varphi] &\db = &\db {\rr \alpha_0} + \int_x {\rr \alpha_1(x)} \varphi(x)
+ \frac{1}{2} \int_{x y}\!\! {\rr \alpha_2(x,y)} \varphi(x)\varphi(y) 
+ \frac{1}{3!} \int_{x y z}\!\!\! {\rr \alpha_3(x,y,z)} 
\varphi(x)\varphi(y)\varphi(z)
\nn &&\db
+ \frac{1}{4!} \int_{x y z w} {\rr \alpha_4(x,y,z,w)} 
\varphi(x)\varphi(y)\varphi(z)\varphi(w) + \ldots 
\label{eq:expansiondensity}
\eea
Here $\rr \varphi(0^+,\bx) = \varphi^{(1)}(\bx)$ and 
$\rr \varphi(0^-,\bx)=\varphi^{(2)}(\bx)$. We emphasize that
(\ref{eq:expansiondensity}) employs a compact notation: 
since the density matrix $\rho_D(0)$ is specified at initial time
$t_0 = 0$, all time integrals contribute only at time $t_0$ of the
closed time contour. As a consequence, the coefficients $\alpha_1(x)$,
$\alpha_2(x,y)$, $\alpha_3(x,y,z)$, \ldots vanish identically
for times different than~$t_0$. For instance, up to quadratic 
order one has 
\bea \db
\int_{x} {\rr \alpha_1(x)} \varphi(x) &\db \equiv&\db \int_{\bx}
\left\{{\rr \alpha_1^{(1)}(\bx)} \varphi^{(1)}(\bx)
+ {\rr \alpha_1^{(2)}(\bx)} \varphi^{(2)}(\bx) \right\}\, ,
\nonumber\\ \db
\int_{x y} {\rr \alpha_2(x,y)} 
\varphi(x) \varphi(y) &\db \equiv&\db \int_{\bx \by} \left\{
{\rr \alpha_2^{(1,1)}}(\bx,\by) \varphi^{(1)}(\bx)\varphi^{(1)}(\by)
+ {\rr \alpha_2^{(1,2)}}(\bx,\by)\varphi^{(1)}(\bx)\varphi^{(2)}(\by)
\right. \nonumber\\
&& \db \left.
+ {\rr \alpha_2^{(2,1)}}(\bx,\by)\varphi^{(2)}(\bx)\varphi^{(1)}(\by)
+ {\rr \alpha_2^{(2,2)}}(\bx,\by)\varphi^{(2)}(\bx)\varphi^{(2)}(\by)\right\}
\nonumber \, .
\eea
Note that $\alpha_0$ is an irrelevant normalization constant.
For a physical density matrix the other coefficients
are of course not arbitrary. Hermiticity implies 
$\alpha_1^{(1)} = - \alpha_1^{(2)*}$,
$\alpha_2^{(1,1)} = - \alpha_2^{(2,2)*}$ and 
$\alpha_2^{(1,2)} = - \alpha_2^{(2,1)*}$, which can be
directly compared to the discussion above.

\subsection{Nonequilibrium 2PI effective action}
\label{sec:noneq2PIeffaction}

Using the parametrization (\ref{eq:denpara}) and 
(\ref{eq:expansiondensity}) for the most general initial density matrix,
one observes that the generating functional (\ref{eq:neqgen}) introduced
above can be written as
\beq\db
Z [{\rr J,R;\rho_D}] \,=\, 
\int \mathscr{D}\! \varphi e^{i  \left(S[\varphi] 
+ \int_{x} {\rr J(x)} \varphi(x)  + \frac{1}{2}
\int_{x y}\! {\rr R(x,y)} 
\varphi(x)\varphi(y) 
+ \frac{1}{3!} \int_{x y z}\! {\rr \alpha_3(x,y,z)} 
\varphi(x)\varphi(y)\varphi(z) + \ldots\right)} \, .
\label{eq:Zneqgen}
\eeq
Here we have neglected an irrelevant normalization constant and  
rescaled the sources in (\ref{eq:neqgen})
as $J(x) \to {\rr J(x) - \alpha_1(x)}$ and 
$R(x,y) \to {\rr R(x,y) - \alpha_2(x,y)}$.
The sources can therefore be conveniently used to absorb the 
lower linear and quadratic contributions
coming from the density matrix specifying the initial state.
This absorbtion of $\alpha_1$ in $J$ and $\alpha_2$ in $R$
exploits the fact that {\rr\em the initial density matrix 
is completely encoded in terms of initial-time sources} 
$\alpha_1$, $\alpha_2$, $\alpha_3$, \ldots for the functional 
integral.  

The generating functional (\ref{eq:Zneqgen}) can be
used to describe situations involving arbitrarily complex initial
density matrices. However, often the initial conditions of an experiment 
may be described by only a few lowest $n$--point functions.
For many practical purposes the initial
density matrix is well described by a Gaussian one.
For instance, the initial 
conditions for the reheating dynamics in the early universe at the
end of inflation are described by a Gaussian density matrix to high 
accuracy. Clearly, the subsequent evolution builds up 
higher correlations which are crucial for the process of thermalization.
These have to be taken into account by the quantum dynamics, which is not
approximated by the specification of an initial density matrix!
 
From (\ref{eq:Zneqgen}) one observes that for Gaussian initial
density matrices, for which $\alpha_3 = \alpha_4 = \ldots = 0$, one has 
\beq 
Z\left[{\rr J,R;\rho_D^{\rm (gauss)}}\right] \,\,\equiv\,\, Z[{\rr J, R}] \, .
\eeq
As a consequence, in this case
the nonequilibrium generating functional corresponds to the 
2PI generating functional introduced in Eq.~(\ref{modZ})
for a closed time path. We can therefore directly take over
all steps from Sec.~\ref{sec:genfunc1} in order to construct the
nonequilibrium 2PI effective action and obtain the important result: 

\vspace*{0.3cm}
\centerline{\framebox{\rr$\,$ Nonequilibrium 2PI effective action
$=$ $\Gamma [\phi,G]$ with closed time path $\C\,$
}}

\vspace*{0.3cm}

All previous discussions of Sec.~\ref{sec:genfunc1} remain 
unchanged except that for nonequilibrium the time integrals 
involve a closed time path! We emphasize again that the use
of the nonequilibrium 2PI effective action represents
no approximation for the dynamics --- higher irreducible 
correlations can build up corresponding to a non-Gaussian 
density matrix for times $t > t_0$. It only restricts 
the ``experimental'' setup described by the initial conditions
for correlation functions. 
Non-Gaussian initial density matrices pose no problems in principle
but require taking into account additional initial-time sources.
This is most efficiently described in terms of $n$PI effective actions 
for $n > 2$.

\subsection{Exact evolution equations}
\label{sec:exactevoleq}

Without approximation the 2PI effective action $\Gamma[\phi,G]$ contains
the complete information about the quantum theory.  
The functional representation of the nonequilibrium
$\Gamma[\phi,G]$ employs a one-point ($\phi$) and a 
two-point field ($G$), whose physical values have to be computed 
for all times of interest.
The equations of motion for these fields are given by the 
stationarity conditions (\ref{stationphi}) and (\ref{eq:station}). 

We will consider first the scalar field theory in the  
symmetric regime, where $\mbox{$\Gamma[\phi=0,G]$} \equiv \Gamma[G]$ 
is sufficient. We come back
to the case of a nonvanishing field expectation
value as well as to fermionic and gauge fields below.
The equation of motion (\ref{SchwingerDysonR}) for $G$ reads 
\bea \db
G^{-1}(x,y) = G_{0}^{-1}(x,y)  - \Sigma (x,y;G) -i R(x,y) \, ,
\label{eq:eomG}
\eea 
where the self-energy $\Sigma(x,y;G)$ is given by (\ref{exactsigma})
The form of the equation (\ref{eq:eomG}) is suitable for
boundary value problems as e.g.~appear for thermal equilibrium.  
However, nonequilibrium time evolution is an initial value problem.
The equation of motion 
can be rewritten as a {\rr\em partial differential equation suitable
for initial-value problems} by convolution with $G$, using
$\int_z G^{-1}(x,z)G(z,y) = \delta(x-y)$: 
\beq \db
\int_z G_{0}^{-1}(x,z)G(z,y) - \int_z 
\left[\Sigma(x,z) + i R(x,y)\right] G(z,y) \,=\, \delta(x-y) \, .
\label{eq:exactG}
\eeq
Here we employ the notation
$\delta(x-y) \equiv \delta_{\C}(x^0-y^0) \delta(\bx - \by)$.
For the scalar theory the classical propagator 
(cf.~Sec.~\ref{sec:genfunc1}) is
$G_{0}^{-1}(x-y) = i \left[\square_x + m^2\right]\delta(x-y)$
and one obtains the {\rr\em evolution equation for the time-ordered
propagator}:
\bea
\mbox{\framebox{\rr
$\,\, \displaystyle
\left( \square_x + m^2 \right) G(x,y) 
+\, i \int_z \left[ \Sigma(x,z;G) + i R(x,y) \right] G(z,y)
= - i \delta(x-y) $
}}
\label{eq:exactevolG}
\eea

\subsubsection{Spectral and statistical components}
\label{sec:specstat}

In the following we introduce a decomposition of the
two-point function $G$ into spectral and statistical components.
The corresponding evolution
equations for the spectral function and statistical 
propagator are fully equivalent to the evolution equation
for $G$, but have a simple physical interpretation.
While the spectral function encodes the
spectrum of the theory, the statistical 
propagator gives information about occupation numbers.
Loosely speaking, the decomposition makes explicit
what states are available and how often they are
occupied.  

For the real scalar field theory 
there are two independent real--valued two--point functions, which 
can be associated to the expectation value of the commutator and 
the anti-commutator of two fields,
\bea
\mbox{\rr commutator:} 
&&\rr \rho(x,y) = i \langle [\Phi(x),\Phi(y)] \rangle \, ,
\label{eq:comrho}\\
\mbox{\db anti-commutator:} 
&&\db F(x,y) = \frac{1}{2} \langle \{\Phi(x),\Phi(y)\} \rangle \, .
\label{eq:anticomF}
\eea
Here $\rho(x,y)$ denotes the spectral function and 
$F(x,y)$ the statistical two-point function. The
{\rr\em decomposition identity for spectral and statistical 
components} of the propagator reads:
\beq
\mbox{\framebox{\db
$\,\, \displaystyle
G(x,y) = F(x,y) - \frac{i}{2} {\rr\rho(x,y)}\, \sign_{\C}(x^0 - y^0)$
}}\label{eq:decompid}
\eeq
The identity is easily understood by making the
time-ordering for the propagator explicit:
\bea \db 
G(x,y) &\db =&\db \langle \Phi(x) \Phi(y) \rangle \Theta_{\C}(x^0 - y^0) 
+ \langle \Phi(y) \Phi(x) \rangle \Theta_{\C}(y^0 - x^0) \nn
&\db =&\db \frac{1}{2} \langle \{\Phi(x), \Phi(y)\} \rangle
\left( \Theta_{\C}(x^0 - y^0) + \Theta_{\C}(y^0 - x^0) \right) \nn
&\db -& \db   \frac{i}{2}\, {\rr i \langle [\Phi(x), \Phi(y)] \rangle}
\underbrace{\left(\Theta_{\C}(x^0 - y^0) - \Theta_{\C}(y^0 - x^0)\right)}  \nn
&&\db  \qquad\qquad \qquad\qquad \qquad  \sign_{\C}(x^0 - y^0) \, . \nnn
\eea
For real scalar fields the real functions obey 
$F(x,y)=F(y,x)$ and $\rho(x,y)=-\rho(y,x)$. 
We note from (\ref{eq:comrho}) that the spectral function $\rho$ 
encodes the equal-time commutation relations
\beq\rr
\rho(x,y)|_{x^0=y^0} = 0 \quad, \quad 
\partial_{x^0}\rho(x,y)|_{x^0=y^0} = \delta(\bx-\by) \, .
\label{eq:bosecomrel}
\eeq
The spectral function is also directly related to the
retarded propagator $\rho (x,y) \Theta (x^0 - y^0)$, or the
advanced one $ - \rho (x,y) \Theta (y^0 - x^0)$. 
However, it is important to realize that out of equilibrium
there are only two independent two-point functions --- no more.
These can be associated to $F$ and $\rho$, which has the advantage
compared to e.g.~an advanced propagator that the non-analyticity
entering through time-ordering is always explicit.  

To obtain a similar decomposition for the self-energy,
we separate $\Sigma$ in a ``local'' 
and ``nonlocal'' part according to
\beq\db
\Sigma(x,y;G) = - i \Sigma^{(0)}(x;G) \delta(x-y)
+ \ol{\Sigma}(x,y;G) \, . 
\label{eq:sighominh}
\eeq
Since $\Sigma^{(0)}$ just corresponds to a space-time dependent 
mass-shift it is convenient for the following to introduce the notation
\beq\db 
M^2(x;G) = m^2 + \Sigma^{(0)}(x;G) \, .
\label{eq:localself}
\eeq
To make the time-ordering for the non-local part of the self-energy, 
$\ol{\Sigma}(x,y;G)$, explicit we can use the same identity as for
the propagator (\ref{eq:decompid}) to decompose:
\bea\db 
\ol{\Sigma} (x,y) = \Sigma_F(x,y) - \frac{i}{2} {\rr \Sigma_{\rho}(x,y)}\, 
\sign_{\C}(x^0 - y^0) \, .
\label{eq:decompself}
\eea
Though the discussion is given here for a vanishing field expectation
value, it should be emphasized that the same decompositions 
(\ref{eq:decompid}) and (\ref{eq:decompself}) apply also for $\phi \not = 0$. 
The r.h.s.~of (\ref{eq:anticomF}) would then receive an additional contribution
subtracting the disconnected part $\sim \phi \phi$. Equivalently,
one can always view (\ref{eq:decompid}) as defining $F$ and $\rho$
from the connected propagator $G$ irrespective of the value of $\phi$.

We emphasize that the equivalent decomposition can be done
for fermionic degrees of freedom as well. The fermion propagator
in terms of spectral and statistical components reads in the same
way as for bosons:
\beq\db
\Delta(x,y) = F^{(f)}(x,y) - \frac{i}{2} {\rr \rho^{(f)}(x,y)}\, 
\sign_{\C}(x^0 - y^0) \, .
\label{eq:fermdec}
\eeq
However, in contrast to bosons for fermions the 
field anti-commutator corresponds to the spectral function,
\bea
\mbox{\rr anti-commutator:} 
&&\rr \rho^{(f)}(x,y) = i \langle \{ \Psi(x), \bar{\Psi}(y) \} 
\rangle \, ,
\label{eq:fanticomrho}\\
\mbox{\db commutator:} 
&&\db F^{(f)}(x,y) =   \frac{1}{2} \langle
[\Psi(x), \bar{\Psi}(y)]\rangle \, .
\label{eq:fcomF}
\eea
This can be directly observed from the time-ordered fermion 
propagator
$\Delta (x,y) = \langle \Psi(x) \bar{\Psi}(y) \rangle \Theta_{\C}(x^0 - y^0) 
- \langle \bar{\Psi}(y) \Psi(x) \rangle \Theta_{\C}(y^0 - x^0)$.
Here the minus sign is a consequence of the anti-commutation property
for fermionic fields (cf.~Sec.~\ref{sec:2PIfermion}). In analogy to
the bosonic case, the equal-time
anti-commutation relations for the fields are again 
encoded in $\rho^{(f)}(x,y)$. For instance, for Dirac fermions one has
\beq\rr
\gamma^0 \rho(x,y)|_{x^0=y^0} = i \delta(\bx-\by) \, 
\label{eq:rhoinitial}
\eeq
with the Dirac matrix $\gamma^0$, and the two-point functions
have the hermiticity properties
\beq
{\rr \left(\,\rho(y,x)\,\right)^\dagger
= - \gamma^0 \rho(x,y) \gamma^0} \, , \qquad
{\db \left(\,F(y,x)\,\right)^\dagger
= \gamma^0 F(x,y) \gamma^0} \, .
\label{eq:hermiticity}
\eeq
The corresponding decomposition for the fermion self-energy 
reads:\footnote{If there is a local contribution to the proper self-energy,
one separates in complete analogy to the scalar equation (\ref{eq:sighominh}).
The decomposition (\ref{eq:selffermdec}) is taken for the non-local part
of the self-energy, while the local contribution
gives rise to an effective space-time dependent fermion mass term.}
\beq
\db \Sigma^{(f)}(x,y) \,=\, 
\Sigma_{F}^{(f)} (x,y) - \frac{i}{2}\, {\rr \Sigma_{\rho}^{(f)}(x,y)}\, 
\sign_\C(x^0-y^0)  \, .
\label{eq:selffermdec}
\eeq

\subsubsection{Detour: Thermal equilibrium}
\label{sec:detourthermal}

To see the above decomposition in a probably more familiar
context, we consider for a moment {\rr\em thermal equilibrium}. This is
done for illustrational purposes only, and we emphasize that {\rr\em the 
notion of an equilibrium temperature is nowhere implemented in 
nonequilibrium quantum field theory}. If a nonequilibrium evolution 
approaches thermal equilibrium at late times then this a prediction 
of the theory and not put in by hand.

The 2PI effective action in thermal equilibrium 
is given by the same expression (\ref{2PIaction}) if the closed 
time path is replaced by an imaginary path $\C=[0,-i\beta]$. 
Here $\beta$ denotes the inverse temperature. Since thermal 
equilibrium is translation invariant, the two-point functions 
depend only on relative coordinates and it is convenient to 
consider the Fourier transforms $F^{\rm (eq)}(\omega,\bp)$
and $\rho^{\rm (eq)}(\omega,\bp)$ with
\beq\db
F^{\rm (eq)}(x,y) = \int \frac{\rmd \omega \rmd^{d} p}{(2\pi)^{d+1}} 
e^{-i \omega (x^0-y^0) + i \bp (\bx - \by)} F^{\rm (eq)}(\omega,\bp)
\eeq
and equivalently for the thermal spectral function.

The periodicity (``KMS'') condition characterizing thermal
equilibrium for the propagator in imaginary
time is given by $G(x,y)|_{x^0=0}=G(x,y)|_{x^0=-i\beta}$.
Employing the decomposition identity (\ref{eq:decompid}) for the
propagator $G$, one can write the periodicity condition as
\beq\db
F^{\rm (eq)}(\omega,\bp) + \frac{i}{2} {\rr \rho^{\rm (eq)}(\omega,\bp)} 
= e^{-\beta \omega} \left(
F^{\rm (eq)}(\omega,\bp) - \frac{i}{2} {\rr \rho^{\rm (eq)}(\omega,\bp)} 
\right) \, .
\label{eq:periodicityFrho}
\eeq
To see this note that $x^0=0$ comes first on the 
imaginary path $\C=[0,-i\beta]$, while $x^0=-i\beta$ comes latest
such that $\sign_{\C}(x^0 - y^0)$ contributes
opposite signs on the left and on the right of the equation. 
Eq.~(\ref{eq:periodicityFrho}) can be rewritten
in a more standard form as a {\rr\em fluctuation-dissipation relation}
for bosons:\footnote{In our conventions
the Fourier transform of the real-valued antisymmetric function
$\rho(x,y)$ is purely imaginary.}
\beq
\mbox{\framebox{\db 
$\,\,\displaystyle F^{\rm (eq)}(\omega,\bp) = -i
\left(\frac{1}{2} + n_{\rm BE}(\omega)\right) \, 
{\rr \rho^{\rm (eq)}(\omega,\bp)}$}}
\label{eq:flucdissbose}
\eeq
with $\rr n_{\rm BE}(\omega)=(e^{\beta \omega}-1)^{-1}$ 
denoting the Bose-Einstein distribution function.
Eq.~(\ref{eq:flucdissbose}) relates the spectral function to the 
statistical propagator. While $\rho^{\rm (eq)}$ encodes 
the information about the spectrum of the theory, one observes from 
(\ref{eq:flucdissbose}) that the function $F^{\rm (eq)}$ 
encodes the statistical aspects in terms of the particle distribution
function $n_{\rm BE}$. In the same way one obtains for the Fourier
transforms of the spectral and statistical components of the
self-energy the thermal equilibrium relation
\beq\db
\Sigma^{\rm (eq)}_F(\omega,\bp) = -i
\left(\frac{1}{2} + n_{\rm BE}(\omega) \right) 
\, {\rr \Sigma^{\rm (eq)}_{\rho}(\omega,\bp)} \, .
\label{eq:sigmadecom}
\eeq
We note that the ratio $\Sigma^{\rm (eq)}_{\rho}(\omega,\bp)/2 \omega$ 
plays in the limit of a vanishing $\omega$-dependence
the role of the decay rate for one-particle excited states with
momentum $\bp$.

For fermions 
the anti-periodicity condition of the fermionic propagator in 
thermal equilibrium, $\Delta(x,y)|_{x^0=0}= - \Delta(x,y)|_{x^0=-i\beta}$, 
implies a corresponding fluctuation-dissipation relation. 
The difference is that the Bose-Einstein distribution in 
(\ref{eq:flucdissbose}) is replaced by the Fermi-Dirac distribution
$n_{\rm FD}(\omega)=(e^{\beta \omega}+1)^{-1}$ 
according to $1/2+n_{\rm BE}(\omega)$ $\rightarrow$ $1/2-n_{\rm FD}(\omega)\,$
in the respective relation.

It is important to realize that {\rr\em out of equilibrium $F$ and $\rho$ are 
not related by the fluctuation-dissipation relation!$\,$} In contrast to
the nonequilibrium theory, the relation (\ref{eq:flucdissbose}) 
is a manifestation of the tremendous simplification that happens if the
system is in thermal equilibrium. 
An even more stringent reduction occurs for the
vacuum where $n_{\rm BE}(\omega) \equiv 0$. In this respect, nonequilibrium 
quantum field theory is more complicated since it admits the description 
of more general situations. Of course, the nonequilibrium theory
encompasses the thermal equilibrium or vacuum theory as special 
cases. We leave now this equilibrium detour and return to the nonequilibrium
case.
  
\subsubsection{Nonequilibrium evolution equations}
\label{sec:noneqeveq}

Out of equilibrium we have to follow the
time-evolution both for the statistical propagator,
$F$, as well as for the spectral function, $\rho$.
The evolution equations are obtained from (\ref{eq:exactevolG})
with the help of the identities (\ref{eq:decompid}) and
(\ref{eq:decompself}). Most importantly, once expressed in terms
of $F$ and $\rho$ the time-ordering is explicit and the
respective sign-functions appearing in the time-ordered 
propagator can be conveniently 
evaluated along the time contour $\C$.

With the notation (\ref{eq:sighominh}) the time evolution equation for the 
time-ordered propagator (\ref{eq:exactevolG}) reads 
\beq\db
\left[ \square_x + M^2(x;G) \right] G(x,y) 
+\, i \int_z \ol{\Sigma}(x,z;G) G(z,y)
= - i \delta(x-y) \, ,
\label{eq:evoleqGM}
\eeq
where we have set $R \equiv 0$. The influence of the
initial-time sources encoded in $R$ is discussed below. 
For the evaluation along the time contour $\C$ 
involved in the integration with 
$\int_z \equiv \int_{\C} \rmd z^0 \int {\rmd}^d z$
we employ  (\ref{eq:decompid}) and
(\ref{eq:decompself}):
\bea 
&& \db i \int_z \ol{\Sigma}(x,z;G) G(z,y) 
= i \int_z \Big\{ \Sigma^F(x,z) F(z,y)\nn
&&  
{\db - \frac{i}{2} \Sigma^F(x,z) \rho(z,y)\, \sign_{\C}(z^0-y^0)} 
{\db - \frac{i}{2} \Sigma^{\rho}(x,z) F(z,y)\, \sign_{\C}(x^0-z^0)}
\label{eq:evalC}\\
&& 
{\rr -\frac{1}{4} \Sigma^{\rho}(x,z) \rho(z,y)\, 
\sign_{\C}(x^0-z^0) \sign_{\C}(z^0-y^0)} 
\Big\} \nnn \, .
\eea
The first term on the r.h.s.~vanishes because of integration 
along the {\rr\em closed time contour $\C$} (cf.~Sec.~\ref{sec:nonequgenfunc}). 
To proceed for the second term one splits the 
contour integral such that the sign-functions have a definite 
value, for instance
\beq\db
\int_{\C}\! {\rmd} z^0\, \sign_{\C}(z^0-y^0) = \int_0^{y^0}\! {\rmd} z^0 
(-1) + \int_{y^0}^0\! {\rmd} z^0 = - 2 \int_0^{y^0}\! {\rmd} z^0 \, 
\eeq 
for the closed contour with initial time $t_0 = 0$.
To evaluate the last term on the r.h.s.~of Eq.~(\ref{eq:evalC}) it is
convenient to distinguish the cases \newline
(a) \ul{$\Theta_{\C}(x^0 - y^0) = 1$}:
\beq\rr
\int_{\C}\! {\rmd} z^0\,\sign_{\C}(x^0-z^0) \sign_{\C}(z^0-y^0)
= \int_0^{y^0}\! {\rmd} z^0 (-1) + \int_{y^0}^{x^0}\! {\rmd} z^0  
+ \int_{x^0}^{0}\! {\rmd} z^0 (-1) \, , 
\eeq
(b) \ul{$\Theta_{\C}(y^0 - x^0) =1$}:
\beq\rr
\int_{\C}\! {\rmd} z^0\,\sign_{\C}(x^0-z^0) \sign_{\C}(z^0-y^0)
= \int_0^{x^0}\! {\rmd} z^0 (-1) + \int_{x^0}^{y^0}\! {\rmd} z^0  
+ \int_{y^0}^{0}\! {\rmd} z^0 (-1) \, .
\eeq
One observes that (a) and (b) differ only by an overall sign factor
$\sim \sign_{\C}(x^0-y^0)$. Combining the integrals therefore
gives:
\bea\db
i \int_z \ol{\Sigma}(x,z;G) G(z,y)
&\db =&\db  \int {\rmd}^d z \left\{
\int_0^{x^0}\! {\rmd} z^0\, \Sigma_{\rho}(x,z) F(z,y)
- \int_0^{y^0}\! {\rmd} z^0\, \Sigma_F(x,z) \rho(z,y) \right.
\nonumber\\
&&\left. 
\db {\rr - \frac{i}{2} \sign_{\C}(x^0-y^0) \int_{y^0}^{x^0}{\rmd} z^0\,
\Sigma_{\rho}(x,z) \rho(z,y)} \right\}\, .
\label{eq:memdec}
\eea
One finally employs 
\beq\db
\square_x G(x,y) = \square_x F(x,y) 
- \frac{i}{2} \sign_{\C}(x^0 - y^0) \square_x {\rr \rho(x,y)} 
- i \delta(x-y)
\label{eq:canceld}
\eeq
such that the $\delta$-term cancels with the respective one on the r.h.s.~of
the evolution equation (\ref{eq:evoleqGM}). Here we have used
\bea\db
-\frac{i}{2} \partial_{x^0}^2 \left[ {\rr \rho(x,y)}\, 
\sign_{\C}(x^0 - y^0) \right]
&\db =&\db 
-\frac{i}{2} \sign_{\C}(x^0 - y^0) \partial_{x^0}^2 {\rr \rho(x,y)} \nn
&& \db {\color{black}
\underbrace{\db - i \delta_{\C}(x^0-y^0) \partial_{x^0} {\rr \rho(x,y)}}} \nn
&& \db \quad \quad \, - i \delta(x-y) 
\,\, {\color{black},}
\nonumber
\eea
where (\ref{eq:bosecomrel}) is employed for the last line and to observe
that a term $\sim \rho(x,y) \delta_{\C}(x^0-y^0)$ vanishes identically.
Comparing coefficients, which here corresponds to
separating real and imaginary parts, one finds from (\ref{eq:memdec})
and (\ref{eq:canceld}) the equations for $F(x,y)$ and $\rho(x,y)$.
Using the abbreviated notation $\int_{t_1}^{t_2}
{\rm d}z \equiv \int_{t_1}^{t_2} {\rm d}z^0 
\int_{-\infty}^{\infty} {\rm d}^d z$ we arrive at the 
{\rr\em coupled evolution equations for the spectral function
and the statistical propagator:} 
\beq
\framebox{
\begin{minipage}{14.3cm} \vspace*{-0.3cm}
\bea \db
\left[\square_x + M^2(x) 
\right] {\rr \rho(x,y)} &\db\! =\!&\db 
- \int_{y^0}^{x^0}\!\! {\rm d} z\, 
{\rr \Sigma_{\rho}(x,z)} {\rr \rho(z,y)}\, , \nonumber\\[0.1cm]
\db \left[ \square_x 
+ M^2(x) \right] F(x,y) &\db\! =\!&\db 
- \int_0^{x^0}\!\! {\rm d} z\,
{\rr \Sigma_{\rho}(x,z)} F(z,y)
\, + \int_0^{y^0}\!\! {\rm d} z\, 
\Sigma_F(x,z) {\rr \rho(z,y)}  . \nonumber\\
\nonumber
\eea
\end{minipage}}
\label{eq:exactrhoF}
\eeq
These are {\rr\em causal equations} with characteristic 
{\rr\em ``memory'' integrals,} which integrate over the time history of the
evolution. We emphasize that the presence of
memory integrals is a property of the exact theory and in
accordance with all symmetries, in particular time reflection symmetry.
The equations themselves do not single out a direction of time and
they should be clearly distinguished from phenomenological
nonequilibrium equations, where irreversibility is typically put in by hand. 
Since these equations are exact they are fully equivalent to any kind of 
identity for the two-point functions such as Schwinger-Dyson/Kadanoff-Baym 
equations without further approximations. 
For $\phi = 0$ the functional dependence of the self-energy corrections
in (\ref{eq:exactrhoF})
is given by $\db M^2 = M^2(F)$, $\db \Sigma_F= \Sigma_F({\rr \rho},F)$ 
and $\rr \Sigma_{\rho} = \Sigma_{\rho}(\rho,{\db F})$.
The case $\phi \not = 0$ is discussed below.

Note that the initial-time properties of the spectral function
have to comply with the equal-time commutation relations 
(\ref{eq:bosecomrel}). In contrast, for
$F(x,y)$ as well as its first derivatives the full initial conditions
at $t_0=0$ need to be supplied in order to solve these equations.
To make contact with the discussion of initial conditions 
in Sec.~\ref{sec:initialconditions} and 
Eq.~(\ref{eq:equirhoc}), we consider for a moment the spatially
homogeneous case for which $F(x,y) = F(x^0,y^0;\bx - \by) =
\int [{\rm d}^d p/(2 \pi)^d] \exp[i \bp (\bx-\by)] F(x^0,y^0; \bp)$
and equivalently for $\rho(x,y)$. 
In terms of the Fourier components $F(t,t'; \bp)$  
the solution of the integro-differential equations
(\ref{eq:exactrhoF}) requires the following initial 
conditions:
\bea
&&\db F(t,t';\bp)_{|t=t'=0} \,\equiv\, {\rr \xi^2_\bp} \quad , \quad 
\frac{1}{2}\left( \partial_t F(t,t';\bp) + \partial_{t'} F(t,t';\bp) 
\right)_{|t=t'=0} \,\equiv\, {\rr \xi_\bp \eta_\bp}\, , \nonumber\\
&&\db \partial_t \partial_{t'} F(t,t';\bp)_{|t=t'=0} \,\equiv\,   
{\rr \eta^2_\bp + \frac{\sigma^2_\bp}{4 \xi^2_\bp}}  \,\, .
\label{eq:initialcondFp}
\eea
Here we have used that the required correlators at initial time
are identical to those given in Eq.~(\ref{eq:equirhoc}) for the 
considered case $\phi \equiv 0$, 
where we had suppressed the momentum labels in the notation.
Accordingly, these are the very same parameters that have to be specified 
for the corresponding most general Gaussian initial density matrix 
(\ref{eq:GaussianrhoD}). We emphasize that the initial conditions for the
spectral function equation are completely
fixed by the properties of the theory itself:
the equal-time commutation relations (\ref{eq:bosecomrel})
specify $\rho(t,t';\bp)|_{t=t'=0} = 0$, 
$\partial_{t}\rho(t,t';\bp)|_{t=t'=0} = 1$ and
$\partial_{t}\partial_{t'}\rho(t,t';\bp)|_{t=t'=0} = 0$
for the anti-symmetric spectral function. 

We are now in the position to discuss the role of the 
initial-time sources, which contain the information 
about the initial-time density matrix. According to the discussion of 
Secs.~\ref{sec:initialconditions} and \ref{sec:noneq2PIeffaction} 
the initial-time sources are fully described by the Fourier components 
of the bilinear source term $R(t,t';\bp)$ at $t=t'=t_0=0$ for the case
$\phi = 0$. Mathematically, the role of the initial-time sources
for the evolution equations is rather simple: Since these
sources have support only for the initial time $t_0$ and vanish
identically for times $t \not = t_0$, they only fix the initial 
values for the correlators and their first derivatives.\footnote{We
consider additional external sources to be absent.}
For simpler differential equations this property is well documented
in the literature on the theory of Green's functions. 
In order to see that this indeed holds also for 
the case considered here, recall that the time evolution equations 
(\ref{eq:exactrhoF}) are derived from (\ref{eq:exactevolG}) for 
$R \equiv 0$. Eq.~(\ref{eq:initialcondFp}) shows that
there is a one-to-one correspondence between the initial-time 
sources parametrizing the density matrix (\ref{eq:GaussianrhoD}) 
and the required initial conditions for the solution
of the time evolution equations. To check that no further dependencies 
on $R$ remain in the evolution equations for times $t \not = t_0$, one notes
that the $R$-dependence appears in (\ref{eq:exactevolG}) as a term 
$\sim \int_z R(x,z) G(z,y)$, which identically vanishes for 
$x^0 \not = t_0$.  
As a consequence, the equation governing $G(x,y)$ (or $F$ and $\rho$)
cannot explicitly depend on $R$ for its time arguments different 
than $t_0$, i.e.~for all times of interest.   

The clear separation of the dynamical role of spectral and 
statistical components is a generic property of nonequilibrium
field theory. As discussed in Sec.~\ref{sec:detourthermal}, the
nonequilibrium theory encompasses standard
vacuum theory as a special case where this separation is absent. 
This dichotomy for nonequilibrium time evolution equations is
not specific to scalar field degrees of freedom. 
In terms of spectral and statistical components the equations 
for {\rr\em fermionic fields} or {\rr\em gauge fields} 
have a very similar structure as well. For instance, the respective form
of the evolution equations for the fermion spectral function
$\rho^{(f)}(x,y)$ and statistical propagator $F^{(f)}(x,y)$ defined 
in (\ref{eq:fermdec}) can be directly obtained from 
(\ref{eq:exactrhoF}) by the l.h.s.~replacement:
\beq \db
\left[ \square_x
+ M^2 \right] {\rr \rho(x,y)}\,\,\, \longrightarrow\,\,\, 
- [i {\partial\!\slash}_{\!x} - m^{(f)} ]\, {\rr \rho^{(f)}(x,y)} 
\, 
\eeq
and equivalently for $F^{(f)}(x,y)$.
On the r.h.s.~of (\ref{eq:exactrhoF}) then appear
the respective fermion propagators
and self-energies $\Sigma_{\rho}^{(f)}$ and $\Sigma_{F}^{(f)}$
as defined in (\ref{eq:selffermdec}).
This can be directly verified from the equation of motion for
the time-ordered fermion propagator (\ref{eq:fermSD}) by convoluting 
it with~$\Delta$. For a free inverse propagator as in 
(\ref{classscalar}) for Dirac fermions this yields 
\beq\db
\left[i \partial\!\slash_{\!x} - m^{(f)} \right] 
\Delta(x,y) - i \int_z\, \Sigma^{(f)}(x,z) \Delta(z,y)
= i \delta_{\C}(x-y) \, .
\label{eq:evolDel}
\eeq 
Following along the lines of the above discussion for scalars one finds
for the fermion case the coupled evolution equations:
\bea\db  
\left[i \partial\!\slash_{\!x} - m^{(f)} \right]  
{\rr \rho^{(f)}(x,y)} &\!\!\db =\!\!&\db  
 \int_{y^0}^{x^0}\! {\rm d}z\,  {\rr \Sigma^{(f)}_{\rho} (x,z)
\rho^{(f)}(z,y)} \, , 
\nonumber\\[0.2cm] \db 
\left[i \partial\!\slash_{\!x} - m^{(f)} \right]  
F^{(f)}(x,y) &\!\!\db =\!\!&\db  
 \int_{0}^{x^0}\! {\rm d}z\, {\rr \Sigma^{(f)}_{\rho}(x,z)} F^{(f)}(z,y)
- \int_{0}^{y^0}\! {\rm d}z\, \Sigma^{(f)}_{F}(x,z) {\rr \rho^{(f)}(z,y)}  . 
\qquad
\label{eq:Fexact}
\eea
Similarly, the nonequilibrium evolution equations 
for gauge fields can be obtained as well. For instance, denoting the
full gauge field propagator by 
\beq\db
D^{\mu\nu}(x,y) = 
F_D^{\mu\nu}(x,y) - \frac{i}{2} {\rr \rho_D^{\mu\nu}(x,y)}\,
{\rm sign}_{\C} (x^0-y^0)
\label{eq:decompgauge}
\eeq
for a theory with
free inverse gauge propagator given by
\beq \db
i D^{-1}_{0,\mu\nu}(x,y) 
= \left[g_{\mu\nu}\, \square - \left( 1 - \xi^{-1}\right) 
\partial_\mu \partial_\nu \right]_x \delta(x-y) 
\eeq
for covariant gauges with gauge-fixing parameter $\xi$ 
and vanishing ``background'' fields, one 
finds the respective equations from (\ref{eq:exactrhoF}) by 
\beq \db
\left[ \square_x + {M^2} \right] {\rr \rho(x,y)} \,\,\, \longrightarrow\,\,\, 
- \left[{g^\mu}_\gamma \square - (1-\xi^{-1}) \partial^\mu 
\partial_{\gamma} \right]_x {\rr \rho_D^{\gamma\nu}(x,y)} \, 
\eeq
and equivalently for $F_D^{\gamma\nu}(x,y)$. Of course, 
the respective indices have to be attached to the corresponding
self-energies on the r.h.s.~of the equations. The derivation 
of the nonequilibrium gauge field evolution equations
will be discussed in more detail in Sec.~\ref{sec:nPI2} in the context of
higher $n$PI effective actions with $n>2$.

\subsubsection{Non-zero field expectation value}

In the presence of a non-zero field expectation value, 
$\phi \not = 0$, the form of the scalar evolution equations for the spectral 
and statistical function (\ref{eq:exactrhoF}) remain the same. However, the
functional dependence now includes $\db M^2 = M^2(\phi,F)$, 
$\db \Sigma_F= \Sigma_F(\phi,{\rr \rho},F)$ and 
$\rr \Sigma_{\rho} = \Sigma_{\rho}({\db \phi},\rho,{\db F})$.
Note that the local self-energy correction (\ref{eq:localself}) 
described by $M^2$ does not depend on the spectral function because
the latter vanishes for equal-time arguments.    
For the $N$-component scalar field theory (\ref{eq:classical})
one has with $\phi^2 \equiv \phi_a\phi_a$:
\bea\db
M_{ab}^2(x;\phi,F) &\db=&\db \left( m^2 + \frac{\lambda}{6N}\, 
\left[F_{cc}(x,x)
+\phi^2(x) \right]
\right) \delta_{ab} \nonumber\\
&\db+&\db \frac{\lambda}{3N}\, \left[F_{ab}(x,x) 
+ \phi_a(x)\phi_b(x) \right] \, ,
\label{Meff}
\eea
with the respective field indices attached to (\ref{eq:exactrhoF}).
In this case the evolution equations for the
spectral function and statistical two-point function (\ref{eq:exactrhoF}) 
are supplemented by a differential equation for $\phi$ given by 
the stationarity condition (\ref{stationphi}), which yields 
the {\rr\em field evolution equation:}
\beq
\mbox{\framebox{\db$\,\displaystyle
\left(\left[\square_x + \frac{\lambda}{6N}
\phi^2(x)\right] \delta_{ab} + M_{ab}^2(x;\phi \equiv 0,F) 
\right) \phi_b(x) 
= \frac{\delta \Gamma_2}{\delta \phi_a(x)}\,$}}
\label{eq:exactphi}
\eeq
For comparison the corresponding equation for the classical field
theory, $\delta S[\phi]/\delta \phi_a(x)$,  
is obtained from (\ref{eq:exactphi}) by the replacement
$M_{ab}(x;\phi \equiv 0,F) \to m^2 \delta_{ab}$ and $\Gamma_2 \equiv 0$.
For the above field evolution equation
we have used that the one-loop type contribution to the
2PI effective action reads 
\bea\db
\frac{i}{2} \Tr\, G_0^{-1}(\phi) G
= - \frac{1}{2} \int_x \left(\left[\square_x + m^2+ \frac{\lambda}{6N}
\phi^2(x)\right] \delta_{ab} + \frac{\lambda}{3N}\phi_a(x)\phi_b(x)\right)
F_{ba}(x,x) \, 
\eea
for the classical inverse propagator (\ref{classprop}), such that one can 
write
\beq\db
m^2 - \frac{\delta [i \Tr\, G_0^{-1}(\phi) G /2]}{\delta \phi_a(x)}
\,=\, M^2_{ab}(x;\phi\equiv 0,F) \,\, .
\eeq
The solution of (\ref{eq:exactphi}) requires specifying the field 
and its first derivative at initial time. In the context of the above 
discussion for homogeneous fields this just corresponds to specifying 
(\ref{eq:equirhoa}), which together with the required
initial conditions (\ref{eq:initialcondFp}) for the two-point function
completes the correspondence with the initial-time density matrix
(\ref{eq:GaussianrhoD}).

\subsubsection{Lorentz decomposition for fermion dynamics}
\label{sec:lorentz}

The evolution equations for the fermion spectral function and 
statistical propagator (\ref{eq:Fexact}) are rather complicated 
in general. For Dirac fermions each
two-point function contains 16 complex components,
which is often supplemented by additional field attributes
such as ``flavor'' or ``color''. However, it is often not necessary 
in practice to consider all components due to the 
presence of symmetries, which require certain components to vanish 
identically without loss of generality. Depending on the symmetry properties of
the initial state and interaction these terms remain 
zero under the nonequilibrium time evolution, which can 
dramatically simplify the analysis. 
It is typically very useful to decompose the fields $\rho^{(f)}(x,y)$ 
and $F^{(f)}(x,y)$ of Eq.~(\ref{eq:Fexact})
into terms that have definite transformation properties under Lorentz
transformation. In order to ease the notation, in this section we
will write $\rho^{(f)} \mapsto \rho$, $F^{(f)} \mapsto F$,
$\Sigma_{\rho}^{(f)} \mapsto A$ and $\Sigma_F^{(f)} \mapsto C$.
Using a standard basis we write
\beq\db
\rho = \rho_S + i \gamma_5 \rho_P + 
\gamma_\mu \rho_V^\mu + \gamma_\mu \gamma_5 \rho_A^\mu
+ \frac{1}{2} \sigma_{\mu\nu} \rho_T^{\mu\nu} \, ,
\label{rhodecomp}\\[0.1cm]
\eeq
where $\sigma_{\mu\nu} = \frac{i}{2}[\gamma_\mu,\gamma_\nu]$
and $\gamma_5=i\gamma^0\gamma^1\gamma^2\gamma^3$.
The 16 \mbox{(pseudo-)}scalar, (pseudo-)vector and 
tensor components 
\beq\db
 \rho_S = \tilde{\tr}\, \rho \, , \quad
 \rho_P = - i\, \tilde{\tr}\, \gamma_5 \rho \, , \quad
 \rho_V^\mu = \tilde{\tr}\, \gamma^\mu \rho \, , \quad
 \rho_A^\mu = \tilde{\tr}\, \gamma_5 \gamma^\mu \rho \, , \quad
 \rho_T^{\mu\nu} = \tilde{\tr}\, \sigma^{\mu\nu} \rho  \, , 
\label{project}
\eeq 
are complex two-point functions.  
Here we have defined $\tilde{\tr} \equiv \frac{1}{4}
\tr$ where the trace acts in Dirac space. Equivalently, there are  16 complex
components for $F$, $A$ and $C$ for given other field attributes.
Using the hermiticity properties (\ref{eq:hermiticity}) they obey
\beq\db
\label{herm}
\rho^{(\Gamma)}_{ij}(x,y)= - \left(\rho^{(\Gamma)}_{ji}(y,x) \right)^*
\quad , \quad F^{(\Gamma)}_{ij}(x,y)= 
\left(F^{(\Gamma)}_{ji}(y,x)\right)^* \, ,
\eeq
where $\Gamma = \{S,P,V,A,T\}$. Inserting the above decomposition 
into the evolution equations (\ref{eq:Fexact})  
one obtains the respective equations for the various
components displayed in~(\ref{project}). 
 
For a more detailed discussion, we first consider the l.h.s.~of the evolution
equations (\ref{eq:Fexact}). In fact, the approximation of a
vanishing r.h.s.~corresponds to the standard mean--field or Hartree--type
approaches frequently discussed in the literature.
However, to discuss thermalization we have to go beyond such a ``Gaussian''
approximation: it is crucial to include direct scattering which is described
by the nonvanishing contributions from the r.h.s.~of the evolution equations.
This is discussed in detail in Sec.~\ref{sec:neqevolution}.
Starting with the l.h.s.~of (\ref{eq:Fexact}) for the evolution
equation for the spectral function one finds:
\bea\db 
 \tilde{\tr}\left[(i {\partial\!\slash} - m_f ) \rho \right] &\db =&\db 
 \left(i \partial_\mu \rho_V^\mu\right) - m_f\, \rho_S \,\, ,
\label{rhoSL}\nonumber\\[0.3cm]
\db  -i\, \tilde{\tr}\left[\gamma_5 (i {\partial\!\slash} - m_f ) \rho \right] 
&\db =&\db 
 - i \left(i \partial_\mu \rho_A^\mu\right) - m_f\, \rho_P \,\, ,
\label{rhoPL}\nonumber\\[0.3cm]
\db  \tilde{\tr}\left[\gamma^\mu (i {\partial\!\slash} - m_f ) \rho \right] 
&\db =&\db 
 \left(i \partial^\mu \rho_S\right) 
 + i \left(i \partial_\nu \rho_T^{\nu\mu}\right) 
 - m_f\, \rho_V^\mu \,\, ,
\label{rhoVL}\\[0.1cm]
\db  \tilde{\tr}
\left[\gamma_5 \gamma^\mu (i {\partial\!\slash} - m_f ) \rho \right] 
&\db =&\db 
 i \left(i \partial^\mu \rho_P\right) + \frac{1}{2}\, 
 \epsilon^{\mu\nu\gamma\delta} 
\left(i\partial_\nu \rho_{T,\gamma\delta}\right) 
 - m_f\, \rho_A^\mu \,\, ,
\label{rhoAL}\nonumber\\[0.1cm]
\db  \tilde{\tr}\left[\sigma^{\mu\nu} (i {\partial\!\slash} - m_f ) 
\rho \right] &\db =&\db 
 -i \left(i\partial^\mu \rho_V^\nu - i \partial^\nu \rho_V^\mu\right)
 + \epsilon^{\mu\nu\gamma\delta} \left(i\partial_\gamma \rho_{A,\delta}\right)
 - m_f\, \rho_T^{\mu\nu} \, . \qquad
\label{rhoTL}\nonumber
\vspace*{0.1cm}
\eea
The corresponding expressions for the l.h.s.~of the evolution equation 
for the statistical propagator $F$ follow from
(\ref{rhoVL}) with the replacement $\rho \to F$. We turn now  
to the integrand on the r.h.s.~of 
(\ref{eq:Fexact}) for the evolution equation of $\rho$. For the various
components (\ref{project}) one finds:
\bea\db 
 \tilde{\tr}\left[A\, \rho \right] &\db =&\db  
 A_S\, \rho_S - A_P\, \rho_P + A_V^\mu\, \rho_{V,\mu}
 - A_A^\mu\, \rho_{A,\mu} \nonumber\\
 &&\db  + \frac{1}{2} A_T^{\mu\nu}\, \rho_{T,\mu\nu} 
 \,\, ,\quad 
\label{rhoSR}\\[0.1cm]
\db  -i\, \tilde{\tr}\left[\gamma_5 A\, \rho \right] &\db =& \db 
 A_S\, \rho_P + A_P\, \rho_S - i A_V^\mu\, \rho_{A,\mu}
 + i A_A^\mu\, \rho_{V,\mu} \nonumber\\
 &&\db  + \frac{1}{4}\, \epsilon^{\mu\nu\gamma\delta} A_{T,\mu\nu}\, 
 \rho_{T,\gamma\delta} \,\, ,
\label{rhoPR}\\[0.1cm]
\db  \tilde{\tr}\left[\gamma^\mu A\, \rho \right] &\db =& \db 
 A_S\, \rho_V^\mu + A_V^\mu\, \rho_S - i A_P\, \rho_A^\mu 
 + i A_A^\mu\, \rho_P + i A_{V,\nu}\, \rho_T^{\nu\mu}  \nonumber\\
 &&\db  + i A_T^{\mu\nu}\, \rho_{V,\nu}
 + \frac{1}{2} \epsilon^{\mu\nu\gamma\delta} \left( A_{A,\nu}\, 
 \rho_{T,\gamma\delta} + A_{T,\nu\gamma}\, 
 \rho_{A,\delta} \right) \, , \qquad
\label{rhoVR}\\[0.1cm]
\db  \tilde{\tr}\left[\gamma_5 \gamma^\mu A\, \rho \right] &\db =&\db 
 A_S\, \rho_A^\mu + A_{A}^\mu\, \rho_S 
 - i A_P\, \rho_V^\mu + i A_V^\mu\, \rho_P
 + i A_{A,\nu}\, \rho_T^{\nu\mu}
  \nonumber\\
 && 
 \db + i A_T^{\mu\nu}\, \rho_{A,\nu} 
 + \frac{1}{2} \epsilon^{\mu\nu\gamma\delta} \left( A_{V,\nu}\, 
 \rho_{T,\gamma\delta} + A_{T,\nu\gamma}\, 
 \rho_{V,\delta}  \right) \, ,\quad
\label{rhoAR}\\[0.1cm]
 \db \tilde{\tr}\left[\sigma^{\mu\nu} A\, \rho \right] &\db =& \db 
 A_S\, \rho_T^{\mu\nu} + A_T^{\mu\nu}\, \rho_S 
 - \frac{1}{2}\,\epsilon^{\mu\nu\gamma\delta} 
 \left( A_P\, \rho_{T,\gamma\delta} + A_{T,\gamma\delta}\, 
\rho_P \right)
 \nonumber\\ 
&&\db 
 - i \left(A_V^\mu\, \rho_V^\nu - A_V^\nu\, \rho_V^\mu \right) 
 + \epsilon^{\mu\nu\gamma\delta} \left( A_{V,\gamma}\, \rho_{A,\delta}
 - A_{A,\gamma}\, \rho_{V,\delta} \right)
 \nonumber\\[0.15cm] 
&&\db 
 + i \left(A_A^\mu\, \rho_A^\nu - A_A^\nu\, \rho_A^\mu \right)
 + i \left(A_T^{\mu\gamma}\, {\rho_{T,\gamma}}^\nu 
 - A_T^{\nu\gamma}\, {\rho_{T,\gamma}}^\mu \right) .\qquad\,\,
\label{rhoTR}
\eea
With the above expressions one obtains the evolution equations for the various
Lorentz components in a straightforward way using (\ref{eq:Fexact}). We note
that the convolutions appearing on the r.h.s.~of the evolution equation
(\ref{eq:Fexact}) for $F$ are of the same form than those computed above for
$\rho$. The respective r.h.s.~can be read off Eqs.\ 
(\ref{rhoSR})--(\ref{rhoTR}) 
by replacing $\rho \to F$ for the first term and $A \to C$ for the second 
term under the integrals of Eq.\ (\ref{eq:Fexact}) for $F$.  
We have now all the relevant 
building blocks to discuss the most general case of nonequilibrium fermionic 
fields. However, this is often not necessary in practice due to the 
presence of symmetries, which require certain components to vanish 
identically.

In the following, we will exploit symmetries of the action 
(\ref{chiralfermact}) for the chiral quark-meson model described in 
Sec.~\ref{sec:chiralqmm}  
and of initial conditions in order to simplify the fermionic 
evolution equations:

\vspace{.2cm} 
{\rr\em Spatial translation invariance and isotropy:}
We will consider spatially homogeneous and isotropic initial conditions. In
this case it is convenient to work in Fourier space and we write
\beq\db
\rho(x,y) \equiv \rho(x^0,y^0;\bx-\by) = \int \frac{{\rm d}^3p}{(2\pi)^3}\,
e^{i \bp \cdot(\bx-\by)} \rho(x^0,y^0;\bp) \, ,
\eeq
and similarly for the other two-point functions. Moreover, isotropy implies 
a reduction of the number of independent two-point functions: e.g.~the vector
components of the spectral function can be written as 
\beq\db
 \rho_V^0(x^0,y^0;\bp) = \rho_V^0(x^0,y^0;p) \, , \qquad
 \vec{\rho}_V(x^0,y^0;\bp) = \bv\, \rho_V(x^0,y^0;p) \, , 
\label{eq:veccomprho}
\eeq
where $p\equiv |\bp|$ and $\bv=\bp/p$. 

\vspace{.2cm}
{\rr\em Parity:} The vector components 
$\rho_V^0(x^0,y^0;p)$ and $\rho_V(x^0,y^0;p)$ 
are unchanged under a parity transformation, whereas the 
corresponding axial-vector
components get a minus sign.  Therefore, parity together with rotational
invariance imply that
\beq\db
 \rho_A^0(x^0,y^0;p) = \rho_A(x^0,y^0;p) = 0 \, .
\eeq
The same is true for the axial-vector components of $F$, $A$ and $C$.
Parity also implies the pseudo-scalar components of the various two-point
functions to vanish.

\vspace{.2cm}
{\rr\em $CP$--invariance:} For instance, under combined charge conjugation 
and parity transformation the vector component of $\rho$ transforms as
\bea\db 
 \rho_V^0(x^0,y^0;p) &\db \longrightarrow&\db  
\rho_V^0(y^0,x^0;p) \, ,\nonumber\\\db
 \rho_V(x^0,y^0;p) &\db \longrightarrow&\db 
 -\rho_V(y^0,x^0;p) \, ,\nonumber
\eea
and similarly for  $A_V^0$ and $A_V$.  The $F$--components transform as 
\bea\db 
 F_V^0(x^0,y^0;p) &\db \longrightarrow&\db  -F_V^0(y^0,x^0;p) \, , \nonumber\\
 \db F_V(x^0,y^0;p) &\db \longrightarrow&\db  F_V(y^0,x^0;p) \, , \nonumber
\eea
and similarly for $C_V^0$ and $C_V$.  Combining this with the hermiticity
relations~(\ref{herm}), one obtains for these components that
\beq
\label{CP}
 \begin{array}{lll}\db  
  \Re \rho_V^0(x^0,y^0;p) &\db =&\db  \Im \rho_V(x^0,y^0;p) = 0 \, , \\
\db   \Re F_V^0(x^0,y^0;p) &\db =&\db  \Im F_V(x^0,y^0;p) = 0  \, ,\\
\db   \Re A_V^0(x^0,y^0;p) &\db =&\db  \Im A_V(x^0,y^0;p) = 0  \, ,\\
\db   \Re C_V^0(x^0,y^0;p) &\db =&\db  \Im C_V(x^0,y^0;p) = 0 \, ,
 \end{array}
\eeq
for all times $x^0$ and $y^0$ and all individual momentum modes.

\vspace{.2cm}
{\rr\em Chiral symmetry:} 
The only components of the decomposition (\ref{rhodecomp}) allowed by
chiral symmetry are those which anticommute with $\gamma_5$. In particular, 
chiral symmetry forbids a mass term for fermions ($m_f\equiv0$) and we 
have
\beq\db
 \rho_S (x^0,y^0;\bp) = \rho_P (x^0,y^0;\bp) 
 = \rho_T^{\mu \nu} (x^0,y^0;\bp) = 0 \, ,
\eeq
and similarly for the corresponding components of $F$, $A$ and $C$. 
We emphasize, however, that in the presence of
spontaneous chiral symmetry breaking there is no such simplification.\\

In conclusion, for the above symmetry properties we are left with only four
independent propagators: the two spectral functions $\rho_V^0(x^0,y^0;\bp)$ 
and $\rho_V(x^0,y^0;\bp)$ and
the two corresponding statistical functions $F_V^0(x^0,y^0;\bp)$ 
and $F_V(x^0,y^0;\bp)$. They are either
purely real or imaginary and have definite symmetry properties under the
exchange of their time arguments $x^0 \leftrightarrow y^0$. These properties as
well as the corresponding ones for the various components of the self-energy
are summarized below:
\begin{center}
\begin{tabular}{ll}\db 
 $\rho_V^0$, $A_V^0$: & imaginary, symmetric; \\\db
 $\rho_V$, $A_V$: & real, antisymmetric;\\\db
 $F_V^0$, $C_V^0$: & imaginary, antisymmetric;\\\db
 $F_V$, $C_V$: & real, symmetric.
\end{tabular}
\end{center}
The exact evolution equations for the spectral functions 
read (cf.~Eq.~(\ref{eq:Fexact})):\footnote{We note that 
the following equations do not rely on the restrictions
(\ref{CP}) imposed by $CP$--invariance: they have the very same form 
for the case that all two-point functions are complex.}
\bea\db 
\nonumber
\lefteqn{
i \frac{\partial}{\partial x^0}\, \rho_V^0(x^0,y^0;p)
= p\, {\rho}_V(x^0,y^0;p) } \\  
&\db +&\db  \int_{y^0}^{x^0} {\rm d}z^0 \Big[
A_V^0(x^0,z^0;p)\, \rho_V^0(z^0,y^0;p) 
- {A}_V(x^0,z^0;p)\, {\rho}_V(z^0,y^0;p) \Big] \, ,
\label{rhoV0eom}
\\[0.2cm]
\nonumber
\db \lefteqn{
i \frac{\partial}{\partial x^0}\, {\rho}_V(x^0,y^0;p)
= p\, \rho_V^0(x^0,y^0;p) } \\  
&\db +&\db  \int_{y^0}^{x^0} {\rm d}z^0 \Big[
A_V^0(x^0,z^0;p)\, {\rho}_V(z^0,y^0;p) 
- {A}_V(x^0,z^0;p)\, \rho_V^0(z^0,y^0;p) \Big] \, .
\label{rhoVeom}
\eea
Similarly, for the statistical two-point functions we obtain
(cf.~Eq.~(\ref{eq:Fexact})):
\bea\db 
\nonumber
\lefteqn{
i \frac{\partial}{\partial x^0}\, F_V^0(x^0,y^0;p)
= p\, {F}_V(x^0,y^0;p) } \\  
&\db +&\db   \int_0^{x^0} {\rm d}z^0 \Big[
A_V^0(x^0,z^0;p)\, F_V^0(z^0,y^0;p) 
- {A}_V(x^0,z^0;p)\, {F}_V(z^0,y^0;p) \Big] 
\nonumber\\
&\db -&\db  \int_0^{y^0} {\rm d}z^0 \Big[
C_V^0(x^0,z^0;p)\, \rho_V^0(z^0,y^0;p) 
- {C}_V(x^0,z^0;p)\, {\rho}_V(z^0,y^0;p) \Big] \, ,
\label{FV0eom}\\
\nonumber\\[0.2cm]
\nonumber\db 
\lefteqn{
i \frac{\partial}{\partial x^0}\, {F}_V(x^0,y^0;p)
= p\, F_V^0(x^0,y^0;p) } \\  
&\db +&\db  \int_0^{x^0} {\rm d}z^0 \Big[
A_V^0(x^0,z^0;p)\, {F}_V(z^0,y^0;p) 
- {A}_V(x^0,z^0;p)\, F_V^0(z^0,y^0;p) \Big] 
\nonumber \\
&\db -&\db  \int_0^{y^0} {\rm d}z^0 \Big[
C_V^0(x^0,z^0;p)\, {\rho}_V(z^0,y^0;p) 
- {C}_V(x^0,z^0;p)\, \rho_V^0(z^0,y^0;p) \Big] \, .
\label{FVeom}
\eea
The above equations are employed in Sec.~\ref{sec:neqevolution} 
to calculate the nonequilibrium
fermion dynamics in a chiral quark-meson model.  

The time evolution for the fermions is described by first-order 
\mbox{(integro-)}differential equations for $F$ and $\rho$: Eqs.\ 
(\ref{rhoV0eom})--(\ref{FVeom}). As pointed out above, the 
initial fermion spectral function is completely 
determined by the equal-time anticommutation relation of fermionic 
field operators (cf.~Eq.~(\ref{eq:rhoinitial})). 
To uniquely specify the time evolution for $F$ we have to set the initial
conditions. The most general (Gaussian) initial conditions for $F$ respecting
spatial homogeneity, isotropy, parity, charge conjugation and chiral symmetry 
can be written as 
\bea\db
 F_V(t,t',p)|_{t=t'=0} &\db =&\db \frac{1}{2}- n_0^{f}(p) \, ,
\label{initialFV}\\\db
 F_V^0(t,t',p)|_{t=t'=0} &\db =&\db 0
\label{initialFV0} \, .
\eea
Here $n_0^{f}(p)$ denotes the initial particle number distribution, whose
values can range between 0 and 1.

\subsection{Nonequilibrium dynamics from 
the 2PI loop expansion}
\label{Subsect2PIloop}

In Sec.~\ref{sec:noneqeveq} we have derived coupled evolution
equations (\ref{eq:exactrhoF}) for the statistical 
propagator $F$ and the spectral function $\rho$. A systematic 
approximation to the exact equations can be obtained from the
loop or coupling expansion of the 2PI effective action, as discussed in 
Sec.~\ref{sec:loopexp}. This determines all the required self-energies
using (\ref{eq:station}) and the decomposition identities (\ref{eq:decompid})
and (\ref{eq:decompself}) for scalars, and ({\ref{eq:fermdec}) for
fermions. (Approximations for gauge field theories will be discussed 
in Sec.~\ref{sec:nPI2} in the context of $n$PI effective actions with $n > 2$.)
We emphasize that {\rr\em all classifications of contributions
are done for the effective action}. Once an approximation order
is specified on the level of the effective action, there are no
further classifications on the level of the evolution
equations. This is a crucial aspect, which ensures the ``conserving'' 
properties of 2PI expansions (cf.~the discussion in Sec.~\ref{sec:how}).
The purpose of this section is to present the relevant formulae, whose
physics will be explained in detail in the later sections.

We consider first the case of the {\rr\em scalar $O(N)$ symmetric 
field theory} with classical action (\ref{eq:classical}) and
a vanishing field expectation value. The case $\phi \not = 0$ is
treated in Sec.~\ref{2PINfield} below. From the 2PI effective 
action to three-loop order
(\ref{2PIGfiveloop}) one finds with $F_{ab}(x,y) = F(x,y) \delta_{ab}$
and $\rho_{ab}(x,y) = \rho(x,y) \delta_{ab}$ the two-loop self-energies:
\bea\db
\Sigma^{(0)}(x) &\db =&\db \lambda\frac{N+2}{6N}\,F(x,x),\\
\db\Sigma_{\rho}(x,y) &\db =&\db
-\lambda^2\,\frac{N+2}{6N^2}\,\rho(x,y)\left[F^2(x,y) 
-\frac{1}{12}\rho^2(x,y)\right] \, ,
\label{eqsigmarho3}\\ 
\db \Sigma_F(x,y) &\db =&\db
-\lambda^2\,\frac{N+2}{18N^2}\,F(x,y)\left[F^2(x,y) 
-\frac{3}{4}\rho^2(x,y)\right] \, ,
\label{eqsigmaF3}
\eea
which enter (\ref{eq:localself}) and (\ref{eq:exactrhoF}).
The two-loop contribution to the  
effective action adds only to $\Sigma^{(0)}$ and 
corresponds to a space-time dependent mass shift 
in the evolution equations.
As is discussed in detail below in Sec.~\ref{eq:scatoffmem},
the time evolution obtained from the two-loop 2PI effective action
(Hartree, or similarly leading-order large-N approximations)
suffers from the presence of an infinite number of spurious 
conserved quantities, which are not present in the fully interacting
theory. As an important consequence no thermalization can be observed 
to that order. A crucial ingredient for the description of nonequilibrium
evolution comes from the three-loop contribution to the
2PI effective action as described by (\ref{eqsigmarho3}) and
(\ref{eqsigmaF3}). 

For the {\rr\em chiral Yukawa model} for $N_f=2$ fermion flavors
and $N_f^2$ scalar fields with classical action (\ref{chiralfermact})
as described in Sec.~\ref{sec:chiralqmm}, 
one obtains
from the two-loop 2PI effective action the fermion self-energies
entering Eqs.~(\ref{rhoV0eom})--(\ref{FVeom}):
\bea\db 
A_V^\mu(x^0,y^0;\bp) &\db\!=\!&\db 
- h^2 \int \frac{{\rm d}^3 q}{(2\pi)^3}\, \Big[
F_V^\mu(x^0,y^0;\bq) \,\rho(x^0,y^0;\bp-\bq) \nonumber\\
&&\db + \rho_V^\mu(x^0,y^0;\bq)\, F(x^0,y^0;\bp-\bq)\Big] \, ,
\label{eq:selffermA}\\[0.2cm] \db
C_V^\mu(x^0,y^0;\bp) &\db\!=\!&\db 
- h^2 \int \frac{{\rm d}^3q}{(2\pi)^3}\, \Big[
F_V^\mu(x^0,y^0;\bq) \,F(x^0,y^0;\bp-\bq) \nonumber\\
&&\db
-\frac{1}{4} \rho_V^\mu(x^0,y^0;\bq)\, \rho(x^0,y^0;\bp-\bq)\Big]\, .
\label{selffermionF}
\eea
Here $\rho_V^\mu(x^0,y^0;\bq)$ and $F_V^\mu(x^0,y^0;\bq)$ are
the vector components of the fermion two-point functions
as introduced in Sec.~\ref{sec:lorentz}. The dynamics of the scalar
two-point functions $\rho(x^0,y^0;\bp)$ and $F(x^0,y^0;\bp)$ is 
described by the evolution equations (\ref{eq:exactrhoF}) with 
the scalar self-energies to this loop order: 
\bea\db
\Sigma^{(0)}(x) &\db\! =\!&\db \lambda\frac{N_f^2+2}{6N_f^2}\,
\int \frac{{\rm d}^3 q}{(2\pi)^3}\, F(x^0,x^0;\bq),
\label{eq:purescalar}\\\db
\label{selfscalarrho}
 \Sigma_{\rho} (x^0,y^0;\bp) &\db\!=\!&\db -\frac{8 h^2}{N_f}
\int \frac{{\rm d}^3 q}{(2\pi)^3}\,
 \rho_V^\mu(x^0,y^0;\bq)\,F_{V,\mu}(x^0,y^0;\bp-\bq) \, ,\\\db
 \Sigma_F (x^0,y^0;\bp) &\db\!=\!&\db - \frac{4 h^2}{N_f}\int 
\frac{{\rm d}^3 q}{(2\pi)^3}\,
 \Big[F_V^\mu(x^0,y^0;\bq)\,F_{V,\mu}(x^0,y^0;\bp-\bq) \nonumber\\
&&\db
 - \frac{1}{4} \rho_V^\mu(x^0,y^0;\bq)\,\rho_{V,\mu}(x^0,y^0;\bp-\bq) 
 \Big] \, .
\label{selfscalarF}
\eea

\subsection{Nonequilibrium dynamics from 
the 2PI $1/N$ expansion}
\label{nonequiN}

We consider first the case of the {\rr\em scalar $O(N)$ symmetric 
field theory} with classical action (\ref{eq:classical}) and
a vanishing field expectation value such that 
$F_{ab}(x,y) = F(x,y) \delta_{ab}$
and $\rho_{ab}(x,y) = \rho(x,y) \delta_{ab}$. The case $\phi \not = 0$ is
treated in Sec.~\ref{2PINfield}.
In the $1/N$--expansion of the 2PI effective action to next-to-leading
order, as discussed in Sec.~\ref{sec:2PIN}, 
the effective mass term $M^2(x;G)$ appearing in the 
evolution equations (\ref{eq:exactrhoF}) is given by
\beq
\db M^2(x;F) \,=\,  m^2 + \lambda \frac{N+2}{6 N}\, F(x,x) \, .
\label{eq:massNLO}
\eeq
One observes that this local self-energy 
part receives LO and NLO contributions.
In contrast, the non-local part of the self-energy (\ref{eq:sighominh})
is nonvanishing 
only at NLO: $\overline{\Sigma}(x,y;G) = - \lambda/(3 N)\, G(x,y)
I(x,y)$ and using the decomposition identities (\ref{eq:decompid})
and (\ref{eq:decompself}) one finds
\bea
{\db \Sigma_F(x,y)} &\db=&\db  - \frac{\lambda}{3 N}\, 
\Big( {\db F(x,y) I_F(x,y)} 
-\frac{1}{4} {\rr \rho(x,y) I_{\rho}(x,y)} \Big) \, ,
\label{SFFR} \\[0.2cm]
{\rr \Sigma_{\rho}(x,y)} &\db=&\db  - \frac{\lambda}{3 N}\, \Big( 
{\db F(x,y)} {\rr I_{\rho}(x,y)} 
+ {\rr \rho(x,y)} {\db I_{F}(x,y)} \Big)  \, . 
\label{SR}
\eea
Here the summation function (\ref{eq:Ifunc}) reads in terms
of its statistical and spectral components:\footnote{This follows
from using the decomposition identity for the propagator 
(\ref{eq:decompid}) and $I(x,y) = I_F(x,y) - \frac{i}{2} 
I_\rho(x,y)\, \sign_{\C}(x^0 - y^0)$. Cf.~also the detailed discussion
in Sec.~\ref{sec:noneqeveq}.} 
\bea
{\db I_{F}(x,y)} &\db=&\db  \frac{\lambda}{6}\, 
\Big( {\db F^2(x,y)} - \frac{1}{4} {\rr \rho^2(x,y)} \Big)  
\nonumber \\\db
- \frac{\lambda}{6}\!\!\!\!\! &\db\!\!\!\Bigg\{\!\!\!&\db \!\!\!\!\!\!
{\db \int\limits_{0}^{x^0} {\rm d}z}\,
{\rr I_{\rho}(x,z)} \Big( {\db F^2(z,y)} - \frac{1}{4}\, {\rr \rho^2(z,y)} 
\Big) -  2 {\db \int\limits_{0}^{y^0} {\rm d}z}\,  
{\db I_F(x,z)  F(z,y)} {\rr \rho(z,y)}  \!  \Bigg\}\!  ,\quad
\label{IFFR}\\
[0.3cm]  
{\rr I_{\rho}(x,y)} &\db=&\db \frac{\lambda}{3}\, {\db F(x,y)} {\rr \rho(x,y)} 
- \frac{\lambda}{3}\, 
{\db \int\limits_{y^0}^{x^0} {\rm d}z}\,  
{\rr I_{\rho}(x,z)} {\db F(z,y)} {\rr \rho(z,y)} \, ,
\label{IRFR}
\eea
using the abbreviated notation $\int_{t_1}^{t_2}
{\rm d}z \equiv \int_{t_1}^{t_2} {\rm d}z^0 
\int_{-\infty}^{\infty} {\rm d}^d z$.
Note that $F(x,y)$, $\rho(x,y)$, $\Sigma_F(x,y)$, $\Sigma_{\rho}(x,y)$, 
$I_F(x,y)$ and $I_{\rho}(x,y)$ are real functions.

For the {\rr\em Yukawa model} of Sec.~\ref{sec:chiralqmm} 
the results from the $1/N_f$ expansion of the 2PI effective action
to NLO can be directly inferred from (\ref{eq:selffermA})--(\ref{selfscalarF}). 
Recall that all classifications are done on the level of the
effective action, as explained for fermions at the
end of Sec.~\ref{sec:classofdiag}. The only difference between 
the NLO evolution equations and the result 
(\ref{eq:selffermA})--(\ref{selfscalarF}) from the two-loop
effective action is that for the former the NNLO part $\sim N_f^{-2}$
in (\ref{eq:purescalar}) is dropped.

\subsubsection{Nonvanishing field expectation value} 
\label{2PINfield}

We consider the {\rr\em scalar $N$-component 
field theory} with classical action (\ref{eq:classical}).
In the presence of a nonzero field expectation value $\phi_a$ the 
most general propagator can no longer be evaluated for the diagonal
configuration (\ref{eq:symmetricG}). Restoring all field
indices the evolution equations (\ref{eq:exactrhoF}) read
\bea \db 
\left[\square_x \delta_{ac} + M_{ac}^2(x) \right] {\rr \rho_{cb} (x,y)} 
& \db \!=\!&  \db 
 -\int_{y^0}^{x^0} {\rm d}z\, {\rr\Sigma^{\rho}_{ac} (x,z)\rho_{cb} (z,y)}\,,
\nonumber\\ \db 
\left[ \square_x \delta_{ac} + M_{ac}^2(x) \right] F_{cb}(x,y) 
& \db \!=\!&  \db 
- \int_0^{x^0} {\rm d}z\, {\rr \Sigma^{\rho}_{ac}(x,z)} F_{cb}(z,y)
\nonumber \\
&& \db  + \int_0^{y^0} {\rm d}z\, \Sigma^{F}_{ac}(x,z) {\rr \rho_{cb}(z,y)}\, .
\label{eq:exactbrok}
\eea
The statistical 
propagator and spectral function components 
have the properties $F^*_{ab}(x,y)=F_{ab}(x,y)=F_{ba}(y,x)$ 
and $\rho^*_{ab}(x,y)=\rho_{ab}(x,y)=-\rho_{ba}(y,x)$. 
At NLO in the 2PI $1/N$ expansion 
$M_{ab}^2(x) \equiv M_{ab}^2(x;\phi,F)$ is 
given by (\ref{Meff}) and the self-energies 
$\Sigma^{F}_{ab}(x,y) \equiv \Sigma^{F}_{ab}(x,y;\phi,\rho,F)$ 
and $\Sigma^{\rho}_{ab}(x,y) \equiv \Sigma^{\rho}_{ab}(x,y;\phi,\rho,F)$
are obtained from (\ref{eq:NLObrok}) as
\bea \db 
\Sigma^{F}_{ab}(x,y) & \db  = & \db   - \frac{\lambda}{3 N}\Big\{
 I_F(x,y)\left[ \phi_a (x)\phi_b(y) + F_{ab}(x,y) \right] -
 \frac{1}{4} {\rr I_{\rho}(x,y)\rho_{ab}(x,y)}  
 \nonumber\\
 && \db  \qquad \quad + P_F(x,y)F_{ab}(x,y) -
 \frac{1}{4} {\rr P_{\rho}(x,y)\rho_{ab} (x,y)} \Big\},
 \label{ASFFR}\\
 \rr \Sigma^{\rho}_{ab} (x,y) & \db  = & \db   - \frac{\lambda}{3 N}\Big\{ 
 {\rr I_{\rho}(x,y)} \left[ \phi_a (x)\phi_b(y) + F_{ab} (x,y) \right] 
 + I_{F}(x,y){\rr \rho_{ab} (x,y)}  
 \nonumber\\
 && \db  \qquad \quad + {\rr P_{\rho}(x,y)} F_{ab}(x,y) + 
 P_{F}(x,y) {\rr \rho_{ab} (x,y)} \Big\}. 
 \label{ASRFR}
\eea
The functions $I_{F}(x,y) \equiv I_{F}(x,y;\rho,F)$ and 
$I_{\rho}(x,y) \equiv I_{\rho}(x,y;\rho,F)$ satisfy the
same equations as for the case of a vanishing macroscopic field given by
(\ref{IFFR}) and (\ref{IRFR}). 
The respective $\phi$-dependent summation functions 
$P_{F}(x,y) \equiv P_{F}(x,y;\phi,\rho,F)$ and 
$P_{\rho}(x,y) \equiv P_{\rho}(x,y;\phi,\rho,F)$
are given by  
\bea \db  
 P_{F} (x,y) & \db =& \db  - \frac{\lambda}{3N} \Bigg\{ H_F (x,y)
   - \int_0^{x^0} {\rm d}z\, \left[ {\rr H_{\rho}} (x,z)I_F (z,y) +
 {\rr I_{\rho} (x,z)} H_F (z,y) \right] 
 \nonumber\\
 && \db  + \int_0^{y^0} {\rm d}z\, \left[  H_F (x,z) {\rr I_{\rho} (z,y)} +
 I_F (x,z) {\rr H_{\rho} (z,y)}\right] 
 \nonumber\\
 && \db  - \int_0^{x^0} {\rm d}z\, \int_0^{y^0} {\rm d}v\,
 {\rr I_{\rho} (x,z)} H_F (z,v) {\rr I_{\rho} (v,y)}
 \nonumber\\
 && \db  + \int_0^{x^0} {\rm d}z\, \int_0^{z^0} {\rm d}v\,
 {\rr I_{\rho} (x,z) H_{\rho} (z,v)} I_F (v,y)
 \nonumber\\
 && \db  + \int_0^{y^0} {\rm d}z\, \int_{z^0}^{y^0} {\rm d}v\,
 I_F (x,z) {\rr H_{\rho} (z,v) I_{\rho} (v,y)} \Bigg\},
\label{eqB11}
\eea
\bea \rr
 P_{\rho} (x,y) & \db  = & \db  - \frac{\lambda}{3N} \Bigg\{ 
{\rr H_{\rho} (x,y)}  - \int_{y^0}^{x^0} {\rm d}z\, 
 \left[ {\rr H_{\rho} (x,z) I_{\rho} (z,y)} + 
 {\rr I_{\rho} (x,z) H_{\rho} (z,y)} \right] 
 \nonumber\\
 && \db  +  \int_{y^0}^{x^0} {\rm d}z\, \int_{y^0}^{z^0} {\rm d}v\,
 {\rr I_{\rho} (x,z) H_{\rho} (z,v) I_{\rho} (v,y)} \Bigg\},
\label{eqB12}
\eea
with 
\beq\db
H_F(x,y) \equiv -\phi_a(x) F_{ab}(x,y) \phi_b(y) \,\, , \qquad 
{\rr H_\rho(x,y) \equiv -\phi_a(x) \rho_{ab}(x,y) \phi_b(y)} \,.
\eeq
The time evolution equation for the field 
(\ref{eq:exactphi}) for the 2PI effective action
to NLO (\ref{eq:NLObrok}) is given by 
\bea\db
\lefteqn{\left(\left[\square_x + \frac{\lambda}{6N}
\phi^2(x)\right] \delta_{ab} +  M_{ab}^2(x;\phi = 0,F) 
\right) \phi_b(x)} \qquad\qquad  \nonumber\\ 
&\db\! =\!&\db \frac{\lambda}{3N} \int_0^{x^0} {\rm d}y \, 
 \left[ {\rr I_{\rho} (x,y)} F_{ab} (x,y) + 
 I_F (x,y) {\rr \rho_{ab} (x,y)} \right] \phi_b (y)
\nonumber\\
&\db\! =\!&\db - \int_0^{x^0}\! {\rm d}y \, 
 {\rr \Sigma^{\rho}_{ab}(x,y;{\db \phi = 0,F},\rho)}\, \phi_b (y)\, .
\label{eq:NLOphi}
\eea
The second equality follows from (\ref{ASRFR}).

\subsection{Numerical implementation}
\label{Sectnumimp}

Beyond the leading-order approximation, the time evolution equations 
(\ref{eq:exactrhoF}) and (\ref{eq:exactphi}) are 
nonlinear integro-differential equations. The approximate self-energies
obtained from a loop expansion or $1/N$ expansion of the
2PI effective action are described in Secs.~\ref{Subsect2PIloop} 
and \ref{nonequiN}.
Though these equations are in general too complicated to be solved
analytically\footnote{Cf.\ Sec.\ \ref{sec:resonance} for an 
approximate analytical solution in the context of parametric 
resonance.} without additional approximations, 
they can be very efficiently implemented and solved on a computer.
Here it is important to note that all equations are explicit
in time, i.e.\ all quantities at some later time $t_f$ can be obtained
by integration over the explicitly known functions for times $t<t_f$
for given initial conditions. This is a direct consequence of
causality. In this respect, solving the nonequilibrium evolution
equations can be technically {\rr\em simpler} than solving the
corresponding theory in thermal equilibrium. The reason is that
for the study of thermal equilibrium the equation of the form 
(\ref{eq:eomG}) is typically employed, which has to be solved
self-consistently since on its l.h.s.~and r.h.s.~the same
variables appear. It involves the propagator for the full range 
of its arguments. In contrast, the time-evolution
equations (\ref{eq:exactrhoF}) for $\rho(x,y)|_{x^0 = t_1,y^0=t_2}$ 
and $F(x,y)|_{x^0 = t_1,y^0=t_2}$
do not depend on the r.h.s.~on $\rho(x,y)$ and $F(x,y)$ for
times $x^0 \ge t_1$ and $y^0 \ge t_2$. To see this note that
the integrands vanish identically for the upper time-limits
of the memory integrals because of the anti-symmetry of the
spectral components, with $\rho(x,y)|_{x^0 = y^0} \equiv 0$ and 
$\Sigma_{\rho}(x,y)|_{x^0 = y^0} \equiv 0$. As a consequence, only
explicitly known quantities at earlier times determine the
time evolution of the unknowns at later times. {\rr\em The numerical
implementation therefore only involves sums over
known functions.} 

To be more explicit we consider first a {\rr\em scalar field
theory}. For spatially homogeneous fields it is
sufficient to implement the equations for the Fourier
components $F(t,t'; \bp)$ and $\rho(t,t'; \bp)$ and we consider $\phi = 0$. 
The numerical implementation with $\phi \not = 0$ follows 
along the very same lines for the equation (\ref{eq:exactphi}).
Spatially inhomogeneous fields pose no complication in principle
but are computationally more expensive. As an example we 
consider a time discretization $t=n a_t$, $t'=m a_t$ with stepsize 
$a_t$ such that $F(t,t') \mapsto F(n,m)$, and  
\bea\db
 \partial_{t}^2 F(t,t') &\db \mapsto&\db \frac{1}{a_t^2}
\Big( F(n+1,m) + F(n-1,m) - 2 F(n,m) \Big),
\label{deriv} \\[0.2cm]
\db \int\limits_0^{t} dt F(t,t') &\db \mapsto&\db
a_t \Big(F(0,m)/2 + \sum\limits_{l=1}^{n-1} F(l,m) + F(n,m)/2
\Big) \label{integral}  \, ,
\eea 
where we have suppressed the momentum labels in the notation.
The second derivative is replaced by a finite-difference expression,
which is symmetric in $a_t \leftrightarrow -a_t$. It is obtained
from employing subsequently the lattice ``forward derivative''
$[F(n+1,m) - F(n,m)]/a_t$ and ``backward derivative''
$[F(n,m) - F(n-1,m)]/a_t$. The integral is approximated
using the trapezoidal rule with the 
average function value $[F(n,m)+F(n+1,m)]/2$ in an interval 
of length $a_t$. The above simple discretization leads already 
to stable numerics for small enough stepsize $a_t$, but the 
convergence properties may be easily improved with more sophisticated 
standard estimators if required. 

As for the continuum the propagators obey 
the symmetry properties $F(n,m)=F(m,n)$ and $\rho(n,m)=-\rho(m,n)$.
Consequently, only ``half'' of the $(n,m)$--matrices have to be computed
and $\rho(n,n) \equiv 0$. Similarly, since the self-energy
$\Sigma_{\rho}$ and the summation function
$I_{\rho}$ appearing in the $1/N$ expansion to NLO (cf.\ Eq.~(\ref{IRFR})) 
are antisymmetric in time one can exploit that 
$\Sigma_{\rho}(n,n)$ and $I_{\rho}(n,n)$ vanish identically. 
For the case of the NLO time evolution the equations (\ref{eq:exactrhoF}) with 
(\ref{SFFR})--(\ref{IRFR}) are used to 
advance the matrices $F(n,m)$ and
$\rho(n,m)$ stepwise in the ``$n$-direction'' for each given $m$. 
As initial conditions one has to specify $F(0,0;\bp)$, $F(1,0;\bp)$
and $F(1,1;\bp)$, while $\rho(0,0;\bp)$, $\rho(1,0;\bp)$ and
$\rho(1,1;\bp)$ are fixed by the equal-time commutation relations
(\ref{eq:bosecomrel}). The time discretized versions of 
(\ref{eq:exactrhoF}) read:
\bea\db \lefteqn{
F({\rr n+1},m;\bp) = 2 F(n,m;\bp) - F(n-1,m;\bp) } \nonumber\\[0.1cm]
&&\db - a_t^2 \left\{ \bp^2 + m^2 + \lambda\, \frac{N+2}{6 N}
\int_{\bk} F(n,n;\bk)
\right\} F(n,m;\bp) 
\nonumber\\
&&\db - a_t^3\, \Bigg\{
\Sigma_{\rho}(n,0;\bp)\, F(0,m;\bp)/2 - \Sigma_F(n,0;\bp)\, \rho(0,m;\bp)/2
\\
&&\db \qquad +\sum\limits_{l=1}^{m-1} \Big(
\Sigma_{\rho}(n,l;\bp)\, F(l,m;\bp) - \Sigma_F(n,l;\bp)\, \rho(l,m;\bp)
\Big) \nonumber\\
&&\db \qquad +\sum\limits_{l=m}^{n-1}\, 
\Sigma_{\rho}(n,l;\bp)\, F(l,m;\bp) \Bigg\}\, , \nonumber 
\eea 
$(n \ge m)$ \footnote{For the discretization of the time integrals it is 
useful to distinguish the cases $n \ge m$ and $n \le m$. We compute the 
entries $F(n+1,m)$ from the discretized
equations for $n \ge m$ except for $n+1=m$ where we have to use 
the equations for $n \le m$.} and similarly for $\rho$. These
equations are explicit in time: 
Starting with $n=1$, for the time step $n+1$ one computes successively
all entries with $m=0,\ldots,n+1$ 
from known functions at earlier times. 
At first sight this property
is less obvious for the non-derivative expressions (\ref{IFFR}) for $I_F$
and (\ref{IRFR}) for $I_{\rho}$ whose form is reminiscent
of a gap equation. However, the discretized 
equation for $I_{\rho}$,
\bea\db \lefteqn{
I_{\rho}(n,m;\bq)=\frac{\lambda}{3} \int_{\bk}
\Bigg\{ F(n,m;\bq-\bk) \rho(n,m;\bk) } 
\nonumber\\
&&\db - a_t \sum\limits_{l=m+1}^{n-1}
I_{\rho}(n,l;\bq) F(l,m;\bq-\bk) \rho(l,m;\bk)  \Bigg\} \, ,
\eea
shows that all expressions for $I_{\rho}(n,m)$ are explicit as well: 
Starting with $m=n$ where $I_{\rho}$ vanishes one should lower 
$m=n,\ldots,0$ successively. For $m=n-1$ one obtains an explicit 
expression in terms of $F(n,m)$ and $\rho(n,m)$ known from the
previous time step in $n$. For $m=n-2$ the
r.h.s.~then depends on the known function $I_{\rho}(n,n-1)$  
and so on. Similarly, for given $I_{\rho}(n,m)$
it is easy to convince oneself that the discretized
Eq.\ (\ref{IRFR}) specifies $I_F(n,m)$ completely
in terms of $I_F(n,0),\ldots,I_F(n,m-1)$, which constitutes
an explicit set of equations by increasing $m$ successively
from zero to $n$.

It is crucial for an efficient numerical 
implementation that each step forward in time does not involve 
the solution of a self consistent or gap equation. This is 
manifest in the above discretization. The main
numerical limitation of the approach is set by the time
integrals (``memory integrals'') 
which grow with time and therefore slow down the 
numerical evaluation. Typically, the influence of early times
on the late time behavior is suppressed and can be neglected 
numerically in a controlled way. In this case, it is often
sufficient to only take into account the contributions from 
the memory integrals
for times much larger then the characteristic inverse damping 
rate (cf.~Sec.~\ref{sec:latetimeuni} below). An error estimate
then involves a series of runs with increasing memory time.   

For scalars one can use a standard lattice discretization for a
spatial volume with periodic boundary conditions. 
For a spatial volume $V=(N_s a)^d$ with lattice spacing $a$ 
one finds for the momenta:
\beq\db
\bp^2 \,\mapsto\, \sum\limits_{i=1}^d \frac{4}{a^2} \sin^2 
\left(\frac{a p_i}{2} \right) 
\, ,
\qquad p_i=\frac{2 \pi n_i}{N_s a}  \, ,
\eeq
where $n_i = 0,\ldots, N_s-1$. This can be easily understood from
acting with the corresponding finite-difference expression (\ref{deriv})
for space-components: $\partial^2_x e^{-i p x} \mapsto 
e^{-i p x} [e^{ipa} + e^{-ipa} - 2]/a^2 =
- e^{-i p x}\, 4 \sin^2(p a/2)/a^2$. On the lattice there 
is only a subgroup of the rotation symmetry generated 
by the permutations of $p_x,p_y,p_z$ and the reflections $p_x
\leftrightarrow -p_x$ etc.~for $d=3$. Exploiting these
lattice symmetries reduces the number of
independent lattice sites to $(N_s+1)(N_s+3)(N_s+5)/48$.
We emphasize that the self-energies are calculated in coordinate space,
where they are given by products of coordinate-space
correlation functions, and then transformed back to
momentum space. The coordinate-space correlation functions
are available by {\rr\em fast Fourier transformation} routines. 

The lattice introduces a 
momentum cutoff $\pi / a$, however, the renormalized
quantities are insensitive to cutoff variations for 
sufficiently large $\pi / a$. We emphasize here again
that {\rr\em it is often convenient to carry out the 
numerical calculations using unrenormalized equations}, or equations
where only the dominant (quadratically) divergent contributions
in the presence of scalars are subtracted. Cf.~the discussion
at the end of Sec.~\ref{sec:renormalize}.   
In order to study the infinite volume limit one has to remove finite 
size effects. Here this is done by increasing the volume until
convergence of the results is observed.\footnote{    
For time evolution problems
the volume which is necessary to reach the infinite volume limit 
to a given accuracy can depend on the time scale. 
This is, in particular, due to the fact that finite systems 
can show characteristic
recurrence times after which an initial effective
damping of oscillations can be reversed. The observed damping can 
be viewed as the result of a superposition 
of oscillatory functions with differing phases or frequencies.  
The recurrence time is given by the time after which 
the phase information contained in the initial oscillations 
is recovered. Then the damping starts again
until twice the recurrence time is reached and so on.
In the LO approximation one can explicitly verify that 
the observed recurrence times e.g.~for the correlation 
$F(x,x)$ scales with the volume or the number of 
lattice sites to ``infinity''. 
We emphasize that the phenomenon of complete recurrences, 
repeating the full initial oscillation pattern after some characteristic time,
is not observed once scattering is taken into account. 
Periodic recurrences can occur with smaller amplitudes as time proceeds and
are effectively suppressed in the large-time
limit.} We finally note that the nonequilibrium
equations are very suitable for parallel computing on
PC clusters using the MPI standard.\\

For {\rr\em fermionic field theories} 
the numerical implementation is more involved than
for purely scalar theories. 
Consider the evolution equations for the Yukawa model
(\ref{rhoV0eom})--(\ref{FVeom}), together with the self-energies 
(\ref{eq:selffermA})--(\ref{selffermionF}).
The structure of the fermionic equations is reminiscent of the 
form of classical
canonical equations. In this analogy, $F_V(t,t')$ plays the role 
of the canonical coordinate and $F_V^0(t,t')$ is analogous to the canonical 
momentum. This suggests 
to discretize $F_V(t,t')$ and $\rho_V(t,t')$ at $t-t'=2n a_t$ (even) and 
$F_V^0(t,t')$ and $\rho_V^0(t,t')$ at $t-t'=(2n+1)a_t$ (odd) time-like lattice 
sites with spacing $a_t$. This is a generalization of the ``leap-frog'' 
prescription for temporally inhomogeneous two-point functions. This implies
in particular that the discretization in the time direction is coarser
for the fermionic two-point functions than for the bosonic ones. This 
``leap-frog'' prescription may be easily extended to the memory integrals 
on the r.h.s.~of Eqs.\ (\ref{rhoV0eom})--(\ref{FVeom}) as well.

We emphasize that the discretization does not suffer from the
problem of so-called fermion doublers. The spatial 
doublers do not appear since (\ref{rhoV0eom})--(\ref{FVeom})
are effectively second order in $\vec{x}$-space. Writing the equations
for $\vec F_V(t,t',\vec x)$ and $\vec \rho_V(t,t',\vec x)$
starting from (\ref{rhoV0eom})--(\ref{FVeom}) one realizes that instead of
first order spatial derivatives there is a Laplacian appearing.
Hence we have the same Brillouin zone for the fermions and scalars.
Moreover, time-like doublers are easily avoided by using a sufficiently 
small stepsize in time $a_t/a_s$.

The fact that Eqs.~(\ref{rhoV0eom})--(\ref{FVeom}) with
(\ref{eq:selffermA})--(\ref{selffermionF}) and the respective
scalar ones contain memory integrals makes numerical
implementations expensive. Within a given numerical precision it is typically
not necessary to keep all the past of the two-point functions in the memory.
A single PIII desktop workstation with
2GB memory allows us to use a memory array with 470 time-steps (with 2 temporal
dimensions: $t$ and $t'$). For instance, we have checked for the presented 
runs in Sec.~\ref{sec:prethermal} that a 30\%
change in the memory interval length did not alter the results.  For a typical
run 1-2 CPU-days were necessary. Much shorter times can be achieved
with parallel computing on PC clusters using the MPI standard. 
The shown plots in that section are calculated on a 
$470\times470\times32^3$ lattice.
(The dimensions refer to the $t$ and $t'$ memory arrays and the momentum-space
discretization, respectively.) By exploiting the spatial symmetries described 
in Sec.~\ref{sec:lorentz} the memory need could be reduced by a factor
of~30.
We have checked that the infrared cutoff is well below any other mass scales
and that the UV cutoff is greater than the mass scales at least by a factor of
three. In the evolution equations we analytically subtract 
only the respective quadratically divergent terms. 
To extract physical quantities we follow the time evolution
of the system for a given lattice cutoff and 
measure e.g.~the renormalized scalar mass from the oscillation frequency
of the correlator zero-modes to set the scale.

\section{Nonequilibrium phenomena}
\label{sec:neqevolution}

Our aim in this section is to study quantum field theories which capture 
important aspects of nonequilibrium physics and which are simple enough that 
one can perform a precise quantitative treatment.
We employ first the two-particle irreducible $1/N$ expansion
for the scalar $N$-component field theory (\ref{eq:classical}).
Considering subsequent orders in the expansion corresponds
to include more and more aspects of the dynamics. This allows
one to study the influence of characteristic ingredients
such as scattering, off-shell and memory effects on the
time evolution. We note that time-reflection invariance for the
equations and conservation of energy is preserved at each order in the
$1/N$--expansion.\footnote{These properties are also taken care of
by the employed numerical techniques.}     
It should be stressed that during the nonequilibrium time evolution 
there is no loss of information in any strict sense.
Here we consider closed systems without coupling to an
external heat bath or external fields. There is no
course graining or averaging involved in the dynamics.
The important process of thermalization is a nontrivial question in
a calculation from first principles. Thermal equilibrium itself 
is time-translation invariant and cannot be reached from
a nonequilibrium evolution on a fundamental level. It is striking 
that we will observe below that scattering drives the evolution 
very closely towards thermal equilibrium results without
ever deviating from them for accessible times.

Here we study the time evolution for various initial condition scenarios away
from equilibrium. One corresponds to a ``quench'' often employed 
to mimic the situation of a 
rapidly expanding hot initial state which cools on time scales much 
smaller than the relaxation time of the fields. 
Initially at high temperature we consider the relaxation processes 
following an instant ``cooling'' described by a sudden drop in the 
effective mass.
Another scenario is characterized by initially densely populated modes in
a narrow momentum range around $\pm \bp_{\rm ts}$.
This so-called ``tsunami'' initial condition is reminiscent 
of colliding wave packets moving with opposite and equal momentum. 
A similar non-thermal and radially
symmetric distribution of highly populated modes may also be encountered in  
the so-called ``color glass'' state at saturated gluon density. Of course,
a sudden change in the two-point function of a previously equilibrated system
or a peaked initial particle number distribution are general enough 
to exhibit characteristic properties
of nonequilibrium dynamics for a large variety
of physical situations. We will first consider
results for the $1+1$ dimensional quantum field theory,
since technical aspects such as the infinite volume limit,
renormalization and large times can be all implemented with
great rigour. The study also has the 
advantage that the physics of off-shell effects 
can be very clearly discussed since they play a more important 
role for one spatial dimension as compared to three dimensions.
The conclusions we will draw are not specific to scalar theories
in low dimensions, and we will compare below with the 
corresponding results in
$3+1$ dimensions for scalars interacting with fermions. 

For the 2PI $1/N$ expansion we will first consider the time
evolution at leading order (LO) and compare it with
the next-to-leading order dynamics (NLO). 
The ``mean-field type'' LO approximation has a long history 
in the literature. However, a drastic consequence of this 
approximation is the appearance of an infinite number of conserved 
quantities, which are not present in the interacting  
finite-$N$ theory. These additional constants of motion can have a 
substantial impact on the time evolution, since they strongly constrain 
the allowed dynamics. As an important consequence of the LO
approximation, the late-time behavior 
depends explicitly on the details of the initial conditions and 
the approach to thermal equilibrium cannot be observed in this case
as is shown below. The extensive use of this approximation was based 
on the hope that deviations from the LO behavior are not too sizeable 
for not too late times. By taking into account the NLO contributions 
it turns out, 
however, that for many important questions there are substantial
corrections long before thermalization sets in. 
In contrast, we will find that various approximations of the 2PI
effective action which go beyond ``mean-field type'' 
(LO, Hartree or any Gaussian) dynamics show thermalization 
and give comparable 
\mbox{answers}. The 2PI approximations beyond mean-field share the property
that the spurious constants of motion are absent, which emphasizes
the important role of conservation laws for the nonequilibrium
dynamics: fake conserved quantities keep the information about the
initial conditions for all times and can spoil any effective
loss of details about the early-time behavior.

\subsection{Scattering, off-shell and memory effects} 
\label{eq:scatoffmem}

\subsubsection{LO fixed points}   
\label{sec:lofixedpoints}

For simplicity we consider spatially homogeneous field expectation 
values $\phi_a(t) = \langle \Phi_a(t,\bx) \rangle$, such that
we can use the Fourier modes $F_{ab}(t,t';\bp)$ and
$\rho_{ab}(t,t';\bp)$.\footnote{Note that 
a spatially {\em homogeneous} field expectation value contains 
all fluctuations arising from an {\em inhomogeneous} $\Phi(x)$ as well. 
From a practical point of view the extraction of physics related
to inhomogeneous field fluctuations is, of course, more difficult 
since it is contained in higher correlation functions.} 
The solution of (\ref{eq:exactrhoF})  
in the limit of a free field theory ($\lambda \equiv 0$) describes modes which 
oscillate with frequency $\sqrt{{\bf p}^2+m^2}/2 \pi$ for
unequal-time functions $F(t,0;\bp)$ and 
$\rho(t,0;\bp)$, unless they are not identically zero. 
Equal-time correlation modes $F(t,t;\bp)$ oscillate either 
with twice that frequency or they are constant in time
($\rho(t,t;\bp)\equiv 0$ 
according to Eq.\ (\ref{eq:bosecomrel})). 
The latter corresponds to
solutions which are translationally invariant in time, i.e.\
$F(t,t';\bp)=F(t-t';\bp)$ and 
$\rho(t,t';\bp)=\rho(t-t';\bp)$. 

The LO contribution to the 2PI 
effective action (\ref{LOcont}) adds a time-dependent mass shift 
to the free field evolution equation. The resulting effective mass term, 
given by (\ref{Meff}) for $N \to \infty$, is the 
same for all Fourier modes and consequently each mode propagates 
``collisionlessly''. There are no further corrections 
since according to (\ref{eq:NLOcont}) and (\ref{eq:NLObrok}) 
the self-energies $\rr \Sigma_{F}$ and $\Sigma_{\rho}$ 
are $\Or (1/N)$ and vanish in this limit! The evolution 
equations (\ref{eq:exactrhoF}) for this
approximation read:
\bea 
\db\left[\partial_t^2 + \bp^2 
+ M^2(t;\phi,F) 
\right]\, F_{ab}(t,t';\bp) &{\db =}& 
{\db 0}\,\, , \nonumber\\ 
{\db \left[\partial_t^2 + \bp^2 
+ M^2(t;\phi,F) \right]}\, {\rr \rho_{ab}(t,t';\bp)} 
&{\db =}& {\db 0}\,\, , 
\label{eq:LOdyn}\\
\db \left[ \partial_t^2 + \frac{\lambda}{6N}
\phi^2(t) + M^2(t;\phi \equiv 0,F) \right] \phi_b(t) 
&{\db =}& {\db 0} \,\, ,\nonumber
\eea
with
\beq\db
M^2(t;\phi,F) \, \equiv \, m^2 + \frac{\lambda}{6N}\, 
\left[\int_{\bp} F_{cc}(t,t;\bp)
+\phi^2(t) \right]  \, , 
\label{eq:LOmass}
\eeq
where $\int_{\bp} \equiv \int {\rm d}^dp/(2\pi)^d$.
In this case one observes that the evolution of $F$ and $\phi$ is completely 
decoupled from~$\rho$. Similar to the free 
field theory limit, to LO the spectral function does not influence 
the time evolution of the statistical propagator. 
The reason is that in this approximation the
spectrum consists only of ``quasiparticle'' modes of energy
$\omega_\bp(t) = \sqrt{\bp^2 + M^2(t)}$ with an
infinite life-time. The associated mode particle numbers are conserved 
for each momentum separately. In contrast, in the interacting quantum 
field theory particles decay, get created, annihilated and the notion of a 
conserved particle number is clearly absent for real scalar fields.   
Note that there are also no memory integrals appearing on the
r.h.s.~of (\ref{eq:LOdyn}), which incorporate in particular 
direct scattering processes. 

As an example we consider a ``tsunami'' type initial condition in $1+1$
dimension. The initial statistical propagator is
given by 
\bea
&&\db F(0,0;p) \, = \, \frac{n_p(0)+1/2}{\sqrt{p^2+M^2(0)}} \quad , \quad
\partial_{t}F(t,0;p)|_{t=0} \, =\, 0  \,\, , 
\nonumber\\
&&\db F(0,0;p)\partial_{t}\partial_{t'} 
F(t,t';p)|_{t=t'=0} \,=\, [n_p(0)+1/2]^2  \, ,
\label{eq:initialtsF}\\[0.3cm]
&&\db \phi(0) \,=\, \partial_t\phi(t)|_{t=0} \,=\, 0 \, .
\label{eq:initialtsrho}
\eea
Here $F_{ab}(t,t';\bp) \equiv F(t,t';\bp) \delta_{ab}$, which is
valid for all times with these initial conditions.
The mass term $M^2(0)$ is given by the gap equation (\ref{eq:LOmass}) 
in the presence of the initial non-thermal 
particle number distribution 
\beq \db
n_p(0)={\cal A} e^{-\frac{1}{2 \sigma^2}(|p|-p_{\rm ts})^2} \, .
\label{eqtsunamidist}
\eeq
As the renormalization condition we choose the initial renormalized mass in
vacuum, $m_R\equiv M(0)|_{n(0)=0}=1$, as our dimensionful scale. There is
no corresponding coupling renormalization in $1+1$ dimension. In these
units the particle number is peaked around $|p|=p_{\rm ts}=5 m_R$ with 
a width determined by $\sigma=0.5 m_R$ and amplitude ${\cal A}=10$. 
We consider the effective coupling $\lambda/(6 m_R^2) = 1$.

\begin{figure}[t]
\epsfig{file=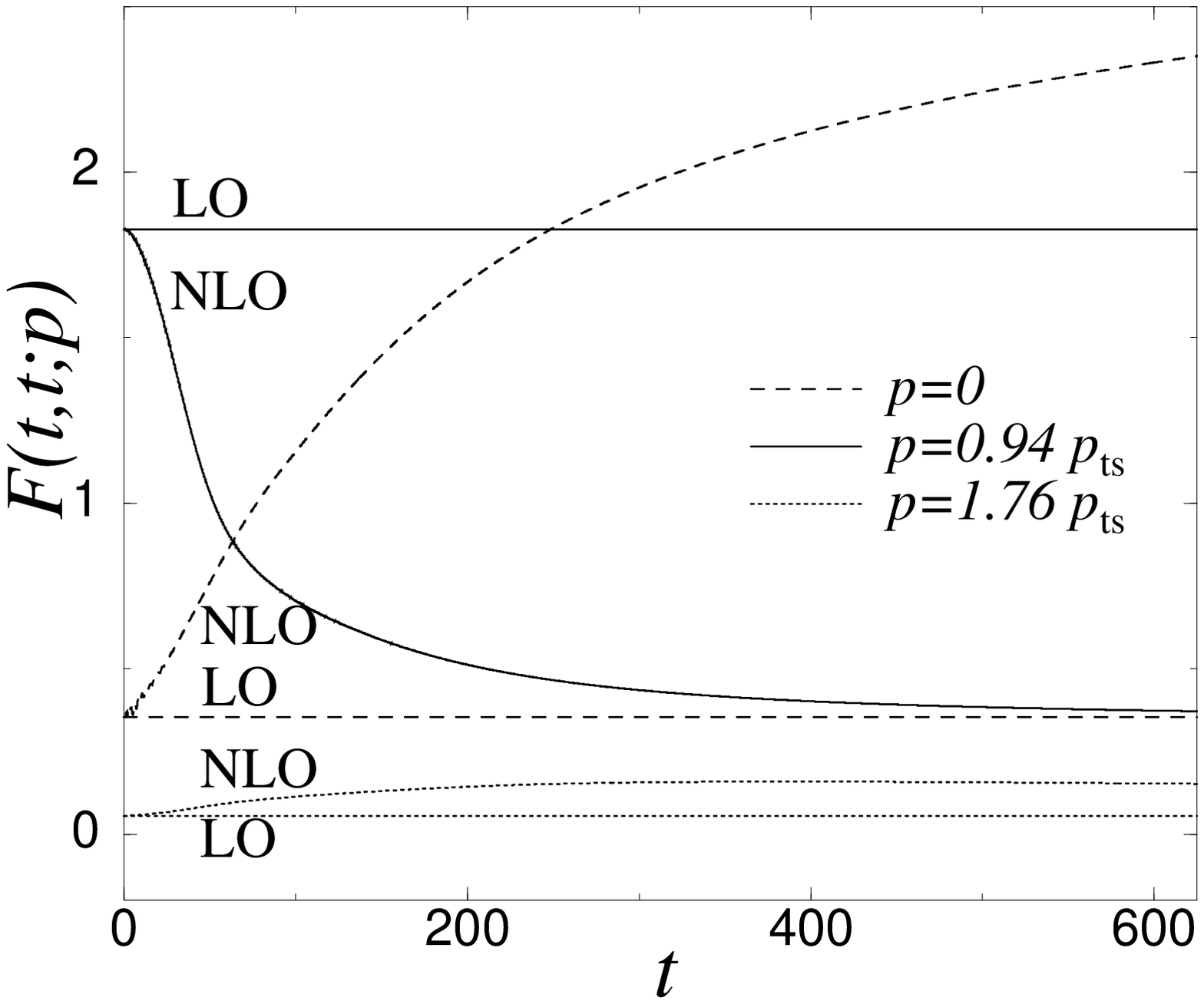,width=6.7cm}
\hspace*{0.4cm}
\epsfig{file=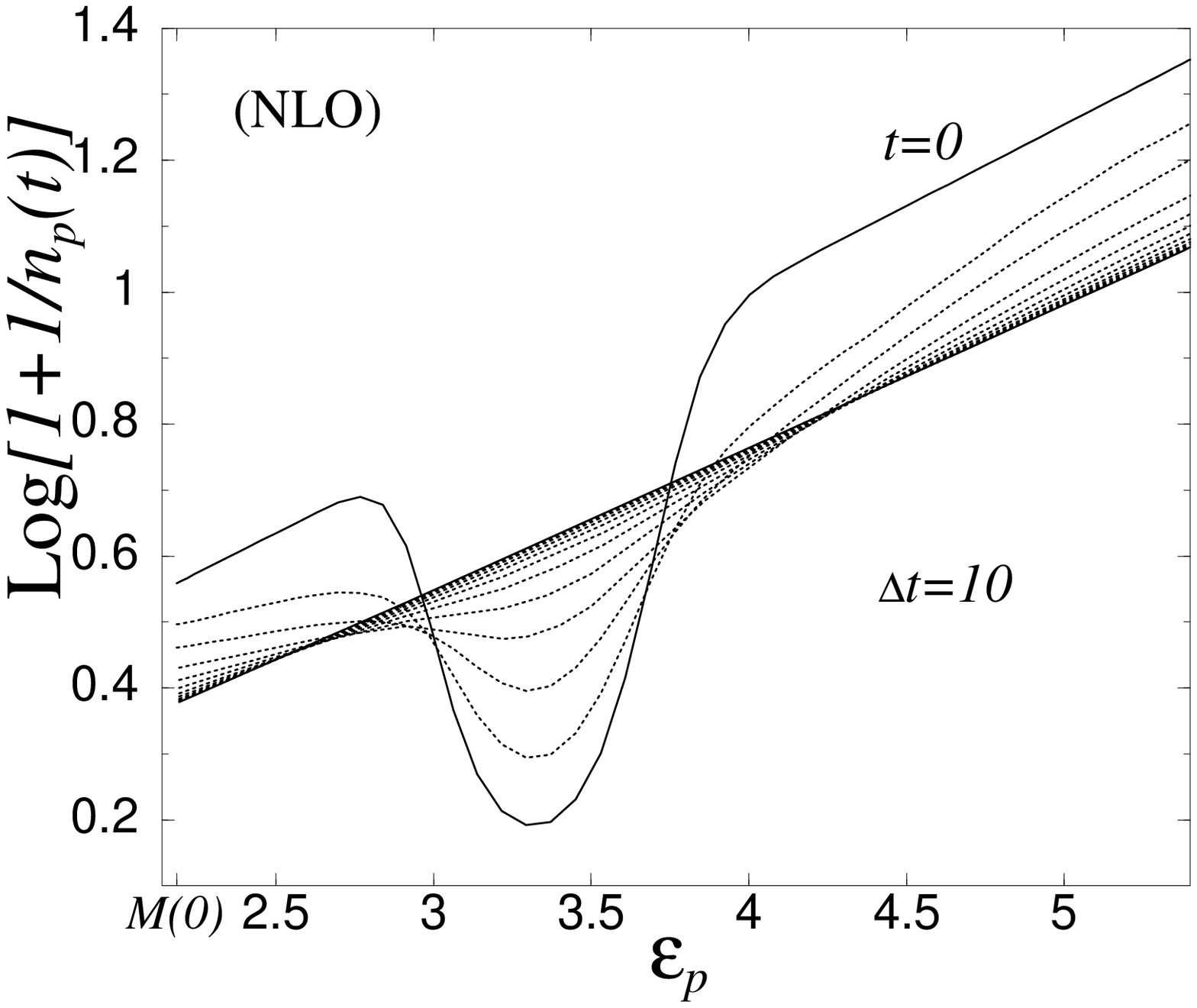,width=7.cm}
\caption{{\bf LEFT:} Comparison of the LO and NLO time dependence of the 
equal-time correlation modes
$F(t,t;p)$ for the ``tsunami'' initial condition 
(\ref{eq:initialtsF})--(\ref{eqtsunamidist}). 
The importance of scattering included in the NLO approximation
is apparent: the non-thermal
LO fixed points become unstable and the ``tsunami'' decays,
approaching thermal equilibrium at late times. 
{\bf RIGHT:} Effective particle number distribution for a ``tsunami''
in the presence of a thermal background. The solid line shows the
initial distribution which for low and for
high momenta follows a Bose-Einstein distribution,
i.e.\ ${\rm Log}[1+1/n_p(0)] \simeq \epsilon_p(0) / T_0$. At late times 
the non-thermal distribution
equilibrates and approaches a straight line with inverse slope
$T_{\rm eq} \, >\, T_0$.}  
\label{Figpeakphi}
\end{figure}
On the left of Fig.\ \ref{Figpeakphi} we present the time evolution of the
equal-time correlation modes $F(t,t;p)$
for different momenta: zero momentum, a momentum close to the 
maximally populated momentum $p_{\rm ts}$ and about twice $p_{\rm ts}$. 
One observes that the equal-time correlations at LO are strictly constant
in time. This behavior can be understood from the fact that for the 
employed ``tsunami'' initial condition 
the evolution starts at a time-translation invariant non-thermal 
solution of the LO equations. There is an infinite number of 
so-called fixed point solutions which are constant in time. 
If the real world would be well approximated by the LO dynamics 
then this would have dramatic consequences. The thermal equilibrium solution,
which is obtained for an initial Bose-Einstein distribution instead
of (\ref{eqtsunamidist}), is not at all particular in this case!
Therefore, everything depends on the chosen initial condition details.
This is in sharp contrast to the well-founded expectation that 
the late-time behavior may be well described by thermal equilibrium 
physics. Scattering effects included in the NLO approximation indeed
drive the evolution away from the LO non-thermal fixed points 
towards thermal equilibrium, which is discussed in detail 
in Sec.~\ref{sec:NLOtherm} below. 

A remaining question is what happens at LO if the time evolution 
does not start from a LO fixed point, as was the case for the 
``tsunami'' initial condition above. 
In the left graph of Fig.~\ref{FigLOM} we plot the evolution of 
$M^2(t)$ in the LO approximation as a function of time $t$,
following a ``quench'' described by an instant drop in the
effective mass term from $2 M^2(0)$ to $M^2(0)$.
The initial particle number distribution is 
$n_p(0)=1/(\exp[\sqrt{p^2+2 M^2(0)}/T_0]-1)$ with
$T_0 = 2 M(0)$ and $\phi(0) = \dot{\phi}(0) = 0$.
The sudden change in the effective mass
term drives the system out of equilibrium and one can study its
relaxation. We present $M^2(t)$ for three different couplings 
$\lambda=\lambda_0 \equiv 
0.5 \, M^2(0)$ (bottom), $\lambda = 10\lambda_0$ (middle)
and $\lambda = 40\lambda_0$ (top).   
All quantities are given in units of appropriate powers 
of the initial-time mass $M(0)$.
Therefore, all curves in the left graph of Fig.~\ref{FigLOM}
start at one. 
The time-dependent mass squared $M^2(t)$ shoots up in response 
to the ``quench'' and stays below the value $2 M(0)^2$ 
of the initial thermal distribution. The amplitude of initial oscillations
is quickly reduced and, averaged over the
oscillation time-scale, the evolution is rapidly driven
towards the asymptotic values.
\begin{figure}[t]
\epsfig{file=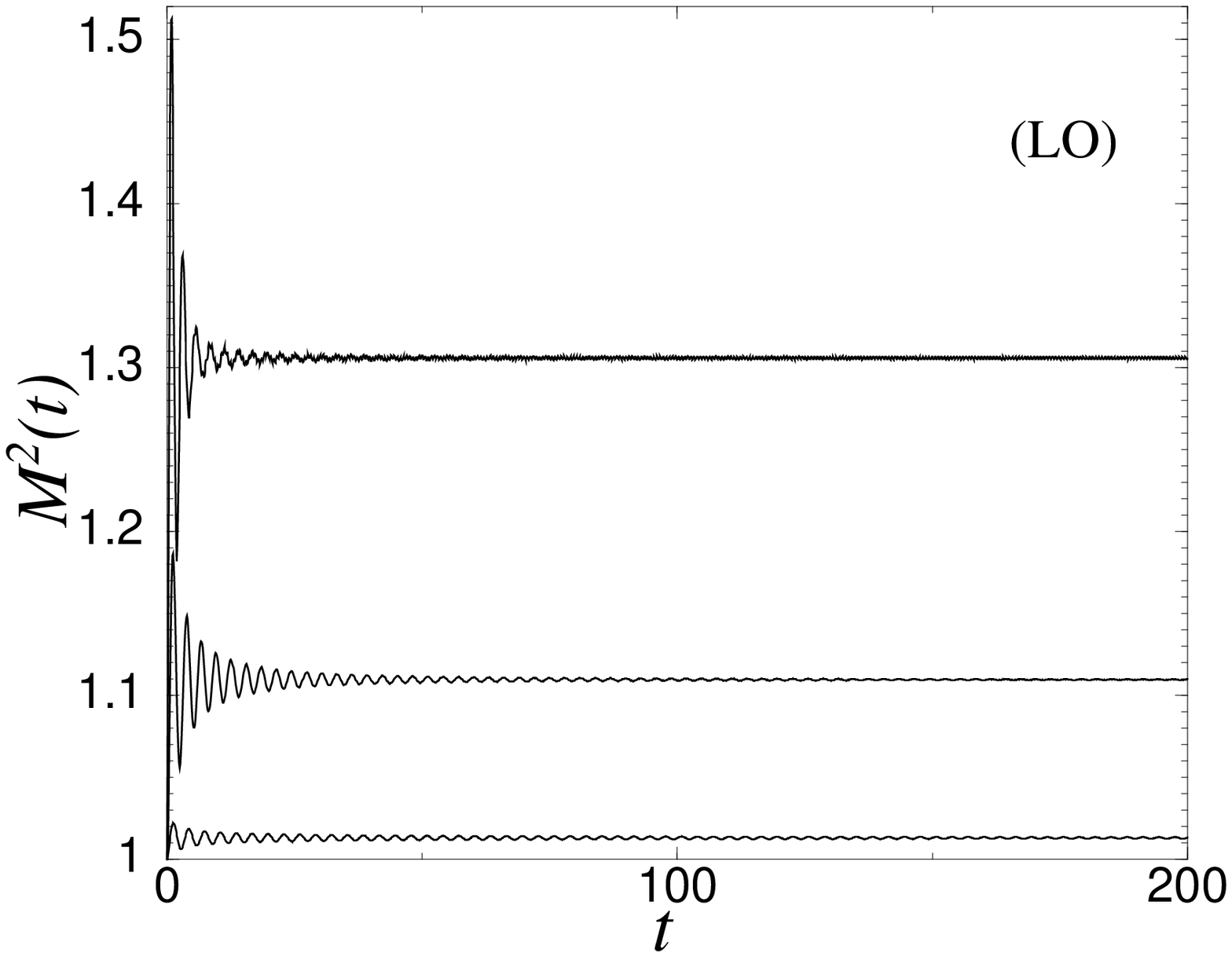,width=8.2cm}
\hspace*{-0.7cm}
\epsfig{file=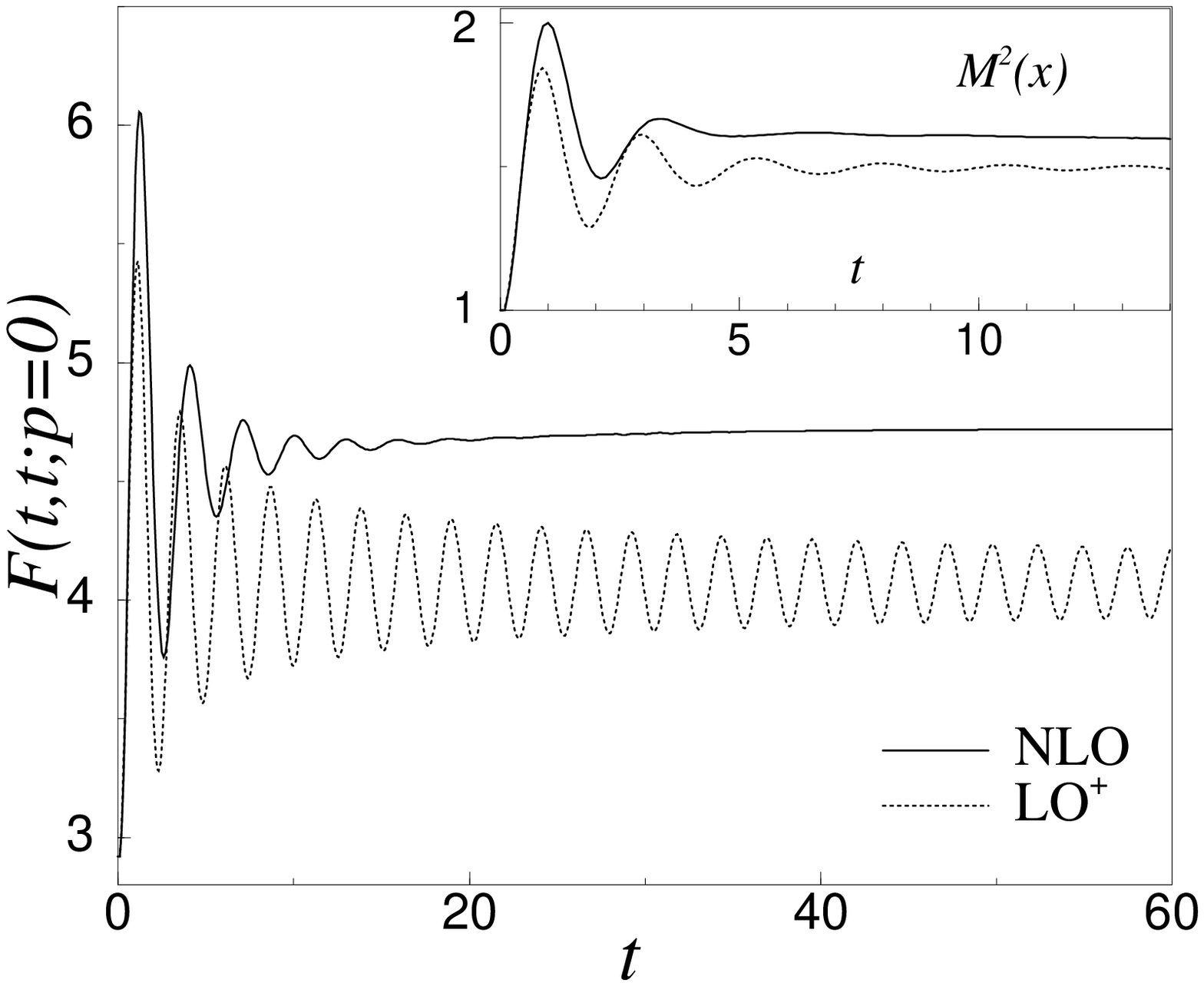,width=6.9cm}
\caption{{\bf LEFT:} Shown is the time-dependent mass term $M^2(t)$ 
in the LO approximation for three different couplings following
a ``quench''. All quantities are given in units of appropriate powers 
of the initial-time mass $M(0)$.
{\bf RIGHT:} Time dependence of the equal-time zero-mode  $F(t,t;p=0)$
after a ``quench'' (see text for details). 
The inset shows the mass term $M^2(t)$, which
includes a sum over all modes. The dotted lines represent the
Hartree approximation (LO$^+$), while the solid lines give
the NLO results. The coupling is $\lambda/6N = 0.17 \, M^2(0)$ for $N=4$.}  
\label{FigLOM}
\end{figure}

We compare the asymptotic values with the self-consistent
solution of the LO mass equation (\ref{eq:LOmass}) 
for constant mass squared $M_{\rm gap}^2$ and given  
particle number distribution $n_p(0)$:\footnote{
Here the logarithmic divergence of the one-dimensional integral 
is absorbed into the bare mass parameter $m^2$ using the same 
renormalization condition as for the dynamical evolution in
the LO approximation, i.e.\ 
$m^2+\frac{\lambda}{6} \int \frac{{\rm d}p}{2 \pi}
\left( n_p(0)+\frac{1}{2} \right) (p^2+M^2(0))^{-1/2}
= M^2(0)$.}
\beq\db
M^2_{\rm gap}\,=\, m^2 +\frac{\lambda}{6} \int \frac{{\rm d}p}{2 \pi}
\left( {\rr n_p(0)} + \frac{1}{2} \right) 
\frac{1}{\sqrt{p^2+M^2_{\rm gap}}} \, .
\label{LOgapequ}
\eeq
The result from this gap equation is 
$M^2_{\rm gap} = \{1.01,1.10,1.29\} M^2(0)$ for the three values
of $\lambda$, respectively.  
For this wide range of couplings the values are in 
good numerical agreement with the corresponding dynamical large-time 
results inferred from Fig.~\ref{FigLOM} as 
$\{1.01,1.11,1.31\}M^2(0)$ at $t = 200/M(0)$. We explicitly checked 
that at $t=400/M(0)$ these values are still the same.  
One concludes that the asymptotic behavior
at LO is well described in terms of the initial particle number distribution
$n_p(0)$. The latter is not a thermal distribution for the
late-time mass terms with values smaller than $2 M^2(0)$. 

The LO approximation for $F(t,t';p)$ and $\rho(t,t';p)$ becomes 
exact\footnote{This is due to the fact that we choose to
consider Gaussian initial conditions.} 
for $t,t' \to 0$, since the memory integrals on the r.h.s.~of
the exact Eqs.~(\ref{eq:exactrhoF}) vanish at initial time. 
For very early times one therefore expects
the LO approximation to yield a quantitatively valid description.
However, from Fig.~\ref{FigLOM} one observes that the time evolution is 
dominated already at early times by the non-thermal
LO fixed points. As discussed above, the latter are artefacts
of the approximation. Though the precise numerical values of the LO
fixed points are pure artefacts, we emphasize  that the presence of 
approximate fixed points governing the early-time behavior is a 
qualitative feature that can be observed also beyond LO (cf.~below). 
The question of how strongly the LO late-time results 
deviate from thermal equilibrium depends of course crucially on the 
details of the initial conditions. 
Typically, time- and/or momentum-averaged quantities 
are better determined by the LO approximation
than quantities characterizing a specific
momentum mode. This is exemplified on the right of Fig.~\ref{FigLOM},
which shows the equal-time zero-mode $F(t,t;p=0)$ along
with $M^2(t)$ including the sum over all modes.
Here we employ a ``quench'' with a larger drop in the
effective mass term from $2.9 M^2(0)$ to $M^2(0)$. The initial 
particle number distribution is $n_p(0)=1/(\exp[\sqrt{p^2+M^2(0)}/T_0]-1)$
with $T_0=8.5 M(0)$. In the figure the dotted curves show the dynamics
obtained from an ``improved'' LO (Hartree) approximation, LO$^+$, 
that takes into account the
local part of the NLO self-energy contribution. 
The resulting equations have the very same structure as
the LO ones, however, with the LO and NLO contribution to 
the mass term $M^2(t)$ included as given by Eq.~(\ref{eq:massNLO})
below. The large-time limit of the mass term in the LO$^+$ approximation
is determined by the LO$^+$ fixed point solution in complete analogy
to the above discussion. We also give in Fig.~\ref{FigLOM} the NLO results,
which are discussed below.
\begin{figure}[t]
\epsfig{file=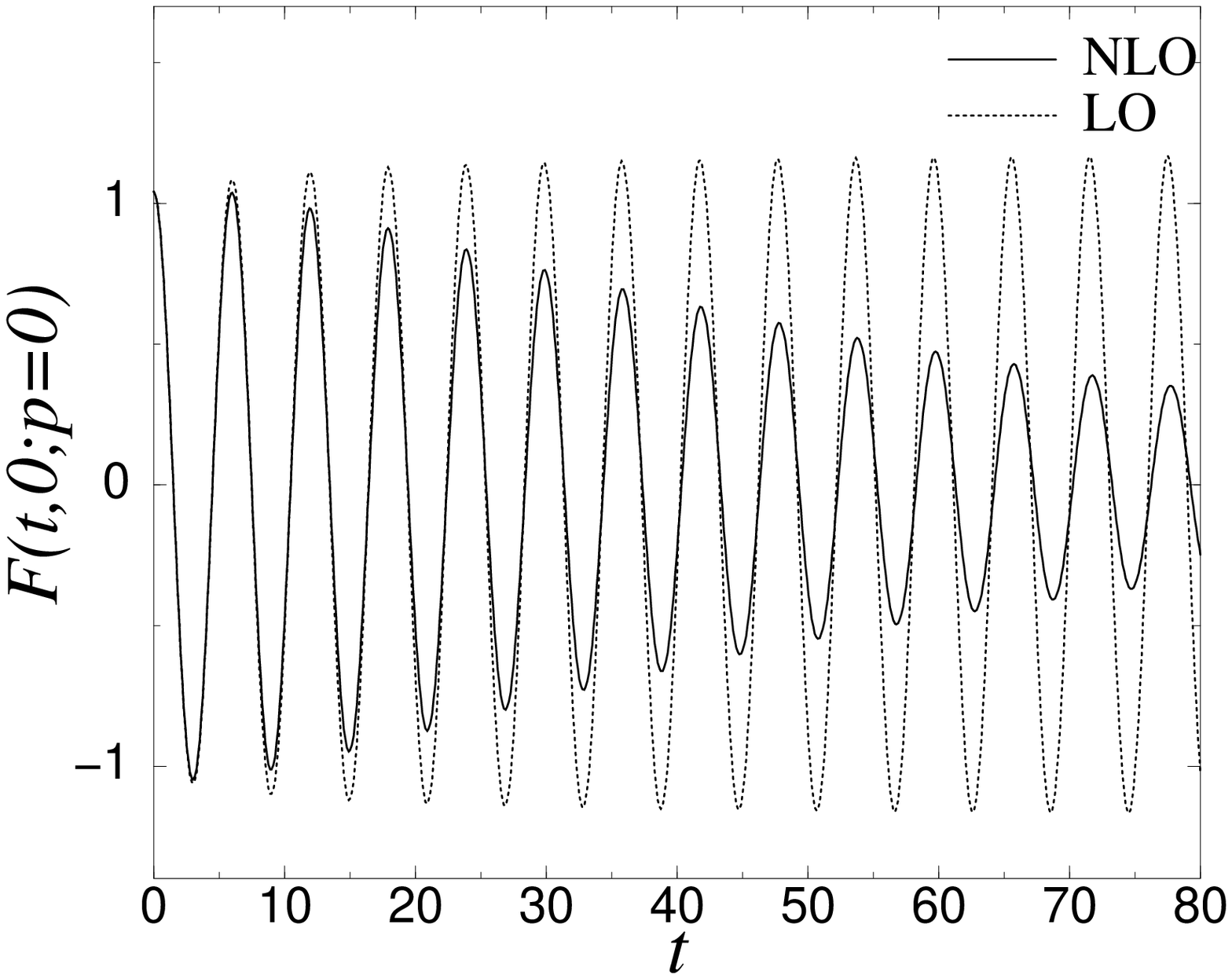,width=7.2cm}
\hspace*{0.3cm}
\epsfig{file=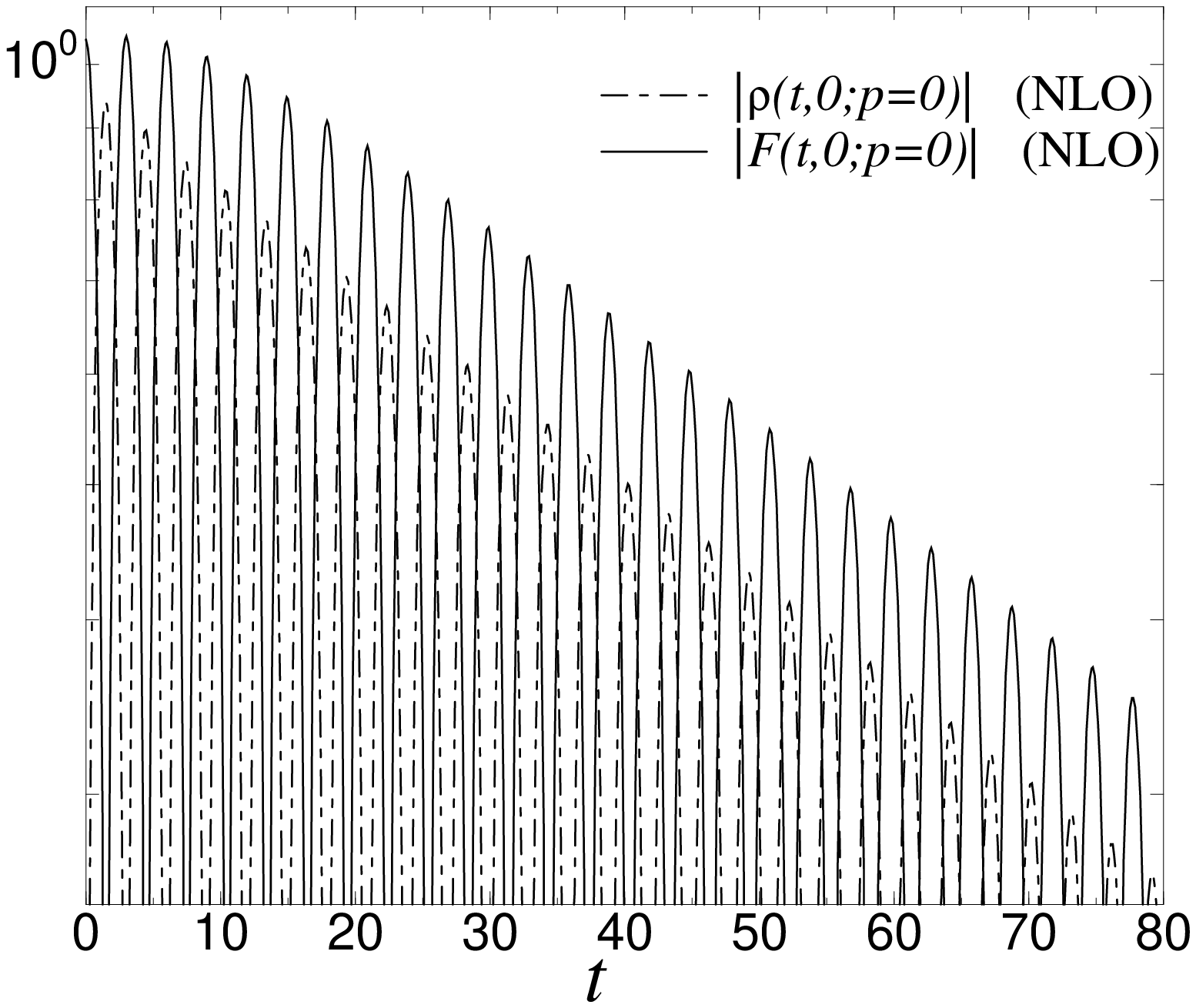,width=6.9cm}
\caption{{\bf LEFT:} Shown is the evolution of the unequal-time correlation 
\mbox{$F(t,0;p=0)$} after a ``quench''. Unequal-time 
correlation functions approach zero in the NLO approximation
and correlations with early times are effectively suppressed
($\lambda/6N = (5/6N\simeq 0.083) \, M^2(0)$ for $N=10$).
In contrast, there is no decay of correlations with earlier times
for the LO approximation.  
{\bf RIGHT:} The logarithmic plot of $|\rho(t,0;p=0)|$
and $|F(t,0;p=0)|$ as a function of time $t$ shows an oscillation
envelope which quickly approaches a straight line.
At NLO the correlation modes therefore approach an exponentially
damped behavior. (All in units of $M(0)$.)}  
\label{FigLogFrho}
\end{figure}

The effective loss of details about the initial conditions is 
a prerequisite for thermalization. At LO this is obstructed by
an infinite number of spurious conserved quantities (mode
particle numbers), which keep initial-time information.
An important quantity in this context is the unequal-time
two-point function $F(t,0;p)$, which characterizes the correlations
with the initial time. Clearly, if thermal equilibrium is
approached then these correlations should be damped.
On the left of Fig.~\ref{FigLogFrho} the dotted line shows the
unequal-time zero-mode $F(t,0;p=0)$ following the same ``quench''
at LO as for Fig.~\ref{FigLOM} left. One observes no decay of correlations 
with earlier times for the LO approximation. Scattering effects
entering at NLO are crucial for a sufficient effective loss of memory 
about the initial conditions, which is discussed next. 

\subsubsection{NLO thermalization}
\label{sec:NLOtherm}

In contrast to the LO approximation at NLO the self-energies
$\Sigma_{F}$ and $\Sigma_{\rho} \sim {\cal O}(1/N)$ do not vanish.
For the initial conditions (\ref{eq:initialtsF}) all correlators
are diagonal in field index space and $\phi \equiv 0$ for all times.
In this case the evolution equations derived from the NLO 2PI effective
action (\ref{eq:NLOcont}) are given by (\ref{eq:exactrhoF}) with
the self-energies (\ref{eq:massNLO})--(\ref{IRFR}). 
As discussed in Sec.~\ref{sec:noneqeveq}, the NLO evolution equations
are {\rr\em causal equations} with characteristic 
{\rr\em ``memory'' integrals}, which integrate over the time history of the
evolution taken to start at time $t_0 = 0$ without loss of generality. 

We consider first the same ``tsunami''
initial condition (\ref{eq:initialtsF}) as for the LO case 
discussed above. The result is shown on the left of Fig.~\ref{Figpeakphi}
for $N=10$. One observes a dramatic effect of the NLO corrections!
They quickly lead to a decay of the
initially high population of modes around $p_{\rm ts}$.
On the other hand, low momentum modes get populated such that
thermal equilibrium is approached at late times.   
In order to make this apparent one can plot the results
in a different way. For this we note that according to
(\ref{eq:initialtsF}) the statistical propagator
corresponds to the ratio of the following
{\rr\em particle number at initial time $t = 0$:}  
\beq\db
n_{\bp}(t) + \frac{1}{2} \,=\, 
\left[ F(t,t;\bp) K(t,t;\bp) - Q^2(t,t;\bp)\right]^{1/2} \, ,
\label{eq:effpart}
\eeq
and the corresponding mode energy. Here we have defined:
\beq\db
K(t,t';\bp) \,\equiv\, \partial_t \partial_{t'} F(t,t';\bp) \quad , \quad
Q(t,t';\bp) \,\equiv\, \frac{1}{2} \left[ \partial_t F(t,t';\bp)
+ \partial_{t'} F(t,t';\bp) \right] \, .
\eeq
Since we employed $Q(0,0;\bp) = 0$ for the initial conditions
(\ref{eq:initialtsF}) the {\rr\em initial mode energy}
is given by
\beq\db
\eps_{\bp}(t) \,=\,  \left(\frac{K(t,t;\bp)}{F(t,t;\bp)}\right)^{1/2} \, ,
\label{eq:effen}
\eeq
such that $F(t,t;p) = [n_\bp(t) + 1/2]/\eps_\bp(t)$.
For illustration of the results we may use (\ref{eq:effpart}) and
(\ref{eq:effen}) for times $t > 0$ in order to {\rr\em define an
effective mode particle number and energy}, where we set 
$Q(t,t';\bp) \equiv 0$ in (\ref{eq:effpart}) for the moment. 
The behavior of the effective particle number is illustrated 
on the right of Fig.~\ref{Figpeakphi}, where we 
plot ${\rm Log}(1+1/n_{\bp}(t))$ as
a function of $\eps_{\bp}(t)$. Note that for a Bose-Einstein
distributed effective particle number this is proportional
to the inverse temperature: ${\rm Log}(1+1/n_{\rm BE}(T)) \sim 1/T$. 
For the corresponding plot of Fig.~\ref{Figpeakphi}
we have employed an initial ``tsunami'' at $p_{\rm ts}/m_R = 2.5$, 
where we added an initial ``thermal background'' distribution
with temperature \mbox{$T_0/m_R = 4$}.\footnote{The initial mass term
is $M(0)/m_R = 2.24$ and $\lambda/6N = 0.5 m_R^2$ for $N=4$.} 
Correspondingly, from the solid
line ($t=0$) in the right figure one observes 
the initial ``thermal background'' as a straight line distorted
by the non-thermal ``tsunami'' peak. 
The curves represent snapshots at equidistant time steps   
$\Delta t\, m_R = 10$. After rapid changes in 
$n_p(t)$ at early times the subsequent curves converge to a 
straight line to high accuracy, with inverse slope 
$T_{\rm eq}/T_0 = 1.175$. The initial high 
occupation number in a small momentum range decays quickly
with time. More and more low momentum modes get populated
and the particle distribution approaches a thermal shape.

The crucial importance of the NLO corrections for the nonequilibrium
dynamics can also be observed for other initial conditions. 
The right graph of Fig.~\ref{FigLOM} shows the equal-time zero-mode 
$F(t,t;p=0)$, along with $M^2(t)$ including the sum over all modes,
following a ``quench'' as described in Sec.~\ref{sec:lofixedpoints}.
While the dynamics for vanishing self-energies $\Sigma_F$ and
$\Sigma_{\rho}$ is quickly dominated by the spurious
LO fixed points, this is no longer the case once the NLO self-energy 
corrections are included. In particular, one observes a very
efficient damping of oscillations at NLO. This becomes even more 
pronounced for unequal-time correlators as shown for $F(t,0;p=0)$
in the left graph of Fig.~\ref{FigLogFrho}. We find that the unequal-time 
two-point functions approach zero for the NLO approximation
and correlations with early times are effectively suppressed.  
Of course, time-reversal invariance implies that the 
oscillations can never be damped out to zero
completely during the nonequilibrium time evolution,
however, zero can be approached arbitrarily closely.
In the NLO approximation we find that 
all modes $F(t,t';p)$ and $\rho(t,t';p)$
approach an approximately exponential damping behavior for both
equal-time and unequal-time correlations. 
On the right of Fig.\ \ref{FigLogFrho} the approach to an exponential behavior 
is demonstrated for $|\rho(t,0;p=0)|$ and $|F(t,0;p=0)|$ with the
same parameters as for the left figure.
The logarithmic plot shows that after a non-exponential 
period at early times the envelope of oscillations
can be well approximated by a straight line. From an asymptotic envelope 
fit of $F(t,0;p=0)$ to an exponential 
form $\sim \exp(-\gamma_0^{\rm (damp)} t)$
we obtain here a damping rate $\gamma_0^{\rm (damp)}=0.016 M(0)$.
For comparison, for the parameters employed for the
right graph of Fig.~\ref{FigLOM} we find $\gamma_0^{\rm (damp)}=0.11 M(0)$
from an exponential fit to the asymptotic behavior of $F(t,0;p=0)$.
The oscillation frequency of $F(t,0;p=0)$ is found to quickly stabilize 
around $1.1 M(0)/2\pi \simeq 0.18 M(0)$, which is of the 
same order than the damping rate.
Correspondingly, one observes in Fig.~\ref{FigLOM} an 
equal-time zero mode which is effectively damped out at NLO
after a few oscillations.
The oscillation frequency is always found to stabilize very quickly
and to be quantitatively well described by $\eps_{\bp}(t)/2\pi$ for
the unequal-time modes ($\eps_{\bp}(t)/\pi$ for the
equal-time modes) with the effective mode energy $\eps_{\bp}(t)$
given by (\ref{eq:effen}). We emphasize that the parameter $M(t)$ 
given by (\ref{eq:massNLO}) cannot be used in general to 
characterize well the oscillation frequencies of the correlation 
zero-modes. Beyond LO or Hartree the renormalized mass, which can
be obtained from the zero-mode oscillation frequency, receives
corrections from the non-vanishing self-energies $\Sigma_F$ 
and $\Sigma_{\rho}$.

The strong qualitative difference between LO and NLO appears because 
an infinite number of spurious conserved quantities is removed once 
scattering is taken into account. It should be emphasized that the
step going from LO to NLO is qualitatively very different than the 
one going from NLO to NNLO or further. In order to understand better 
what happens going from LO to NLO we consider again the
effective particle number (\ref{eq:effpart}). It is straightforward
by taking the time derivative on both sides of (\ref{eq:effpart})
to obtain an {\rr\em exact evolution equation for $n_{\bp}(t)$} with the
help of the exact relations (\ref{eq:exactrhoF}):
\beq
\framebox{
\begin{minipage}{13.cm} \vspace*{-0.3cm}
\bea \db
\lefteqn{
\left( n_{\bp}(t) +\frac{1}{2} \right) \partial_t n_{\bp}(t) =} \nonumber\\
&& \db \!\!\! \int_{t_0}^{t}\!\! d t'' \, \Big\{
\left[ {\rr \Sigma_{\rho}(t,t'';\bp)}F(t'',t;\bp) 
- {\rr \Sigma_{F}(t,t'';\bp)}
\rho(t'',t;\bp) \right] \partial_{t}F(t,t';\bp)|_{t=t'}
\nonumber\\ 
&&\,\qquad \db 
- \left[ {\rr \Sigma_{\rho}(t,t'';\bp)} \partial_{t} F(t'',t;\bp) 
- {\rr \Sigma_{F}(t,t'';\bp)} \partial_{t} \rho(t'',t;\bp) \right] 
F(t,t;\bp) \Big\} \nonumber
\eea
\end{minipage}}
\label{eq:exactn}
\eeq
Here $t_0$ denotes the initial time which was set to zero in
(\ref{eq:exactrhoF}) without loss of generality. 
Since $\rr \Sigma_F \sim \Or (1/N)$ as well as $\rr \Sigma_{\rho}$,
one directly observes that at LO, i.e.~for $N \to \infty$, the
particle number for each momentum mode is strictly conserved:
{\rr\em $\partial_t n_{\bp}(t) \equiv 0$ at LO.} 
Stated differently, eq.~(\ref{eq:effpart}) just 
specifies the infinite number of additional constants of motion 
which appear at LO. In contrast, once corrections beyond LO 
are taken into account then (\ref{eq:effpart}) no longer
represent conserved quantities.
The first non-zero contribution to the self-energies occurs
at NLO. As a consequence $\partial_t n_{\bp}(t) \not \equiv 0$ in general. 
 
For the LO approximation we have seen that there is an infinite 
number of time-translation invariant solutions (\ref{eq:effpart})
to the nonequilibrium evolution equations. Thermal equilibrium
plays no particular role at LO. However, this is very different
at NLO and beyond. 
It is very instructive to consider what is required to find  
solutions of (\ref{eq:exactn}) which are homogeneous in time.
Of course, any time-translation 
invariant solution cannot be achieved if one respects
all symmetries of the quantum field theory such
as time reflection symmetry. However, we can consider 
the limit $t_0 \to - \infty$ and ask under what conditions 
time-translation invariant $F(t,t';\bp) = F^{\rm (hom)}(t-t';\bp)$ 
etc.~represent solutions of (\ref{eq:exactn}). Using the \mbox{(anti-)}symmetry
of the (spectral function) statistical propagator we have
\bea\rr
\rho^{\rm (hom)}(t-t';\bp) &\rr \!=\!&\rr - i \int \frac{{\rm d} \omega}{2\pi}
\sin[\omega(t-t')]\, \rho^{\rm (hom)}(\omega,\bp) \, ,
\nonumber\\\db
F^{\rm (hom)}(t-t';\bp) &\db \!=\!&\db \int \frac{{\rm d} \omega}{2\pi}
\cos[\omega(t-t')]\, F^{\rm (hom)}(\omega,\bp) \, .
\label{eq:FouriertimeFrho}
\eea  
Note that in our conventions $\rho^{\rm (hom)}(\omega;\bp) 
= - \rho^{\rm (hom)}(- \omega;\bp)$ is
purely imaginary, while $F^{\rm (hom)}(\omega;\bp) 
= F^{\rm (hom)}(-\omega;\bp)$ 
is real. 
Since $\partial_{t}F^{\rm (hom)}(t-t';\bp)|_{t=t'} \equiv 0$,  
time-translation invariant solutions of (\ref{eq:exactn}) 
require:
\bea
&\db\lim\limits_{(t-t_0) \to \infty}&\db \!\!\!\!
\int_{t_0}^{t}\!\! d t'' \,
\left[ {\rr \Sigma^{\rm (hom)}_{\rho}(t-t'';\bp)} \partial_{t} 
F^{\rm (hom)}(t''-t;\bp) \right.  \nonumber\\ 
&& \qquad  \db \left. - {\rr \Sigma^{\rm (hom)}_{F}(t-t'';\bp)} 
\partial_{t} \rho^{\rm (hom)}(t''-t;\bp) \right] = 0
\, . 
\label{eq:homcondition}
\eea
Inserting the Fourier transforms (\ref{eq:FouriertimeFrho}) and
the respective ones for the self-energies and using 
\beq
\db
\lim_{(t-t_0) \to \infty} \frac{\sin \left[ \omega (t-t_0) \right]}{\omega}
\,=\, \pi \delta (\omega) \, ,
\label{eq:delta}
\eeq
one finds that the condition (\ref{eq:homcondition}) 
can be equivalently written as
\beq\db
\int {\rm d}\omega \, \omega \left[
{\rr \Sigma^{\rm (hom)}_{\rho}(\omega,\bp)} F^{\rm (hom)}(\omega,\bp)
- \Sigma^{\rm (hom)}_F(\omega,\bp) {\rr \rho^{\rm (hom)}(\omega,\bp)} 
\right] = 0 \, .
\label{eq:condfourier}
\eeq
In contrast to the LO approximation, at NLO and beyond
we find with $\Sigma^{\rm (hom)} \not = 0$ 
that time-translation invariant solutions are highly nontrivial.
However, one observes
from the equilibrium fluctuation-dissipation relation 
(\ref{eq:flucdissbose}) with (\ref{eq:sigmadecom}) that
indeed {\rr\em thermal equilibrium singles out a solution which
fulfills this condition.$\,$} We emphasize that in equilibrium 
the fluctuation-dissipation relation is fulfilled at all orders in the 
2PI $1/N$ expansion. In this respect there is no qualitative change 
going from NLO to NNLO or beyond. In particular, very similar results
as shown in Fig.~\ref{Figpeakphi} are obtained from the loop expansion
of the 2PI effective action beyond two-loop order, which is
demonstrated in Sec.~\ref{sec:latetimeuni}. Three-loop is 
required since the two-loop order corresponds to a Gaussian 
approximation, which suffers from the same infinite number
of spurious conserved quantities as the LO approximation.

We emphasize that the above discussion does not imply that the evolution
always has to approach thermal equilibrium at late times. There are many
initial conditions for which the system {\rr\em cannot thermalize}. 
Obvious examples are initial conditions inferred from statistical
ensembles which do not obey the clustering property. These
do not exhibit thermal correlation functions in general.

\subsubsection{Detour: Boltzmann equation}
\label{sec:detourboltzmann}

The nonequilibrium 2PI effective action can be employed to obtain
{\rr\em effective kinetic or Boltzmann-type descriptions for quasiparticle 
number distributions}. The latter are widely used in the literature
and it is important to understand their merits and limits. 
The ``derivation'' of such an equation from the 2PI effective action
requires a number of additional approximations which 
go beyond a loop or $1/N$ expansion of the 2PI effective action.
At the end of this procedure we will find an irreversible equation,
which therefore lost part of the symmetries of the underlying quantum field
theory. The comparison will allow us, in particular, to clearly
point out {\rr\em the role of neglected off-shell effects as compared to 
the quantum field theory description}.

Kinetic descriptions concentrate on the behavior of particle
number distributions. The latter contain information about the
statistical propagator $F(x,y)$ at equal times $x^0=y^0$.
Correlations between unequal times are outside the scope
of a Boltzmann equation. We can therefore start the discussion
from the {\rr\em exact equation for the effective 
particle number (\ref{eq:exactn})}.\footnote{For this it
is useful to rewrite the following characteristic term as:
\bea
-i \left(\Sigma_{\rho} F  - \Sigma_{F} \rho\right)  = 
\left( \Sigma_{F} -i \Sigma_{\rho}/2 \right)
\left( F +i \rho/2 \right) -
\left( \Sigma_{F} +i  \Sigma_{\rho}/2 \right)
\left( F  -i \rho/2 \right) \, .\nonumber
\eea
The difference of the
two terms on the r.h.s.~can be directly interpreted as the
difference of a ``loss'' and a ``gain'' term in
a Boltzmann type description.} 
For our current purposes it
is irrelevant that we will consider a spatially homogeneous one-component
field theory with vanishing field expectation value ($\phi = 0$). The steps
leading to the Boltzmann equation are summarized as follows:

{\rr 1.} 
Consider a {\rr\em 2PI loop expansion of $\Gamma[G]$ to three-loop order.}
This leads to self-energy corrections up to two loops,
diagrammatically given by \hspace*{0.1cm}
\parbox{1.cm}{
\centerline{\epsfig{file=2loopSELF.eps,width=1.cm}}
}$\,\,$ , $\,\,$ 
\parbox{1.cm}{
\centerline{\epsfig{file=SELF.eps,width=1.cm}}
}, which lead to the time-dependent mass term 
\beq\db
M^2(t) = m^2 + \frac{\lambda}{2}\, \int_{\bp} F(t,t;\bp)
\label{eq:loopM}
\eeq
and 
\bea \db
\Sigma_{F}(t,t';\bp) &\db\!\!\!=\!\!\!&\db -\frac{\lambda^2}{6} 
\int_{\bq,\bk} 
F(t,t';\bp-\bq-\bk)\! \left[F(t,t';\bq)F(t,t';\bk) 
- \frac{3}{4}\, {\rr \rho(t,t';\bq)\rho(t,t';\bk)} \right], 
\nonumber\\
\rr \Sigma_{\rho}(t,t';\bp) &\db\!\!\! =\!\!\!&\db -\frac{\lambda^2}{2} 
\int_{\bq,\bk}
{\rr \rho(t,t';\bp-\bq-\bk)}\!
\left[F(t,t';\bq)F(t,t';\bk) 
- \frac{1}{12}\, {\rr \rho(t,t';\bq)\rho(t,t';\bk)} \right]. 
\nonumber\\
\label{eq:3loopsigma}
\eea

{\rr 2.} 
Choose a {\rr\em quasiparticle ansatz}, i.e.~a free-field type form for
the two-point functions and their derivative. This corresponds to the
replacements on the r.h.s.~of (\ref{eq:exactn}):
\bea \db
F(t'',t;\bp) &\db \, \to \,&\db \left(n_{\bp} + 1/2\right) 
\cos \left[\omega_{\bp} (t''-t) \right]/ \omega_{\bp} \, ,
\nn
\rr \rho (t'',t;\bp) &\rr\, \to \,&\rr  
\sin \left[\omega_{\bp} (t''-t) \right]/ \omega_{\bp} \, ,
\label{eq:freefieldFrho}\\[0.2cm]
\db \partial_{t} F(t'',t;\bp) &\db \, \to \,&\db \left( n_{\bp} + 1/2 \right) 
\sin \left[\omega_{\bp} (t''-t) \right] \, ,
\nn
\rr \partial_{t} \rho (t'',t;\bp) &\rr\, \to \,&\rr 
- \cos \left[\omega_{\bp} (t''-t) \right] \, .
\eea 

{\rr 3.} Here the particle number $n_{\bp}$ and mode energy
$\omega_{\bp}$ may still depend weakly on time: 
One assumes a {\rr\em separation of scales}, with a
sufficiently slow time variation of $n_{\bp}(t)$  
such that one can pull all factors of $n_\bp(t)$
out of the time integral on the r.h.s.~of (\ref{eq:exactn}).
The particle numbers are then evaluated at the latest time
of the memory integral.

{\rr 4.} {\rr\em Send the initial time $t_0$ to the remote past: 
$t_0 \to -\infty$.} Of course, by construction the resulting equation 
is not meant to describe the detailed early-time
behavior since $t_0 \to -\infty$. In the context of kinetic
descriptions, one finally specifies the initial condition for the effective 
particle number distribution at some finite time and approximates
the evolution by the equation with $t_0$ in the
remote past.

A standard alternative derivation employs first steps {\rr 1.} 
and {\rr 4.}, then a first-order gradient expansion 
of the two-point functions in the center coordinate $(t+t')/2$
and then a quasiparticle ansatz. However, for the typically
employed first-order gradient approximation both approaches are 
fully equivalent. The current procedure has the advantage
that one can send the initial time $t_0 \to - \infty$ last,
which allows one to discuss a few finite-time effects that are
typically discarded.
Applying the assumptions {\rr 1.--3.} to the exact equation
(\ref{eq:exactn}) leads after some lengthy shuffling of terms
to the expression: 
\bea\lefteqn{
\db \partial_t n_{\bp}(t) \,\,=\,\, \frac{\lambda^2}{3} \int_{\bs\bq\bk}
(2\pi)^d \delta(\bp - \bq - \bk - \bs) 
\frac{1}{2 \omega_{\bp}2 \omega_{\bq}2 \omega_{\bk}2 \omega_{\bs}} }\nn
&& \db \Bigg\{
{\db \left[ (1+n_{\bp})(1+n_{\bq})(1+n_{\bk})(1+n_{\bs}) 
- n_{\bp}n_{\bq}n_{\bk}n_{\bs} \right] \qquad\quad \mbox{\bf (I)} }
\nn
&&{\db \int_{t_0}^t dt'' \cos 
\left[ (\omega_{\bp} + \omega_{\bq} + \omega_{\bk} + \omega_{\bs})(t-t'') 
\right]}
\nn 
&\db +& 
{\db 3 \left[ (1+n_{\bp})(1+n_{\bq})(1+n_{\bk}) n_{\bs} 
- n_{\bp}n_{\bq}n_{\bk} (1+n_{\bs}) \right]\qquad\quad\,\, \mbox{\bf (II)}}
\nn
&&{\db \int_{t_0}^t dt'' \cos 
\left[ (\omega_{\bp} + \omega_{\bq} + \omega_{\bk} - \omega_{\bs})(t-t'') 
\right]}
\nn 
&\rr +& 
{\rr 3 \left[ (1+n_{\bp})(1+n_{\bq}) n_{\bk} n_{\bs} 
- n_{\bp}n_{\bq} (1+n_{\bk}) (1+n_{\bs}) \right] \qquad\quad\,\, 
\mbox{\bf (III)}}
\nn
&&{\rr \int_{t_0}^t dt'' \cos 
\left[ (\omega_{\bp} + \omega_{\bq} - \omega_{\bk} - \omega_{\bs})(t-t'') 
\right]}
\nn 
&\db +&
{\db \left[ (1+n_{\bp}) n_{\bq} n_{\bk} n_{\bs} 
- n_{\bp} (1+n_{\bq}) (1+n_{\bk}) (1+n_{\bs}) \right] \qquad\quad\,\,\,\,\,\, 
\mbox{\bf (IV)}}
\nn
&&{\db \int_{t_0}^t dt'' \cos 
\left[ (\omega_{\bp} - \omega_{\bq} - \omega_{\bk} - \omega_{\bs})(t-t'') 
\right]} \db
\Bigg\} \, .
\label{eq:preboltzmann}
\eea
The contributions on the r.h.s.~of this equation have a typical 
``gain \& loss'' structure with a simple interpretation: 
\begin{itemize}
\item[{\db{(I)}}] \vspace*{-0.1cm}
 describes production and annihilation of four
``quasiparticles'' ($0 \to 4$, $4 \to 0$).
\item[{\db{(II)}}] \vspace*{-0.1cm}
\rr and {\db (IV)} describe $1 \to 3$ and $3 \to 1$
processes.  
\item[{\rr{(III)}}] \vspace*{-0.1cm}
 describes $2 \leftrightarrow 2$ scattering 
processes, which are the well-known contributions to
the standard Boltzmann equation.
\end{itemize}
We emphasize that {\rr\em only the processes described by {\rr (III)} 
lead to a non-zero contribution in the limit $\rr (t-t_0) \to \infty$}.
For time-independent mode energies $\omega_\bp$ one can perform
the time integrations in (\ref{eq:preboltzmann}) explicitly. This leads using 
(\ref{eq:delta}) to $\delta$-functions, which 
enforce the strict conservation of the mode energies involved
in the scattering. At this point the description of the
quantum field theory has been fully reduced to the physics of
colliding ``on-shell'' quasiparticles. Of course, strict energy and momentum 
conservation for these quasiparticles leads to
vanishing contributions {\db (I)}, {\db (II)} and
{\db{(IV)}}. The latter describe processes, which change
the total particle number such as $1 \to 3$ processes. 
We will see below that {\rr\em in a quantum field theory ``off-shell'' particle
number changing processes such as described by {\db{(II)}} can
play a very important role!} An obvious example is given by the
fact that the Boltzmann equation including the contribution {\rr (III)} 
admits grand canonical
thermal solutions with nonzero chemical potential $\mu$: 
\beq
\rr n_\bp(t) \to \frac{1}{e^{(\omega_\bp - \mu)/T}-1} \,\, .
\eeq
However, the real scalar quantum field theory is charge neutral and
$\mu \equiv 0$! We conclude that taking into account 
only $2 \leftrightarrow 2$ scattering processes of ``on-shell''
quasiparticles obviously fails to describe the quantum world
if for a given initial condition the chemical potential turns
out to be large. We emphasize that total particle number
changing processes are included
in the original quantum field theory equation before applying the 
additional steps~{\rr 2.--4.} In particular, they do not only
appear as contributions from finite initial-time effects
as discussed here for illustration. The main ``off-shell'' effects 
are removed in step~{\rr 2.}, when the quasiparticle assumption
enters the description. 

The physics of ``off-shell'' processes can be very nicely
observed in the $1+1$ dimensional quantum field theory.
Note that for one spatial dimension the contribution {\rr{(III)}}
vanishes as well in the limit $\rr (t-t_0) \to \infty$! 
In this case both momentum and energy conservation,
$p + q - k - s = 0$ $\wedge$ $\sqrt{p^2 + M^2} + \sqrt{q^2 + M^2} 
- \sqrt{k^2 + M^2} - \sqrt{s^2 + M^2} = 0$, lead to an
``ineffective'' two-to-two scattering where the incoming and outcoming
modes are unaffected: for instance with $p = k$ and $q = s$
the contribution {\rr{(III)}} in Eq.~(\ref{eq:preboltzmann}) 
vanishes identically. The equation becomes trivial in this case.
We conclude that for the $1+1$ dimensional quantum field theory there is 
{\rr\em no dynamics from the Boltzmann equation!}
In contrast, from the same three-loop approximation of the 2PI effective
action without the steps~{\rr 2.--4.} we will see in the next 
section that the $1+1$ dimensional quantum field theory thermalizes.

\subsubsection{Characteristic time scales}
\label{sec:latetimeuni}

\begin{figure}[t]
\epsfig{file=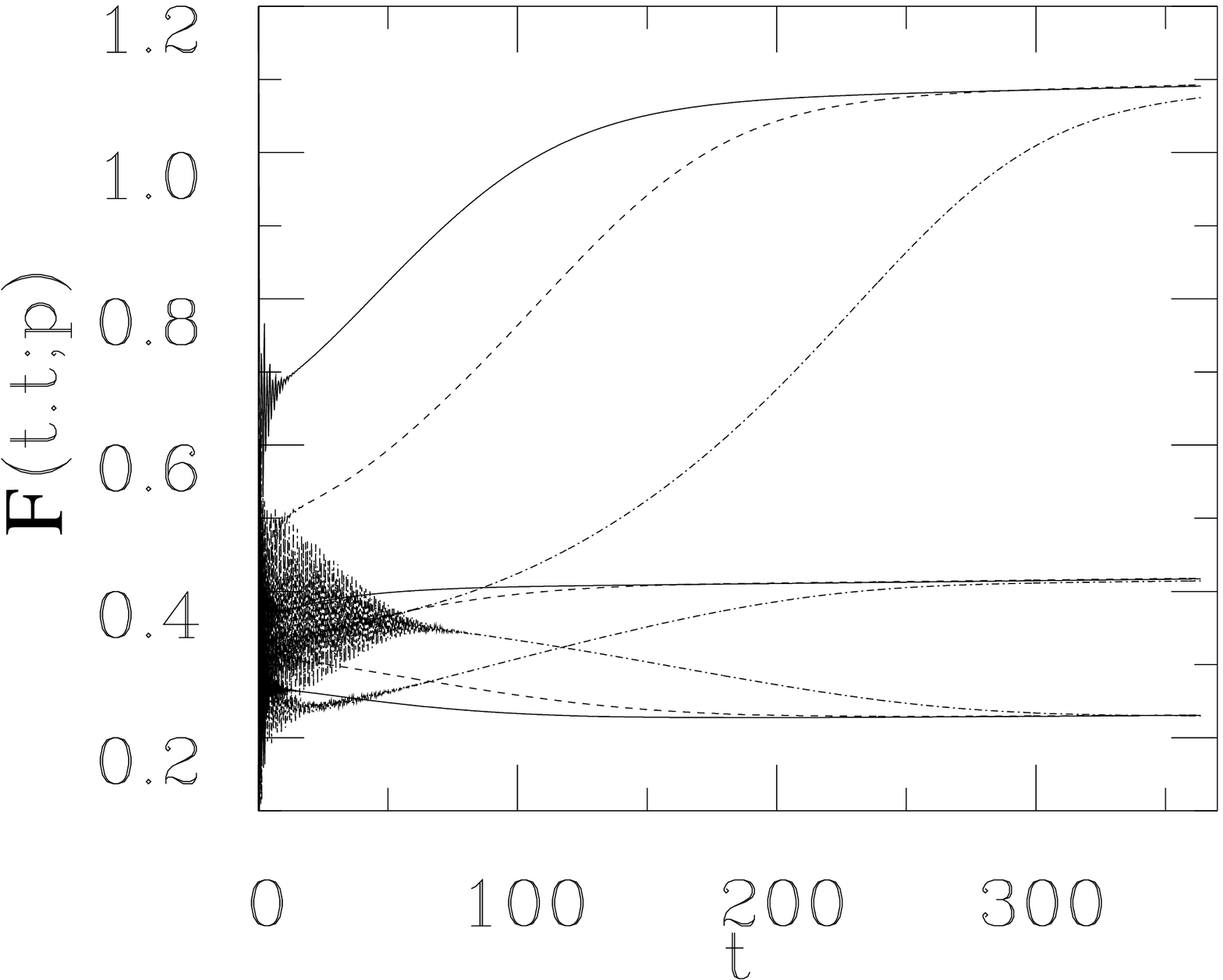,width=7.cm}
\hspace*{0.4cm}
\epsfig{file=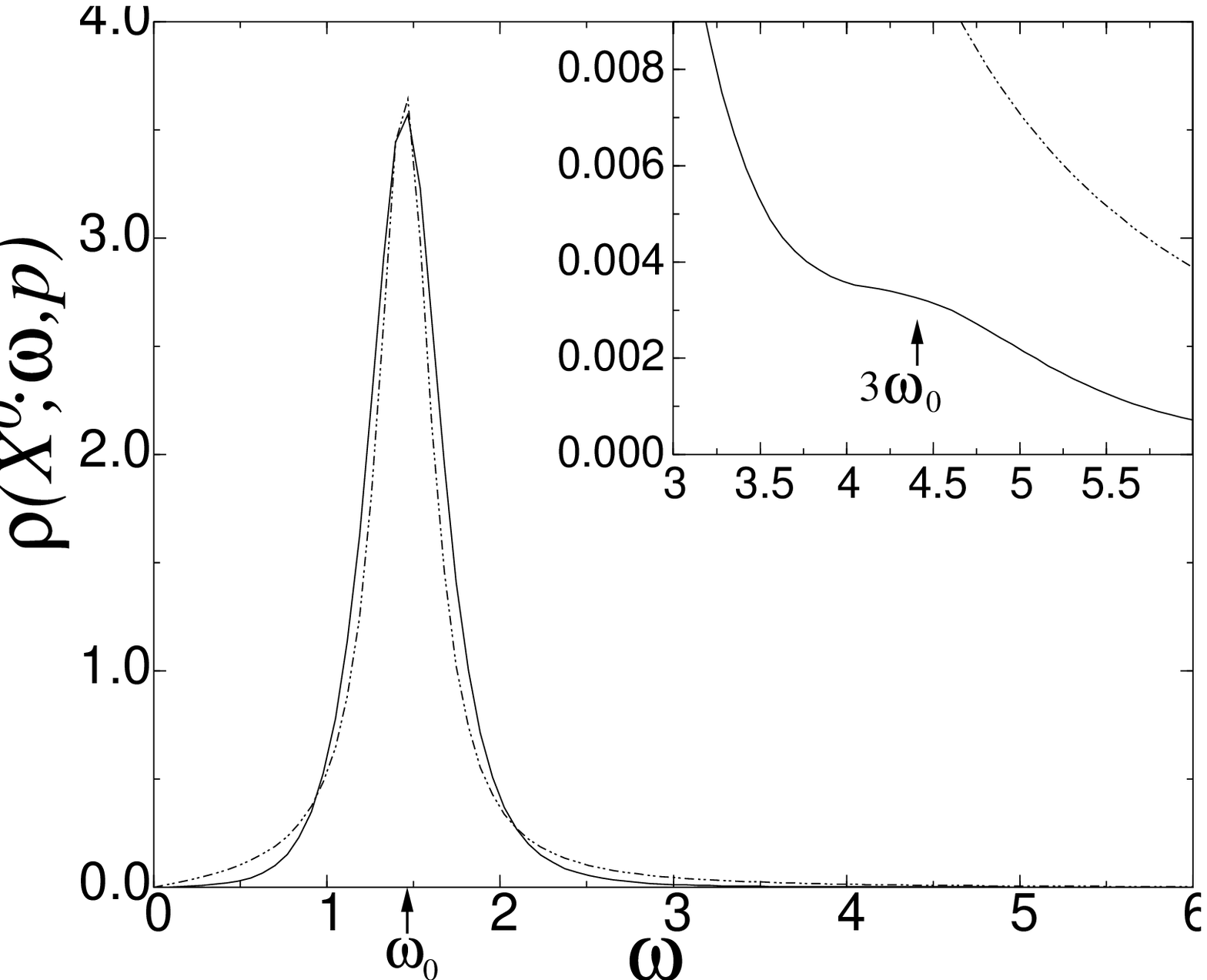,width=7.cm}
\caption{{\bf LEFT:} Examples for the time dependence of the equal-time 
propagator $F(t,t;\bp)$ with Fourier modes $p=0,3,5$ 
from the 2PI three-loop effective action. 
The evolution is shown for three very different nonequilibrium
initial conditions with the same energy density (all in initial-mass units). 
{\bf RIGHT:} Wigner transform $\varrho(X^0;\om,p)$ of the spectral
function as a function of $\om$ at $X^0=35.1$ for $p=0$ (in units of $m_R$). 
Also shown are fits to a Breit-Wigner function (dotted) with 
$(\omega_{0},\Gamma_{0}) = (1.46,0.37)$. The inset shows a blow-up around 
$3\omega_{0}$. The expected bump from ``off-shell'' 
$1 \leftrightarrow 3$ processes is small but visible.}
\label{fig:lateuni}
\end{figure}
Nonequilibrium dynamics requires the
specification of an initial state. A crucial question of thermalization is
how quickly the nonequilibrium system effectively
looses the details about the initial conditions, and 
what are the characteristic stages of a partial loss of information.
Thermal equilibrium keeps no memory about 
the time history except for the values of a few conserved charges.
As a consequence, for the real scalar field theory thermalization
requires that the late-time result 
is uniquely determined by energy density.

In Fig.~\ref{fig:lateuni} we show the
time dependence of the equal-time propagator $F(t,t;\bp)$
for three Fourier modes $p =0,3,5$ and three very {\rr\em different 
initial conditions with the same energy density.} All quantities
are given in appropriate units of the mass at initial time,
and we consider $1+1$ dimensions.
For the solid line the initial conditions are close
to a mean field thermal solution, the initial mode 
distribution for the dashed and the dashed-dotted lines deviate 
more and more substantially from a thermal equilibrium distribution.  
It is striking to observe that propagator modes with very different
initial values but with the same momentum $p$ 
approach the same large-time value. The asymptotic behavior of the 
two-point function modes are universal and uniquely determined by the initial 
energy density. 

One observes that after an effective damping of rapid oscillations 
the modes are still far from equilibrium. 
All correlation functions quickly approach an 
exponentially damped
behavior. A characteristic rate $\gamma^{\rm (damp)}_0$ can be obtained 
from the zero mode of the unequal-time two-point function
$F(t,0;p=0)$ with a 
corresponding time scale proportional to the inverse rate. 
In this early-time range correlations with the initial time are effectively
suppressed and asymptotically $F(t,0;p=0) \to 0^+$. 
The early-time range is followed by a smooth ``drifting'' of modes,
which is characterized by a slow dependence of $F(t,t';p=0)$ on
$(t+t')/2$. The presence of such a regime is a prerequisite 
for descriptions based on gradient expansions in the center coordinate 
$(t+t')/2$. Clearly, the early-time behavior for times 
$\sim 1/\gamma^{\rm (damp)}_0$ is beyond the scope of
a gradient expansion. Though the exponential damping at early times
is crucial for an effective loss of details of the initial conditions,
it does not determine the time scale for thermalization.
One typically finds very different
rates for damping and for the late-time approach of $F(t,t;p=0)$ 
to thermal equilibrium. 
\begin{figure}[b]
\centerline{\epsfig{file=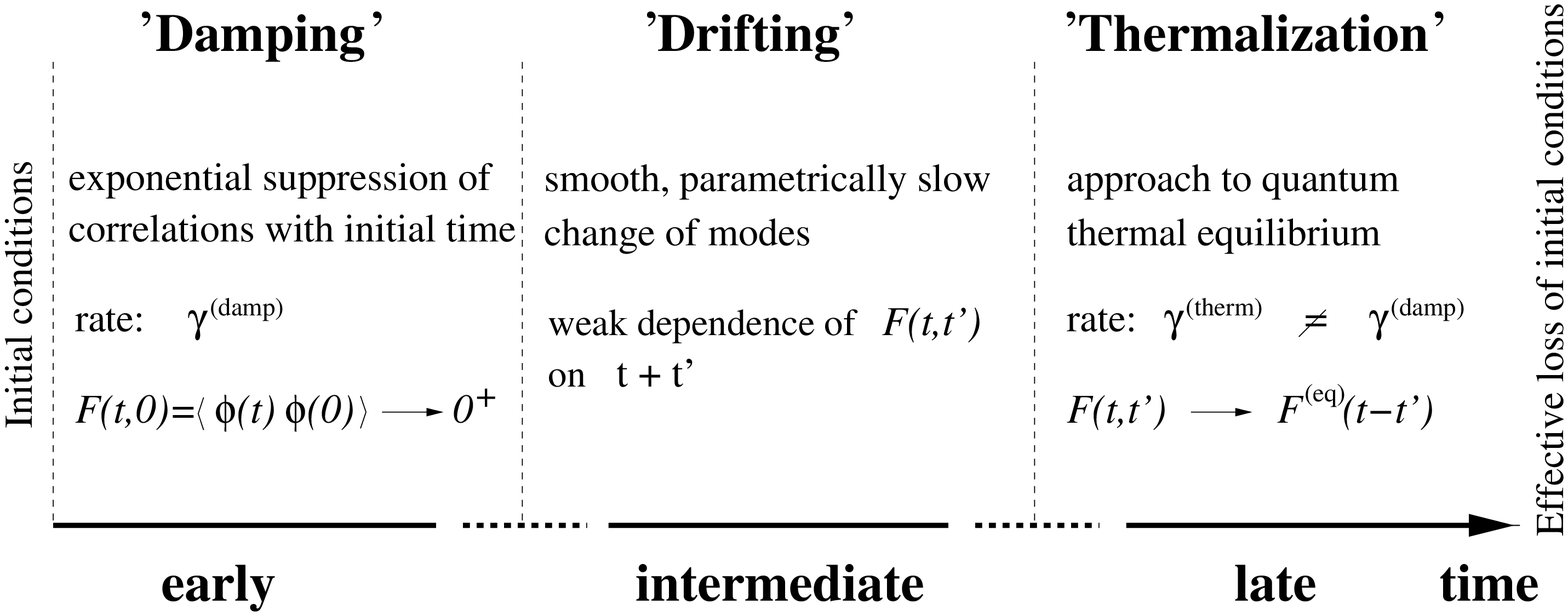,
width=14.cm}}
\end{figure}

One of the ``key'' properties
for the success of the quantum field theoretical description
is the nontrivial (i.e.~not the free-field or ``$\delta$''-type) 
{\rr\em dynamical spectral function $\rho\,$}. To analyze the 
spectral function (not to solve the dynamics!), we perform a 
Wigner transformation for the
modes $\rho(t,t';\bp)$ and write with $X^0=(t+t')/2$, $s^0= t-t'$: 
\beq\db
i\varrho(X^0; \omega, p) = \int_{-2X^0}^{2X^0}ds^0\, e^{i\omega s^0}
\rho(X^0+s^0/2, X^0-s^0/2;p) \, .
\label{eqW}
\eeq
The $i$ is introduced such that $\varrho(X^0;\om,\bp)$ is real. Because the 
spectral function is antisymmetric, $\varrho(X^0;\om, \bp) = 
-\varrho(X^0;-\om, \bp)$, and we will present the positive-frequency part
only. Since we consider an initial-value problem with  
$t,t'\geq 0\,$, the time integral over $s^0=t-t'$ is bounded by
$\pm 2X^0$.

On the right of Fig.~\ref{fig:lateuni} we display the Wigner
transform for the zero momentum mode as a function of $\omega$.
One clearly observes that the interacting theory has a
continuous spectrum described by a peaked spectral function with a nonzero
width. The peak is located at $\omega_0/m_R = 1.46$ in units
of the initial renormalized mass (cf.~Sec.~\ref{sec:lofixedpoints}),
and the results are shown for $m_R X^0=35.1$ with 
$\lambda/m_R^2=4$.\footnote{Here we used a ``tsunami''
similar to the one discussed in Sec.~\ref{sec:lofixedpoints}, however,
the shown results are insensitive to the initial condition details.}
The inset shows a blow-up around $3\omega_0/m_R=4.38$. 
The expected bump in the spectral function is
small but visible. We stress that this bump in the spectral function is
kinematically forbidden for the ``on-shell'' approximation and arises from
``off-shell'' $1 \leftrightarrow 3$ processes. 
In Fig.\ \ref{fig:lateuni} we also present fits to a Breit-Wigner spectral
function
\beq\db
\varrho_{\rm BW}(X^0;\om,\bp) =
\frac{2\om\Gamma_\bp(X^0)}{[\om^2-\omega_\bp^2(X^0)]^2
+ \om^2\Gamma^2_\bp(X^0)}
\label{BreitWigner} \, ,
\eeq
with a width $\Gamma_\bp(X^0)=2\gamma^{\rm (damp)}_\bp(X^0)$.
While the position of the peak can be fitted
easily, the overall shape and width are only qualitatively captured.
In particular, the slope of $\varrho(X^0;\om,p)$ for small $\om$ is 
quantitatively different. We also see that the Breit-Wigner fits give
a narrower spectral function (smaller width) and therefore would predict a
slower exponential relaxation in real time. 

The {\rr\em characteristic time scales} observed for the nonequilibrium 
evolution of modes in Fig.~\ref{fig:lateuni} can be associated to 
\begin{enumerate}
\item rapid oscillations of correlation functions with period
 $\sim 1/\omega_{\bp}$ {\rr\em described by the ``peak'' of the
spectral function.}
\item damping of oscillations with inverse 
rate $1/\gamma_{\bp}^{\rm (damp)}$ {\rr\em described by a nonzero ``width'' 
$\Gamma_\bp = 2\gamma^{\rm (damp)}_\bp$ 
of the spectral function.} In equilibrium 
the ``width'' is given by
\bea \db
2 \omega \Gamma^{\rm (eq)}(\omega,\bp) \equiv 
- \Sigma_{\rho}^{\rm (eq)}(\omega,\bp) 
\,\sim\, \Or (\lambda^2/N) \, . 
\eea  
\item {\rr\em late-time thermalization with inverse 
rate $1/\gamma_{\bp}^{\rm (therm)}$ because of ``off-shell'' number 
changing processes.} For the three-loop or NLO in $1/N$ approximation of the
2PI effective action the processes changing the total particle number 
are perturbatively of order $\sim \lambda^4/N^2$ (``slow!''). 
This can be understood
from the fact that the total particle number changing processes require
a nonzero ``width'' $\sim \lambda^2/N$, and this width enters
in $\Or (\lambda^2/N)$ evolution equations for the two-point functions. 
\end{enumerate}
One finds qualitatively the same characteristic ranges for the
corresponding nonequilibrium evolution in $3+1$ dimensions, 
and we will consider results for the chiral ``quark-meson'' model
below. An important quantitative difference between one and three
spatial dimensions results from the fact that 
``on-shell'' $2 \leftrightarrow 2$ processes
contribute for the latter. However, as for the $1+1$ dimensional case 
they do not change the total ``quasiparticle'' 
number. Total number changing ``off-shell'' or ``on-shell'' processes 
are required in general to reach thermal equilibrium. As we have seen
above, ``off-shell'' number 
changing processes appear at NLO in the 2PI $1/N$ expansion. In contrast,
they can be achieved ``on-shell'' first at NNLO. E.g.~for $\phi^4$-theory 
the lowest order ``on-shell'' ($2 \to 4$) contribution appears
from the five-loop diagram (cf.~Sec.~\ref{sec:loopexp}): 

\vspace*{0.2cm}

\centerline{\epsfig{file=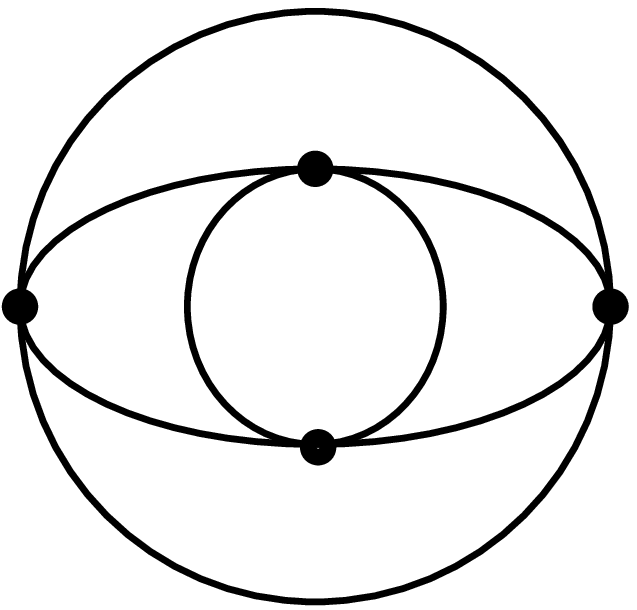,width=1.5cm}}

\vspace*{0.2cm}

\noindent
Perturbatively, this contribution is of the same order $\sim \lambda^4/N^2$
than the ``off-shell'' particle number changing processes arising 
at three-loop or NLO in the $1/N$ expansion of the 2PI effective action.
In this respect it is interesting to observe that precision tests
for the nonequilibrium dynamics as discussed in Sec.~\ref{sec:precision} 
indicate accurate results already at NLO in the 2PI expansion.

\subsection{Prethermalization}
\label{sec:prethermal}

\begin{figure}[t]
\centerline{\epsfig{file=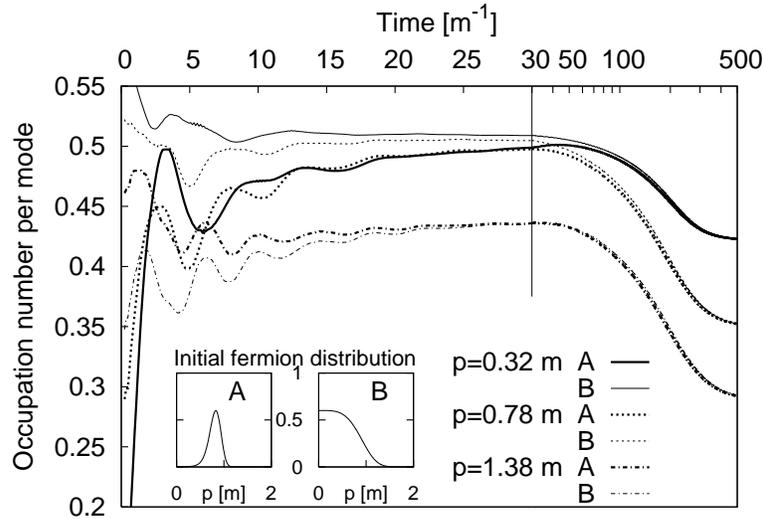,width=10.5cm}}
\caption{Fermion occupation number $n^{(f)}(t;p)$
for three different momentum modes as a function of time. 
The evolution is shown for two different initial
conditions with {\em same} energy density. 
The long-time behavior is shown on a 
logarithmic scale for $t \ge 30\, m^{-1}$.}
\label{fig:join_fn}
\end{figure}
Prethermalization is a universal far-from-equilibrium phenomenon
which describes the very rapid establishment of an almost
constant ratio of pressure over energy density (equation
of state), as well as a kinetic temperature based on
average kinetic energy. The phenomenon occurs on time scales 
dramatically shorter than the thermal equilibration time.
As a consequence, prethermalized quantities approximately
take on their final thermal values already at a time
when the occupation numbers of individual momentum 
modes still show strong deviations from the late-time
Bose-Einstein or Fermi-Dirac distribution. 

Here we consider the nonequilibrium evolution of
quantum fields for a low-energy  
quark-meson model, which is described in Sec.~\ref{sec:chiralqmm}.
It takes into account two quark flavors with a Yukawa coupling $\sim h$ 
to a scalar $\sigma$-field and a triplet of pseudoscalar pions, $\vec{\pi}$. 
The theory corresponds to the well-known ``linear $\sigma$-model'',
which incorporates the chiral symmetries of massless two-flavor QCD.  
The employed couplings in the action (\ref{chiralfermact})
with (\ref{eq:potterm}) are taken to be of order one, and if
not stated otherwise $h=\lambda=1$.
We emphasize that the main results about
prethermalization are independent
of the detailed values of the couplings.
Here we employ the 2PI effective action (\ref{totaleffact}) to 
two-loop order given by (\ref{eq:chiraltwoloop}).~\footnote{For
the relation to a nonperturbative expansion of the 2PI effective action to 
next-to-leading order in $N_f$ see~Secs.~\ref{sec:chiralqmm}
and \ref{sec:2PIN}).} All quantities
will be given in units of the scalar thermal
mass $m$.\footnote{The thermal mass $m$ is evaluated in equilibrium.
It is found to prethermalize very rapidly. The employed spatial
momentum cutoff is $\Lambda/m = 2.86$.}  

\begin{figure}[t]
\centerline{\epsfig{file=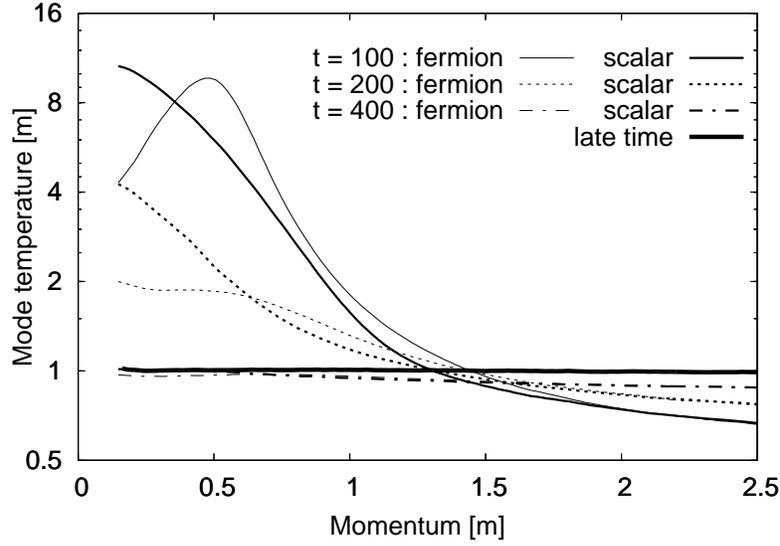,width=10.5cm}}
\caption{Fermion and scalar mode temperatures $T^{(f,s)}_p(t)$ as 
a function of momentum $p$ for various times.}
\label{fig:Tdist}
\end{figure}
{\em Thermalization:} 
In Fig.~\ref{fig:join_fn} we show the effective occupation
number density of fermion momentum modes, $n^{(f)}(t;p)$, 
as a function of time for three different 
momenta~\footnote{This quantity is directly related to the expectation value
of the vector component of the field commutator 
$\langle [\psi,\bar{\psi}] \rangle$ in Wigner
coordinates and fulfills $0 \le n^{(\rm f)}(t;p) \le 1$.}.
The plot shows two runs denoted as (A) and (B) 
with different initial conditions but same energy density.
Run (A) exhibits a high initial particle
number density in a narrow momentum range around
$\pm p$. This situation is reminiscent of  
two colliding wave packets with opposite 
and equal momentum. We emphasize, however, that we 
are considering a spatially homogeneous and isotropic
ensemble with a vanishing net charge density.   
For run (B) an initial particle number density is employed which
is closer to a thermal distribution. 

One observes that for a given momentum the mode numbers
of run (A) and (B) approach each other at early times. 
The characteristic time scale for
this approach is well described by the damping
time $t_{\rm damp}(p)$~\footnote{The rate $1/t_{\rm damp}(p)$ is 
determined by the spectral component of the 
self-energy.}. 
Irrespective of the initial distributions (A) or (B), we find 
(for $p/m\simeq 1$)
$t_{\rm damp}^{(f)} \simeq 25\, m^{-1}$ for fermions
and $t_{\rm damp}^{(s)} \simeq 28\, m^{-1}$ for 
scalars. In contrast to the initial rapid changes, 
one observes a rather slow or ``quasistationary'' subsequent
evolution. The equilibration time
$t_{\rm eq} \simeq 95\, m^{-1}$ 
is substantially larger than
$t_{\rm damp}$ and is approximately the same for fermions and 
scalars. Thermal equilibration is a collective phenomenon which is,
in particular, rather independent of the momentum.
As we have also observed for the $1+1$ dimensional scalar
theory in Sec.~\ref{sec:latetimeuni}, mode quantities such as effective
particle number distribution functions show a characteristic 
two-stage loss of initial conditions: after the 
damping time scale much of the 
details about the initial conditions are effectively lost. 
However, the system is still far from equilibrium and  
thermalization happens on a much larger time scale.

\begin{figure}[t]
\centerline{\epsfig{file=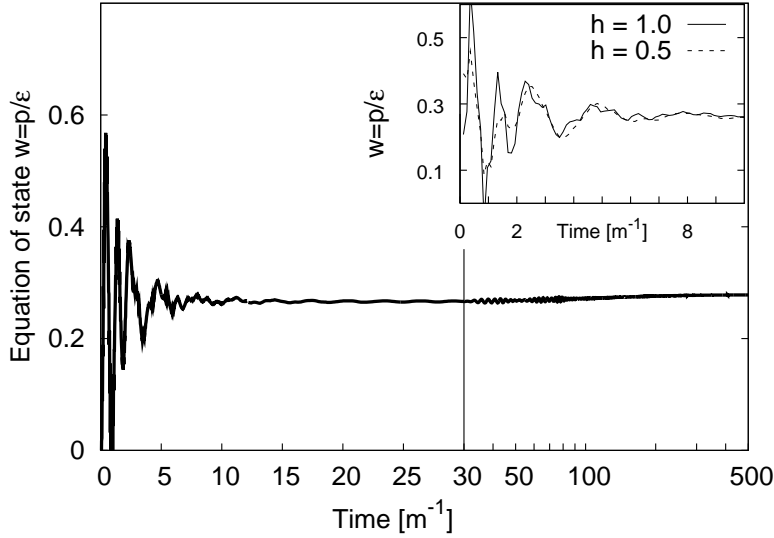,width=10.5cm}}
\caption{The ratio of pressure over energy density $w$ as a function of
time. The inset shows the early stages 
for two different couplings and demonstrates
that the prethermalization
time is independent of the interaction details.}
\label{fig:wevol}
\end{figure}
We define mode temperatures $T^{(f,s)}_p(t)$
by equating the mode particle numbers $n^{(f,s)}_p(t)$ with a 
time and momentum dependent Bose-Einstein or Fermi-Dirac distribution, 
respectively:
\beq\db
n_p(t) \stackrel{!}{=} 
\left[ \exp\left(\omega_p(t)/T_p(t) \right) 
\pm 1 \right]^{-1} \, .
\label{eq:occup}
\eeq
This definition is a quantum mechanical
version of its classical counterpart as defined by the
squared ``generalized velocities''.
In thermal equilibrium with $\omega_p \simeq \sqrt{p^2 + M^2}$
and $T_p = T_{\rm eq}$
equation (\ref{eq:occup}) yields the familiar occupation numbers ($\mu=0$).
Here the mode frequency $\omega^{(f,s)}_p(t)$ is determined by the peak of the
spectral function for given time and momentum, as detailed for the
scalar theory in Sec.~\ref{sec:latetimeuni}. 
In Fig.~\ref{fig:Tdist} we show the fermion and 
scalar mode temperature as a function of momentum for various 
times $t \gg t_{\rm damp}$. 
One observes that at late times, when thermal equilibrium is approached,
all fermion and scalar mode temperatures become constant 
and agree: $T_p^{(f)}(t) = T_p^{(s)}(t) = T_{\rm eq}$. In contrast,
there are sizeable deviations from the thermal result
even for times considerably larger than the characteristic damping 
time.  

\begin{figure}[t]
\centerline{\epsfig{file=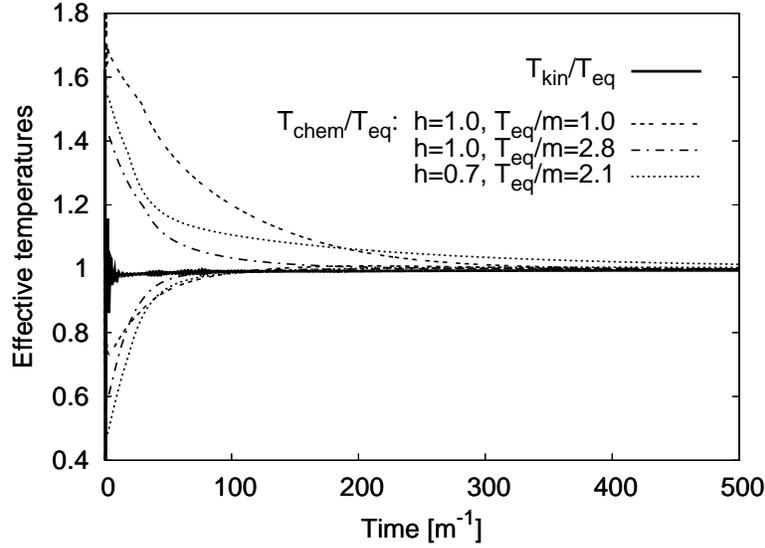,width=10.5cm}}
\caption{Chemical temperatures for scalars (upper curves) and
fermions (lower curves) for different 
values of the coupling $h$ and $T_{\rm eq}$.
We also show the kinetic temperature $T_{\rm kin}(t)$ (solid line), 
which prethermalizes on a very short time scale as compared to
chemical equilibration.}
\label{fig:Tkinevolrat}
\end{figure}
{\em Kinetic prethermalization:}
In contrast to the rather long thermalization time,
prethermalization sets in extremely rapidly. 
In Fig.~\ref{fig:wevol} 
we show the ratio of pressure over energy density,
$w = p/\epsilon$, as a function of time. One observes that
an almost time-independent equation of state builds up 
very early, even though the system is still far from equilibrium!
The prethermalization time $t_{\rm pt}$ is here 
of the order of the characteristic
inverse mass scale $m^{-1}$. This is a typical consequence of
the loss of phase information by summing over oscillating functions 
with a sufficiently dense frequency spectrum. In order to see that this 
phenomenon is not related to scattering or 
to the strength of the interaction, we compare
with a smaller coupling in the inset and observe good agreement
of both curves. The dephasing phenomenon is 
unrelated to the scattering-driven process of thermalization.

Given an equation of state, the question arises 
whether there exists a suitable definition of a global 
kinetic temperature $T_{\rm kin}$. In contrast to a mode quantity such 
as $T_p(t)$, a temperature
measure which averages over all momentum modes may prethermalize.
Building on the classical association of temperature with
the mean kinetic energy per degree of freedom, we use here
a definition based on the total kinetic 
energy $E_{\rm kin}(t)$:
\beq\db
T_{\rm kin}(t) = E_{\rm kin}(t)/c_{\rm eq}\,  . 
\eeq
Here the extensive dimensionless proportionality constant  
$c_{\rm eq} = E_{\rm kin,eq}/T_{\rm eq}$ is given 
solely in terms of equilibrium
quantities\footnote{For a relativistic plasma
one has $E_{\rm kin}/N = \epsilon/n = \alpha T$.
As alternatives, one may consider the weighted average
$\bar{T}(t) = \sum n(t;p) T(t;p)/ \sum n(t;p)$ where the sum is over 
all modes, or a definition analogous to Eq.~(\ref{eq:chem}) below.}. 
Since total energy is conserved, the time scale 
when ``equipartition'' is reached (i.e.~$E_{\rm kin}/E$ is 
approximately constant) also corresponds to a time-independent 
kinetic temperature. The latter equals the equilibrium temperature
$T_{\rm eq}$ if $E_{\rm kin}/E$ has reached the thermal value.

The solid line of Fig.~\ref{fig:Tkinevolrat} shows $T_{\rm kin}(t)$
normalized to the equilibrium temperature (for $T_{\rm eq}/m = 1$). 
One observes that an almost time-independent kinetic temperature is
established after the short-time scale $t_{\rm pt} \sim m^{-1}$. 
The time evolution of bulk quantities such
as the ratio of pressure over energy density $w$, or the
kinetic temperature $T_{\rm kin}$, are dominated by
a single short-time scale. These quantities approximately
converge to the thermal equilibrium values already at early
times and can be used for an efficient ``quasi-thermal'' 
description in a far-from-equilibrium situation!

{\em Chemical equilibration:}
In thermal equilibrium the relative particle numbers of different species 
are fixed in terms of temperature and particle masses. A system
has chemically equilibrated if these ratios are reached, as observed for 
the hadron yields in heavy ion collisions.
Obviously, the chemical equilibration time $t_{\rm ch}$ will
depend on details of the particle number changing interactions in a 
given model and $t_{\rm ch} \le t_{\rm eq}$. In our
model we can study the ratio between the numbers of fermions
and scalars. For this purpose we introduce the 
chemical temperatures $T_{\rm ch}^{(f,s)}(t)$ 
by equating the integrated number density
of each species, $n^{(f,s)}(t) = g^{(f,s)}
\int {\rm d}^3 p/(2 \pi)^3\, n^{(f,s)}_p(t)$, 
with the integrated Bose-Einstein/Fermi-Dirac form of distributions:
\beq\db
n(t) \stackrel{!}{=} 
\frac{g}{2 \pi^2} \int_{0}^{\infty}\! {\rm d} p p^2
 \left[ \exp\left(\omega_p(t)/T_{\rm ch}(t) \right) 
\pm 1 \right]^{-1} \, .
\label{eq:chem}
\eeq      
Here $g^{(f)}=8$ counts the number of fermions and 
$g^{(s)}=4$ for the scalars. 

The time evolution of the ratios 
$T_{\rm ch}^{(s,f)}(t)/T_{\rm eq}$ is shown
in Fig.~\ref{fig:Tkinevolrat} 
for different values of the coupling constant $h$ and  
the equilibrium temperature $T_{\rm eq}$. 
One observes that chemical equilibration with 
$T_{\rm ch}^{(s)}(t) = T_{\rm ch}^{(f)}(t)$ does not
happen on the prethermalization time scale, in contrast
to the behavior of $T_{\rm kin}(t)$.
Being bulk quantities, the scalar and fermion 
chemical temperatures can approach each other rather quickly at first. 
Subsequently, a slow evolution towards equilibrium sets in.   
For the late-time chemical equilibration we find for
our model $t_{\rm ch} \simeq t_{\rm eq}$. However, the
deviation from the thermal result can become relatively small
already for times $t \ll t_{\rm eq}$. 

Let us finally consider our findings in view of collisions of heavy
nuclei and try to estimate the prethermalization time. Actually,
$t_{\rm pt}$ is rather independent of the details of the model
like particle content, values of couplings etc. It mainly reflects
a characteristic frequency of the initial oscillations. 
If the ``temperature'' (i.e.~average kinetic energy per mode)
sets the relevant scale one expects
$T\, t_{\rm pt} = {\rm const}$.
(For low $T$ the scale will be replaced by the mass.)
For our model we indeed find 
$T\, t_{\rm pt} \simeq 2 - 2.5$.~\footnote{We
define $t_{\rm pt}$ by $|w(t_{\rm pt}) - w_{\rm eq}|/w_{\rm eq}
< 0.2$ for $t > t_{\rm pt}$.} We expect
such a relation with a similar constant to hold for the quark-gluon
state very soon after the collision. For $T \gtrsim 400 - 500\,$MeV we
obtain a very short prethermalization
time $t_{\rm pt}$ of somewhat less than $1\,$fm. This is consistent 
with observed very early hydrodynamic behavior in collision
experiments, however, one further has to
investigate the required isotropization of pressure.

\subsection{Far-from-equilibrium field dynamics: Parametric resonance}
\label{sec:resonance}

In classical mechanics 
parametric resonance is the phenomenon of resonant amplification of 
the amplitude of an oscillator having a time-dependent periodic frequency.
In the context of quantum field theory a similar phenomenon describes 
the {\rr\em amplification of quantum fluctuations, which can be interpreted as 
particle production.} It provides an important building block for our 
understanding of the (pre)heating of the early universe at the end of an 
inflationary period, and may also be operative 
in various situations in the context of relativistic heavy-ion 
collision experiments. Here we will consider the phenomenon
as a ``paradigm'' for far-from-equilibrium dynamics of macroscopic 
fields or one-point functions. Dynamics of correlation functions
for vanishing --- or similarly for constant --- macroscopic 
fields has been described in detail above. There is a wealth 
of phenomena associated to non-constant fields such as 
parametric resonance, spinodal decomposition or in general the dynamics
of phase transitions where the field can play the role of an order 
parameter. 

The example of parametric resonance is particularly
challenging since it is a nonperturbative phenomenon even in the 
presence of arbitrarily small couplings. 
Despite being a basic phenomenon that can occur in a large variety
of quantum field theories, parametric resonance is a rather complex 
process, which in the past defied most attempts for a complete 
analytic treatment
even for simple theories. It is a far-from-equilibrium phenomenon that 
involves densities inversely proportional to the coupling. 
The nonperturbatively large occupation numbers cannot be described by 
standard kinetic descriptions. So far, classical statistical 
field theory simulations on the lattice have been the only quantitative 
approach available. These are expected to be valid for 
not too late times, before the approach to quantum thermal 
equilibrium sets in (cf.~also Sec.~\ref{sec:precision}). 
Studies in quantum field 
theory have been mainly limited to linear or mean-field type 
approximations (leading-order 
in large-$N$, Hartree), which present a valid description
for sufficiently early times. However, they are known 
to fail to describe thermalization and miss important rescattering 
effects (cf.~also Sec.~\ref{eq:scatoffmem}).
The 2PI $1/N$ expansion
provides for the first time a quantitative nonperturbative 
approach in quantum field theory taking into account rescattering. 

Recall for a moment the classical mechanics example of 
resonant amplitude amplification 
for an oscillator with time-dependent periodic frequency.
The amplitude $y(t)$ is described by the differential
equation $\ddot{y} + \omega^2(t) y = 0$ with 
periodic $\omega(t+T)=\omega(t)$ of period $T$.
Since the equation is invariant under $t \to t + T$  
there are periodic solutions $y(t+T) = c\, y(t)$.
This can be expressed as $y(t) = c^{t/T} \Pi(t)$ with 
periodic $\Pi(t+T)=\Pi(t)$. One concludes that for
real $c > 1$ there is an {\em\rr instability with an 
exponential growth.} For small elongations a physical 
realization of this situation is a pendulum with a periodically 
changing length as displayed:  

\vspace*{0.2cm}

\centerline{
\epsfig{file=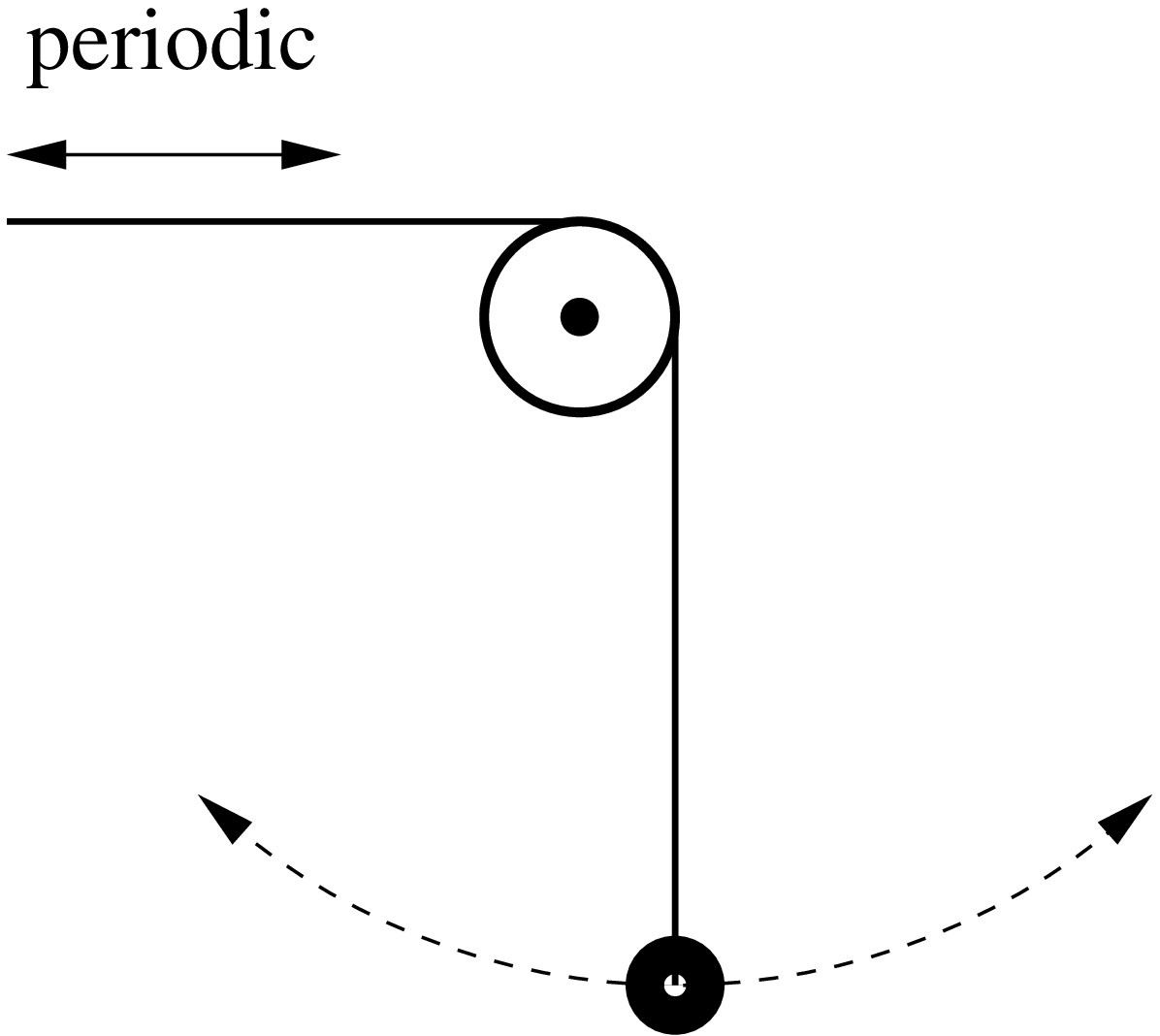,width=3.cm}
}}

\vspace*{0.2cm}

\noindent
In contrast to the mechanics example, in quantum field theory
there will be no external periodic source. A large {\rr\em coherent 
field amplitude coupled to its own quantum fluctuations}
will trigger the phenomenon of parametric resonance. 
Mathematically, however, important aspects are very similar 
to the above classical example for sufficiently early times:
The mechanical oscillator amplitude $y$ plays the role of the statistical 
two-point function $F$ in quantum field theory, and the periodic
$\omega^2(t)$ plays the role of an effective mass
term $M^2(\phi(t))$ whose time dependence is induced by an
oscillating macroscopic field $\phi(t)$. Simple linear
approximations to the problem, which have been much employed 
in the literature, are even mathematically equivalent to the
above mechanics example. Accordingly, the well-known Lam{\'e}--type
solutions of the mechanics problem will also play a role 
in the quantum field theory study. Substantial deviations
do, however, quickly set in with important non-linear effects.

\subsubsection{Parametric resonance in the $O(N)$ model}

We consider the scalar $N$-component quantum
field theory with classical action (\ref{eq:classical}), 
and employ the 2PI $1/N$-expansion to next-to-leading order.
The relevant equations of motion are given by 
(\ref{eq:exactbrok})--(\ref{eq:NLOphi}) as described 
in Sec.~\ref{2PINfield}. We will describe {\rr\em numerical solutions}
of these equations in $3+1$ dimensions without further
approximations. Moreover, the
approach allows us to identify the relevant contributions to the dynamics
at various times and to obtain an {\rr\em approximate analytic solution of the 
nonlinear dynamics} for the entire amplification range.
It should be emphasized that the approach solves the problem of an 
analytic description of the dynamics at nonperturbatively large densities.
It is necessary to take
into account the infinite set of NLO 2PI diagrams as described
in Sec.~\ref{sec:classofdiag}. As we will show 
in the following, each of these eventually contributes to the same order in 
the coupling $\lambda$ such that any finite order in loops or
couplings is not sufficient. 
In this sense, the 2PI $1/N$-expansion
to NLO represents a minimal approach for the controlled 
description of the phenomenon. This justifies the rather
involved complexity of the approximation.

We have in mind a situation reminiscent of that in
the early universe after a period of chaotic inflation, driven by a 
macroscopic (inflaton) field.
We consider a weakly coupled system that is initially 
in a pure quantum state, characterized by a large field amplitude
\beq\db    
\phi_a(t) \,=\, \sigma(t) M_0 \sqrt{\frac{6 N}{\lambda}}\, \delta_{a1} \, .
\label{eq:rescaledfield}
\eeq
and small quantum fluctuations, corresponding to vanishing particle numbers at 
initial time. Here $M_0$ sets our unit of mass, $\sigma(0)=\sigma_0$ is 
$\sim {\mathcal O}(1)$ 
and $\partial_t\sigma(t)|_{t=0}=0$. The initial statistical
propagator contains a ``longitudinal'' component $F_{\pa}$ and
$(N-1)$ ``transverse'' components $F_{\pe}$:
\beq\db
F_{ab} \,=\, {\rm diag}\{F_{\pa},F_{\pe},\ldots,F_{\pe}\} \, ,
\eeq 
with $\partial_{t}F(t,0;\bp)|_{t=0} = 0$,
and $\partial_{t}  \partial_{t'} F(t,t';\bp)|_{t=t'=0} \equiv
F^{-1}(0,0;\bp)/4$ for a pure-state initial density matrix 
(cf.~Sec.~\ref{sec:initialconditions}).
\begin{figure}[t]
\centerline{
\epsfig{file=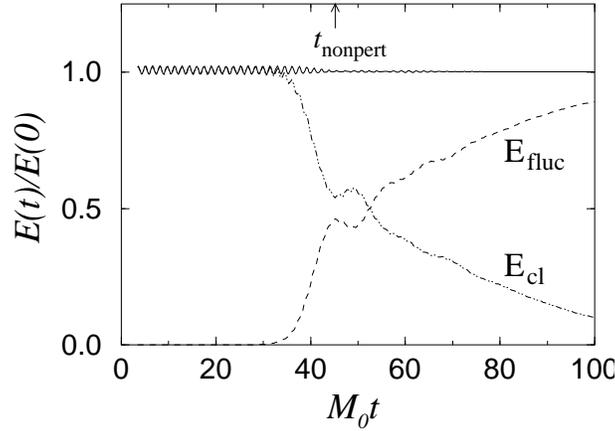,width=8.cm}}
\caption{Total energy (solid line) and 
  classical-field energy $E_\cl$ (dotted line) as a function of time. 
  The dashed line represents the fluctuation part $E_{\rm fluc}$,
  showing a transition from a classical-field to a
  fluctuation dominated regime.}
  \label{fig:energy}
\end{figure}  

\begin{figure}[t]
\centerline{
\epsfig{file=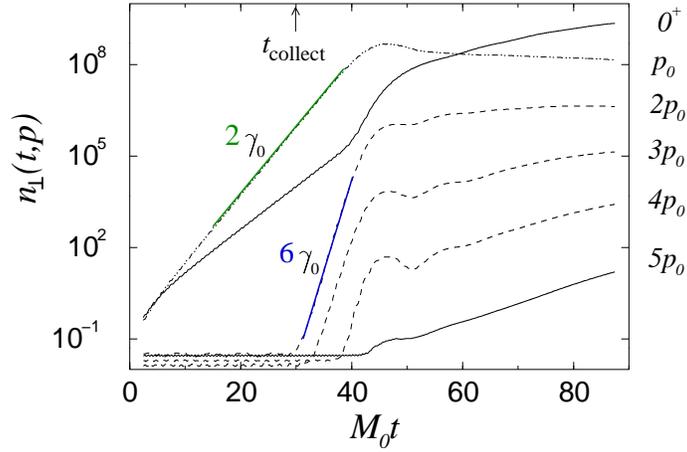,width=9.cm}} 
\caption{Effective particle number density for the 
 transverse modes as a function of time for various momenta 
 $p \le 5 p_0$. At early times, modes with $p \simeq p_0$ are  
 exponentially amplified with a rate $2 \gamma_0$. Due to 
 nonlinearities, one observes subsequently an enhanced growth
 with rate~$6\gamma_0$ for a broad momentum range.}
 \label{fig:number_tr}
\end{figure}

\begin{figure}[t]
\centerline{
\epsfig{file=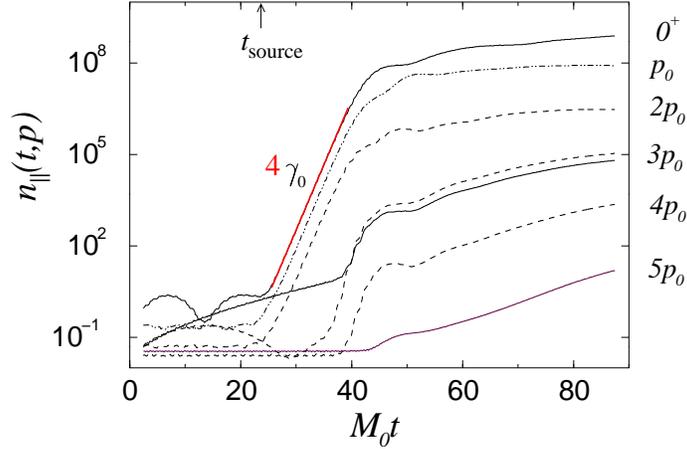,width=9.cm}}  
\caption{Same as in Fig.~\ref{fig:number_tr}, for the longitudinal 
  modes. Nonlinear source effects trigger an exponential growth with 
  rate $4\gamma_0$ for $p \lesssim 2p_0$. 
  The thick line corresponds to a mode in the parametric resonance band, 
  and the long-dashed line for a similar one outside the band. The resonant 
  amplification is quickly dominated by source-induced particle 
  production.}
  \label{fig:number_lg}
\end{figure}

To get an overview, in Fig.~\ref{fig:energy} we show the contributions from the
macroscopic field and the from the fluctuations to the (conserved) total
energy as functions of time for $N=4$ and $\lambda=10^{-6}$.~\footnote{Typical 
volumes $(N_s a_s)^3$ of $N_s = 36$--$48$ 
with $a_s = 0.4$--$0.3$ lead to results that are rather
insensitive to finite-size and cutoff 
effects.} The total energy $E_{\rm tot}$ 
is initially dominated by the classical part $E_{\rm cl}(t=0)$, with
reads in terms of the rescaled field (\ref{eq:rescaledfield}):
\bea\db
\frac{E_{\rm cl}(t)}{L_3}  
\,=\,  \frac{3N}{\lambda} M_0^2\left[ 
\partial_t \partial_{t'} + m^2 +  \frac{1}{2} M_0^2 \sigma^2(t)
\right] \sigma(t) \sigma(t')\Big|_{t=t'} \, ,
\eea
where $L_3$ denotes the spatial volume, and
$E_{\rm fluc} = E_{\rm tot} - E_{\rm cl}$. More and more energy 
is converted into fluctuations as the system evolves.
A characteristic time --- denoted as $t_{\rm nonpert}$ 
in Fig.~\ref{fig:energy} ---
is reached when both contributions become of the same size,
i.e.~$E_{\rm fluc} \simeq E_{\rm cl}$. Before this time,
the coherent oscillations of the field $\phi$ lead to a resonant enhancement 
of the statistical propagator Fourier modes 
\beq\rr
 F_{\pe} (t,t';\bp_0) \,\sim\, e^{\gamma_0 (t+t')} \, .
\label{earlytime}
\eeq
in a narrow range of momenta around 
a specific value $|\bp |\simeq p_0$: this is parametric resonance. 
Apart from the resonant amplification in the linear regime, we identify 
two characteristic time scales --- denoted as $t_{\rm source}$ and 
$t_{\rm collect}$ below ---, which signal strongly enhanced particle 
production in a broad momentum range.  
Nonlinear interactions between field modes cause the resonant amplification to 
spread to a broad range of momenta. More specifically, 
the initially amplified modes act as a source for other modes. The rate
of the source-induced exponential amplification exceeds the characteristic 
rate $\gamma_0$ for the resonant amplification. This is illustrated in 
Figs.~\ref{fig:number_tr} and~\ref{fig:number_lg}, 
where the effective particle numbers are shown
for various momenta as a function of time in the transverse and the
longitudinal sector. Before the transition to a 
fluctuation dominated regime around $t_{\rm nonpert}$, 
one observes a rapid change of the particle numbers due to resonant as well as
source-induced amplification. The source-induced amplification is crucial 
for the rapid approach to the subsequent, quasistationary regime, where 
direct scattering drive a very 
slow evolution towards thermal equilibrium. The 
transition to this slow regime around $t_{\rm nonpert}$ can be
very well observed from Figs.~\ref{fig:number_tr} and~\ref{fig:number_lg}.
 
\begin{figure}[t]
\centerline{\epsfig{file=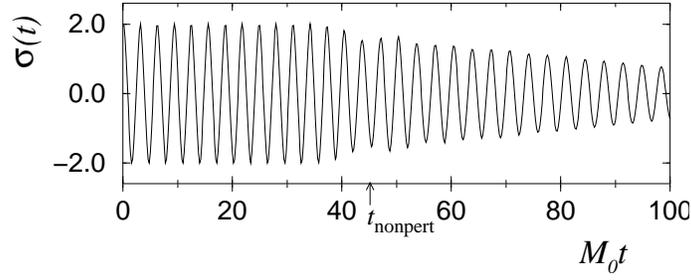,width=9.cm}}
 \caption{The rescaled field $\sigma$ as a function of time.}
  \label{fig:field}
\end{figure}

The fluctuation dominated regime is characterized
by strong nonlinearities. For instance, from Fig.\ \ref{fig:energy} one infers
for $t \simeq t_{\rm nonpert}$ that the classical field decay ``overshoots''
and is temporarily reversed by feed-back from the modes.
This can be directly seen in the evolution for the rescaled field shown in 
Fig.~\ref{fig:field}, which shows a dip around $t_{\rm nonpert}$. 
(The particle numbers of Figs.~\ref{fig:number_tr} 
and~\ref{fig:number_lg} exhibit correspondingly a reverse behavior.)  
The oscillations in the envelope of $\sigma(t)$ damp out 
with time, and one observes a slow decay of the field and the associated 
energy at later times. The latter phenomena 
cannot be seen in leading-order or Hartree--type approximations and
it is crucial to include the next-to-leading order contributions
(cf.~also Sec.~\ref{sec:lofixedpoints}). 
They are important for a reliable 
description of the system at the end of the resonance stage for finite 
$N \lesssim 1/\lambda$. For realistic inflationary models with typically 
$\lambda \ll 1$ this is, in particular, crucial to determine whether there 
are any radiatively restored symmetries. 

\begin{figure}[t]
\centerline{
\epsfig{file=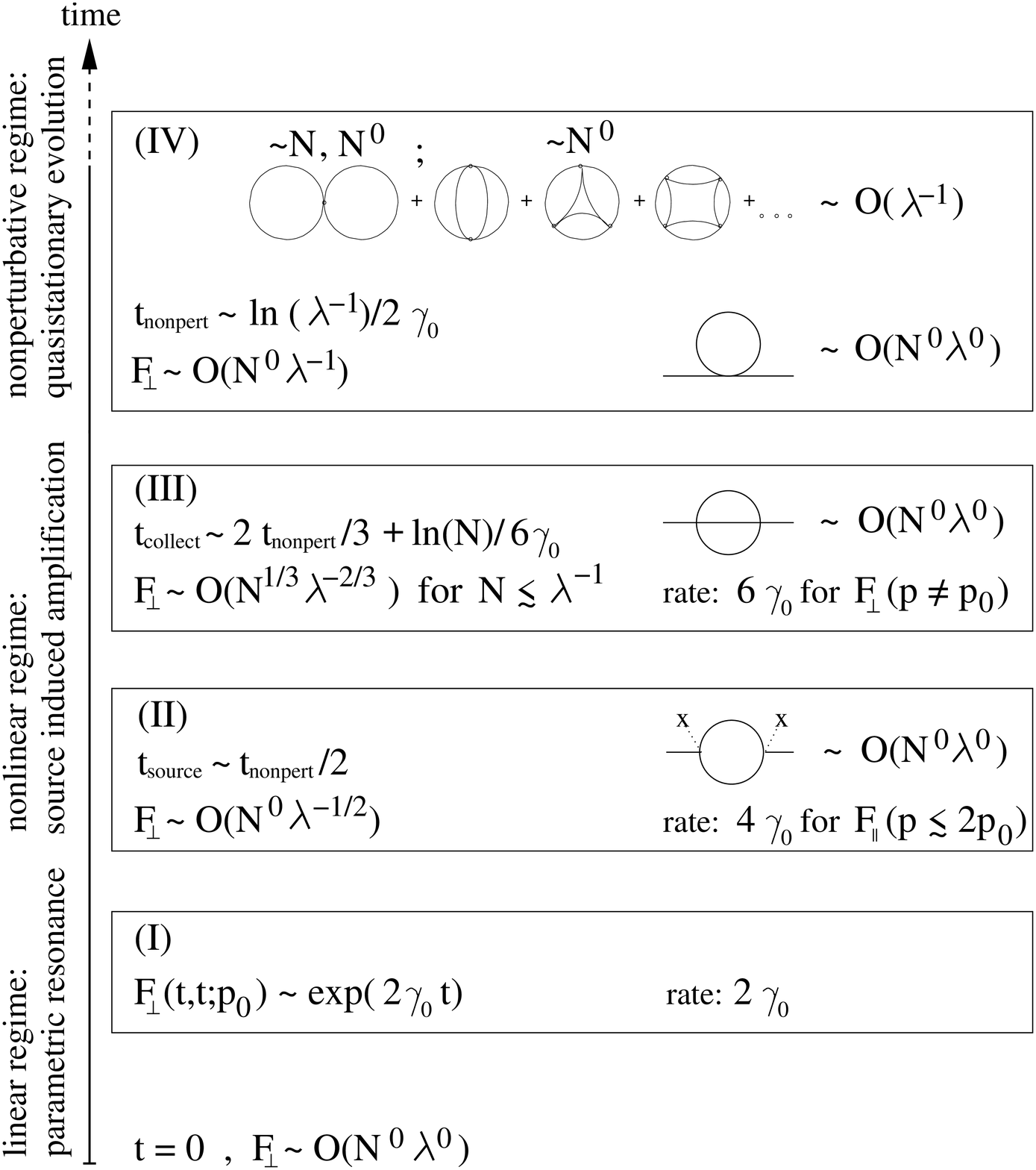,width=11.cm}}
\caption{Schematic overview of the characteristic time scales and the
respective relevant diagrammatic contributions (see text).}
\label{fig:timescheme}
\end{figure}

The characteristic properties described above can be understood analytically
from the evolution equations for the one- and two-point functions. 
To set the scale we use the initial longitudinal mass squared
$M_0^2 \equiv M^2(t=0)$ with (cf.~Eq.~(\ref{Meff})):
\bea\db 
 M^2(t) &\db =&\db  m^2 + \frac{\lambda}{6N} 
 \Big[3 T_\pa(t) + (N-1) T_\pe(t)\Big] \, , \\
\label{eq:massscale}
\db  T_{\pa,\pe}(t) &\db =&\db 
 \int^{\Lambda} \frac{{\rm d}^3 p}{(2\pi)^3} F_{\pa,\pe}(t,t;\bp) \, ,
\label{eq:tadpoles}
\eea
where we denote the ``tadpole'' contributions by 
$T_{\pa}$ and $T_{\pe}$ from the longitudinal and transverse
propagator components, respectively, and some \mbox{$\Lambda \gg p_0$}.
Initially, $F_{\pa}(0,0;\bp) = 1/\left[2 \om_{\pa}(\bp)\right]$ with frequency
$\om_{\pa}(\bp) = [\bp^2 + M_0^2\, (1 + 3 \sigma^2_0)]^{1/2}$, 
and similarly for the transverse components
$F_{\pe}(0,0;\bp)$ where the frequency $\om_{\pe}(\bp)$ 
contains a mass term $\sigma^2_0$ instead of $3 \sigma^2_0$.
Parametrically the initial statistical propagator is of order
one, i.e.~$\sim {\mathcal O}(N^0 \lambda^0)$. Parametric resonance
leads to the dominant amplification of $F_{\pe}(t,t;\bp_0)$ with rate
$2 \gamma_0$. As a consequence of the exponential amplification
of the statistical propagator there is a characteristic time
at which loop corrections will become of order one as well.
This is schematically summarized in Fig.~\ref{fig:timescheme}.
The time when $F_{\pe}(t,t;\bp_0) 
\sim \Or (N^0 \lambda^{-1/2})$ is denoted by $t = t_{\rm source}$.
At this time the one-loop diagram with two field insertions
indicated by crosses as depicted in Fig.~\ref{fig:timescheme}
will give a contribution of order one to the evolution equation for 
$F_{\pa}(t,t;\bp)$. For instance, the two powers of the coupling coming 
from the vertices of that diagram are canceled by the field amplitudes
(\ref{eq:rescaledfield}) and by propagator lines associated
to the amplified $F_{\pe}(t,t';\bp_0)$. Similarly, at the time
$t = t_{\rm collect}$ the maximally amplified transverse propagator 
mode has grown to $F_{\pe}(t,t';\bp_0) 
\sim \Or (N^{1/3}\lambda^{-2/3})$. As a consequence, the 
``setting sun'' diagram in Fig.~\ref{fig:timescheme} becomes
of order one and is therefore of the same order as the classical
contributions. Though the loop corrections become of order one
later than the initial time, they induce amplification rates
that are multiples of the rate $\gamma_0$ which lead to
a very rapid growth of modes in a wide momentum range. 
Finally, when the fluctuations have grown 
nonperturbatively large with $F_{\pe}(t,t';\bp_0) 
\sim \Or (N^0\, \lambda^{-1})$
any loop correction will no longer be suppressed by
powers of the small coupling $\lambda$. In this case
the nonperturbative $1/N$ expansion becomes of crucial
importance for a quantitative description of the
dynamics to later times. In the following, we derive
these results from the evolution equations
(\ref{eq:exactbrok})--(\ref{eq:NLOphi}) in more detail.

\paragraph{(I) Early-time (linear) regime: Resonant amplification} 
 
At early times the $\sigma$--field evolution equation receives the
dominant, i.e.\ ${\mathcal O} (\lambda^0)$, 
contributions from the classical action 
$S$ given in (\ref{eq:classical}). As a consequence, the field dynamics can be 
described by the classical equation of motion. 
In the following, all quantities are 
rescaled with appropriate powers of $M_0$ to become 
dimensionless and we set $M_0 \equiv 1$. The classical field equation reads
\beq\rr
\partial_t^2 \sigma(t) + \sigma(t) + \sigma^3(t) = 0 \, . 
\label{eq:rescfieldequation}
\eeq
Differential equations of this time have been extensively 
studied in the literature. For the initial condition considered 
here it has the (anti-)periodic solution 
$\sigma(t+\pi/\omega_0) = - \sigma(t)$, which can be
expressed in term of the Jacobian cosine ${\rm cn}$ as
$\sigma(t) = \sigma_0\, 
{\rm cn}[t \, \sqrt{1+\sigma_0^2},\sigma_0/\sqrt{2(1+\sigma_0^2)}]$.
Therefore, the solution shows a rapidly oscillating behavior
with characteristic frequency 
\beq\rr
\omega_0 \simeq 2 \sqrt{1 + \sigma_0^2} \, .
\eeq
The period average of the field amplitude can be also expressed 
in terms of the initial amplitude $\sigma_0$ as
$\ol{\sigma^2(t)}\simeq \sigma_0^2/2$.

The evolution equations for the two-point functions
to order ${\mathcal O} (\lambda^0)$ correspond to
free-field equations with the addition of a
time-dependent mass term $\sim 3 \sigma^2(t)$ for the longitudinal
and $\sim \sigma^2(t)$ for the transverse modes:
\bea\db
\Big[ \partial_t^2 + \bp^2 + 1 + {\rr \sigma^2(t)} \Big]F_{\pe}(t,t';\bp) 
&\db =& \db 0 \nn
\db \Big[ \partial_t^2 + \bp^2 + 1 
+ {\rr 3 \sigma^2(t)} \Big]F_{\pa}(t,t';\bp) 
&\db =& \db 0 \, ,
\eea
and equivalently for the spectral functions $\rho_{\pe}(t,t';\bp)$ 
and $\rho_{\pa}(t,t';\bp)$. As a consequence of this approximation, 
the two-point functions can be factorized
as products of single-time functions:
\bea \db 
 F_\pe(t,t';\bp) &\db =&\db  \frac{1}{2} 
 \left[f_\pe(t;\bp)f_\pe^*(t';\bp)+f_\pe^*(t;\bp)f_\pe(t';\bp)\right] \, , 
\nonumber\\\db 
 \rho_\pe(t,t';\bp) &\db =& \db  
 i \left[f_\pe(t;\bp)f_\pe^*(t';\bp)-f_\pe^*(t;\bp)f_\pe(t';\bp)\right] \, , 
\label{rhomodef} 
\eea
and similarly for the longitudinal components. We emphasize that the
simple decomposition (\ref{rhomodef}) is no longer correct at higher
orders in the coupling. In terms of these so-called 
mode functions the equations of motion read, e.g.~for the
transverse modes: 
\bea 
\db \Big[ \partial_t^2 + \bp^2 + 1 + {\rr \sigma^2(t)} \Big]
f_\pe(t;\bp) \,=\, 0 \, .
\label{eq:lametype}
\eea
Up to an overall arbitrary phase,
the above initial conditions for the two-point functions translate into  
$f_\pe(0,\bp)=1/\sqrt{2\omega_\pe(\bp)}$,  
$\partial_t f_\pe(t,\bp)|_{t=0}= -i \sqrt{\omega_\pe(\bp)/2}$ 
and equivalently for $f_\pa$.
For this approximation the quantum field theory problem becomes mathematically 
equivalent to the well-known classical mechanics problem described
in the above introduction. The
analytical solution of the Lam{\'e}--type equation (\ref{eq:lametype})
is well known and can be summarized for our purposes as follows:
There is an exponential amplification of $F_\pe(t,t';\bp)$ for a
{\rr\em bounded momentum range} $0 \le \bp^2 \le \sigma_0^2/2$.
This corresponds to a maximum momentum for amplification
$\bp^2_{\rm max} + 1 + \ol{\sigma^2(t)} \simeq
1 + \sigma_0^2 = (\omega_0/2)^2$. A further important result is
that there is a separation of scales: $\omega_0 \gg \gamma_0$,
with the maximum amplification rate $\gamma_0 \simeq 2 \delta \omega_0$ for 
$\bp^2=\bp_0^2 \simeq \sigma_0^2/4$ with the small number
$\delta \le e^{-\pi} \simeq 0.043$.
There is much smaller growth in a narrow momentum range for $F_\pa$, which
is of no importance here since it is quickly dominated by loop corrections
as is shown below. Time-averaged over $\sim \omega_0^{-1}$, for 
$t,t' \gg \gamma_0^{-1}$ one finds the result of Eq.~(\ref{earlytime}).
Note here that also leading-order large-$N$ or Hartree--type
approximations show the same {\em\rr bounded amplification regime},
such that at all times no higher momentum modes would get populated.
This is, of course, an artefact of the approximation which
is absent once NLO corrections are taken into account as is
described below.

The analytic results for the Lam{\'e} regime agree precisely with the
NLO numerical results for sufficiently early times. 
Figs.~\ref{fig:number_tr} and \ref{fig:number_lg} show 
the transverse and longitudinal particle numbers for various momenta
in the range $0 < p \le 5 p_0$, averaged over 
the rapid oscillation time $\sim 1/\omega_0$. For the transverse
modes the lowest two momenta shown are inside the resonance band. 
We define the effective particle numbers as in 
Eq.~(\ref{eq:effpart}).\footnote{We have employed $Q \equiv 0$ as
in Sec.~\ref{sec:NLOtherm} which is, however, irrelevant.}
As shown in Sec.~\ref{sec:NLOtherm} for the $O(N)$ model
these definitions yield an efficient description 
approaching a Bose-Einstein distributed particle number 
at sufficiently late times.

\paragraph{(II) Source-induced (nonlinear) amplification regime:
Strongly enhanced particle production for longitudinal modes} 

The $\Or (\lambda^0)$ approximation for longitudinal modes breaks down at 
the time 
\beq
\qquad t\simeq t' ={\rr t_{\rm source}}\,\,:
\hspace*{0.3cm} {\rr F_{\pe}(t,t';\bp_0) 
\,\sim\, \Or (N^0 \lambda^{-1/2})}  \, .
\eeq
This can be derived from the ${\mathcal O} (\lambda)$ evolution equations
to which one-loop self-energies contribute, which diagrammatically
are given by
\epsfig{file=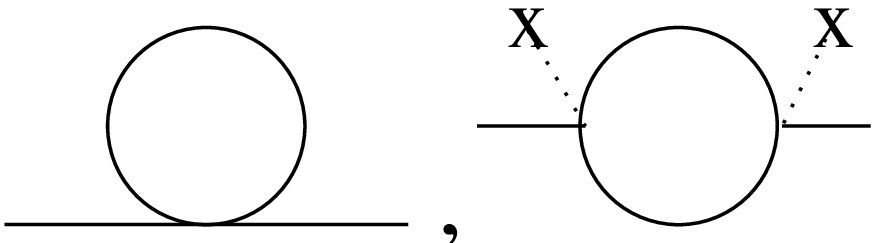,width=2.cm}.
The approximate evolution equation reads:
\bea\db
\lefteqn{
\Big(\partial_t^2 + {\bp}^2 +  M^2(t) + 3 \sigma^2(t) \Big) 
F_\pa(t,t';{\bp}) \simeq   } \nonumber \\
&&\db  { \frac{2 \lambda (N-1)}{3N} \sigma(t) \Bigg\{
\int_0^t {\rm d} t'' \sigma(t'')
\Pi^\rho_\pe(t,t'';{\bp}) F_\pa(t'',t';{\bp})} \nonumber\\ 
&&\db { -\frac{1}{2}\int_0^{t'} {\rm d} t'' \sigma(t'')
\Pi^F_\pe(t,t'';{\bp})\rho_\pa(t'',t';{\bp})\Bigg\}} \,\equiv\, RHS  \, .
\label{eq:orderlambda}
\eea
Here we have abbreviated $\Pi^A_\pe(t,t'';{\bp}) 
= \int {\rm d}^3 q/(2 \pi)^3\, F_\pe(t,t'';{\bp}-\bq)\, A_\pe(t,t'';\bq)$
with \mbox{$A = \{F, \rho\}$}, and we used that
$\Pi_\pe \gg \Pi_\pa$ and $F_{\pe}^2 \gg \rho_{\pe}^2$.\footnote{The latter
inequality will be discussed in detail in Sec.~\ref{sec:precision},
where it is shown to indicate the validity of classical statistical
field theory approximations.} One observes that indeed for $F_{\pe} 
\sim \Or (N^0 \lambda^{-1/2})$ the 
r.h.s.~of Eq.~(\ref{eq:orderlambda}) becomes $\sim \Or (1)$ 
and cannot be neglected.

In order to make analytical progress, one has to evaluate the ``memory 
integrals'' in the above equation. This is dramatically simplified
by the fact that the integral is approximately local in time,
since the exponential growth lets the latest-time contributions 
dominate the integral. (This will be the case for times
$t \lesssim t_{\rm nonpert}$ after which exponential amplification
stops, cf.~below.) For the approximate evaluation of the memory integrals we
consider time-averages over $\omega_0^{-1} \ll \gamma_0^{-1}\,$:
\beq
\db \int_0^t {\rm d} t'' \;\;\longrightarrow\;\;  
\int_{t - c/\omega_0}^t {\rm d} t'' \qquad (c \sim 1)
\eeq
and perform a Taylor expansion around the latest time $t$ $(t')$:
\bea\db
\rho_{\pa,\pe}(t,t'';\bp) &\db \simeq& \db
\partial_{t''}\rho_{\pa,\pe}(t,t'';\bp)|_{t=t''}
(t'' - t) \equiv (t - t'') \, , \nn
\db F_{\pa,\pe}(t,t'';\bp) &\db \simeq& \db F_{\pa,\pe}(t,t;\bp) \, ,
\eea
where we have used the equal-time commutation relations (\ref{eq:bosecomrel}).
With these approximations 
the r.h.s.~of Eq.~(\ref{eq:orderlambda}) can be evaluated as:
\bea \db
RHS
&\db\simeq&\db
{\db \lambda}\, \sigma^2(t)\,  
\frac{c^2}{\omega_0^2} \frac{(N-1)}{3N}\, {\db T_{\pe}(t)} \, 
{\db F_{\pa}(t,t';{\bp})}\,
\!\!\!\!\!\qquad {\db\rm (mass\; term)}
\label{eq:massesta}
\\
&\rr +&\rr {\rr \lambda}\, 
\sigma(t)\sigma(t')\, \frac{c^2}{\omega_0^2}\,\frac{(N-1)}{6N}
{\rr \Pi^F_\pe(t,t';{\bp})} \!\!\!\!\!\qquad\;\; {\rr\rm (source\; term)} 
\label{eq:sourceasta}
\eea
The first term is a contribution from NLO to the mass, whereas the second 
term represents a source. Note that both the LO mass term 
and the above correction to this mass are of the same order in 
$\lambda$, however, with opposite sign.
To evaluate the momentum integrals,
we use a saddle point approximation around the dominant
$p\simeq p_0$, valid for $t,t'\gg \gamma_0^{-1}$, 
with $F_\pe(t,t',\bp) \simeq F_\pe(t,t',\bp_0) 
\exp [-|\gamma_0^{\prime\prime}|(t+t')(p-p_0)^2/2]$.\footnote{Here
$\gamma(p) \simeq \gamma_0 + \gamma_0^{\prime\prime} (p-p_0)^2/2$
with $\gamma_0^{\prime\prime} \simeq - 64 \delta (1 - 6 \delta)
\sqrt{1+\sigma_0^2}/\sigma_0^2$.}
From this one obtains for the above mass term:
\beq\db
 \lambda\, T_{\pe}(t) \,\simeq\, 
\lambda\, \frac{p_0^2\,F_{\pe} (t,t;\bp_0)}{2 
(\pi^3 |\gamma_0^{\prime\prime}| t)^{1/2}} \, .
\label{eq:mest}
\eeq
The result can be used to obtain an estimate at what time $t$
this loop correction becomes an important contribution to the
evolution equation. Note that 
to this order in $\lambda$ it is correct to use 
$F_{\pe} (t,t';\bp_0) \sim e^{\gamma_0 (t+t')}$ on the
r.h.s.~of (\ref{eq:mest}). The condition 
\mbox{$\lambda\, T_{\pe}(t=t_{\rm nonpert})$} $\sim \Or(1)$
can then be written for $\lambda \ll 1$ as:   
\beq
\mbox{\framebox{\db$\displaystyle t_{\rm nonpert} \simeq 
\frac{1}{2\gamma_0}\, \ln \lambda^{-1}$}}
\label{eq:tnonpert}
\eeq
The same saddle point approximation can be performed to
evaluate the source term (\ref{eq:sourceasta}):
\beq \rr
\lambda\, \Pi^F_\pe(t,t';0) \,\simeq \, \lambda\,
\frac{p_0^2\,F_{\pe}^2(t,t';\bp_0) }{4 (\pi^3 |\gamma_0^{\prime\prime}| 
(t+t'))^{1/2}} \, .
\eeq
Here we only wrote the source term for $\bp = 0$ where it has its maximum, 
although it affects all modes with $p \lesssim 2p_0$. Again this can
be used to estimate the time $t = t_{\rm source}$ at which 
$\lambda\, \Pi^F_{\pe} \sim \Or(1)$:
\beq
\mbox{\framebox{\rr $\displaystyle t_{\rm source} \simeq \frac{1}{2}\,
t_{\rm nonpert}$}}
\eeq
One arrives at the important conclusion that the source term 
(\ref{eq:sourceasta}) becomes {\rr\em earlier} of order one than
the mass term (\ref{eq:massesta}): For $t_{\rm source} \lesssim t 
\lesssim t_{\rm nonpert}$ the source term dominates the dynamics! 
Using these estimates in
(\ref{eq:orderlambda}) one finds that the longitudinal modes
with $0 \lesssim p \lesssim 2p_0$ get amplified with twice the 
rate $2\gamma_0$:
\beq\rr 
F_{\pa}(t,t;\bp) \, \sim\,  \lambda\, F_{\pe}^2(t,t;\bp_0) \sim 
\lambda\, e^{4\gamma_0 t} \, . 
\label{eq:Fpaamp}
\eeq
Though the non-linear contributions start later, they grow
{\rr\em twice as fast!} The analytical estimates for $t_{\rm source}$ 
and rates agree very well with the numerical results 
shown in Fig.~\ref{fig:number_lg}.

\paragraph{(III) Collective amplification regime:
explosive particle production in a broad momentum range for
transverse modes}

A similar analysis can be made for the transverse modes. 
Beyond the Lam{\'e}--type ${\mathcal O}(\lambda^0)$ description, the 
evolution equation for $F_{\pe}$ receives contributions from the 
feed-back of the longitudinal modes at ${\mathcal O} (\lambda)$ 
as well as from 
the amplified transverse modes at ${\mathcal O} (\lambda^2)$. 
They represent {\rr\em source terms} in the evolution equation
for $F_{\pe}(t,t';\bp)$ which are both 
parametrically of the form $\sim \lambda^2 F_\pe^3/N$ as is depicted
below: 

\vspace*{0.3cm}

\parbox{7.6cm}{
\hspace*{0.3cm}\parbox{3cm}{
\centerline{
\epsfig{file=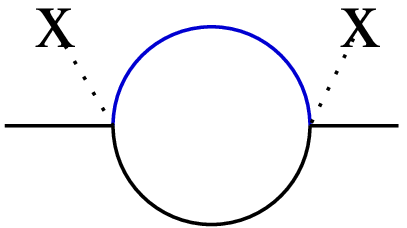,width=2.cm}
}}
{\rr $\displaystyle \sim \frac{\lambda}{N} {\db F_{\pa}} F_{\pe}
\,\,\stackrel{\mbox{\footnotesize \db cf.~(\ref{eq:Fpaamp})}}{\sim}\,\,
 \frac{\lambda^2}{N} F_{\pe}^3$}

\hspace*{0.3cm}\parbox{3cm}{
\centerline{
\epsfig{file=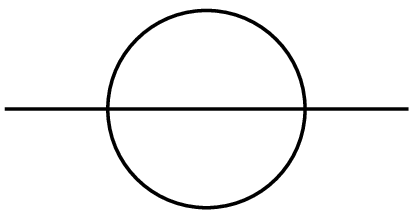,width=2.cm}
}}
{\rr $\displaystyle \sim \frac{\lambda^2}{N} F_{\pe}^3$}
}
\parbox{2.5cm}{$\displaystyle\Bigg\}$ \, \, 
$\displaystyle\rr \sim \frac{\lambda^2}{N}
\, e^{6 \gamma_0 t}$}

\vspace*{0.3cm}

\noindent
Following along the lines of the above paragraph and using
(\ref{eq:tnonpert}) this leads to 
the characteristic time $t =  t_{\rm collect}$ at which
these source terms become of order one:
\beq
\mbox{\framebox{\rr$\displaystyle t_{\rm collect} \simeq \frac{2}{3}\, 
t_{\rm nonpert} + \frac{\ln N}{6 \gamma_0}$}} 
\label{eq:tcollect}
\eeq
For $t\simeq t' \simeq  t_{\rm collect}$ the dominant 
transverse mode has grown to 
\beq\rr
F_{\pe}(t,t';\bp_0) 
\,\sim\, \Or (N^{1/3}\lambda^{-2/3})  \, .
\eeq
Correspondingly, for $t_{\rm collect } \lesssim t \lesssim t_{\rm nonpert}$
one finds a large particle production rate $\sim\! 6 \gamma_0$ 
in a momentum range $0 \lesssim p \lesssim 3 p_0$, in 
agreement with the full NLO results
shown in Fig.\ \ref{fig:number_tr}. In this time range
the longitudinal modes exhibit an enhanced amplification as well 
(cf.~Fig.\ \ref{fig:number_lg}). It is important to realize that the phenomenon
of source-induced amplification repeats itself: the newly amplified modes, 
together with the primarily amplified ones, act as a source for other 
modes, and so on. In this way, even higher growth rates of multiples of 
$\gamma_0$ can be observed
and the ``explosive'' amplification rapidly propagates towards
higher momentum modes.
We emphasize that the collective amplification regime is absent 
in the LO large-$N$ approximation. Consequently, even for the transverse 
sector the latter does not give an accurate description if 
$t_{\rm collect }\le t_{\rm nonpert}$, that is for $N \lesssim \lambda^{-1}$.

\paragraph{Behavior of the field}

Around $t  \lesssim t_{\rm nonpert}$ is the earliest 
time when sizeable corrections
to the classical field equation (\ref{eq:rescfieldequation}) appear,
which give (cf.~Eq.~\ref{eq:NLOphi}):
\beq\db
\left[ \partial_t^2 + 1 + {\rr \delta M^2(t)} + \sigma^2(t)\right] \sigma(t) = 
- {\rr \int_0^t \rmd t' \Sigma^{\rho}_{\pa}(t,t';\bp=0)|_{\sigma=0}}\,
\sigma(t') \, ,
\label{eq:fieldeq}
\eeq
where $\rr \delta M^2 \simeq \frac{\lambda (N-1)}{6 N} T_{\pe}$.
Before $t_{\rm nonpert}$, where the ``memory expansion'' discussed above
is valid, we can discuss the behavior of the field
$\sigma = \sigma^{(0)} + \delta \sigma$ perturbatively in terms of 
a slowly varying small correction $\delta \sigma$ to the classical
solution $\sigma^{(0)}$ of (\ref{eq:rescfieldequation}) 
and a small $\delta M^2$. If we neglect for a
moment the NLO contributions on the r.h.s.~the linearized
equation (\ref{eq:fieldeq}) then yields 
\bea \db
\sigma(t) \simeq \left( 1  - 
\frac{1}{1+3 \sigma_0^2}
\frac{\lambda (N-1)}{6 N} T_{\pe}(t) 
\right) \sigma^{(0)}(t) \, . 
\eea
One concludes that there is an exponential decrease of the field 
amplitude at LO   
since $T_{\pe}(t) \sim e^{2\gamma_0 t}$ for $t  \lesssim t_{\rm nonpert}$ 
according to (\ref{eq:mest}). The dominant NLO corrections read 
\bea \db
 \int_{t-c/\omega_0}^t \!\!\!\!\!\!\!\!\!\!\!\!dt' \, 
 \Sigma^\rho_\pa(t,t';\bp=0)|_{\sigma=0}\, {\db \sigma (t')}
 \simeq \frac{c^2}{2\omega_0^2} \Big( \partial_t 
 \Sigma^\rho_\pa(t,t';\bp=0)|_{\sigma=0} \Big)|_{t=t'} \, {\db \sigma (t)} 
&& \nn \db
 \stackrel{{\db \Or(\lambda^2)}}{\simeq} 
- \frac{c^2}{2\omega_0^2} \frac{\lambda^2}{18 N}
\int^\Lambda\!\!\! \frac{\rmd^3 q}{(2\pi)^3}\frac{\rmd^3 k}{(2\pi)^3}
F_{\pe}(t,t;- \bq - \bk)F_{\pe}(t,t;\bk)\, {\db \sigma (t)} &&\nn
\db \simeq {\rr - \frac{c^2}{2\omega_0^2}
\frac{\lambda^2}{18 N} T_{\pe}^2(t)\,} \sigma (t) \, . &&
\eea
Note that the NLO correction to the effective mass term
comes with an opposite sign than the LO correction.  
For $t \to t_{\rm nonpert}$ one has $\delta M^2 \sim
\Or (N^0 \lambda^0)$ and $\Sigma^{\rho}_{\pa} \sim \Or (N^{-1} \lambda^0)$,
i.e.~they become of the same order in $\lambda$. As a consequence,
cancellations may lead (temporarily) to reverse field decay. This is
indeed what one observes from the full numerical NLO solution for
$N=4$ and $\lambda=10^{-6}$ shown in Fig.~\ref{fig:field}.
Around $t_{\rm nonpert}$ strong nonlinearities appear, where 
the field decay ``overshoots'' and is shortly reversed by 
feedback from modes, overshoots again etc.

\paragraph{(IV) Fluctuation dominated regime: 
nonperturbatively large densities $\sim 1/\lambda$ 
and quasistationary evolution}

When the system evolves in time, 
more and more energy is converted into fluctuations.
The description in terms of $\Or (\lambda^2)$ evolution 
equations break down at
\beq
\rr t \simeq t' = t_{\rm nonpert}\,\,: 
\hspace*{0.5cm} F_{\pe}(t,t';\bp_0) 
\sim \Or (N^0\, \lambda^{-1}) \, .
\label{eq:parametricnp}
\eeq
Transverse and longitudinal modes have grown to $\Or (N^0\, \lambda^{-1})$
in a wide momentum range for times $t \gtrsim t_{\rm nonpert}$. 
This corresponds
to nonperturbatively large particle number densities $n_{\pe}(\bp)$
and $n_{\pa}(\bp)$ inversely proportional to the coupling. 
Because of this parametric dependence there are leading 
contributions to the dynamics coming from all loop orders. 
As a consequence, shortly after $t_{\rm nonpert}$ 
a comparably slow, quasistationary evolution sets.
It is important to note that descriptions based on a 
finite-order loop expansion of the 2PI effective 
action cannot be applied.  
To describe this regime it is crucial to employ 
a nonperturbative approximation as provided by the 2PI
$1/N$ expansion at NLO.

In order to discuss this in more detail, we consider the self--energies 
appearing in the evolution 
equations (\ref{eq:exactrhoF}) for $F_{\pe}(t,t';\bp)$
and $\rho_{\pe}(t,t';\bp)$. From the NLO expressions given in (\ref{ASFFR})
and (\ref{ASRFR}) one finds for $\Sigma^{F}_{\pe}$: 
\bea\db 
\Sigma^{F}_{\pe}(t,t';\bp) &\db  = &\db  
- \frac{\lambda}{3 N}\int \frac{{\rm d}^3 q}{(2\pi)^3} \Big\{
\\
&&\db  {I}_F(t,t';\bq) F_{\pe}(t,t';\bp - \bq) 
- \frac{1}{4}\, {I}_{\rho}(t,t';\bq) \rho_{\pe}(t,t';\bp - \bq)  
 \nonumber\\
 &\db +&\db  {P}_F(t,t';\bq)F_{\pe}(t,t';\bp - \bq) -
 \frac{1}{4}\, {P}_{\rho}(t,t';\bq)\rho_{\pe}(t,t';\bp - \bq) \Big\}\,.
\nonumber \label{ASFFRtrans}
\eea
The functions ${I}_{F,\rho}$ and ${P}_{F,\rho}$ contain
the summation of an infinite number of ``chain'' graphs, 
where each additional element
adds another loop to the graph (cf.~also the figures in 
Sec.~\ref{sec:classofdiag}). We will argue in the following
that each loop order of this infinite number of graphs contributes
with the same order in the coupling $\lambda$. 
 
According to 
(\ref{IFFR}) and (\ref{IRFR}) the sum of ``chain'' graphs described
by ${I}_{F,\rho}$ can be iteratively generated by the relations
\bea\db 
{I}_{F}(t,t';\bq) =  -\frac{\lambda}{3N} 
 \Pi_F (t,t';\bq) +   \frac{\lambda}{3N}\int_{0}^{t} dt''\,
 {I}_{\rho}(t,t'';\bq)\Pi_F (t'',t';\bq) &&  
\nonumber\\
\db  - \frac{\lambda}{3N}\,\int_0^{t'} dt''\,  
 {I}_F(t,t'';\bq) \Pi_{\rho} (t'',t';\bq), &&
 \label{AIFFRmode}\\
\db {I}_{\rho}(t,t';\bq) = - \frac{\lambda}{3N}\Pi_{\rho} (t,t';\bq)
 + \frac{\lambda}{3N} \int_{t'}^{t} dt''\,  
 {I}_{\rho}(t,t'';\bq) \Pi_{\rho} (t'',t';\bq) \, ,
 &&\label{AIRFRmode}
\eea
and corresponding expressions for ${P}_{F,\rho}$ (cf.\
Eqs.~(\ref{eqB11}) and (\ref{eqB12})).
Here the ``chain elements'' $\Pi_{F,\rho}$ are given by
\bea
\db  \Pi_F (t,t';\bq) & \db = & \db 
- \frac{1}{2}\int \frac{{\rm d}^3 k}{(2\pi)^3}
\Bigg\{ F_{\pa}(t,t';\bq-\bk) F_{\pa}(t,t';\bk)
\nonumber\\
&&\db  +(N-1) F_{\pe}(t,t';\bq-\bk) F_{\pe}(t,t';\bk) -
 \frac{1}{4}\Big[\rho_{\pa}(t,t';\bq-\bk) 
\nonumber\\
&&\db  \rho_{\pa}(t,t';\bk)
+(N-1) \rho_{\pe}(t,t'';\bq-\bk) \rho_{\pe}(t,t'';\bk) \Big]\Bigg\},
 \label{PiFmode}\\
\db  \Pi_{\rho}(t,t';\bq)  &\db  = &\db  - \int \frac{{\rm d}^3 k}{(2\pi)^3}
\Big\{ F_{\pa}(t,t';\bq-\bk) \rho_{\pa}(t,t';\bk)
\nonumber\\
&&\db  + (N-1) F_{\pe}(t,t';\bq-\bk) \rho_{\pe}(t,t';\bk) \Big\}\, .
\label{PiRmode}
\eea 
Denoting a given loop order $l$ by ${I}_{F}^{(l)}$ we can write
\bea
\db {I}_{F}(t,t';\bq) &\db =&\db  
\sum_{l=1}^{\infty}{I}_{F}^{(l)}(t,t';\bq)\, ,
\label{loopdecomp}\\
\db {I}_{F}^{(1)}(t,t';\bq) &\db =&\db  -\frac{\lambda}{3N} 
 \Pi_F (t,t';\bq) \, ,
\nonumber\\
\db {I}_{F}^{(2)}(t,t';\bq) &\db =&\db  - \left(\frac{\lambda}{3N}\right)^2
\int_{0}^{t} dt''\,
 {\Pi}_{\rho}(t,t'';\bq)\Pi_F (t'',t';\bq)
\nonumber\\
&&\db  + \left(\frac{\lambda}{3N}\right)^2 \int_0^{t'} dt''\,  
 {\Pi}_F(t,t'';\bq) \Pi_{\rho} (t'',t';\bq),
\nonumber\\
\ldots \nonumber
\eea
and similarly for ${I}_{\rho} = \sum_{l=1}^{\infty}{I}_{\rho}^{(l)}$.
Concentrating on the 
transverse modes, one observes from (\ref{PiFmode}) and (\ref{PiRmode}) that
\beq\db 
\Pi_F (t,t';0) \sim {\mathcal O}(N \lambda^{-2}) \, , \quad
\Pi_{\rho} (t,t';0) \sim {\mathcal O}(N \lambda^{-1}) \, .
\eeq
Inserting this into (\ref{AIFFRmode}) or (\ref{loopdecomp}) for
${I}_{F}$, and into the corresponding expression (\ref{AIRFRmode}) for 
${I}_{\rho}$, one finds
\beq\db 
{I}_{F}^{(l)}(t,t';0) \sim {\mathcal O}(N^0 \lambda^{-1})\, , \quad
{I}_{\rho}^{(l)}(t,t';0) \sim {\mathcal O}(N^0 \lambda^{0}) \, ,
\eeq
irrespective of the loop order $l$. Here it is important to note that
averaged over the rapid oscillation time the spectral functions are
of order one:
\beq\rr
\rho_{\pe}(t,t';\bp) \sim \rho_{\pa}(t,t';\bp) 
\sim {\mathcal O}(N^0 \lambda^{0}) \, .
\label{estimate}
\eeq
As a consequence, with (\ref{estimate})
the self--energy (\ref{ASFFRtrans}) receives leading contributions 
in the coupling from all loop orders. Since these 
contributions are  proportional $N^{-1} \lambda^0$, they are 
subleading in $1/N$. In particular, the expansion in powers
of $1/N$ employed here remains a valid nonperturbative approximation
scheme. It is interesting to note that the 2PI $1/N$ expansion to NLO
can be understood as a ``three-loop'' approximation with an effective
vertex. This is denoted schematically below for the graphs appearing at 
NLO: 

\vspace*{0.3cm}

\centerline{
\begin{minipage}{6cm}
\centerline{\epsfig{file=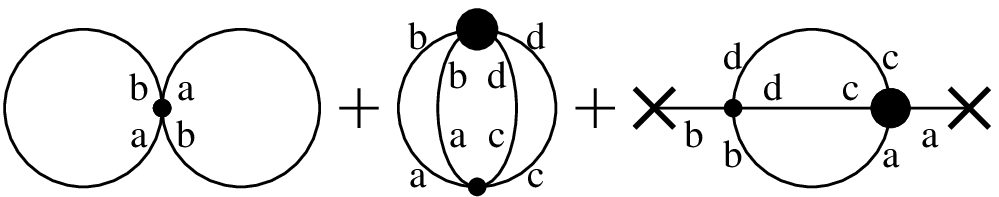,width=5.5cm}}
\centerline{\epsfig{file=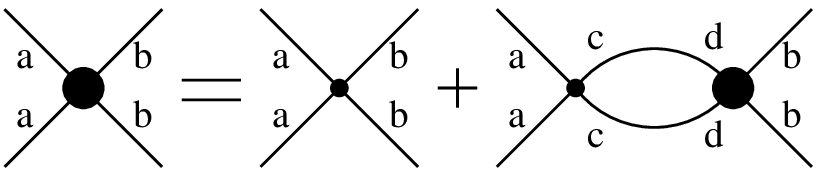,width=5.5cm}}
\end{minipage}
}

\vspace*{0.3cm}

\noindent
The effect of the nonperturbatively large densities is taken into
account by a self-consistent vertex correction as indicated by
the diagrammatic equation for the four-vertex. The question of how
self-consistent vertex corrections can be systematically described 
for cases where no $1/N$ expansion is applicable will be treated in
Sec.~\ref{sec:nPI2}.
We finally note that for times $t \gtrsim t_{\rm nonpert}$
the evolution equations are no
longer characterized by the effective time-locality
in the sense described above such that the ``memory expansion''
cannot be used for an approximate description at late times.

\section{Classical aspects of nonequilibrium quantum fields: Precision tests}
\label{sec:precision}

It is an important question to what extend nonequilibrium quantum field
theory can be approximated by classical statistical field theory.  
It is a frequently employed strategy in the literature to consider
nonequilibrium classical dynamics instead of quantum dynamics since the
former can be simulated numerically up to controlled statistical errors.
Classical statistical field theory
indeed gives important insights when the number of field quanta per mode 
is sufficiently large such that quantum fluctuations 
are suppressed compared to statistical fluctuations. 
We will derive below a {\rr\em sufficient condition for the
validity of classical approximations to nonequilibrium
dynamics.} The description in terms of spectral and statistical
correlation functions as introduced in Sec.~\ref{sec:specstat}
is particularly suitable for comparisons 
since these correlation functions possess a well-defined
classical limit.
However, classical Rayleigh-Jeans divergences and the lack of genuine quantum 
effects --- such as the approach to quantum thermal equilibrium
characterized by Bose-Einstein or Fermi-Dirac statistics --- limit
the use of classical statistical field theory. To find out its
use and its limitations we perform below direct {\em comparisons
of nonequilibrium classical and quantum evolution} for same initial 
conditions. One finds
that classical methods can give an accurate description of
quantum dynamics for the case of large enough initial occupation numbers
and not too late times, before the approach to quantum thermal
equilibrium sets in. Classical approaches are unsuitable, in 
particular, to determine thermalization rates. 

Classical methods have been extensively used in the past
to rule out ``candidates'' for approximation schemes applied 
to nonequilibrium quantum field theory. Approximations that 
fail to describe classical nonequilibrium dynamics should be
in general discarded also for the quantum case. If the dynamics is
formulated in terms of correlation functions then approximation
schemes for the quantum evolution can be straightforwardly implemented
as well for the respective classical theory. For instance,
the 2PI $1/N$-expansion introduced in Sec.~\ref{sec:2PIN}
can be equally well implemented in
the classical as in the quantum theory. Therefore, in the
classical statistical approach one can compare NLO results 
with results including 
all orders in $1/N$. This gives a rigorous answer to
the question of what happens at NNLO or beyond in this case.
In particular, for increasing occupation numbers per mode the classical and
the quantum evolution can be shown to approach each other 
if the same initial conditions are applied and for not too late times.
For sufficiently high particle number densities one
can therefore strictly verify how rapidly the $1/N$ series
converges for far-from-equilibrium dynamics! The possibility
of {\rr\em precision tests} are an important aspect of 
classical statistical field theory methods.

\subsection{Exact classical time-evolution equations}
\label{seccl}

In this section we define the basic classical correlation functions and 
derive exact dynamical equations for them. The evolution equations
for the classical spectral function and the 
classical statistical propagator will turn out to be of
the same form as Eqs.~(\ref{eq:exactrhoF}) for the corresponding 
quantum correlators --- 
with the only difference that the quantum self-energies are replaced by
the respective classical ones. The comparison between quantum
and classical evolution can be very clearly discussed using the
the language of correlation functions, since the quantum spectral function
$\rho(x,y)$ and the statistical propagator $F(x,y)$ of Eq.~(\ref{eq:decompid}) 
both have a well-defined classical equivalent.
 
We consider the classical $N$-component scalar field $\varphi_a$ with
action (\ref{eq:classical}). The classical field equation of motion 
is then given by
\bea \db
\left[\square_x +m^2 + \frac{\lambda}{6N} 
\varphi_b(x)\varphi_b(x)\right]\varphi_a(x) \, =\, 0 \, ,
\label{eqmotion} 
\eea
whose solution requires specification of the initial conditions
$\varphi_a(0,\bx) = \varphi_a(\bx)$ and
$\pi_a(0,\bx) = \pi_a(\bx)$ with $\pi_a(x) \equiv \partial_{x^0}\varphi_a(x)$.
We define the {\rr\em macroscopic or average classical field:}
\bea\db
\phi_{a,\cl} (x) = \langle \varphi_a(x)\rangle_\cl \equiv 
\int D\pi D\varphi\, W[\pi,\varphi]\varphi_a(x) \, .
\label{eq:phicl}
\eea
Here $W[\pi,\varphi]$ denotes the normalized
probability functional at initial time. The measure indicates integration
over classical phase-space:
\beq\db
\int\! D\pi D\varphi  = \int\!
{\prod\limits_{a=1}^N\prod\limits_{\bx}} 
{\rm d}\pi_a(\bx) {\rm d}\varphi_a(\bx) \, .
\eeq
Here the theory will be defined on a spatial lattice to regulate 
the Rayleigh-Jeans divergence of classical statistical field theory. 
The connected {\rr\em classical statistical propagator} $F_{ab,\cl} (x,y)$
is defined by
\bea\db
F_{ab,\cl} (x,y) + \phi_{a,\cl} (x) \phi_{b,\cl} (y) 
= \langle \varphi_a(x)\varphi_b(y)\rangle_\cl \equiv 
\int D\pi D\varphi\, W[\pi,\varphi]\varphi_a(x)\varphi_b(y) \, .
\label{eqFcl}
\eea
The classical equivalent of the quantum spectral function  
is obtained by replacing
$-i$ times the commutator with the Poisson bracket:\footnote{
Recall that the Poisson bracket with respect to the initial fields is
\beq
\{ A(x), B(y) \}_{\rm PoissonBracket} 
= \sum_{a=1}^N\int {\rm d}^d z\,\left[ 
 \frac{\delta A(x)}{\delta\varphi_a(\bz)} 
 \frac{\delta B(y)}{\delta\pi_a(\bz)} 
-\frac{\delta A(x)}{\delta\pi_a(\bz)} 
 \frac{\delta B(y)}{\delta\varphi_a(\bz)}\right] \, .
\eeq} 
\beq \db
 \rho_{ab,\cl}(x,y) = -\langle\,\, 
\{\varphi_a(x),\varphi_b(y)\}_{\rm PoissonBracket} \,\,\rangle_\cl \, .
\label{eq:classspec}
\eeq
As a consequence, one finds the {\rr\em equal-time relations for the
classical spectral function:}
\beq
\db \rho_{ab,\cl}(x,y)|_{x^0=y^0} = 0,\;\;\;\;
 \partial_{x^0}\rho_{ab,\cl}(x,y)|_{x^0=y^0} = \delta_{ab} \delta(\bx-\by) \, .
\label{eq:cleqtime}
\eeq
Though their origin is different, note that they are 
in complete correspondence with the respective quantum relations  
(\ref{eq:bosecomrel}). 

The evolution equations for the classical statistical correlators 
can be obtained by functional methods in a similar way as for
the quantum theory.
In order not to be too redundant, we employ here a different approach 
starting from the differential equation for the free correlators.
We will choose here $W[\pi,\varphi]$ to be invariant under the $O(N)$ 
symmetry with $\phi_a \equiv 0$ such that
\beq\db
F_{ab,\cl}(x,y)=F_\cl(x,y)\delta_{ab} \, ,  
\eeq
and equivalently for $\rho_{ab,\cl}(x,y)$.\footnote{An explicit example
for $W[\pi,\varphi]$ would be
$W[\pi,\varphi] = Z_0^{-1}\,  \exp\left\{\int {\rm d}^d \bx \left( 
\pi^2 + (\nabla \varphi)^2 + m^2 \varphi^2
\right)/2\right\}$
where $Z_0 \equiv \int D\pi D\varphi\, W[\pi,\varphi]$. We emphasize
that none of the following derivations will make use of 
a specific form for $W[\pi,\varphi]$.} We will discuss
$\phi_a \not = 0$ below. The unperturbed spectral 
function is a solution of the homogeneous equation\footnote{This is the
case both in the quantum and the classical theory.} 
\beq\db
\left[\square_x +m^2\right]\rho_0(x,y)=0\, ,
\label{eqrho0}
\eeq
with initial conditions determined by the equal-time canonical relations
(\ref{eq:cleqtime}). Let $F_0$ denote the solution of the
unperturbed homogeneous problem
\beq\db
\left[\square_x+m^2\right]F_0(x,y) = 0 \, ,
\label{eqF0}
\eeq
with initial conditions determined by the initial probability functional.
The derivation of the time evolution equations of classical correlation 
functions is conveniently
formulated by introducing two additional two-point functions: 
the classical retarded and advanced Green functions, which are
related to the classical spectral function $\rho_\cl$ by
\beq\db
G^R_{\cl}(x,y) = \Theta(x^0-y^0)\rho_\cl(x,y) = G^A_{\cl}(y,x) \, .
\label{eqdefGR}
\eeq
The classical retarded self-energy $\Sigma_{R,\cl}(x,y)$ can be defined as
the difference between the full and free inverse retarded Green functions
\beq\db
\Sigma_{R,\cl}(x,y) = {\left(G^R_{\cl}\right)}^{-1}(x,y) 
- {\left(G^R_0\right)}^{-1}(x,y) \, ,
\label{eq:clSD}
\eeq 
where the free retarded Green function
\beq\db
G^R_0(x,y) = \Theta(x^0-y^0)\rho_0(x,y) \, ,
\eeq
solves the inhomogeneous equation
\beq\db
\left[\square_x +m^2\right]G^R_0(x,y) = \delta^{d+1}(x-y) \, ,
\label{eqGR0}
\eeq
with retarded boundary conditions. In the same way we define the advanced
self-energy $\Sigma_{A,\cl}$. Retarded (advanced) Green functions and
self-energies vanish when \mbox{$x^0<y^0$} \mbox{($x^0>y^0$)}. 
With these definitions we can rewrite (\ref{eq:clSD}) and the
respective equation for the advanced Green function as:
\bea
\label{eqGR}
&&\db G^R_{\cl} = G^R_{0} - G^R_{0}\cdot\Sigma_{R,\cl}\cdot G^R_{\cl}\, ,\\
\label{eqGA}
&&\db G^A_{\cl} = G^A_{0} - G^A_{0}\cdot\Sigma_{A,\cl}\cdot G^A_{\cl}\, ,
\eea
where we use a compact notation 
\beq\db
A\cdot B = \int {\rm d}^{d+1}z\, A(x,z)B(z,y) \, .
\eeq
A combination of Eqs.\ (\ref{eqGR}) and (\ref{eqGA}) gives the 
{\rr\em Schwinger-Dyson equation for the classical spectral function} 
$\rho_\cl = G^R_{\cl}-G^A_{\cl}$ with
\beq\rr
\rho_\cl = \rho_0 - G^R_{0}\cdot\Sigma_{R,\cl}\cdot G^R_{\cl} 
+ G^A_{0}\cdot\Sigma_{A,\cl}\cdot G^A_{cl} \, .
\label{eqrhocl}
\eeq
There is a similar identity for the classical statistical propagator $F_\cl$,
defined in (\ref{eqFcl}). Using the definitions (\ref{eqGR}) and (\ref{eqGA}) 
one can write the following
identity for the statistical function
\beq\db
F_{\rm cl} = F_0 -
G^R_{0}\cdot \left[ \Sigma_{R,\cl} - {G^R_{\cl}}^{-1}\right] \cdot F_{\rm
cl} 
- F_0 \cdot\left[\Sigma_{A,\cl} + {G^A_{0}}^{-1}\right]\cdot
G^A_{\cl} \, .
\eeq
We can now define a classical statistical self-energy as
\beq\db 
\Sigma_{F,\cl} = -{G^R_\cl}^{-1}\cdot F_\cl\cdot 
{G^A_\cl}^{-1} + {G^R_0}^{-1}\cdot F_0 \cdot {G^A_0}^{-1} \, ,
\eeq
and find the {\rr\em Schwinger-Dyson equation for the statistical propagator:}
\beq\rr
F_{\rm cl} = F_0 - G^R_0\cdot \Sigma_{R,\cl}\cdot F_\cl 
- F_0\cdot \Sigma_{A,\cl}\cdot G^A_\cl - G^R_0\cdot
\Sigma_{F,\cl} \cdot G^A_\cl \, .
\label{eqSDFcl}
\eeq
Acting with $(\Box + m^2)$ on (\ref{eqrhocl}) and (\ref{eqSDFcl}) brings 
the classical Schwinger-Dyson equations in a form which is more
suitable for initial-value problems. We make the retarded nature of 
$\Sigma_{R,\cl}$ manifest by writing
\beq\db
\Sigma_{R,\cl}(x,y) = \Sigma^{(0)}_\cl(x)\delta^{d+1}(x-y) +
\Theta(x^0-y^0) \Sigma_{\rho,\cl}(x,y) \, ,
\eeq
and similarly for $\Sigma_{A,\cl}(x,y)$. The spectral component of the 
classical self-energy is $\Sigma_{\rho,\cl}(x,y) = \Sigma_{R,\cl}(x,y) - 
\Sigma_{A,\cl}(x,y) = -\Sigma_{\rho,\cl}(y,x)$. After properly
taking into account all $\Theta$--functions as well as 
(\ref{eqrho0}), (\ref{eqGR0}) and (\ref{eqF0}), 
one finds for the {\rr\em exact time evolution
equations for $\rho_\cl$ and $F_\cl$:}
\bea\rr
\left[\square_x +M^2_\cl(x)\right]\rho_\cl(x,y) &\rr =&\rr
- \int_{y^0}^{x^0}\!\! {\rm d} z \,\,
\Sigma_{\rho, \cl}(x,z)\rho_\cl(z,y) \, ,
\label{eqrho1cl}
\\\rr
\left[\square_x +M^2_\cl(x)\right]F_\cl(x,y)\! &\rr =&\rr
- \int_0^{x^0}\!\! {\rm d} z
\,\, \Sigma_{\rho, \cl}(x,z)F_\cl(z,y) \nonumber \\
&&\rr + \int_0^{y^0}\!\! {\rm d} z \,\, \Sigma_{F,\cl}(x,z)\rho_\cl(z,y) \, ,
\label{eqF1cl}
\eea        
where we use the abbreviated notation $\int_{t_1}^{t_2}
{\rm d}z \equiv \int_{t_1}^{t_2} {\rm d}z^0 
\int_{-\infty}^{\infty} {\rm d}^d z$ and 
\beq\db
M^2_\cl(x) = m^2 + \Sigma^{(0)}_\cl(x).
\eeq
We emphasize that the form of these classical time evolution equations 
is the same as for the quantum evolution described by (\ref{eq:exactrhoF}). 
If the initial conditions
are chosen to be the same then the only difference between classical
and quantum theory originates from the self-energies entering
the evolution equations.   
As a direct consequence, one concludes that {\rr\em LO large-$N$ or any 
Gaussian/Hartree type approximation exhibits purely classical
dynamics:} since the quantum self-energies  
$\Sigma_F$ and $\Sigma_\rho$ as well as the classical $\Sigma_{F,\cl}$
and $\Sigma_{\rho,\cl}$ vanish for these approximations the evolution 
equations derived from the
quantum theory are identical to those derived from the classical
theory. The tadpole contributions $\Sigma^{(0)}$ and $\Sigma^{(0)}_\cl$ 
entail no difference if the initial conditions
are chosen to be the same.
In the following we consider approximations with
non-vanishing spectral and statistical self-energies in order to 
discuss the differences between classical and quantum dynamics.

\subsection{Classicality condition}
\label{sec:classicality}

The classical self-energies 
$\Sigma_{\rho, \cl}$, $\Sigma_{F,\cl}(x,z)$
and $\Sigma^{(0)}_\cl$ could be approximated by 
perturbation theory to a given order in the coupling $\lambda$.
However, as for the case of the quantum evolution the perturbative
classical self-energies lead to a secular time-evolution and 
fail to provide a reliable description for the
classical nonequilibrium dynamics. The analytical description
can, however, be based on the same 2PI summation techniques
as introduced for the quantum theory above. We will only state here
the result of the respective calculation within classical statistical 
field theory, since we will then discuss how the same result can be directly
obtained from the classical limit of the known quantum self-energies. 

One finds for the classical statistical $O(N)$ model employing the 
2PI $1/N$ expansion to next-to-leading order:
\bea\db
M^2_\cl(x) &\db =&\db m^2 + \lambda\frac{N+2}{6N} F_\cl(x,x) \, ,
\nonumber
\\
\db \Sigma_{F,\cl}^{\rm (NLO)}(x,y) 
&\db =&\db -\frac{\lambda}{3N} F_\cl(x,y)I_{F,\cl}(x,y) \, ,
\label{eqSigmacl}\\
\db \Sigma_{\rho,\cl}^{\rm (NLO)}(x,y) &\db =&\db -\frac{\lambda}{3N}
\Big(F_\cl(x,y)I_{\rho,\cl}(x,y)+\rho_\cl(x,y)I_{F,\cl}(x,y)\Big) \, ,
\nonumber
\eea
with the classical summation functions 
\bea
\db I_{F,\cl}(x,y) &\db=&\db  \frac{\lambda}{6}\, F_\cl^2(x,y)  
\nonumber \\
&\db-&\db  \frac{\lambda}{6}\, \int {\rm d}^d z
\left\{  \int\limits_{0}^{x^0} dz^0 \, I_{\rho,\cl}(x,z) F_\cl^2(z,y) 
-\, 2 \int\limits_{0}^{y^0} dz^0\, I_{F,\cl}(x,z)  F_\cl(z,y) 
\rho_\cl(z,y) \right\} \, ,
\nonumber\\\db
I_{\rho,\cl}(x,y) &\db=&\db \frac{\lambda}{3}\, F_\cl(x,y) \rho_\cl(x,y) 
- \frac{\lambda}{3}\, \int {\rm d}^d z
{\db \int\limits_{y^0}^{x^0} dz^0}\,  
I_{\rho,\cl}(x,z) F_\cl(z,y) \rho_\cl(z,y) \, .
\label{eq:clIRFR}
\eea
At this stage we can compare the result from 
the 2PI $1/N$ expansion for the quantum theory, 
as discussed in Sec.~\ref{sec:2PIN}, with the resummed
classical expressions obtained here. One observes that the classical 
self-energies  (\ref{eqSigmacl})---(\ref{eq:clIRFR}) 
are obtained from the corresponding quantum expressions 
(\ref{eq:massNLO})---(\ref{IRFR}) by dropping
the terms with a product of two spectral functions $\rho$
compared to a product of two statistical
propagators $F$. Accordingly, the classical self-energies
are obtained from the respective
quantum ones as
\beq \db
\Sigma_{F,\cl} \,=\,  \Sigma_{F} (F^2 \gg {\rr \rho^2}) \quad ,\quad
\Sigma_{\rho,\cl} \,=\,  
\Sigma_{\rho} (F^2 \gg {\rr \rho^2}) \nonumber\, .
\label{eq:classlimit}
\eeq

The same analysis can be done employing the 2PI loop expansion. For the
following analytical discussion we consider the spatially homogeneous
case and employ the Fourier modes $\Sigma_{F,\cl}(t,t';\bp)$
and $\Sigma_{\rho,\cl}(t,t';\bp)$. For
the classical statistical $O(N)$ symmetric theory one obtains 
to two-loop order the self-energies: 
\bea\db
\Sigma_{F,\cl}^{\rm (2loop)}(t,t';\bp) &\db\!=\!& 
\db - \lambda^2\frac{N+2}{18N^2}
\int_{\bq,\bk} 
F_\cl(t,t';\bp-\bq-\bk) F_\cl(t,t';\bq)F_\cl(t,t';\bk)  \, ,
\nonumber\\\db
\Sigma_{\rho,\cl}^{\rm (2loop)}(t,t';\bp) &\db\!=\!&
\db - \lambda^2\frac{N+2}{6N^2}
\int_{\bq,\bk} 
\rho_\cl(t,t';\bp-\bq-\bk) F_\cl(t,t';\bq)F_\cl(t,t';\bk) \, .
\label{eq:cl2loop}
\eea
Let us directly confront this with the respective two-loop 
self-energies for the respective {\rr\em quantum} $O(N)$ theory: 
\bea\db 
\Sigma_{F}^{\rm (2loop)}(t,t';\bp) &\db\!=\!&\db - \lambda^2\frac{N+2}{18N^2} 
\int_{\bq,\bk} 
F(t,t';\bp-\bq-\bk)\nonumber \\ &&\db \left[F(t,t';\bq)F(t,t';\bk) 
{\rr - \frac{3}{4}\, \rho(t,t';\bq)\rho(t,t';\bk)} \right] \, , 
\nonumber\\\db
\Sigma_{\rho}^{\rm (2loop)}(t,t';\bp) &\db\!=\!&
\db - \lambda^2\frac{N+2}{6N^2} 
\int_{\bq,\bk} \rho(t,t';\bp-\bq-\bk)
  \nonumber\\ &&\db \left[F(t,t';\bq)F(t,t';\bk) 
{\rr - \frac{1}{12}\, \rho(t,t';\bq)\rho(t,t';\bk)} \right] \, .
\label{eq:qm2loop}
\eea
From this one infers a {\rr\em sufficient 
condition for classical evolution:}
\beq
\fbox{\db
$\displaystyle |F(t,t';\bq)F(t,t';\bk)|\, \gg \,
 {\rr \frac{3}{4}\, |\rho(t,t';\bq)\rho(t,t';\bk)|}$
}\label{eq:classicality}
\eeq
This condition has to be fulfilled for {\rr\em all times and all momenta}
in order to ensure that classical and quantum evolution agree.
However, we will observe below that it can be typically only achieved 
for a limited range of time and/or momenta. 

One expects that the classical description
becomes a reliable approximation for the quantum theory if
the number of field quanta in each mode is sufficiently high. 
The classicality condition (\ref{eq:classicality}) entails
the justification of this expectation. In order to illustrate the 
condition in terms of a more intuitive picture of occupation numbers,
we employ the free-field theory type form of the spectral function and
statistical propagator given in Eq.~(\ref{eq:freefieldFrho}).
From this one obtains the following estimates for the
{\rr\em time-averaged} correlators:
\bea \db
\overline{F^2}(t,t';\bp) &\db\!\equiv\!&\db
\frac{\omega_{\bp}}{2 \pi} \int_{t-2 \pi/\omega_{\bp}}^{t}  
\!\!\!\!\!\!\!\!\!\!\!\!\!\! \rmd t' \,
F^2(t,t';\bp) \,\, \leadsto\,\, 
\frac{(n_{\bp}(t)+1/2)^2}{2 \omega_{\bp}^2(t)} \, ,  
\nonumber\\
\db \overline{\rho^2}(t,t';\bp) &\db\!\leadsto\!&
\db \frac{1}{2 \omega_{\bp}^2(t)} \,\, .
\eea
Inserting these estimates in (\ref{eq:classicality}) for equal momenta 
yields 
\beq
{\db\left[n_{\bp}(t)+\frac{1}{2} \right]^2 \, \gg \, \frac{3}{4} }
\qquad {\rm or} \qquad {\db n_{\bp}(t) \, \gg \, 0.37 } \, .
\label{eq:estimatencl}
\eeq
This limit agrees rather well with what is obtained for the
case of thermal equilibrium.
For a Bose-Einstein distributed particle number $n_{\rm BE} = 
(e^{\omega/T}-1)^{-1}$ with temperature $T$ one finds 
$n_{\rm BE}(\omega = T) = 0.58$, below which deviations from the 
classical thermal distribution become sizeable. We emphasize 
that in thermal equilibrium the fluctuation-dissipation 
relation (\ref{eq:flucdissbose}) ensures $|F^{\rm (eq)}(\omega,\bp)| \gg 
|\rho^{\rm (eq)}(\omega,\bp)|$ for high-temperature modes $T \gg \omega $:
\beq \db
F^{\rm (eq)}(\omega,\bp) = -i
\Big({\rr n_{\rm BE}(\omega)}+\frac{1}{2} \Big) \, 
\rho^{\rm (eq)}(\omega,\bp)  
\stackrel{{\rr T \gg \omega}}{\simeq} -i\, {\rr \frac{T}{\omega}}\, 
\rho^{\rm (eq)}(\omega,\bp)  \, .
\label{eq:flucdissestimate}
\eeq
The nonequilibrium estimate (\ref{eq:estimatencl}) as well as the
equilibrium result (\ref{eq:flucdissestimate}) leads one to expect
that the quantum evolution described by $F$ and $\rho$ 
is well approximated by the classical evolution in terms of 
$F_{\cl}$ and $\rho_{cl}$ for
large initial occupation numbers \mbox{$n_{\bp}(t = 0)$}. 
However, note that the nonequilibrium classicality condition 
(\ref{eq:classicality}) is typically not fulfilled at all times,
since the unequal-time correlator $F(t,t';\bp)$ can oscillate
around zero with a phase difference to the oscillations of 
$\rho(t,t';\bp)$ (cf.~also the free-field 
expressions (\ref{eq:freefieldFrho})). In particular, we will find
that characteristic quantities for the quantum late-time behavior 
such as thermalization times are very badly approximated
by the respective classical estimates. Nevertheless, for 
time-averaged quantities and not too late times $t \ll t_{\rm eq}$,
i.e.~before the approach to quantum thermal equilibrium sets in,
the classical evolution can give an accurate description for
sufficiently high initial occupation numbers as will be demonstrated 
below.

A similar analysis can be performed in the presence of
a non-vanishing classical macroscopic field $\phi_{a,\cl}(x)$ defined
in (\ref{eq:phicl}). For instance, to NLO in the 2PI $1/N$
expansion the {\rr\em classical macroscopic field} is described by the
evolution equation  
\bea\db
\left(\left[\square_x + \frac{\lambda}{6N}
\phi_\cl^2(x)\right] \delta_{ab} 
+ M_{\cl,ab}^2(x;\phi_\cl \equiv 0,F_\cl) 
\right) \phi_{\cl,b}(x) \qquad\qquad  \nonumber\\
\db = -  \int_0^{x^0}\!\! {\rm d}y \, 
 \Sigma^{\cl}_{\rho,ab}(x,y;\phi_\cl \equiv 0,F_\cl,\rho_\cl)\,\, 
\phi_{\cl,b} (y)\, ,
\label{eq:clNLOphi}
\eea
where $M_{\cl,ab}^2(x;\phi_\cl \equiv 0,F_\cl)$ is given by
(\ref{Meff}) with the replacement $F \to F_\cl$, and 
$\Sigma^{\cl}_{\rho,ab}(x,y;\phi_\cl \equiv 0,F_\cl,\rho_\cl)$
is obtained from the respective quantum self-energy
$\Sigma_{\rho,ab}(x,y;\phi \equiv 0,F,\rho)$ given in
(\ref{ASRFR}) by employing (\ref{eq:classlimit}). This can be
directly compared to the corresponding evolution
equation (\ref{eq:NLOphi}) for the case of the quantum field theory.

\subsection{Precision tests and the role of quantum corrections}

The nonequilibrium evolution of classical
correlation functions in the $O(N)$ model can be obtained numerically
up to controlled statistical errors.   
Initial conditions for the nonequilibrium evolution are
determined from a probability functional on classical phase-space. 
The subsequent time evolution is solved numerically using the
classical equation of motion for the field (\ref{eqmotion}).
The results presented below have been obtained from sampling
50000-80000 independent initial conditions to approximate the exact
evolution of correlation functions. 
The statistical propagator is constructed from these
individual runs according to (\ref{eqFcl}) and using 
(\ref{eq:classspec}) for the spectral function.
Since these results include all orders 
in $1/N$, they can be used for a precision test of the 2PI $1/N$
expansion implemented in classical statistical field theory.
We emphasize that this compares two very different calculational
procedures: the results from the simulation involve thousands of 
individual runs from which the correlators are constructed, while the 
corresponding results employing the 2PI $1/N$ expansion involve
only a single run solving directly the evolution equation for the
correlators. The accuracy of the simulations manifests itself 
also in the fact that the time-reversal invariant dynamics can 
be explicitly reversed in practice for not too late times.
The close agreement between
full and 2PI NLO results, which will be observed below, is 
therefore also indicative of the numerical precision of the
latter. 

We consider a system that is invariant under space translations and work
in momentum space.
We choose a Gaussian initial state such that a specification of the
initial two-point functions is sufficient. 
As mentioned above,
the classical spectral function at initial time is completely
determined from the equal-time relations (\ref{eq:cleqtime}). 
For the statistical propagator we take $F(0,0;p) = [n_p(0)+\half]/\om_p$,
with the initial particle number    
representing a peaked ``tsunami'' 
(cf.\ Sect.~\ref{sec:lofixedpoints}).

\begin{figure}[t]
 \centerline{
\epsfig{file=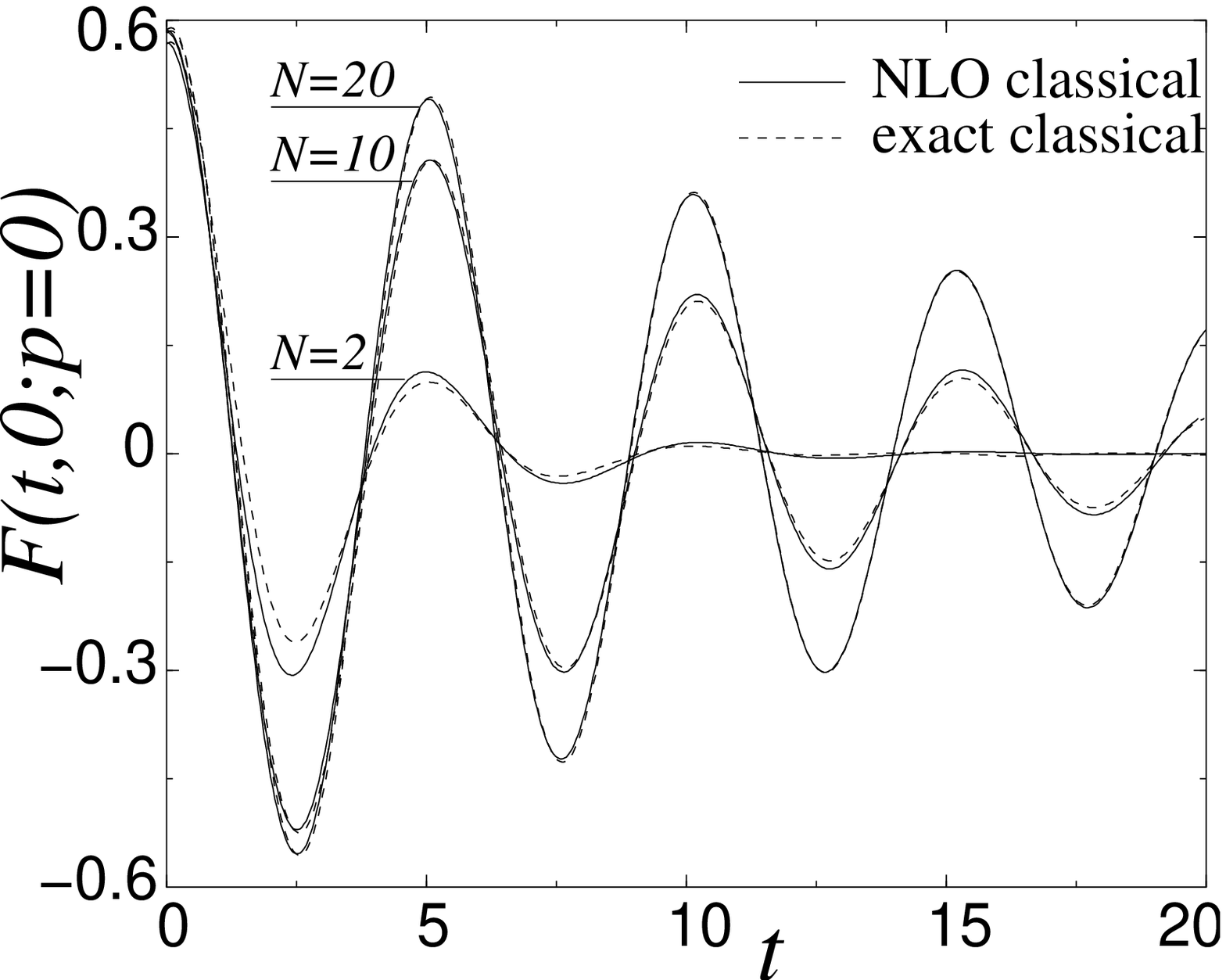,width=7.3cm}
 \hspace{.cm}
\epsfig{file=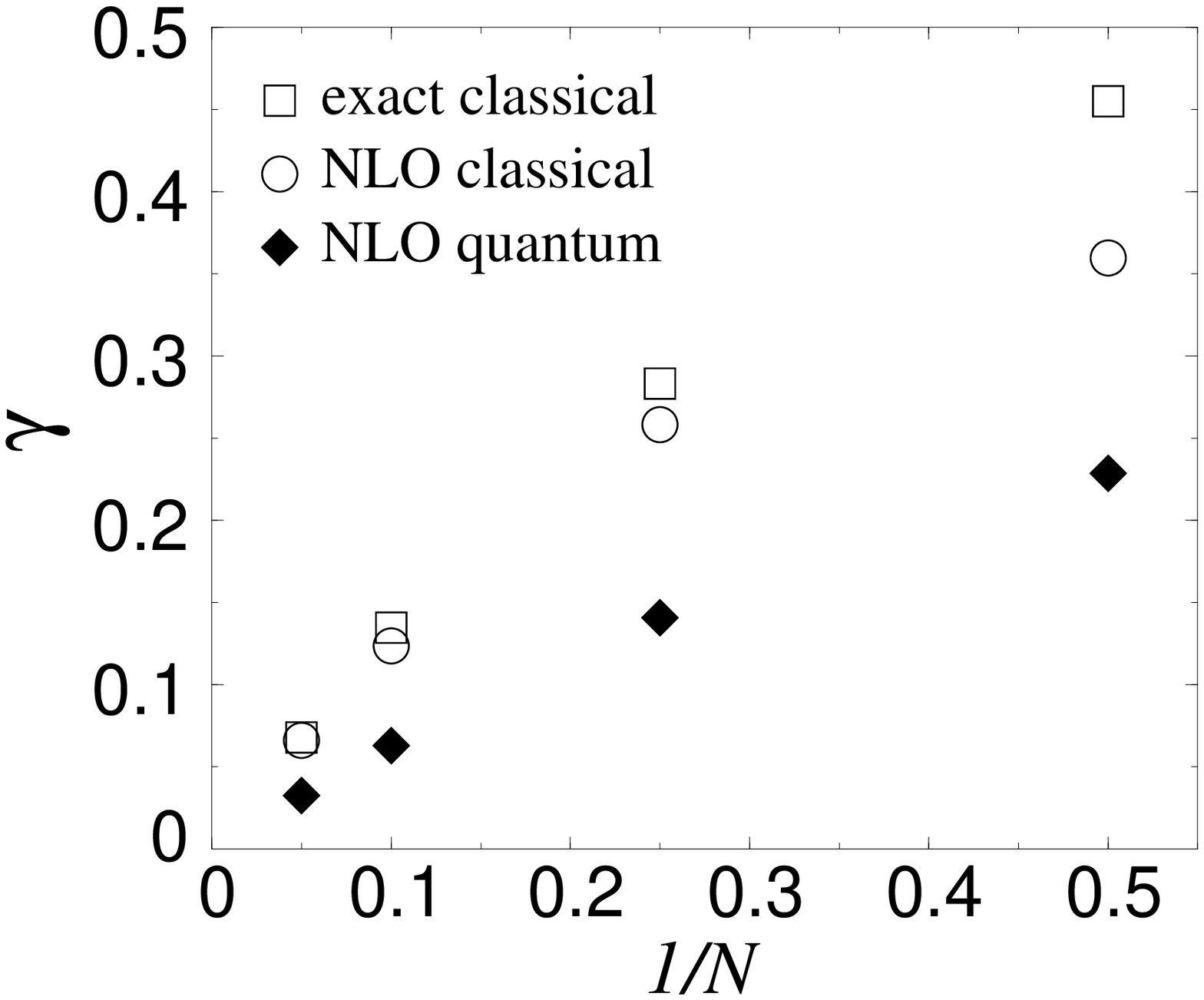,width=7.2cm}}
 \caption{\label{fig:classearly} {\bf Left:} Unequal-time two-point 
function $F(t,0;p=0)$ at zero
momentum for $N=2,10,20$. The full lines show
results from the NLO classical evolution and the dashed lines from the
exact classical evolution (MC). One observes a convergence of 
classical NLO and exact results already for moderate values of $N$.
 {\bf Right:} Damping rate extracted 
 from $F(t,0;p=0)$ as a function of $1/N$. 
 Open symbols represent NLO and exact classical evolution.
 The quantum results are shown with full symbols for comparison.
 The initial conditions are characterized by low occupation 
 numbers so that quantum effects become sizeable. One observes
 that in the quantum theory
 the damping rate is reduced compared to the classical theory. 
 (All in units of $m_R$.)}
\end{figure}
The initial mode energy is given by $\om_p = (p^2+M_\cl^2)^{1/2}$
where $M_\cl$ is the one-loop renormalized mass in presence of the
nonequilibrium medium, determined from the one-loop ``gap equation''
for $M_\cl$ in Eq.~(\ref{eqSigmacl}). 
As a renormalization condition we choose the one-loop renormalized mass in
vacuum $m_R\equiv M|_{n(0)=0}=1$ as our dimensionful scale.
The results shown below are obtained using a fixed coupling constant 
$\lambda/m_R^2=30$. 

On the left of Fig.~\ref{fig:classearly} the classical statistical propagator 
$F_\cl(t,0;p=0)$ is presented for three values of $N$. 
All other parameters
are kept constant. The figure compares the time evolution using the 2PI
$1/N$ expansion to NLO and the full Monte Carlo (MC)
calculation.  One observes that the
approximate time evolution of the correlation function shows a rather good
agreement with the exact result even for small values of $N$ (note that
the effective four-point coupling is strong, $\lambda/6N=2.5 m_R^2$ for
$N=2$). For $N=20$ the exact and NLO evolution can hardly be
distinguished. A very sensitive quantity for comparisons is the damping rate
$\gamma$, which is obtained from an exponential fit to the envelope of
$F_\cl(t,0;p=0)$.  The systematic convergence of the NLO and the Monte Carlo
result as a function of $1/N$ can be observed from the right graph of
Fig.~\ref{fig:classearly}. The
quantitatively accurate description of far-from-equilibrium processes
within the NLO approximation of the 2PI effective action is manifest.

The right graph of Fig.~\ref{fig:classearly} 
also shows the damping rate from the quantum
evolution, using the same initial conditions and parameters. One observes
that the damping in the quantum theory differs and, in particular, is
reduced compared to the classical result. The effective loss of details 
about the initial conditions takes more time for the quantum system
than for the corresponding classical one. In the limit $N\to \infty$
damping of the unequal-time correlation function goes to zero
since the nonlocal part of the self-energies
vanishes identically at LO large-$N$ and scattering is absent. In
this limit there is no difference between evolution in a quantum and
classical statistical field theory. (Cf.~Sec.~\ref{sec:lofixedpoints} 
and the discussion above.)

\begin{figure}[t]
 \centerline{
\epsfig{file=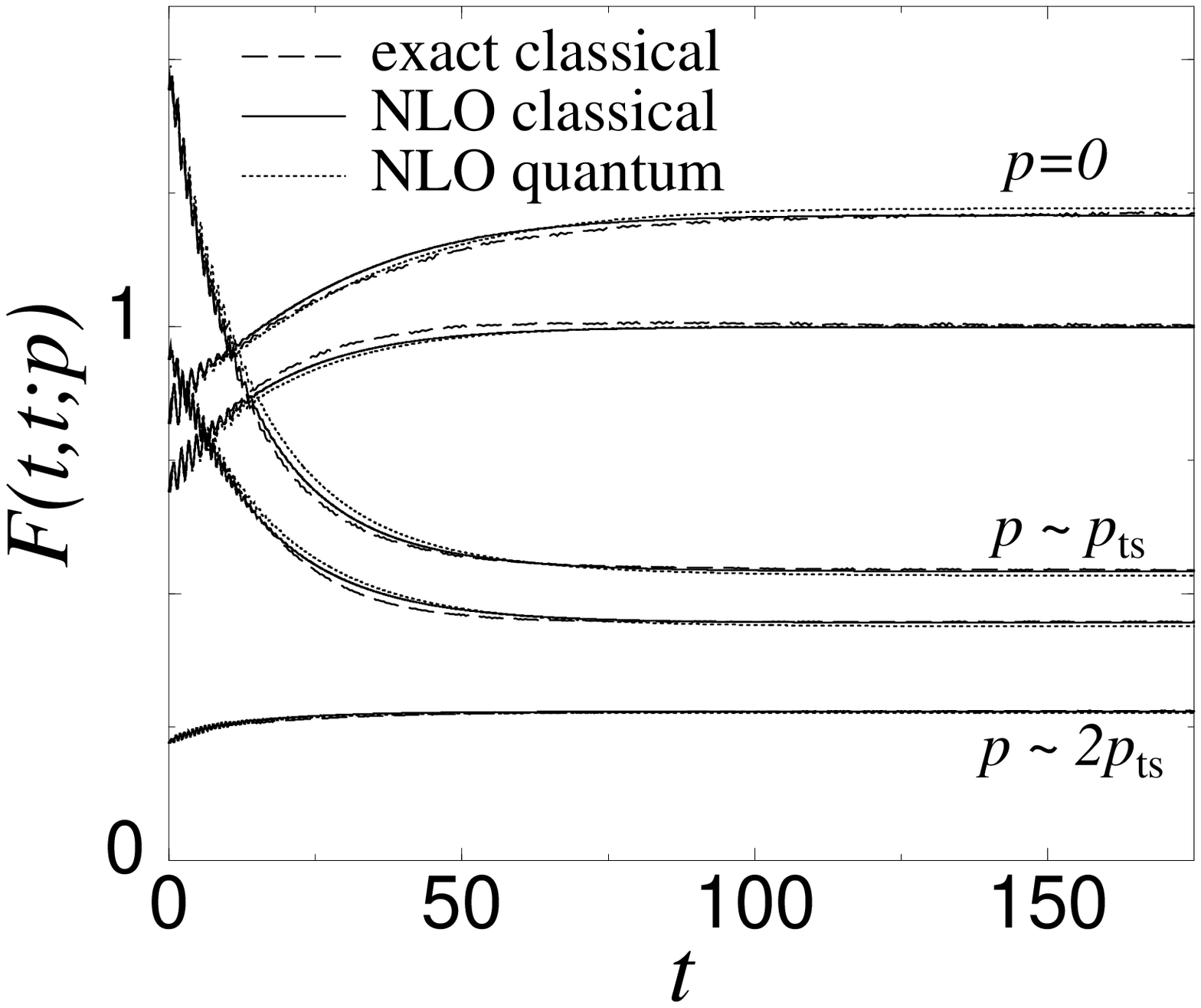,width=7.3cm}
 \hspace{.1cm}
\epsfig{file=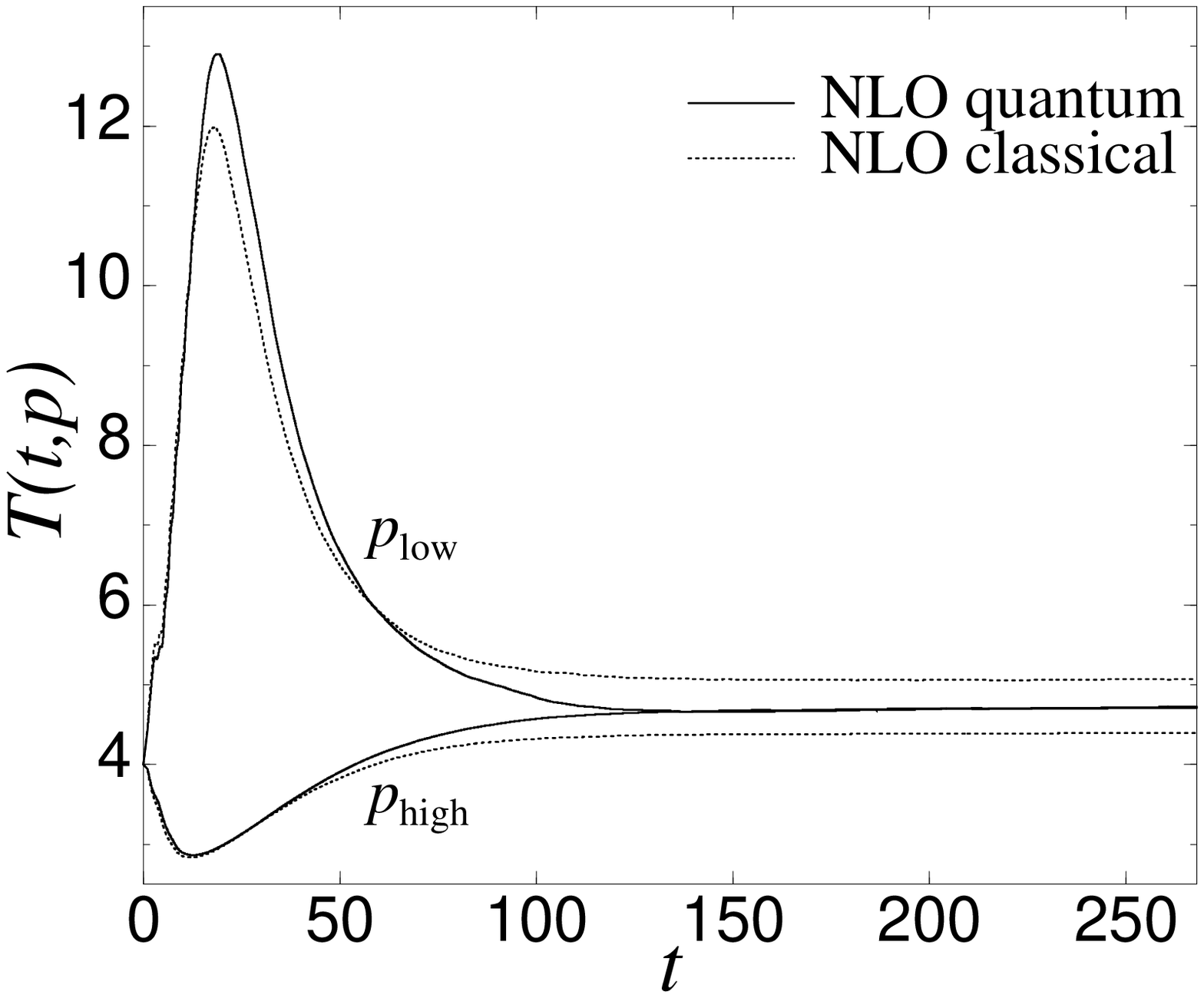,width=7.cm}}
 \caption{\label{fig:classlate} {\bf Left:} Nonequilibrium evolution of the 
equal-time two-point function $F(t,t;p)$ for $N=10$ for various momenta $p$. 
One observes a good agreement between the exact MC (dashed) and
the NLO classical result (full). The quantum evolution is shown with
dotted lines. The integrated initial particle density is six times as high as
in Fig.\ \ref{fig:classearly}. 
 {\bf Right:} A very sensitive quantity to 
study deviations is the time dependent inverse slope 
$T(t,p)$ defined in the text. When quantum thermal equilibrium 
with a Bose-Einstein distributed particle number is approached all 
modes get equal $T(t,p)=T_{\rm eq}$, as can be observed to high 
accuracy for the quantum evolution. For classical 
thermal equilibrium the defined inverse slope remains momentum dependent.}
\end{figure}

For finite $N$ scattering is present and quantum and classical
evolution differ in general. However, as discussed in 
Sec.~\ref{sec:classicality}, the classical field approximation
may be expected to become a reliable description for the quantum theory if
the number of field quanta in each field mode is sufficiently high. 
We observe that increasing the initial particle number density leads
to a convergence of quantum and classical time evolution at not too late
times. In Fig.~\ref{fig:classlate} (left) the time evolution of 
the equal-time
correlation function $F(t,t;p)$ is shown for several momenta $p$ and $N=10$.
Here the integrated particle density $\int \frac{dp}{2\pi} n_p(0)/M_\cl=1.2$ 
is six times as high as in Fig.~\ref{fig:classearly}. 
At $p = 2 p_{\rm ts}$ one finds  
$n_{p = 2 p_{\rm ts}}(0) \simeq 0.35$ and a slightly larger value 
at this momentum of about $\simeq 0.5$ for late times. 
For these initial conditions the estimate (\ref{eq:estimatencl}) 
for the classicality condition (\ref{eq:classicality}) is therefore 
approximately valid up to momenta $p \simeq 2 p_{\rm ts}$, and
one indeed observes from the left of Fig.~\ref{fig:classlate} 
a rather good agreement of quantum and classical evolution 
in this range. For an estimate of 
the NLO truncation error we also give the full (MC)
result for $N=10$ showing a quantitative agreement 
with the classical NLO evolution both at early and later times.

\subsubsection{Classical equilibration and quantum thermal equilibrium}  

From the left of Fig.~\ref{fig:classlate} one observes that the initially 
highly occupied ``tsunami'' modes ``decay'' as time proceeds and the
low momentum modes become more and more populated. At late times
the classical theory and the quantum theory approach their 
respective equilibrium distributions. Since classical and quantum 
thermal equilibrium are distinct, the classical and quantum time evolutions
have to deviate at sufficiently late times.
Figure~\ref{fig:classlate} shows the time dependent inverse slope 
parameter 
\beq\db
T(t,p) \equiv - n_p(t) \mbox{$[n_p(t)+1]$} 
\left(\frac{{d} n_p}{{d} \epsilon_p}\right)^{-1}  \, ,
\label{alsoinvslope}
\eeq
which has been introduced in Sec.~\ref{sec:NLOtherm} to
study the approach to quantum thermal equilibrium.\footnote{Note
that $\,{d} {\rm Log}(n_p^{-1}(t) + 1)/{d} \eps_p(t) = T^{-1}(t,p)$.} 
It employs the effective particle number $n_p(t)$ defined in
(\ref{eq:effpart})\footnote{Here we employ $Q(t,t';\bp) = 0$,
cf.~Sec.~\ref{sec:NLOtherm}.} and mode energy $\epsilon_p(t)$ given by
(\ref{eq:effen}).
Initially one observes a very different behavior of $T(t,p)$ for the low
and high momentum modes, indicating that the system is far from
equilibrium. The quantum evolution approaches
quantum thermal equilibrium with a momentum independent inverse slope 
$T_{\rm eq}=4.7\, m_R$ to high accuracy. 
In contrast, in the classical limit the slope parameter remains momentum
dependent since the classical dynamics does of course not reach a 
Bose-Einstein distribution.

To see this in more detail we note that 
for a Bose-Einstein distribution, $n_{\rm BE}(\ep_p) =
1/[\exp(\eps_p/T_{\rm eq})-1]$, the inverse slope (\ref{alsoinvslope})
is independent of the mode energies and
equal to the temperature $T_{\rm eq}$.  During the nonequilibrium evolution
effective thermalization can therefore be observed if $T(t,p)$ 
becomes time and momentum independent, $T(t,p) \to T_{\rm eq}$. This is 
indeed seen on the left of Fig.~\ref{fig:classlate} for the quantum 
system. If the system is approaching classical equilibrium 
at some temperature $T_\cl$ and is weakly coupled, the following behavior 
is expected.  From the definition (\ref{eq:effpart}) of $n_p(t)$ in
terms in two-point functions, we expect to find approximately
\beq\db
T(t,p)  \to  T_\cl\left( 1- \ep_p^2/T_\cl^2\right) \, ,
\label{eqinvsl}
\eeq
i.e.~a remaining momentum dependence with $T(t,p) <
T(t,p')$ if $\ep_p>\ep_p'$. Indeed, this is
what one observes for the classical field theory 
result in Fig.~\ref{fig:classlate}.

For a classical theory a very simple test 
for effective equilibration
is available. An exact criterion can be obtained 
from the classical counterpart of the
``KMS'' condition for thermal equilibrium discussed in 
Sec.~\ref{sec:detourthermal}. In coordinate space the 
classical equilibrium ``KMS'' 
condition reads
\beq\db
\frac{1}{T_\cl} \frac{\partial}{\partial x^0}F^{\rm (eq)}_\cl(x-y) = 
-\rho^{\rm (eq)}_\cl(x-y),
\label{eqclKMS}
\eeq
and in momentum space 
\beq\db
F^{\rm (eq)}_\cl(k) = -in_\cl(k^0)\rho^{\rm (eq)}_\cl(k), \;\;\;\;\;\;\;\;
n_\cl(k^0) = \frac{T_\cl}{k^0}.
\eeq
Differentiating Eq.\
(\ref{eqclKMS}) with respect to $y^0$ at $x^0=y^0=t$ gives
\beq\db
\frac{1}{T_\cl}\frac{\partial}{\partial y^0}\frac{\partial}{\partial x^0}
F^{\rm (eq)}_\cl(x-y)\Big|_{x^0=y^0=t} 
= -\frac{\partial}{\partial y^0}\rho^{\rm (eq)}_\cl(x-y)\Big|_{x^0=y^0=t}. 
\eeq
Combining this KMS relation with the equal-time condition
(\ref{eq:cleqtime}) for the spectral function leads to
\beq\db
\partial_t\partial_{t'}F_\cl^{\rm (eq)}(t,t';\bx-\by)|_{t=t'} 
= T_\cl\delta(\bx-\by).
\eeq
In terms of the fluctuating classical fields 
$\pi_a(x) \equiv \partial_{x^0}\varphi_a(x)$ 
this is of course the well-known equilibrium relation
$\langle\pi_a(t,\bx)\pi_b(t,\by)\rangle_\cl^{\rm (eq)} =
T_\cl\delta(\bx-\by)\delta_{ab}$.
Out of equilibrium one can define an effective classical mode temperature
\beq\db
T_\cl(t,p) = \partial_t\partial_{t'}F_\cl(t,t';p)|_{t=t'} \, ,
\eeq
and effective classical equilibration is observed if $T_\cl(t,p)$
becomes time and momentum independent, $T_\cl(t,p)\to T_\cl$.

\begin{figure}[t]
\centerline{\epsfig{file=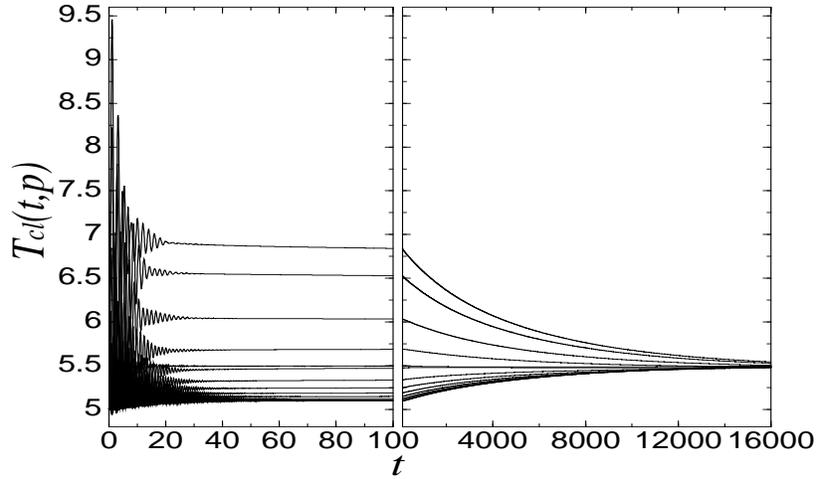,height=6.5cm,width=10.5cm}}
\caption{
Nonequilibrium time evolution in $1+1$ dimensions in the
classical limit of the three-loop approximation of the 2PI effective
action for one scalar field, $N=1$. Shown is the effective mode 
temperature $T_\cl(t,p)$, defined in the text, with initial
$T_\cl(0,p)/m_R = 5$ for all $p$. The classical field theory 
approaches classical  
equilibrium, $T_\cl(t,p) \to T_{\rm cl}$, after a very long time.
}
\label{fixedpointclassical}
\end{figure}
Apart from the 2PI $1/N$ expansion
the late-time behavior can also be studied from the 2PI loop expansion. 
For the quantum theory this has been demonstrated in 
Sec.~\ref{sec:latetimeuni}   
for the three-loop approximation of the 2PI
effective action (cf.~Fig.~\ref{fig:lateuni}, left).
The equivalent approximation in the classical theory is given 
by the evolution equations (\ref{eqrho1cl}) and (\ref{eqF1cl}) 
with the two-loop self-energies (\ref{eq:cl2loop}). In order
to demonstrate this,
we will use the loop approximation to verify explicitly the above 
statements about classical equilibration.

In Fig.\ \ref{fixedpointclassical} the nonequilibrium evolution of the
effective ``mode temperature'' $T_\cl(t,p)$ is shown 
for various momentum modes in the
$(1+1)$-dimensional classical scalar field theory for $N=1$.
The equations are solved by a lattice discretization with spatial
lattice spacing $m_R a=0.4$, time step $a_0/a=0.2$, and $N_s=24$ sites.
Without loss of generality we use $\lambda/m_R^2=1$. For the initial
ensemble we take $F_\cl(0,0;p) = T_0/(p^2+m_R^2)$ and 
$\partial_t\partial_{t'}F_\cl(t,t';p)|_{t=t'=0} = T_0$ with
$T_0/m_R=5$. We have observed that at sufficiently late times the
contributions from early times to the dynamics are effectively suppressed.
This fact has been
employed in Fig.\ \ref{fixedpointclassical} to reach the very late times.
One sees that at sufficiently late times the system relaxes towards
classical equilibrium with a final temperature $T_{\rm cl}/m_R \approx 5.5$.  
For the zero-momentum mode we find an exponential late-time relaxation of
$T_\cl(t,p=0)$ towards $T_{\rm cl}$ with a small rate ($\sim 2
\times 10^{-4} m_R$). It should be emphasized that typical 
classical equilibration times are substantially
larger than the times required to approach thermal equilibrium
in the respective quantum theory. In contrast to the
quantum theory statements about equilibration times are,
however, in general not insensitive to the employed momentum 
regularization for the classical theory because of the
classical Rayleigh-Jeans divergences.

\section{$n$-Particle irreducible generating functionals~II: Equivalence
hierarchy}
\label{sec:nPI2}

To understand the success and, more importantly, the limitations 
of expansion schemes based on the 2PI effective action we 
consider in this section $n$-particle irreducible ($n$PI) effective actions
for $n > 2$. Recall that the description of the 2PI effective 
action $\Gamma[\phi,D]$ employs a self-consistently dressed 
one-point function $\phi$ and two-point function, which for
notational purposes we denote here by $D$:\footnote{$G$ will denote
the ghost propagator in gauge theories below.} 
The one-point and two-point functions are 
dressed by solving the equations of motion 
$\delta \Gamma/\delta \phi = 0$ and $\delta \Gamma/\delta D = 0$
for a given order in the (e.g.~loop) expansion of 
$\Gamma[\phi,D]$ (cf.~Sec.~\ref{sec:genfunc1}). 
However, the 2PI effective action does not treat the higher 
$n$-point functions with $n>2$ on the same footing as the lower ones:
The three- and four-point function etc.~are not self-consistently 
dressed in general, i.e.~the corresponding proper three-vertex 
$V_3$ and four-vertex $V_4$ are given by the classical ones. In contrast, 
the $n$PI effective action $\Gamma[\phi,D,V_3,V_4,\ldots,V_n]$ 
provides a dressed description 
for the proper vertices $V_3,V_4,\ldots,V_n$ as well,
with $\delta\Gamma/\delta V_3 =0,\delta\Gamma/\delta V_4 =0, 
\ldots ,\delta\Gamma/\delta V_n =0$. 

The use of $n$PI effective actions with higher $n > 2$ is not entirely 
academic. They are relevant in the presence of initial-time sources describing 
a non-Gaussian initial density matrix for nonequilibrium evolutions
(cf.~the discussion in Secs.~\ref{sec:initialconditions} and 
\ref{sec:noneq2PIeffaction}). They are also known to be relevant in
high-temperature gauge theories for a quantitative 
description of transport coefficients in the context
of kinetic theory. 
As an example, the calculation of shear viscosity in a 
theory like QCD can be based on the inclusion of an infinite series of 
2PI ``ladder'' diagrams in order to recover the leading order 
``on-shell'' results in the gauge coupling $g$. 
Further examples where approximation
schemes based on higher $n$PI effective actions are relevant  
include critical phenomena near second-order phase transitions.
For instance, the quantitative description of the universal behavior near the 
second-order phase transition of scalar $\phi^4$ theory goes beyond a 
2PI loop expansion.\footnote{Critical phenomena can be
described using the $1/N$ expansion of
the 2PI effective action beyond leading 
order. These approximations indeed resum an infinite number of loop diagrams
and, for instance, the NLO result can be rewritten as 
a ``loop approximation in the presence of an 
effective four-vertex'' (cf.~Secs.~\ref{sec:loopexp} and
\ref{sec:resonance}).} It requires taking into account vertex 
corrections that start with the 4PI effective action to four-loop
order. The latter agrees with the most general $n$PI loop expansion
to that order, which is shown below.

The evolution equations, which are obtained by variation of 
the \mbox{$n$PI} effective action, are closely related to known
exact identities for correlation functions, i.e.~Schwinger-Dyson 
(SD) equations. Without approximations 
the equations of motion obtained from an exact \mbox{$n$PI} 
effective action and the exact SD equations have to agree since one 
can always map identities onto each other. However, in general
this is no longer the case for a given order in the
loop or coupling expansion of the $n$PI effective action.
By construction, SD equations are expressed in terms of loop diagrams 
including both classical and dressed vertices, which leads to 
ambiguities of whether classical or dressed ones should be 
employed at a given truncation order.     
In particular, SD equations are not closed a priori in the sense that
the equation for a given $n$-point function always involves
information about $m$-point functions with $m > n$. 
These problems are absent using effective action 
techniques. In turn, the $n$PI results can be used to resolve 
ambiguities of whether classical or dressed vertices should be employed 
for a given truncation of a SD equation. For instance, in QCD
the three-loop effective action result leads to evolution equations, which 
are equivalent to the SD equation for the two-point 
function and the one-loop three-point function {\em if}$\,$ all
vertices in loop-diagrams for the latter are replaced by the full 
vertices at that order.  
As mentioned in previous sections, the ``conserving'' property 
of using an effective action truncation can have important 
advantages, in particular if 
applied to nonequilibrium time evolution problems, where the 
presence of basic constants of motion such as energy conservation 
is crucial. 

We will derive below the 4PI effective
action for a nonabelian $SU(N)$ gauge theory with fermions
up to four-loop or ${\mathcal O} (g^6)$ corrections,
starting from the 2PI effective action and doing subsequent
Legendre transforms   
(Secs.~\ref{sec:highereffective actions} and \ref{sec:nonabgauge}). 
The class of models include gauge 
theories such as QCD or abelian theories as QED, as well as simple 
scalar field theories with cubic or quartic interactions. 
In Sec.~\ref{sec:selfcomplete} we derive an equivalence
hierarchy for $n$PI effective actions, which implies that the 
4PI results to this order
are equivalent to those from the $n$PI effective action
up to four-loop or ${\mathcal O} (g^6)$ corrections for
arbitrary $n > 4$. We derive the
non-equilibrium gauge field and fermion evolution equations
(Sec.~\ref{sec:nonequilibrium}), and
discuss the connection to kinetic theory in Sec.~\ref{sec:kinetic}.

\subsection{Higher effective actions}
\label{sec:highereffective actions}

Recall that all information about  
the quantum theory can be obtained from the effective 
action, which is a generating functional
for Green's functions evaluated in
the absence of external sources, i.e.~at the physical or 
stationary point. All functional representations of the
effective action are equivalent in the sense that they are
generating functionals for Green's functions including all 
quantum/statistical fluctuations and, in the absence of
sources, have to agree by construction:\footnote{Of course,
the nonequilibrium $n$PI effective action in the presence
of initial-time sources representing an initial
density matrix can differ in general. However, these pose no
additional complications since they vanish identically
for times different than the initial time. Cf.~the discussions
in Secs.~\ref{sec:noneq2PIeffaction} and \ref{sec:noneqeveq}.}
\bea\db
\Gamma[\phi] = \Gamma[\phi,D] = \Gamma[\phi,D,V_3] 
= \Gamma[\phi,D,V_3,V_4] = \ldots
= \Gamma[\phi,D,V_3,V_4,\ldots,V_n] 
\label{eq:equalitynPI}
\eea  
for arbitrary $n$ without further approximations. However, e.g.~loop expansions
of the 1PI effective action to a given order in the presence of 
the ``background'' field $\phi$ differ in general from a loop expansion
of $\Gamma[\phi,D]$ in the presence of $\phi$ and $D$. A similar
statement can be made for expansions of higher functional
representations. As mentioned in the introduction in 
Sec.~\ref{sec:introuniv},
for applications it is often desirable to obtain a self-consistently 
complete description, which to a given order in the expansion determines 
$\Gamma[\phi,D,V_3,V_4,\ldots,V_n]$ for arbitrarily high $n$. 
For this it is important to realize that there 
exists an {\rr\em equivalence hierarchy} 
as displayed in the introduction in Eq.~(\ref{eq:hierarchy}), which is derived
in Sec.~\ref{sec:selfcomplete}. For instance at three-loop order one has:
\bea\db
\Gamma^{\rm (3loop)}[\phi] \not = \Gamma^{\rm (3loop)}[\phi,D]
\not = \Gamma^{\rm (3loop)}[\phi,D,V_3] 
&\db\! =\!&\db \Gamma^{\rm (3loop)}[\phi,D,V_3,V_4] \\ 
&\db\! =\!&\db \Gamma^{\rm (3loop)}[\phi,D,V_3,V_4, \ldots,V_n]\, , \nonumber 
\eea
for arbitrary $n > 4$ in the absence of sources. 
As a consequence, there is
no difference between $\Gamma^{\rm (3loop)}[\phi,D,V_3]$ and
$\Gamma^{\rm (3loop)}[\phi,D,V_3,V_4]$ etc.~at the stationary point,
and the 3PI effective action captures already the 
complete answer for the self-consistent description
to this order. More explicitly,
the equality reads $\Gamma^{\rm (3loop)}[\phi,D(\phi),V_3(\phi)] =
\Gamma^{\rm (3loop)}[\phi,D(\phi),V_3(\phi),V_4(\phi)] = \ldots$
since at the stationary point of the effective action all $n$-point
correlations become functions of the field expectation value $\phi$.
At four-loop order the 4PI effective action 
would become relevant etc. For instance, for a theory as quantum 
electrodynamics (QED) or chromodynamics (QCD) the 2PI
effective action provides a self-consistently complete
description to two-loop order or\footnote{Here, 
and throughout the paper, $g$ means the strong gauge coupling $g_s$ for 
QCD, while it should be understood as the electric charge $e$ for QED.
For the power counting we take $\phi \sim {\mathcal O} (1/g)$
(cf.~Sec.~\ref{sec:compu}). The metric is denoted as 
$g^{\mu\nu} = g_{\mu\nu} = \mbox{diag}(1,-1,-1,-1)$. 
} 
${\mathcal O}(g^2)$: For a two-loop approximation
all $n$PI descriptions with $n \ge 2$ are equivalent
in the absence of sources. In contrast, a self-consistently complete 
result to three-loop order or ${\mathcal O}(g^4)$
requires at least the 3PI effective action etc.  
To go to much higher loop-order can become somewhat 
academic from the point of view of calculational feasibility. 

To present the argument we will first consider the 4PI effective action for 
a simple generic scalar model with cubic and
quartic interactions. The formal generalization to fermionic and
gauge fields is straightforward, and in Sec.~\ref{sec:nonabgauge} 
the construction is done for $SU(N)$ gauge theories with fermions. 
We use here a concise notation where Latin indices represent all 
field attributes, numbering real field components and their internal
as well as space-time labels, and sum/integration over repeated 
indices is implied.
We consider the classical action 
\beq\db
S[\varphi] = \frac{1}{2} \varphi_i\, iD_{0,ij}^{-1}\, \varphi_j - \frac{g}{3 !}
V_{03,ijk} \varphi_i\varphi_j\varphi_k
- \frac{g^2}{4 !} V_{04,ijkl} \varphi_i\varphi_j\varphi_k\varphi_l \, ,
\label{eq:action}
\eeq
where we scaled out a coupling constant $g$ for later convenience.
The generating functional for Green's functions in the
presence of quadratic, cubic and quartic source terms is
given by:
\bea\db
Z[J,R,R_3,R_4] &\db=&\db \exp\left(i W[J,R,R_3,R_4]\right) \nonumber\\
&=&\db \int \mathcal{D}\varphi 
\exp\Big\{ i \Big( S[\varphi] + J_i\, \varphi_i
+ \frac{1}{2} R_{ij}\, \varphi_i \varphi_j \label{eq:Z}\\
&&\db
+\frac{1}{3 !} R_{3,ijk}\, \varphi_i \varphi_j \varphi_k
+\frac{1}{4 !} R_{4,ijkl}\, \varphi_i \varphi_j \varphi_k \varphi_l
\Big)\Big\} \, . \nonumber
\eea
The generating functional for connected Green's functions, $W$, 
can be used to define the connected two-point ($D$), three-point ($D_3$) and
four-point function ($D_4$) in the presence of the sources,
\bea\db
\frac{\delta W}{\delta J_i} &\db=&\db \phi_i \, ,  \\
\db\frac{\delta W}{\delta R_{ij}} &\db=&\db \frac{1}{2}
\left( D_{ij} + \phi_i \phi_j \right) \, , \\
\db\frac{\delta W}{\delta R_{3,ijk}} &\db=&\db \frac{1}{6}
\left( D_{3,ijk} + D_{ij}\, \phi_k + D_{ki}\, \phi_j + D_{jk}\, \phi_i 
+ \phi_i\phi_j\phi_k\right) \, , \\
\db\frac{\delta W}{\delta R_{4,ijkl}} &\db=&\db \frac{1}{24}
( D_{4,ijkl} + [D_{3,ijk}\, \phi_l + 3\, {\rm perm.}] 
+ [D_{ij} D_{kl} + 2\, {\rm perm.}] \nonumber \\
&&\db
+ [D_{ij}\, \phi_k \phi_l + 5\, {\rm perm.}] 
+ \phi_i\phi_j\phi_k\phi_l ) \, .
\label{eq:Wderiv}
\eea
We denote the proper three-point and four-point vertices
by $g V_3$ and $g^2 V_4$, respectively, and define 
\bea\db
D_{3,ijk} &\db=&\db 
- i g\, D_{ii'}D_{jj'}D_{kk'} V_{3,i'j'k'}  \label{eq:V3def}
\, , \\\db
D_{4,ijkl} &\db=&\db - i g^2\, D_{ii'}D_{jj'}D_{kk'}D_{ll'} V_{4,i'j'k'l'}
\nonumber\\
&&\db + g^2\, (D_{ii'}D_{jj'}D_{k'u'}D_{w'l}D_{v'k} 
+ D_{ii'}D_{j'u'}D_{k'l}D_{jv'}D_{w'k} 
\nonumber\\
&&\db + D_{ii'}D_{j'u'}D_{k'k}D_{jv'}D_{l'l})
V_{3,i'j'k'} V_{3,u'v'w'} \, . 
\label{eq:V4def}
\eea
The effective action is obtained as the Legendre transform of
$W[J,R,R_3,R_4]$:\footnote{
In terms of the standard one-particle irreducible effective action
$\Gamma[\phi]=W[J]-J \phi$ the proper vertices $V_3$ and
$V_4$ are given by
\bea
g V_{3} = - \frac{\delta^3 \Gamma[\phi]}{\delta \phi\delta \phi\delta \phi} 
\quad , \quad g^2 V_{4} = 
- \frac{\delta^4 \Gamma[\phi]}{\delta \phi\delta \phi\delta \phi\delta \phi}
\nonumber .
\eea
Here it is useful to note that in terms of the connected 
Green's functions $D_n$ one has
\bea
\frac{\delta^2 W[J]}{\delta J \delta J} &=& i\, D \qquad , \qquad
\frac{\delta^2 \Gamma[\phi]}{\delta \phi \delta \phi}\,\, 
= \,\, i\, D^{-1} \, ,
\nonumber \\
\frac{\delta^3 W[J]}{\delta J \delta J \delta J} &=& -\, D_3 
\,\, =\,\, -i\, D^3\, 
\frac{\delta^3 \Gamma[\phi]}{\delta \phi \delta \phi \delta \phi}\, .
\nonumber \\
\frac{\delta^4 W[J]}{\delta J \delta J \delta J \delta J} &=& -i\, D_4 
\,\, =\,\, D^4\,  
\frac{\delta^4 \Gamma[\phi]}{\delta \phi \delta \phi \delta \phi \delta \phi}
+ 3 i\, D^5 
\left(\frac{\delta^3 \Gamma[\phi]}{\delta \phi \delta \phi \delta \phi}
\right)^2\, .
\nonumber
\eea
}
\bea\db
\Gamma[\phi,D,V_3,V_4] &\db=&\db W 
-\frac{\delta W}{\delta J_i} J_i - \frac{\delta W}{\delta R_{ij}} R_{ij}
\nonumber\\
&&\db - \frac{\delta W}{\delta R_{3,ijk}} R_{3,ijk} 
- \frac{\delta W}{\delta R_{4,ijkl}} R_{4,ijkl} \, .
\label{eq:LT}
\eea
For vanishing sources one observes from (\ref{eq:LT}) the stationarity 
conditions
\beq\db
\frac{\delta \Gamma}{\delta \phi} = \frac{\delta \Gamma}{\delta D}
= \frac{\delta \Gamma}{\delta V_3} = \frac{\delta \Gamma}{\delta V_4} = 0 \, ,
\label{eq:statcond}
\eeq
which provide the equations of motion for $\phi$, $D$, $V_3$ and $V_4$.

\subsubsection{4PI effective action up to four-loop order corrections}
\label{sec:compu}

Since the Legendre transforms employed in (\ref{eq:LT}) can be
equally performed subsequently, a most convenient 
computation of $\Gamma[\phi,D,V_3,V_4]$ 
starts from the 2PI effective action $\Gamma[\phi,D]$.
According to (\ref{2PIaction}) 
the exact 2PI effective action can be written as:
\beq\db
\Gamma[\phi,D] = S[\phi] + \frac{i}{2} \Tr \ln D^{-1}
+ \frac{i}{2} \Tr\, D_{0}^{-1}(\phi) D 
+ \Gamma_{2}[\phi,D] 
+ {\rm const} \, ,
\label{eq:2PIdef}
\eeq
with the field-dependent inverse classical propagator
\beq\db
i D_{0}^{-1}(\phi) = \frac{\delta^2 S[\phi]}{\delta\phi \delta\phi} \, .
\eeq
To simplify the presentation, we use in the following a
symbolic notation which suppresses indices and summation or
integration symbols (suitably regularized). 
In this notation the inverse classical propagator
reads
\beq\db
i D_{0}^{-1}(\phi) = i D_{0}^{-1} 
- g \phi V_{03}  
-\frac{1}{2} g^2 \phi^2 V_{04}  \, , 
\eeq
and to three-loop order one has\footnote{Note that
for $\phi \not = 0$, in the phase with spontaneous symmetry breaking,
$\phi \sim \mathcal{O}(1/g)$, and the three-loop result 
(\ref{eq:2PI3loop}) takes into account the contributions up to
order $g^6$.} (cf.~Sec.~\ref{sec:loopexp}) 
\bea\db
\Gamma_2 [\phi,D] &\db=&\db - \frac{1}{8} g^2 D^2 V_{04}  
+ \frac{i}{12} D^3 (g V_{03} + g^2 \phi V_{04})^2 
+ \frac{i}{48} g^4 D^4 V_{04}^2  
\nonumber\\
&&\db + \frac{1}{8} g^2 D^5 (g V_{03} + g^2 \phi V_{04})^2 V_{04} 
- \frac{i}{24} D^6 (g V_{03} + g^2 \phi V_{04})^4 
\nonumber\\[0.13cm]
&&\db +\, \mathcal{O}\left( 
g^n (g^2 \phi)^m |_{n+m=6}
\right)   \, ,
\label{eq:2PI3loop}
\eea
for $n,m=0,\ldots,6$. We emphasize that the exact $\phi$-dependence of 
$\Gamma_2[\phi,D]$ can be written as a function of the combination 
$(g V_{03} + g^2 \phi V_{04})$.
In order to obtain the vertex 2PI effective action $\Gamma[\phi,D,V_3,V_4]$
from $\Gamma [\phi,D]$, one can exploit 
that the cubic and quartic source terms 
$\sim R_3$ and $\sim R_4$ appearing in (\ref{eq:Z}) can be conveniently 
combined with the vertices $g V_{03}$ and $g^2 V_{04}$
by the replacement:
\beq\db
g V_{03} \,\,\to\,\, g V_{03} - R_3 \equiv g \tilde{V}_3 
\quad , \quad g^2 V_{04} \,\,\to\,\, g^2 V_{04} - R_4 
\equiv g^2 \tilde{V}_4 \,\, .
\eeq
The 2PI effective action with the modified interaction is given by
\beq\db
\Gamma_{\tilde{V}}[\phi,D] = W[J,R,R_3,R_4] 
-\frac{\delta W}{\delta J} J - \frac{\delta W}{\delta R} R \, .
\eeq
Since
\beq\db
\frac{\delta \Gamma_{\tilde{V}}}{\delta R_3} = 
\frac{\delta W}{\delta R_3} \qquad ,\qquad
\frac{\delta \Gamma_{\tilde{V}}}{\delta R_4} = 
\frac{\delta W}{\delta R_4} \,\, ,
\eeq
one can express the remaining Legendre transforms, leading to 
$\Gamma[\phi,D,V_3,V_4]$, in terms of the vertices $\tilde{V}_{3}$,
$\tilde{V}_{4}$ and
$V_{03}$, $V_{04}$:   
\bea\db
\lefteqn{\Gamma[\phi,D,V_3,V_4] = \Gamma_{\tilde{V}}[\phi,D] 
- \frac{\delta \Gamma_{\tilde{V}}[\phi,D]}{\delta R_{3}} R_{3}
- \frac{\delta \Gamma_{\tilde{V}}[\phi,D]}{\delta R_{4}} R_{4}}
\nonumber\\
&\db=&\db \Gamma_{\tilde{V}}[\phi,D] 
- \frac{\delta \Gamma_{\tilde{V}}[\phi,D]}{\delta \tilde{V}_{3}} 
(\tilde{V}_3 - V_{03})
- \frac{\delta \Gamma_{\tilde{V}}[\phi,D]}{\delta \tilde{V}_{4}} 
(\tilde{V}_4 - V_{04})  \, .
\label{eq:VGcomp}
\eea
What remains to be done is expressing 
$\tilde{V}_3$ and $\tilde{V}_4$ in terms of $V_3$ and $V_4$.
On the one hand, from (\ref{eq:Wderiv}) and the 
definitions (\ref{eq:V3def}) and (\ref{eq:V4def}) one has
\bea\db
\frac{\delta \Gamma_{\tilde{V}}[\phi,D]}{g \delta \tilde{V}_{3}}
&\db=&\db - \frac{1}{6}
\left( - i g\, D^3 V_3 + 3 D \phi  
+ \phi^3 \right)\, ,
\label{eq:DGV3}\\
\db\frac{\delta \Gamma_{\tilde{V}}[\phi,D]}{g^2 \delta \tilde{V}_{4}}
&\db=&\db - \frac{1}{24}
\left( - i g^2\, D^4 V_4
- 3 g^2 D^5 V_3^2
- 4 i g\, D^3 V_3 \phi 
+ 3 D^2 \right. 
\nonumber\\
&&\db \left. + 6 D \phi^2  
+ \phi^4 \right) \, . 
\label{eq:DGV4}
\eea
On the other hand, from the expansion of the 2PI effective action
to three-loop order with (\ref{eq:2PI3loop}) one 
finds\footnote{Note that since the exact $\phi$-dependence of 
$\Gamma_2[\phi,D]$ can be written as a function of 
$(g V_{03} + g^2 \phi V_{04})$, the parametrical dependence of
the higher order terms in 
the variation of (\ref{eq:2PI3loop}) 
with respect to $(g V_{03})$ is given by 
$\mathcal{O}(g^{n} (g^2 \phi)^m |_{n+m=5})$ (cf.~(\ref{eq:2PIV3})). 
}  
\bea\db
\frac{\delta \Gamma_{\tilde{V}}[\phi,D]}{g \delta \tilde{V}_{3}}
&\db=&\db - \frac{1}{6} \phi^3 - \frac{1}{2} D \phi + \frac{i}{6} D^3 
(g \tilde{V}_3 + g^2 \phi \tilde{V}_4) 
\nonumber \\
&&\db + \frac{1}{4} g^2 D^5 (g \tilde{V}_3 + g^2 \phi \tilde{V}_4) \tilde{V}_4 
- \frac{i}{6} D^6 (g \tilde{V}_3 + g^2 \phi \tilde{V}_4)^3 
\nonumber\\[0.13cm]
&&\db +\, \mathcal{O}\left( 
g^{n} (g^2 \phi)^m |_{n+m=5}
\right) \, ,
\label{eq:2PIV3}\\[0.13cm]
\db \frac{\delta \Gamma_{\tilde{V}}[\phi,D]}{g^2 \delta \tilde{V}_{4}}
&\db=&\db -\frac{1}{24} \phi^4 - \frac{1}{4} D \phi^2 - \frac{1}{8} D^2 
+ \frac{i}{6} D^3 \phi (g \tilde{V}_3 + g^2 \phi \tilde{V}_4)
\nonumber\\
&&\db + \frac{i}{24} g^2 D^4 \tilde{V}_4 
+ \frac{1}{4} g^2 D^5 \phi 
(g \tilde{V}_3 + g^2 \phi \tilde{V}_4) \tilde{V}_4
\nonumber\\
&&\db 
+ \frac{1}{8} D^5 (g \tilde{V}_3 + g^2 \phi \tilde{V}_4)^2 
- \frac{i}{6} D^6 \phi (g \tilde{V}_3 + g^2 \phi \tilde{V}_4)^3
\nonumber\\[0.13cm]
&&\db +\, \mathcal{O}\left( 
g^{n-2} (g^2 \phi)^m |_{n+m=6}
\right) \, .
\label{eq:2PIV4}
\eea
Comparing (\ref{eq:2PIV3}) and (\ref{eq:DGV3}) yields
\bea\db
g V_3 &\db=&\db (g \tilde{V}_3 + g^2 \phi \tilde{V}_4) 
- \frac{3}{2} i g^2 D^2 (g \tilde{V}_3 + g^2 \phi \tilde{V}_4) \tilde{V}_4
- D^3 (g \tilde{V}_3 + g^2 \phi \tilde{V}_4)^3 
\nonumber\\
&&\db +\, \mathcal{O}\left( 
g^{n} (g^2 \phi)^m |_{n+m=5}
\right) \, .
\label{eq:V3pre}
\eea
Similarly, for $V_4$ comparing (\ref{eq:2PIV4}) and (\ref{eq:DGV4}), 
and using (\ref{eq:V3pre}) one finds 
\beq\db
g^2 V_4 = g^2 \tilde{V}_4  + \mathcal{O}\left( 
g^{n-2} (g^2 \phi)^m |_{n+m=6}
\right) \, .
\label{eq:V3res} 
\eeq
This can be used to invert the above relations as 
\bea\db
g \tilde{V}_3 + g^2 \phi \tilde{V}_4 &\db=&\db
g V_3 + \frac{3}{2} i g^3 D^2 V_3 V_4
+ g^3 D^3 V_3^3 
+ \mathcal{O}\big(g^5\big) \, ,  \\[0.14cm]
\db g^2 \tilde{V}_4 &\db=&\db g^2 V_4   + \mathcal{O}\left( g^4 \right) \, .
\eea
Plugging this into (\ref{eq:VGcomp}) and expressing the free, the one-loop and
the $\Gamma_2$ parts in terms of $V_{3}$ and $V_{4}$ as well as
$V_{03}$ and $V_{04}$, one obtains from a straightforward calculation:
\beq\rr
\Gamma[\phi,D,V_3,V_4] = S[\phi] + \frac{i}{2} \Tr \ln D^{-1}
+ \frac{i}{2} \Tr\, D_0^{-1}(\phi) D + \Gamma_2[\phi,D,V_3,V_4] \, ,
\label{eq:V2PIeffact}
\eeq
with 
\bea\rr
\Gamma_2[\phi,D,V_3,V_4] &\rr =&\rr \Gamma_2^0[\phi,D,V_3,V_4]
+ \Gamma_2^{\rm int}[D,V_3,V_4] \, , \\[0.2cm]
\rr \Gamma_2^0[\phi,D,V_3,V_4] &\rr =&\rr - \frac{1}{8} g^2 D^2 V_{04}
+ \frac{i}{6} g D^3 V_3 (g V_{03} + g^2 \phi V_{04})
\nonumber\\
&&\rr
+ \frac{i}{24} g^4 D^4 V_4 V_{04}
+ \frac{1}{8} g^4 D^5 V_3^2 V_{04} \, ,
\label{eq:G20phi} \\
\rr\Gamma_2^{\rm int}[D,V_3,V_4] &\rr=&\rr - \frac{i}{12} g^2 D^3 V_3^2 
- \frac{i}{48} g^4 D^4 V_4^2  
- \frac{i}{24} g^4 D^6 V_3^4 
+ \mathcal{O}\left(g^6\right) . \qquad
\label{eq:G2int}
\eea
The diagrammatic representation of these results is given
in Figs.~\ref{fig:ym3loop1} and~\ref{fig:ym3loop3} of 
Sec.~\ref{sec:calcSUN}. There the equivalent calculation is
done for a $SU(N)$ gauge theory and one has to replace
the propagator lines and vertices of the figures by the corresponding scalar 
propagator and vertices. Note that for the scalar theory
the thick circles represent the dressed three-vertex $g V_3$ 
and four-vertex $g^2 V_4$, respectively, while the small circles denote
the corresponding {\em effective} classical three-vertex 
$g V_{03} + g^2 \phi V_{04}$ and classical four-vertex
$g^2 V_{04}$. As a consequence, the diagrams
look the same in the absence of spontaneous symmetry breaking,
indicated by a vanishing field expectation value $\phi$. 

In (\ref{eq:V2PIeffact}), the action $S[\phi]$
and $D_0$ depend on the classical vertices as before.
The expression for $\Gamma_2^0$, which includes all terms of 
$\Gamma_2$ that depend on the classical vertices, is valid to
all orders: $\Gamma_2^{\rm int}$ contains no explicit dependence
on the field $\phi$ or the classical vertices $V_{03}$ and $V_{04}$,
independent of the approximation for the 4PI effective action. This 
can be straightforwardly observed from (\ref{eq:VGcomp}), where the complete
(linear) dependence of $\Gamma$ on $V_{03}$ and $V_{04}$ is explicit,
together with (\ref{eq:DGV3}) and (\ref{eq:DGV4}).

\subsubsection{Equivalence hierarchy for $n$PI effective actions}
\label{sec:selfcomplete}

As pointed out in Sec.~\ref{sec:introuniv} of the introduction,
for applications it is often desirable to obtain a self-consistently 
complete description, which to a given order of a loop or coupling
expansion determines the $n$PI effective action 
$\Gamma[\phi,D,V_3,V_4,\ldots,V_n]$ for arbitrarily 
high $n$. Despite the complexity of a general $n$PI effective action
such a description can be obtained in practice because 
of the equivalence hierarchy displayed in Eq.~(\ref{eq:hierarchy}):
Typically the 2PI, 3PI or maybe the 4PI effective action  
captures already the complete answer for the self-consistent 
description to the desired/computationally feasible order of 
approximation. Higher effective actions, which are relevant 
beyond four-loop order, may not be entirely irrelevant in the 
presence of sources describing complicated initial conditions 
for nonequilibrium evolutions. However, their discussion would 
be somewhat academic from the point of view of calculational
feasibility and we will concentrate
on up to four-loop corrections or ${\mathcal O}(g^6)$
in the following.
Below we will not explicitly write in addition to the
loop-order the corresponding order of the coupling $g$ for the 
considered theory, which is detailed above in Sec.~\ref{sec:compu}.

To show (\ref{eq:hierarchy}) we will first observe that
to one-loop order all $n$PI effective actions agree in the
absence of sources. The one-loop result for the 1PI effective action 
is given by (\ref{G1Roneloop}) for vanishing source $R$. 
As has been explicitly 
shown in Sec.~\ref{sec:genfunc1} (cf.~Eq.~(\ref{eq:onelopp2PIR})),
the one-loop 2PI effective action agrees with that expression, i.e. 
\beq\db
\Gamma^{\rm (1loop)}[\phi,D] = \Gamma^{\rm (1loop)}[\phi] \, ,
\eeq 
in the absence of sources. The equivalence with the
one-loop 3PI and 4PI effective actions can be explicitly observed from 
the results of Sec.~\ref{sec:compu}. 
In order to obtain the 3PI expressions we could 
directly set the source $R_4\equiv 0$ from the beginning in 
the computation of that section such that there is no
dependence on $V_4$. Equivalently, we can note from 
Eqs.~(\ref{eq:V2PIeffact})--(\ref{eq:G2int}) that already the 4PI 
effective action to this order simply agrees with (\ref{G1Roneloop})
for zero sources.
As a consequence, it carries no dependence on $V_3$ and $V_4$, i.e.
\beq\db
\Gamma^{\rm (1loop)}[\phi,D,V_3,V_4] 
= \Gamma^{\rm (1loop)}[\phi,D,V_3] 
= \Gamma^{\rm (1loop)}[\phi,D] \, .
\eeq  
For the one-loop case it remains to be shown that in addition
\beq\db
\Gamma^{\rm (1loop)}[\phi,D,V_3,V_4,\ldots,V_n]
= \Gamma^{\rm (1loop)}[\phi,D,V_3,V_4]
\label{eq:equiv4vn} 
\eeq
for arbitrary $n \ge 5$. For this we note that the 
number $I$ of internal lines in a given loop diagram is given
by the number $v_3$ of proper 3-vertices, the number $v_4$ of 
proper 4-vertices, \ldots, the number $v_n$ of 
proper n-vertices in terms of the standard relation: 
\beq\db
2 I = 3 v_3 + 4 v_4 + 5 v_5 \ldots + n v_n \, ,
\eeq
where $v_3 + v_5 + v_7 + \ldots$ has to be even.
Similarly, the number $L$ of loops in such a diagram is
\bea\db
L &\db =&\db I - v_3 - v_4 - v_5 \ldots - v_n + 1 \nonumber\\
&\db =&\db \frac{1}{2} v_3 + v_4 + \frac{3}{2} v_5 \ldots
+ \frac{n - 2}{2} v_n + 1  \,\, .
\label{eq:numberofloops}
\eea
The equivalence~(\ref{eq:equiv4vn}) follows from the fact
that for~$L=1$ equation~(\ref{eq:numberofloops}) implies that 
$\Gamma^{\rm (1loop)}[\phi,D,V_3,V_4,\ldots,V_n]$ 
cannot depend in particular on $V_5, \ldots V_n$.\footnote{Note
that we consider here theories where there is no 
classical 5-vertex or higher, whose presence would lead to a trivial
dependence for the classical action and propagator.}

The two-loop equivalence of the 2PI and higher effective
actions follows along the same lines. According to
(\ref{eq:V2PIeffact})--(\ref{eq:G2int}) the 4PI effective
action to two-loop order is given by:
\bea\db
\Gamma^{\rm (2loop)}[\phi,D,V_3,V_4] 
&\db\!=\!&\db S[\phi] + \frac{i}{2} \Tr \ln D^{-1}
+ \frac{i}{2} \Tr\, D_0^{-1}(\phi) D \nonumber\\
&&\db +\, \Gamma_2^{\rm (2loop)}[\phi,D,V_3,V_4] 
\, , \label{eq:4pitwoloop} \\
\db \Gamma_2^{\rm (2loop)}[\phi,D,V_3,V_4] &\db \!=\!&\db
- \frac{1}{8} g^2 D^2 V_{04}
+ \frac{i}{6} g D^3 V_3 (g V_{03} + g^2 \phi V_{04})
- \frac{i}{12} g^2 D^3 V_3^2 \nonumber \, .
\eea
There is no dependence on $V_4$ to this order and, following 
the discussion above, there is no dependence on
$V_5,\ldots,V_n$ according to (\ref{eq:numberofloops}) for
$L=2$. Consequently, 
\beq\db
\Gamma^{\rm (2loop)}[\phi,D,V_3,V_4,\ldots,V_n] =
\Gamma^{\rm (2loop)}[\phi,D,V_3,V_4] =
\Gamma^{\rm (2loop)}[\phi,D,V_3] \, ,
\eeq
for arbitrary $n$ in the absence of sources. The latter yields
\beq\db
\frac{\delta \Gamma^{\rm (2loop)}[\phi,D,V_3]}{\delta V_3} = 
\frac{\delta \Gamma^{\rm (2loop)}_2[\phi,D,V_3]}{\delta V_3} = 0
\quad \Rightarrow \quad 
g V_3 = g V_{03} + g^2 \phi V_{04} \, ,
\eeq
which can be used in (\ref{eq:4pitwoloop}) to show in addition
the equivalence of the
3PI and 2PI effective actions (cf.~Eq.~(\ref{eq:2PI3loop})) to 
this order:
\bea\db
\Gamma_2^{\rm (2loop)} [\phi,D,V_3] &\db=&\db - \frac{1}{8} g^2 D^2 V_{04}  
+ \frac{i}{12} D^3 (g V_{03} + g^2 \phi V_{04})^2 \nonumber\\
&\db=&\db \Gamma_2^{\rm (2loop)} [\phi,D] \, ,
\label{eq:2pitwoloop}
\eea 
for vanishing sources. The {\rr\em inequivalence} of the 2PI with 
the 1PI effective action to this order,
\beq\rr
\Gamma^{\rm (2loop)} [\phi,D] \not = \Gamma^{\rm (2loop)}[\phi] \, ,
\eeq
follows from using the result of 
$\delta\Gamma_2^{\rm (2loop)} [\phi,D]/\delta D = 0$ for $D$
in (\ref{eq:2pitwoloop}) in a straightforward way.\footnote{
Here $\Gamma^{\rm (2loop)} [\phi,D]$ includes e.g.~the 
summation of an infinite series of so-called ``bubble''
diagrams, which form the basis of mean-field or Hartree-type
approximations, and clearly go beyond a perturbative two-loop
approximation $\Gamma^{\rm (2loop)}[\phi]$ (cf.~Sec.~\ref{sec:loopexp}).}

In order to show the three-loop equivalence of the 3PI and
higher effective actions, we first note from 
(\ref{eq:V2PIeffact})--(\ref{eq:G2int}) that the 4PI
effective action to this order yields $V_4 = V_{04}$ in
the absence of sources:
\beq\db
\frac{\delta \Gamma^{\rm (3loop)}[\phi,D,V_3,V_4]}{\delta V_4} = 
\frac{\delta \Gamma^{\rm (3loop)}_2[\phi,D,V_3,V_4]}{\delta V_4} = 
\frac{i}{24} g^4 D^4 \left( V_{04} - V_4 \right) = 0 \, .
\eeq
Constructing the 3PI effective action to three-loop would mean to 
do the same calculation as in Sec.~\ref{sec:compu} but with
$V_4 \to V_{04}$ from the beginning ($R_4 \equiv 0$). 
The result of a classical four-vertex for the 4PI effective action
to this order, therefore, directly implies:
\beq\db
\Gamma^{\rm (3loop)}[\phi,D,V_3,V_4] = \Gamma^{\rm (3loop)}[\phi,D,V_3] \, ,
\label{eq:parteq} 
\eeq
for vanishing sources. To see the equivalence with a 5PI effective
action $\Gamma^{\rm (3loop)}[\phi,D,V_3,V_4,V_5]$,
we note that to three-loop order the only possible diagram
including a five-vertex requires $v_3 = v_5 = 1$ for $L=3$
in Eq.~(\ref{eq:numberofloops}). As a consequence, to this
order the five-vertex corresponds to the classical one, which 
is identically zero for the theories considered here, 
i.e.~$V_5 = V_{05} \equiv 0$. In order to obtain
that (to this order trivial) result along the lines of 
Sec.~\ref{sec:compu}, one can formally include a classical
five-vertex $V_{05}$ and observe that the three-loop
2PI effective action
admits a term $\sim D^4 V_{05} V_3$. After performing the
additional Legendre transform the result then follows
from setting $V_{05} \to 0$ in the end. 
The equivalence with $n$PI effective actions for
$n \ge 6$ can again be observed from the fact
that for~$L=3$ Eq.~(\ref{eq:numberofloops}) implies 
no dependence on $V_6, \ldots V_n$. In addition to
(\ref{eq:parteq}), we therefore have for
arbitrary $n \ge 5$:
\beq\db
\Gamma^{\rm (3loop)}[\phi,D,V_3,V_4,\ldots,V_n] =
\Gamma^{\rm (3loop)}[\phi,D,V_3,V_4] \, .
\eeq
The {\rr\em inequivalence} of the three-loop 3PI and 
2PI effective actions can be readily observed
from (\ref{eq:V2PIeffact})--(\ref{eq:G2int}) and
(\ref{eq:parteq}):
\bea\db
\frac{\delta \Gamma^{\rm (3loop)}[\phi,D,V_3]}{\delta V_3} = 0
\quad \Rightarrow \quad 
g V_3 = g \left(V_{03} + g \phi V_{04}\right) - g^3 D^3 V_3^3 \, .
\eea 
Written iteratively, the above self-consistent equation for $V_3$
sums an infinite number of contributions in terms of the
classical vertices. As a consequence, the three-loop 3PI result 
can be written as an infinite series of diagrams for the
corresponding 2PI effective action, which
clearly goes beyond $\Gamma^{\rm (3loop)}[\phi,D]$ 
(cf.~Eq.~(\ref{eq:2PI3loop})):
\beq\rr
\Gamma^{\rm (3loop)}[\phi,D,V_3] \not = \Gamma^{\rm (3loop)}[\phi,D] \, .
\eeq
The importance of such an infinite
summation will be discussed for the case of gauge theories below.

\subsection{Nonabelian gauge theory with fermions} 
\label{sec:nonabgauge}

We consider a $SU(N)$ gauge theory with $N_f$ flavors of
Dirac fermions with classical action
\bea\db
\lefteqn{
S_{\rm eff} = S + S_{\rm gf} + S_{\rm FPG} }  \nonumber\\
&\db=&\db \int d^4x \Big( 
- \frac{1}{4} F_{\mu\nu}^a F^{\mu\nu\, a}
- \frac{1}{2 \xi} \left({\cal G}^a(A) \right)^2
- \bar{\psi} (-i D\!\slash) \psi
- \bar{\eta}^a \partial_\mu \left(D^\mu \eta \right)^a
\Big)\,\, , 
\label{eq:Sefffirst}
\eea
where $\psi$ ($\bar{\psi}$), $A$ and $\eta$ ($\bar{\eta}$) denote
the (anti-)fermions, gauge and 
\mbox{(anti-)ghost} fields, respectively, with gauge-fixing term
${\cal G}^a(A) = \partial^\mu A_\mu^a$ for covariant gauges. 
The color indices in the adjoint representation are 
$a,b,\ldots = 1,\ldots,N^2-1$, while those for the fundamental representation
will be denoted by $i,j,\ldots$ and run from $1$ to $N$. Here
\bea\db
F_{\mu\nu}^a &\db =&\db  \partial_\mu A_\nu^a - \partial_\nu A_\mu^a
- g f^{abc} A_\mu^b A_\nu^c \, ,  \\ 
\db \left(D^\mu \eta \right)^a &\db=&\db
 \partial^\mu \eta^a - g f^{abc} A^{\mu\, b} \eta^c
\, ,\\\db
D\!\slash &\db =&\db 
\gamma^\mu \left(\partial_\mu + i g A^a_\mu t^a \right) \, , 
\eea
where $[t^a,t^b] = i f^{abc} t^c$, $\tr(t^a t^b) = \delta^{ab}/2$. For QCD,
$t^a = \lambda^a/2$ with the Gell-Mann matrices $\lambda^a$
($a=1,\ldots,8$). We will suppress Dirac and flavor indices in
the following. It is convenient to write (\ref{eq:Sefffirst}) 
as 
\bea\db
S_{\rm eff} &\db=&\db \frac{1}{2} \int_{x y} A^{\mu\, a}(x)\,
i D^{-1\, ab}_{0\, \mu\nu}(x,y) A^{\nu\, b}(y) 
+ \int_{x y} \bar{\eta}^a(x)\, i G_0^{-1\, ab}(x,y) \eta^b(y)
\nonumber\\
&\db+&\db \int_{x y} \bar{\psi}_i(x)\, i \Delta_{0\, ij}^{-1}(x,y) \psi_j(y)
- \frac{1}{6}\, g \int_{x y z} V_{03\, \mu\nu\gamma}^{abc}(x,y,z)
A^{\mu\, a}(x) A^{\nu\, b}(y) A^{\gamma\, c}(z) 
\nonumber\\
&\db-&\db \frac{1}{24}\, g^2 \int_{x y z w} 
V_{04\, \mu\nu\gamma\delta}^{abcd}(x,y,z,w) 
A^{\mu\, a}(x) A^{\nu\, b}(y) A^{\gamma\, c}(z) A^{\delta\, d}(w) 
\nonumber\\
&\db-&\db g \int_{x y z} V_{03\, \mu}^{{\rm (gh)}ab,c}(x,y;z)
\bar{\eta}^a(x)  \eta^b(y) A^{\mu\, c}(z)
\nonumber\\
&\db-&\db g \int_{x y z} V_{03\, \mu\, i j}^{{\rm (f)}a}(x,y;z)
\bar{\psi}_i(x)  \psi_j(y) A^{\mu\, a}(z) \, ,
\label{eq:Seff}
\eea
with the free inverse fermion, ghost and gluon propagator
in covariant gauges given by
\bea\db
i \Delta^{-1}_{0\, ij}(x,y) &\db=&\db i \partial\!\slash_{\!x}\, 
\delta_{ij} \delta_\C (x-y) \, ,
\\[0.14cm]
\db i G^{-1\, ab}_{0}(x,y) &\db=&\db 
- \square_x \delta^{ab} \delta_\C (x-y) \, ,
\\[0.14cm]
\db i D^{-1\, ab}_{0\, \mu\nu}(x,y) 
&\db =&\db \left[g_{\mu\nu}\, \square - \left( 1 - \xi^{-1}\right) 
\partial_\mu \partial_\nu \right]_x \delta^{ab} \delta_\C (x-y) 
 \, ,
\eea
where we have taken the fermions to be massless.
The tree-level vertices read in coordinate space: 
\bea\db
V_{03\, \mu\nu\gamma}^{abc}(x,y,z) &\db =&\db  f^{abc} \Big( 
\nonumber\\
&&\db g_{\mu\nu} [\delta_{\C}(y-z)\, \partial^x_\gamma\delta_{\C}(x-y)
- \delta_{\C}(x-z)\, \partial^y_\gamma\delta_{\C}(y-x)]
\nonumber\\[0.13cm]
&\db +& \db
 g_{\mu\gamma} [\delta_{\C}(x-y)\, \partial^z_\nu\delta_{\C}(z-x)
- \delta_{\C}(y-z)\, \partial^x_\nu\delta_{\C}(x-z)]
\nonumber\\
&\db +&\db 
 g_{\nu\gamma} [\delta_{\C}(x-z)\, \partial^y_\mu\delta_{\C}(y-x)
- \delta_{\C}(x-y)\, \partial^z_\mu\delta_{\C}(z-x)]
\Big) \, ,\nonumber\\
\label{eq:vert1}\\ 
\db V_{04\, \mu\nu\gamma\delta}^{abcd}(x,y,z,w) &\db=&\db \Big(
f^{abe}f^{cde} [ g_{\mu\gamma}g_{\nu\delta} - g_{\mu\delta}g_{\nu\gamma}]
\nonumber\\
&\db+&\db 
f^{ace}f^{bde} [ g_{\mu\nu}g_{\gamma\delta} - g_{\mu\delta}g_{\nu\gamma}]
+ f^{ade}f^{cbe} [ g_{\mu\gamma}g_{\delta\nu} - g_{\mu\nu}g_{\gamma\delta}]  
\Big)
\nonumber\\
&&\db \delta_{\C}(x-y)\delta_{\C}(x-z)\delta_{\C}(x-w) \, , 
\\[0.14cm]
\db V_{03\, \mu}^{{\rm (gh)}ab,c}(x,y;z) &\db =&\db  - f^{abc} \partial_\mu^x 
\delta_{\C}(x-z)\delta_{\C}(y-z) \, ,
\\[0.14cm]
\db V_{03\, \mu\, i j}^{{\rm (f)}a}(x,y;z) &\db =&\db  \gamma_\mu t^a_{ij}
\delta_{\C}(x-z)\delta_{\C}(z-y) \, .
\label{eq:fermvertex}
\eea
Note that $V_{03, abc}^{\mu\nu\gamma}(x,y,z)$
is symmetric under exchange of $(\mu,a,x) \leftrightarrow (\nu,b,y) 
\leftrightarrow (\gamma,c,z)$. Likewise, 
$V_{04, abcd}^{\mu\nu\gamma\delta}(x,y,z,w)$ is symmetric in its space-time
arguments and under exchange of $(\mu,a) \leftrightarrow (\nu,b) 
\leftrightarrow (\gamma,c) \leftrightarrow (\delta,d)$.   

In addition to the linear and bilinear source terms, which
are required for a construction of the 2PI effective action,  
following Sec.~\ref{sec:highereffective actions} 
we add cubic and quartic source terms to
(\ref{eq:Seff}): 
\bea\db 
S'_{\rm source} &\db =&\db  
 \frac{1}{6} \int_{x y z} R_{3\, \mu\nu\gamma}^{abc}(x,y,z)
A^{\mu\, a}(x) A^{\nu\, b}(y) A^{\gamma\, c}(z) 
\nonumber\\
&\db +&\db \db  \frac{1}{24} \int_{x y z w} 
R_{4\, \mu\nu\gamma\delta}^{abcd}(x,y,z,w) 
A^{\mu\, a}(x) A^{\nu\, b}(y) A^{\gamma\, c}(z) A^{\delta\, d}(w) 
\nonumber\\
&\db +&\db  \int_{x y z} R_{3\, \mu}^{{\rm (gh)}ab,c}(x,y;z)
\bar{\eta}^a(x)  \eta^b(y) A^{\mu\, c}(z)
\nonumber\\
&\db +&\db  \int_{x y z} R_{3\, \mu\, i j}^{{\rm (f)}a}(x,y;z)
\bar{\psi}_i(x)  \psi_j(y) A^{\mu\, a}(z)  \, ,
\label{eq:sources}
\eea
where the sources $R_{3,4}$ obey the same symmetry properties
as the corresponding classical vertices $V_{03}$ and $V_{04}$ 
discussed above.
The definition of the corresponding three- and four-vertices 
follows Sec.~\ref{sec:highereffective actions}. In particular, we have
for the vertices involving Grassmann fields:
\bea\db 
\frac{\delta W}{\delta R_{3\, \mu}^{{\rm (gh)}ab,c}(x,y;z)}
&\db \!=\!&\db  - i g \int_{x' y' z'} D^{\mu\mu'\, c c'}(z,z') G^{b a'}(y,x') 
  V_{3\, \mu'}^{{\rm (gh)}a' b' c'}(x',y';z') G^{b' a}(y',x) , 
\nonumber\\[0.2cm]
\db \frac{\delta W}{\delta R_{3\, \mu\, i j}^{{\rm (f)}a}(x,y;z)}
&\db \!=\!&\db  
- i g \int_{x' y' z'} D^{\mu\mu'\, a a'}(z,z') \Delta_{ji'}(y,x') 
 V_{3\, \mu'\, i' j'}^{{\rm (f)}a'}(x',y';z') \Delta_{j'i}(y',x) \, ,\,\, 
\eea
for the case of vanishing ``background'' fields 
$\langle A \rangle = \langle \psi \rangle = \langle \bar{\psi} \rangle
= \langle \eta \rangle = \langle \bar{\eta} \rangle = 0$, which we will
consider in the following.

\subsubsection{Effective action up to four-loop or
${\mathcal O}(g^6)$ corrections}
\label{sec:calcSUN}

\begin{figure}[t]
\centerline{
\epsfig{file=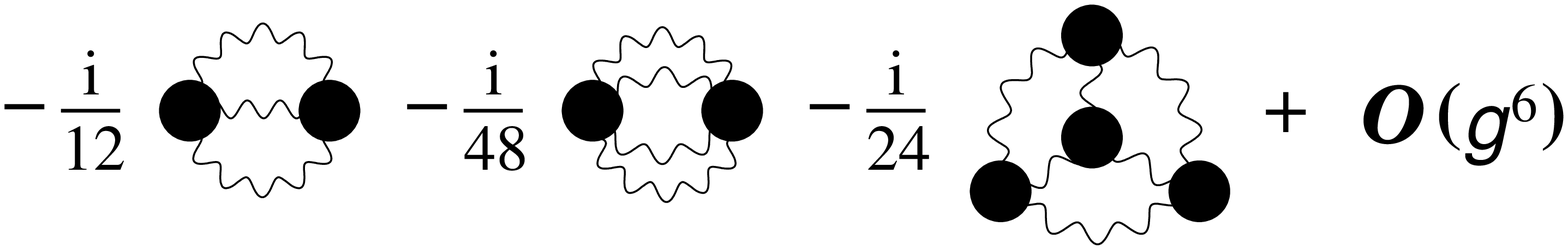,width=9.6cm}
}
\caption{ The figure shows together with Fig.~\ref{fig:ym3loop4}
the diagrammatic representation of 
$\Gamma_2^{\rm int}[D,G,V_{3},V_3^{\rm (gh)},V_{4}]$
to three-loop order as given in Eq.~(\ref{eq:gamma2}). 
For the gauge theory with fermions, to this order there is in 
addition the same contribution as in
Fig.~\ref{fig:ym3loop4} with the unwiggled propagator lines
representing the fermion propagator $\Delta$ and the ghost vertex 
replaced by the corresponding fermion vertex $V_{3}^{\rm (f)}$. 
This functional contains no explicit
dependence on the classical vertices independent of the
order of approximation.}
\label{fig:ym3loop3}
\end{figure}

\begin{figure}[t]
\centerline{
\epsfig{file=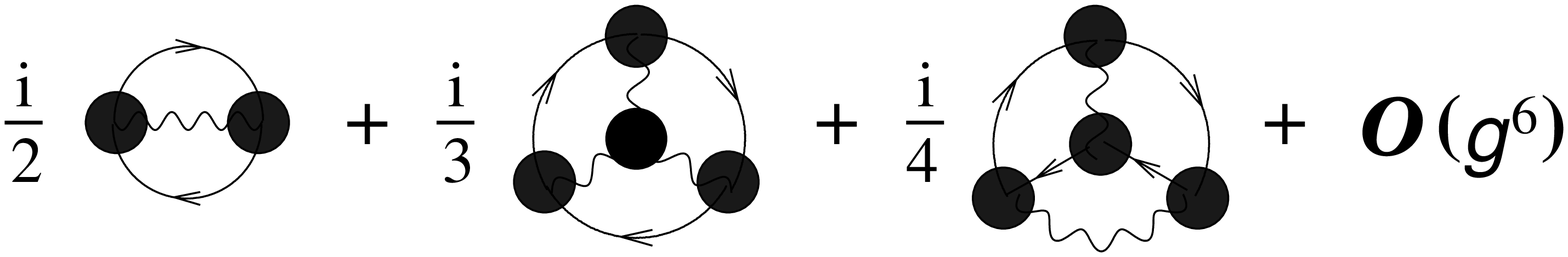,width=9.3cm}
}
\caption{Ghost/fermion part of 
$\Gamma_2^{\rm int}$ to three-loop order.}
\label{fig:ym3loop4}
\end{figure}

Consider first the 2PI effective action with vanishing
``background'' fields,
which according to Sec.~\ref{sec:genfunc1} can be written as
\bea\db 
\Gamma[D,\Delta,G] 
&\db =&\db \frac{i}{2} \Tr \ln D^{-1}
+ \frac{i}{2} \Tr\, D_0^{-1} D 
- i \Tr \ln \Delta^{-1}
- i \Tr\, \Delta_0^{-1} \Delta 
\nonumber\\
&&\db
- i \Tr \ln G^{-1}
- i \Tr\, G_0^{-1} G 
+ \Gamma_2[D,\Delta,G] \, 
\label{eq:exact2PI}.
\eea
Here the trace $\Tr$ includes an integration over the
time path $\C$, as well as integration over spatial 
coordinates and summation over flavor, color and Dirac indices. 
The exact expression for $\Gamma_2$ contains all 2PI diagrams
with vertices described by (\ref{eq:vert1})--(\ref{eq:fermvertex}) 
and propagator lines associated to the full connected two-point functions
$D$, $G$ and $\Delta$. In order to clear up the presentation, we will
give all diagrams including gauge and ghost propagators only.
The fermion diagrams can simply be obtained from the
corresponding ghost ones, since they have the same signs and 
prefactors.\footnote{Note that to three-loop order there 
are no graphs with more than one closed ghost/fermion loop,
such that ghosts and fermions cannot appear in the same diagram
simultaneously.} For the 2PI effective action of the gluon-ghost 
system, $\Gamma[D,G]$, to three-loop order the 2PI effective
action is given by (using the same compact notation as introduced in 
Sec.~\ref{sec:compu}):
\bea\db
\Gamma_2[D,G] &\db =&\db 
- \frac{1}{8} g^2 D^2 V_{04}
+ \frac{i}{12} g^2 D^3 V_{03}^2 
- \frac{i}{2} g^2 D G^2 V_{03}^{\rm (gh)\, 2}
+ \frac{i}{48} g^4 D^4 V_{04}^2
\nonumber\\
&&\db + \frac{1}{8} g^4 D^5 V_{03}^2 V_{04}  
- \frac{i}{24} g^4 D^6 V_{03}^4 
+ \frac{i}{3} g^4 D^3 G^3 V_{03}^{\rm (gh)\, 3}V_{03} 
\nonumber\\
&&\db + \frac{i}{4} g^4 D^2 G^4 V_{03}^{\rm (gh)\, 4} 
+ \mathcal{O}\left(g^6 \right)
\, .
\label{eq:3loop2PISUN}
\eea
The result can be compared with (\ref{eq:2PI3loop}) and taking
into account an additional factor of $(-1)$ for each closed
loop involving Grassmann fields (cf.~Sec.~\ref{sec:2PIfermion}).
Here we have suppressed in the notation the dependence of $\Gamma_2[D,G]$
on the higher sources (\ref{eq:sources}). The desired effective action
is obtained by performing the remaining Legendre
transforms:
\bea\db
\Gamma[D,G,V_{3},V_3^{\rm (gh)},V_{4}] = \Gamma[D,G]
- \frac{\delta W}{\delta R_3} R_3 
- \frac{\delta W}{\delta R_3^{\rm (gh)}} R_3^{\rm (gh)}
- \frac{\delta W}{\delta R_4} R_4  \, .
\eea
The calculation follows the same steps as detailed in Sec.~\ref{sec:compu}.  
\begin{figure}[t]
\centerline{
\epsfig{file=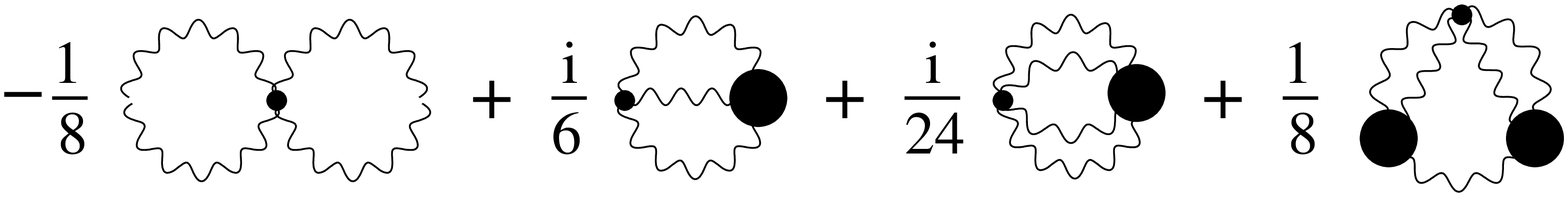,width=10.5cm}
}
\caption{The figure shows together with Fig.~\ref{fig:ym3loop2}
the diagrammatic representation of 
$\Gamma_2^0[D,G,V_{3},V_3^{\rm (gh)},V_{4}]$ as given
in Eq.~(\ref{eq:G20}). Here the wiggled lines denote the
gauge field propagator $D$ and the unwiggled lines the
ghost propagator $G$. The thick circles denote the dressed
and the small ones the classical vertices. This functional 
contains all terms of $\Gamma_2$ that depend on the classical 
vertices $g V_{03}$, $g V_{03}^{\rm (gh)}$ and $g^2 V_{04}$ for an 
$SU(N)$ gauge theory. There are no further contributions to 
$\Gamma_2^0$ appearing at higher order in the expansion. 
For the gauge theory with 
fermions there is in addition the same contribution as in
Fig.~\ref{fig:ym3loop2} with the unwiggled propagator lines
representing the fermion propagator $\Delta$ and the ghost vertices 
replaced by the corresponding fermion vertices $V_{03}^{\rm (f)}$ 
and $V_{3}^{\rm (f)}$ (cf.~Eq.~(\ref{eq:fermvertex})).}
\label{fig:ym3loop1}
\end{figure}
\begin{figure}[t]
\centerline{
\epsfig{file=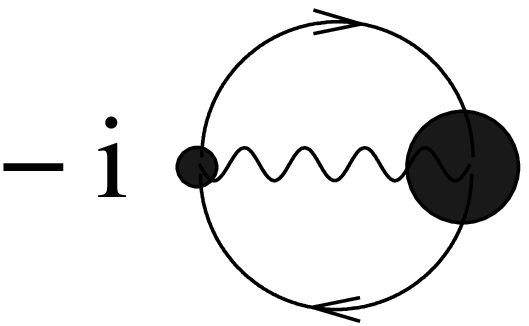,width=1.8cm}
}
\caption{Ghost/fermion part of $\Gamma_2^0$.}
\label{fig:ym3loop2}
\end{figure}
For the effective action to $\mathcal{O}(g^6)$ we obtain:  
\bea\rr 
\Gamma[D,G,V_{3},V_3^{\rm (gh)},V_{4}] &\rr =&\rr  \frac{i}{2} \Tr \ln D^{-1}
+ \frac{i}{2} \Tr\, D_0^{-1} D 
- i \Tr \ln G^{-1}
- i \Tr\, G_0^{-1} G 
\nonumber\\
&&\rr  + \Gamma_2[D,G,V_{3},V_3^{\rm (gh)},V_{4}] \, ,
\label{eq:exact4PI}
\eea
with
\bea\rr 
\Gamma_2[D,G,V_{3},V_3^{\rm (gh)},V_{4}] 
&\rr =&\rr  \Gamma_2^0[D,G,V_{3},V_3^{\rm (gh)},V_{4}]
+ \Gamma_2^{\rm int}[D,G,V_{3},V_3^{\rm (gh)},V_{4}] \, , 
\nonumber\\[0.2cm]
\rr \Gamma_2^0[D,G,V_{3},V_3^{\rm (gh)},V_{4}] 
&\rr =&\rr 
- \frac{1}{8} g^2 D^2 V_{04}
+ \frac{i}{6} g^2 D^3 V_{3} V_{03}
- i g^2 D G^2 V_3^{\rm (gh)} V_{03}^{\rm (gh)}
\nonumber\\
&&\rr  + \frac{i}{24} g^4 D^4 V_{4} V_{04}
+ \frac{1}{8} g^4 D^5 V_{3}^2 V_{04} \, , 
\label{eq:G20} \\[0.2cm]
\rr \Gamma_2^{\rm int}[D,G,V_{3},V_3^{\rm (gh)},V_{4}] &\rr =&\rr 
- \frac{i}{12} g^2 D^3 V_{3}^2 
+ \frac{i}{2} g^2 D G^2 V_{3}^{\rm (gh)\, 2}
- \frac{i}{48} g^4 D^4 V_{4}^2  
\nonumber\\
&&\rr  - \frac{i}{24} g^4 D^6 V_{3}^4 
+ \frac{i}{3} g^4 D^3 G^3\, V_{3}^{\rm (gh)\, 3}V_{3} 
\nonumber\\
&&\rr  + \frac{i}{4} g^4 D^2 G^4\, V_{3}^{\rm (gh)\, 4}
+ \mathcal{O}(g^6) \, .
\label{eq:gamma2}
\eea
The contributions are displayed diagrammatically 
in Figs.~\ref{fig:ym3loop1} and~\ref{fig:ym3loop2} 
for $\Gamma_2^0$, and in Figs.~\ref{fig:ym3loop3} 
and~\ref{fig:ym3loop4} for $\Gamma_2^{\rm int}$.

The equivalence of the 4PI effective action to three-loop order with
the 3PI and $n$PI effective actions for $n \ge 5$ in the absence of 
sources follows along the lines of 
Sec.~\ref{sec:selfcomplete}. As a consequence, to three-loop order the 
$n$PI effective
action does not depend on higher vertices $V_5$, $V_6$, \ldots $V_n$.
In particular with vanishing sources the four-vertex is given by the 
classical one: 
\beq\db
\frac{\delta \Gamma^{\rm (3loop)}[D,G,V_3,V_3^{\rm (gh)},V_4]}{\delta V_4} = 
\frac{\delta
\Gamma^{\rm (3loop)}_2[D,G,V_3,V_3^{\rm (gh)},V_4]}{\delta V_4} = 0 
\,\, \Rightarrow \,\, V_4 = V_{04} .
\label{eq:classical4}
\eeq
If one plugs this into (\ref{eq:G20}) and (\ref{eq:gamma2}) one
obtains the three-loop 3PI effective action,
$\Gamma^{\rm (3loop)}[\phi,D,V_3,V_3^{\rm (gh)}]$. Similarly, to two-loop
order one has 
\bea\db 
\frac{\delta \Gamma^{\rm (2loop)}[D,G,V_3,V_3^{\rm (gh)}]}{\delta V_3} 
&\db \!=\!&\db  
\frac{\delta \Gamma^{\rm (2loop)}_2[D,G,V_3,V_3^{\rm (gh)}]}{\delta V_3} = 0 
\,\, \Rightarrow \,\, V_3 = V_{03} \, , \nonumber\\
\db \frac{\delta 
\Gamma^{\rm (2loop)}[D,G,V_3,V_3^{\rm (gh)}]}{\delta V_3^{\rm (gh)}} 
&\db \!=\!&\db  \frac{\delta \Gamma^{\rm (2loop)}_2[D,G,V_3,V_3^{\rm (gh)}]}
{\delta V_3^{\rm (gh)}} = 0 
\,\, \Rightarrow \,\, V_3^{\rm (gh)} = V_{03}^{\rm (gh)} \, ,
\nonumber
\eea
and equivalently for the fermion vertex $V_3^{\rm (f)}$.
To this order, therefore, the combinatorial factors of the
two-loop diagrams of Fig.~\ref{fig:ym3loop1} and
\ref{fig:ym3loop3} for the gauge part, as well as 
of Fig.~\ref{fig:ym3loop2} and
\ref{fig:ym3loop4} for the ghost/fermion part,
combine to give the result (\ref{eq:3loop2PISUN}) to two-loop order for
the 2PI effective action. 

\begin{figure}[t]
\centerline{
\epsfig{file=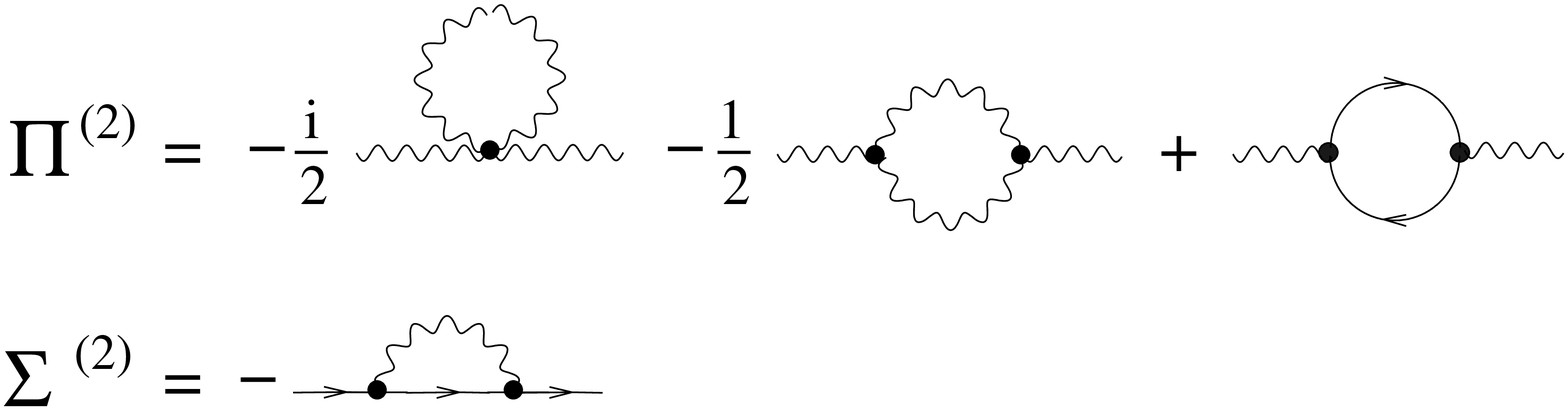,width=10.5cm}
}
\caption{The self-energy for the gauge field ($\Pi$) and
the ghost/fermion ($\Sigma$) propagators as obtained 
from the self-consistently complete 
two-loop approximation of the effective action. 
Note that at this order all vertices correspond to 
the classical ones.}
\label{fig:ymmerge2}
\end{figure}

We have seen above that to two-loop order 
the proper vertices of the $n$PI effective action 
correspond to the classical ones. Accordingly, at this order
the only non-trivial equations of motion in the absence
of ``background'' fields are those for the two-point functions:
\beq\db
\frac{\delta \Gamma}{\delta D} = 0 \, , \quad \, 
\frac{\delta \Gamma}{\delta G} = 0 \, , \quad \,
\frac{\delta \Gamma}{\delta \Delta} = 0\, ,
\label{eq:statprop}
\eeq
for vanishing sources. Applied to an $n$PI effective
action ($n>1$), as e.g.~(\ref{eq:exact4PI}),
one finds for the gauge field propagator:
\beq\db
D^{-1} = D^{-1}_0 
- \Pi \quad , \quad
\Pi = 2 i\, 
\frac{\delta \Gamma_2}{\delta D}  \, .
\label{eq:SDforD}  
\eeq 
The ghost propagator and self-energy are 
\beq\db
G^{-1} = G^{-1}_0 
- \Sigma \quad , \quad 
\Sigma = -i \frac{\delta \Gamma_2}{\delta G} \, ,
\label{eq:SDforG}  
\eeq
and equivalently for the fermion propagator $\Delta$. (Cf.~also
Sec.~\ref{sec:genfunc1} for the same relations in the context of 2PI
effective actions.)
The self-energies to this order are shown in diagrammatic form 
in Fig.~\ref{fig:ymmerge2}. In contrast, for the three-loop 
effective action the three-vertices get dressed and the stationarity 
conditions, 
\beq \db
\frac{\delta \Gamma}{\delta V_{3}} = 0 \, , \quad \, 
\frac{\delta \Gamma}{\delta V_{3}^{\rm (gh)}} = 0 \, , \quad \,
\frac{\delta \Gamma}{\delta V_{3}^{\rm (f)}} = 0 \, , 
\label{eq:statver}
\eeq
applied to (\ref{eq:exact4PI})--(\ref{eq:gamma2})
lead to the equations shown in the left graph of Fig.~\ref{fig:ymver3}.
Here the diagrammatic form of the contributions is always 
the same for the ghost and for the fermion propagators or vertices. We 
therefore only give the expressions for the gauge-ghost
system. If fermions are present, the respective diagrams
have to be added in a straightforward way.  

\begin{figure}[t]
\centerline{
\epsfig{file=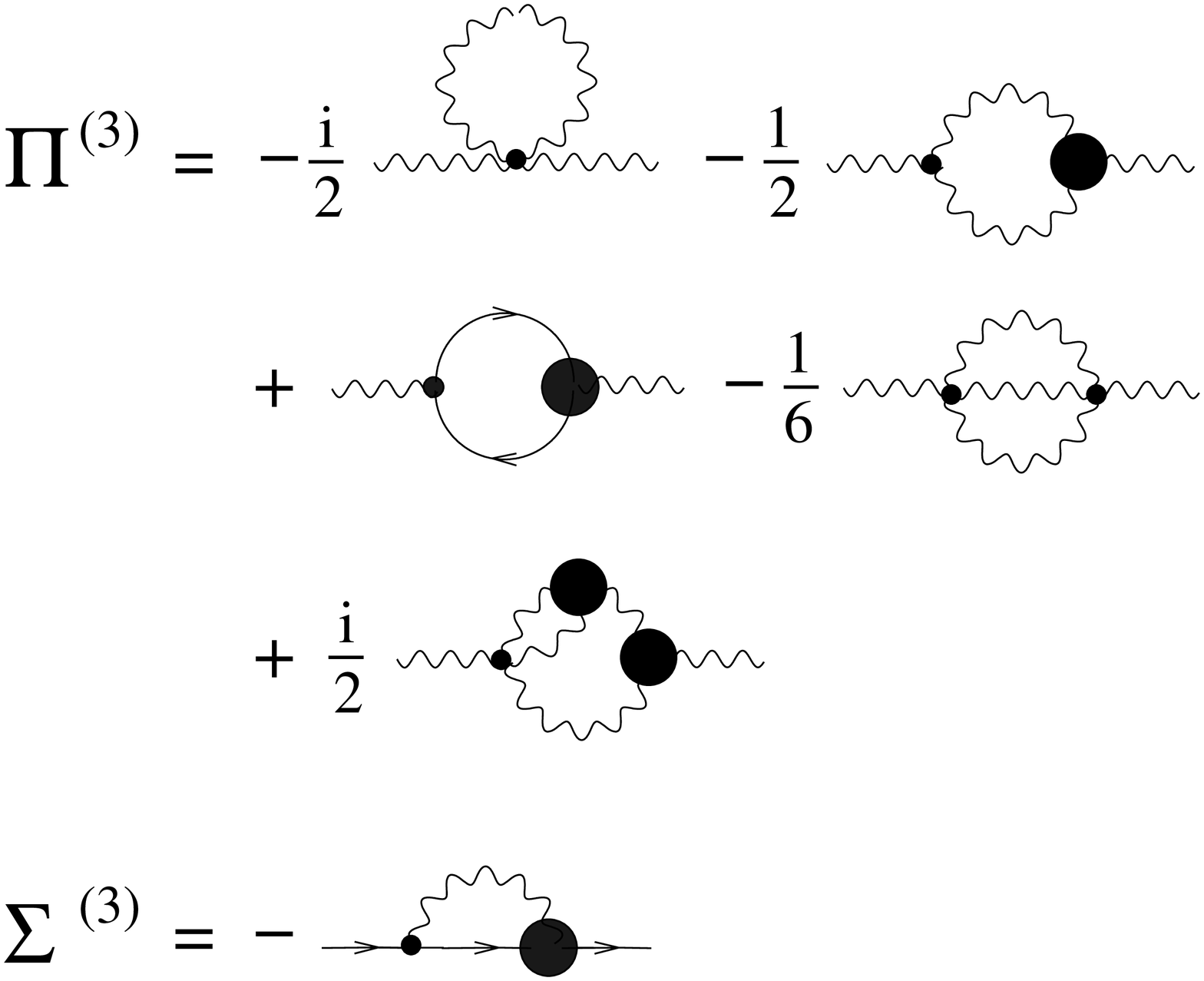,width=6.8cm}
}
\caption{The self-energy for the gauge field ($\Pi$) and
the ghost/fermion ($\Sigma$) propagators as obtained 
from the self-consistently complete 
three-loop approximation of the effective action.
(Cf.~Fig.~\ref{fig:ymver3} for the vertices.)}
\label{fig:ymmerge3}
\end{figure}

The self-energies to this order are displayed 
in Fig.~\ref{fig:ymmerge3}. It should be emphasized that 
their relatively simple form is a consequence of the
equations for the proper vertices, Fig.~\ref{fig:ymver3}.  
To see this we consider first the many terms generated 
by the functional derivative
of (\ref{eq:G20}) and (\ref{eq:gamma2}) with respect to
the gauge field propagator:
\bea\db
\Pi^{(3)}\,\, \equiv\,\, 2i \frac{\delta \Gamma_2^{\rm (3loop)}}{\delta D} 
&\db \!=\!&\db
- \frac{i}{2}\, 
\begin{minipage}{1.2cm}
\epsfig{file=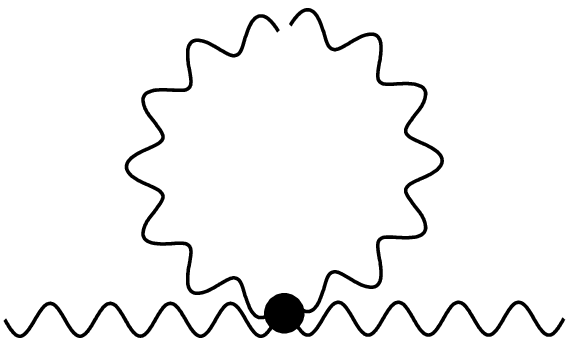,width=1.2cm}
\end{minipage}
- 
\begin{minipage}{1.4cm}
\epsfig{file=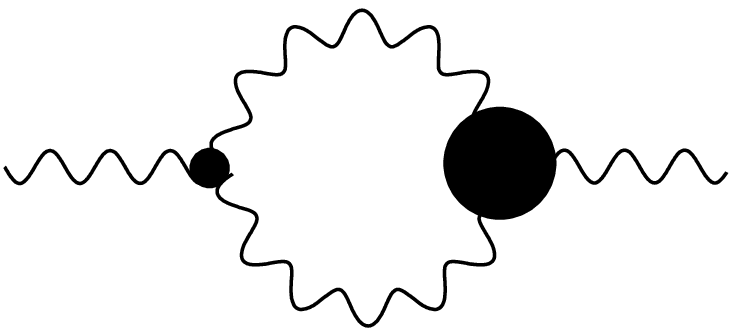,width=1.4cm}
\end{minipage}
+ \frac{1}{2}\, 
\begin{minipage}{1.4cm}
\epsfig{file=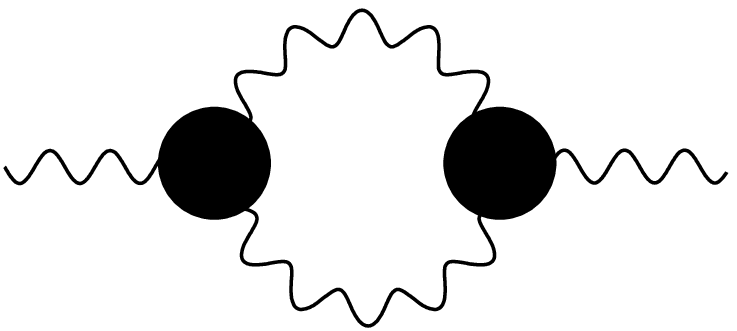,width=1.4cm}
\end{minipage}
+ 2\, 
\begin{minipage}{1.4cm}
\epsfig{file=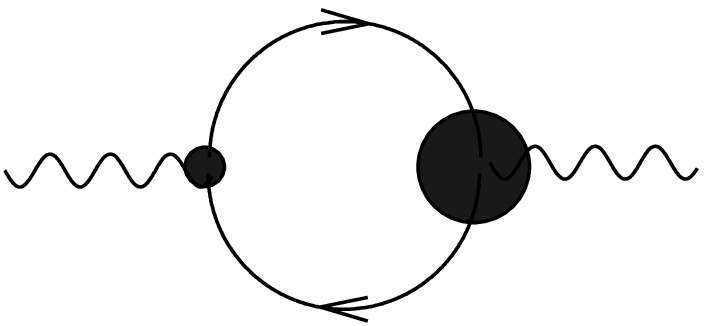,width=1.4cm}
\end{minipage}
\nonumber\\
&&\db
-\,  
\begin{minipage}{1.4cm}
\epsfig{file=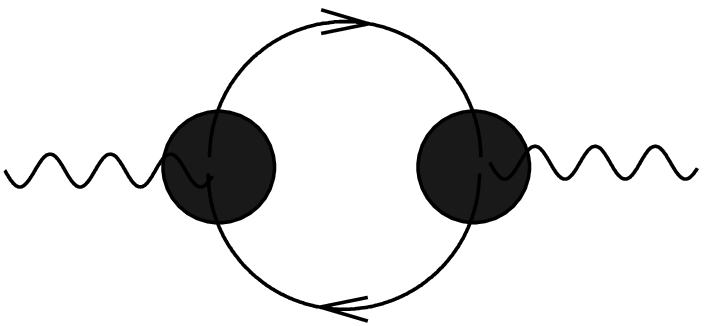,width=1.4cm}
\end{minipage}
- \frac{1}{3}\, 
\begin{minipage}{1.4cm}
\epsfig{file=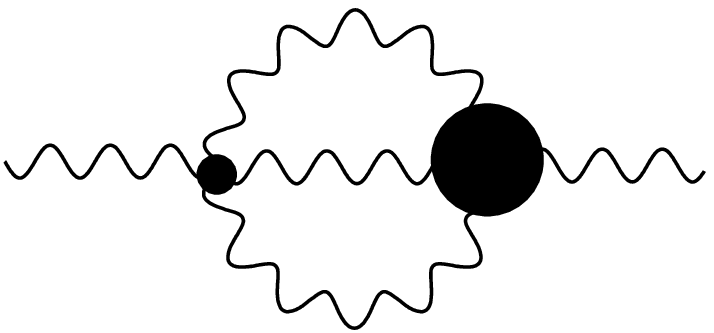,width=1.4cm}
\end{minipage}
+ \frac{1}{6}\, 
\begin{minipage}{1.4cm}
\epsfig{file=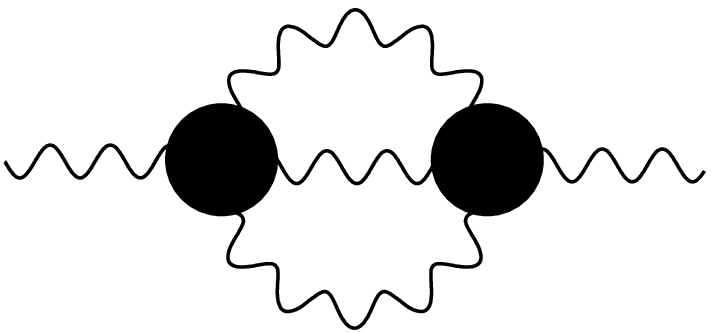,width=1.4cm}
\end{minipage}
+ i\, 
\begin{minipage}{1.4cm}
\epsfig{file=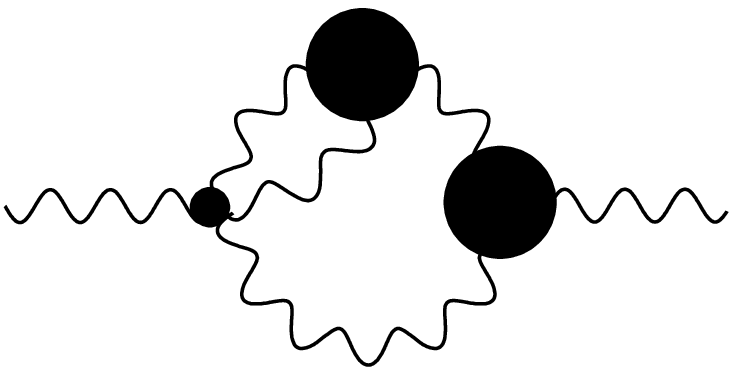,width=1.4cm}
\end{minipage}
\\
&&\db 
\,{\rr + \frac{i}{4}}\, 
\begin{minipage}{1.7cm}
\epsfig{file=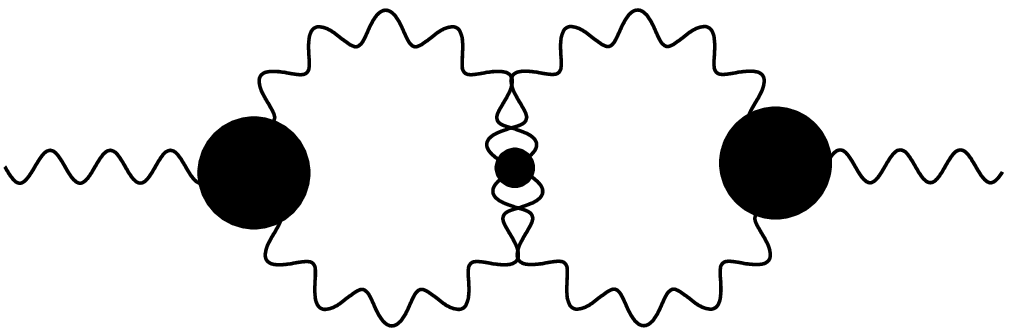,width=1.7cm}
\end{minipage}
\,{\rr + \frac{1}{2}}\, 
\begin{minipage}{1.6cm}
\epsfig{file=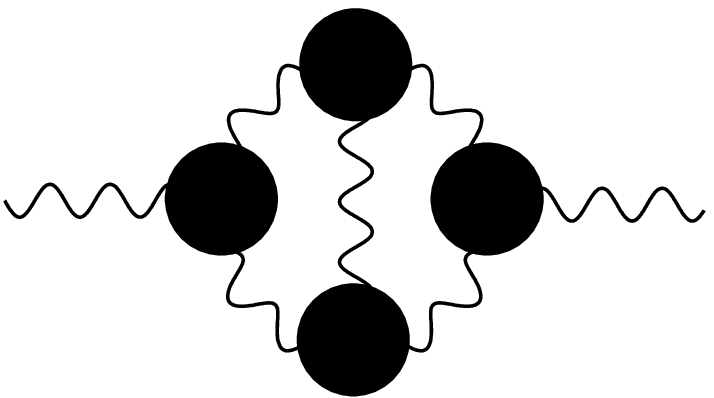,width=1.6cm}
\end{minipage}
\,{\rr - 2}\, 
\begin{minipage}{1.6cm}
\epsfig{file=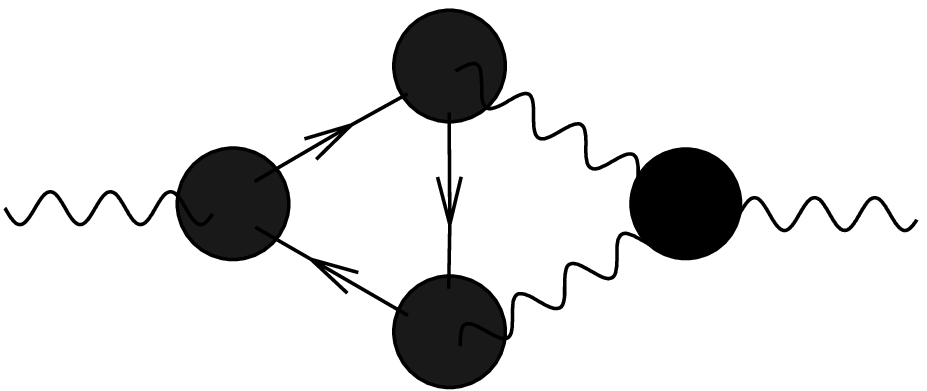,width=1.6cm}
\end{minipage}
\,{\rr -}\, 
\begin{minipage}{1.6cm}
\epsfig{file=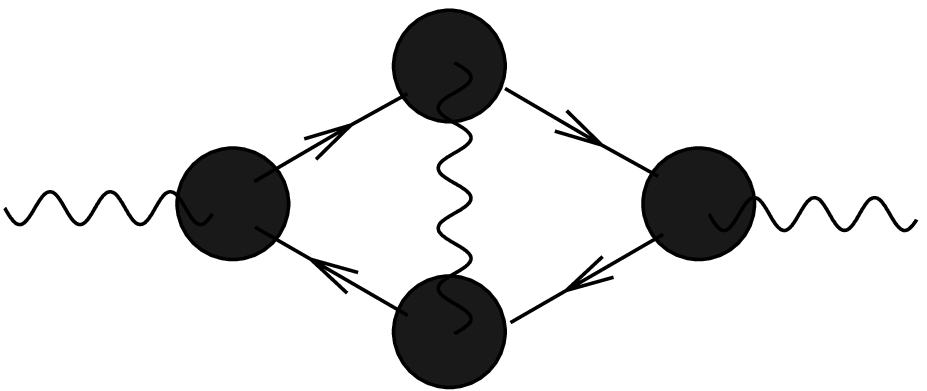,width=1.6cm}
\end{minipage} \, .
\nonumber
\eea
The short form for the self-energy of Fig.~\ref{fig:ymmerge3}
is obtained through cancellations by replacing in the above 
expression 
\bea\db
 \frac{1}{2}\, 
\begin{minipage}{1.4cm}
\epsfig{file=ymsun2.eps,width=1.4cm}
\end{minipage}
&\db \!=\!&\db \frac{1}{2}\, 
\begin{minipage}{1.4cm}
\epsfig{file=ymsun.eps,width=1.4cm}
\end{minipage}
\,{\rr -\frac{1}{2}}\, 
\begin{minipage}{1.6cm}
\epsfig{file=ymdum3.eps,width=1.6cm}
\end{minipage}
\,{\rr -\frac{i}{4}}\, 
\begin{minipage}{1.6cm}
\epsfig{file=ymdum1.eps,width=1.6cm}
\end{minipage}
\nonumber\\
&&\db
-\frac{i}{2}\, 
\begin{minipage}{1.6cm}
\epsfig{file=ymshade.eps,width=1.6cm}
\end{minipage}
\,{\rr +}\, 
\begin{minipage}{1.6cm}
\epsfig{file=ymdum2.eps,width=1.6cm}
\end{minipage}
\eea
as well as
\bea\db
-\, 
\begin{minipage}{1.4cm}
\epsfig{file=ymsunghost2.eps,width=1.4cm}
\end{minipage}
&\db \!=\!&\db -\, 
\begin{minipage}{1.4cm}
\epsfig{file=ymsunghost.eps,width=1.4cm}
\end{minipage}
\,{\rr +}\, 
\begin{minipage}{1.6cm}
\epsfig{file=ymdum4.eps,width=1.6cm}
\end{minipage}
\,{\rr +}\, 
\begin{minipage}{1.6cm}
\epsfig{file=ymdum2.eps,width=1.6cm}
\end{minipage} \, .
\eea
The latter equations follow from inserting 
the expressions for the dressed vertices
of Fig.~\ref{fig:ymver3}. Noting in addition that 
the proper four-vertex to this order 
corresponds to the classical one 
(cf.~(\ref{eq:classical4})) leads
to the result. Along the very same lines a 
similar cancellation
yields the compact form of the 
ghost/fermion self-energy displayed in 
Fig.~\ref{fig:ymmerge3}.

\subsubsection{Comparison with Schwinger-Dyson equations}
\label{sec:compSD}

\begin{figure}[t]
\framebox{\begin{minipage}{6.78cm}
{\small\rr Gauge three-vertex as well as ghost/fermion
vertex from $\Gamma^{\rm (3loop)}$:}

\vspace*{0.3cm}

\epsfig{file=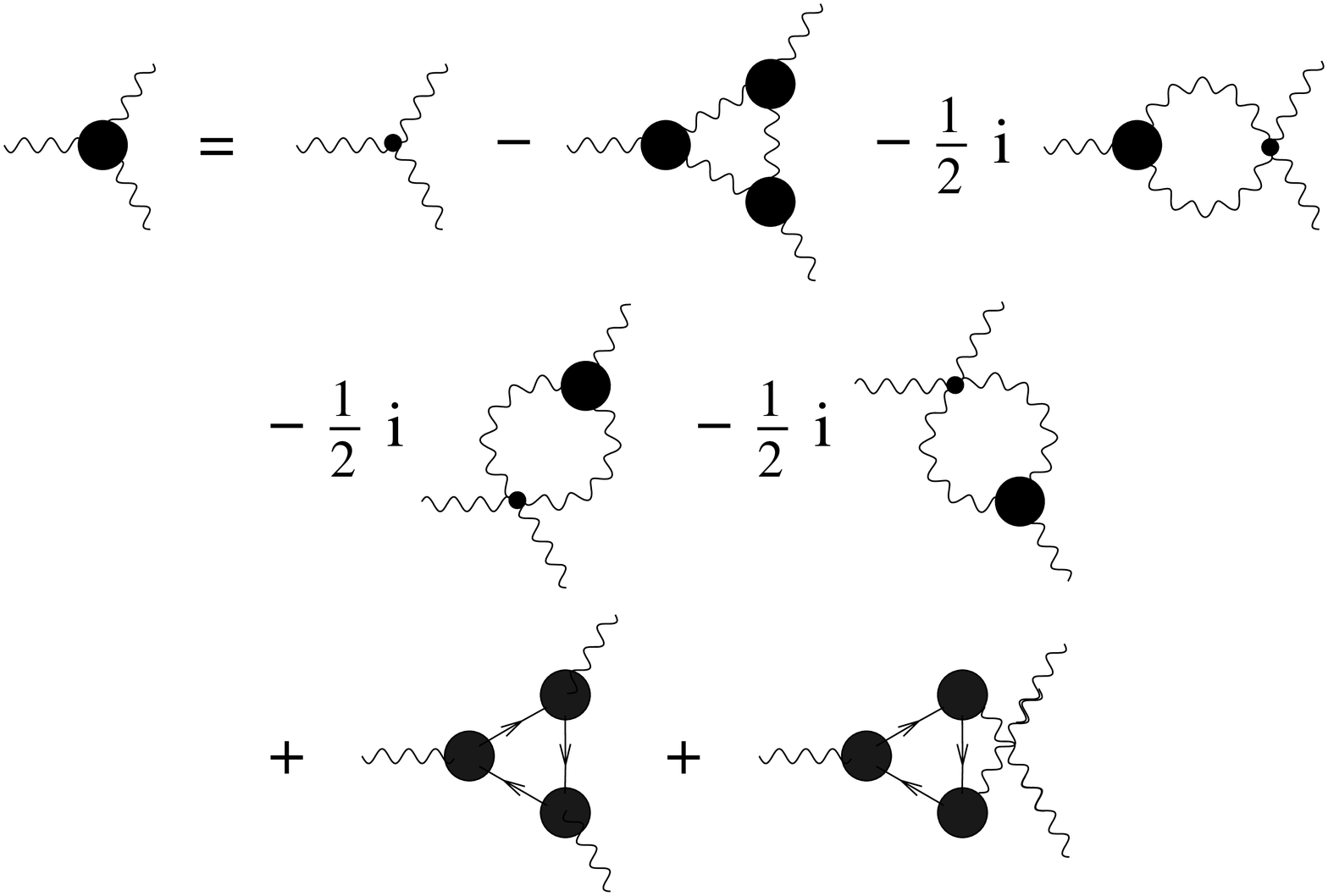,width=6.78cm}
\epsfig{file=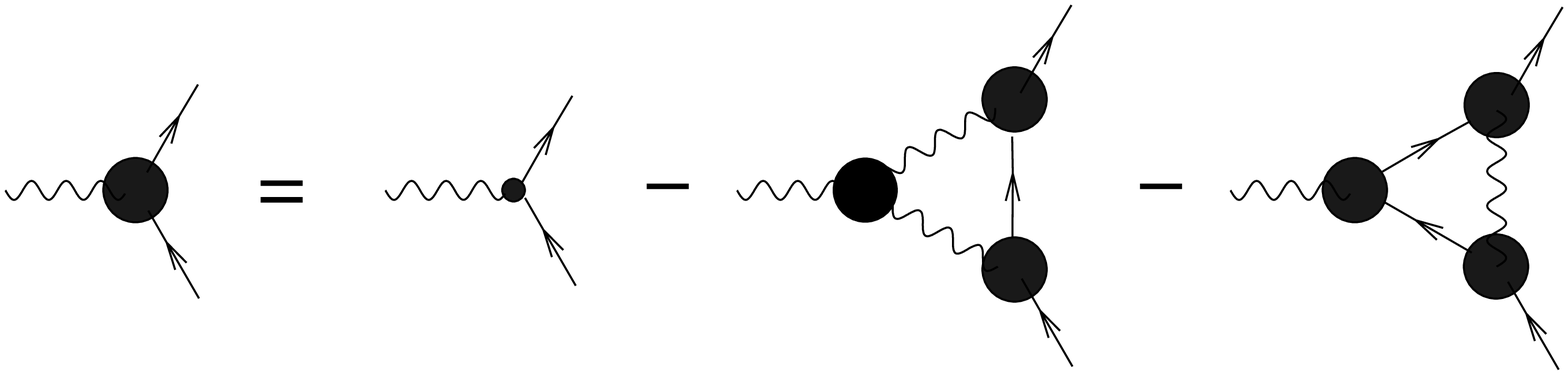,width=6.78cm}
\end{minipage}}
\framebox{\begin{minipage}{7.42cm}
{\small\db Compare: exact SD equation 
(ghost/fermion\newline diagrams not displayed):}

\vspace*{0.3cm}

\epsfig{file=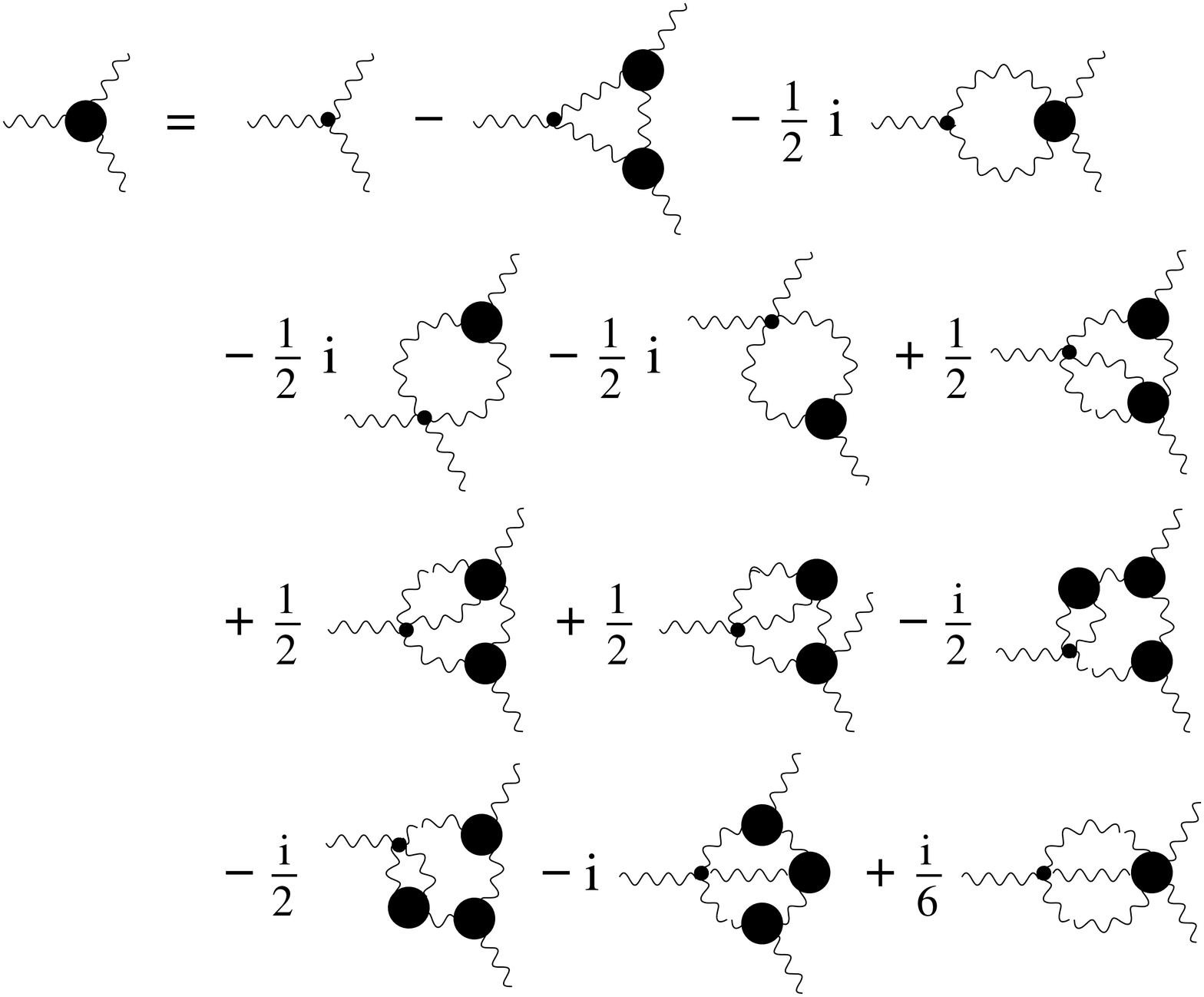,width=7.42cm}
\end{minipage}}
\caption{{\bf Left:} 
The gauge field three-vertex as well as the ghost (fermion)
vertex as obtained from the self-consistently complete 
three-loop approximation of the effective action.
Apart from the isolated classical three-vertex, all 
vertices in the equations correspond to dressed ones since at
this order the four-vertex equals the classical vertex. 
The result reflects the proper symmetry of the three-vertex. 
{\bf Right:} 
Schwinger-Dyson equation for the proper three-vertex $V_3$.
Additional diagrams involving ghost or
fermion vertices are not displayed for brevity.
We show it for comparison with the three-loop effective action result
displayed on the left. A naive
truncation of the Schwinger-Dyson equation at the one-loop level
does not agree with the latter and symmetries (the second and third 
diagram contain a classical three-vertex instead of a dressed one).}
\label{fig:ymver3}
\end{figure}
The equations of motions of the last section are self-consistently
complete to two-loop/three-loop order 
of the $n$PI effective action for arbitrarily large $n$.
We now compare them with Schwinger-Dyson (SD) equations,
which represent identities between $n$-point functions.
Clearly, without approximations the equations of motion
obtained from an exact \mbox{$n$PI} effective action and
the exact (SD) equations have to agree since one can
always map identities onto each other. However, in general
this is no longer the case for a given order in the
loop expansion of the $n$PI effective action.

By construction each diagram in a SD equation contains
at least one classical vertex. In general, this is not the case
for equations obtained from the $n$PI effective action:
The loop contributions of $\Gamma_2^{\rm int}$ in Eq.~(\ref{eq:gamma2})
or Figs.~\ref{fig:ym3loop3}--\ref{fig:ym3loop4} are 
solely expressed in terms of full vertices.
However, to a given loop-order cancellations can occur for
those diagrams in the equations of motion which do not
contain a classical vertex. For the three-loop effective
action result this has been demonstrated above
for the two-point functions. Indeed, the equations for the 
two-point functions shown in Fig.~\ref{fig:ymmerge3} correspond
to the SD equations, if one takes into account that to the considered 
order the four-vertex is trivial and given by the classical one 
(cf.~Eq.~(\ref{eq:classical4})). However, such a correspondence
is not true for the proper three-vertex to that order.

As an example, we show on the right of  
Fig.~\ref{fig:ymver3} the
SD equation for the proper three-vertex,
where for brevity we do not display the additional diagrams
coming from ghost/fermion degrees of 
freedom. One observes that a naive neglection of the two-loop contributions
of that equation would not lead to the effective action result 
for the three-vertex shown in Fig.~\ref{fig:ymver3}.
Of course, the straightforward one-loop truncation of the 
SD equation would not even respect the property of 
$V_3$ being completely symmetric in its space-time and group
labels. This is the well-known problem of loop-expansions
of SD equations, where one encounters the ambiguity of whether
classical or dressed vertices should be employed at 
a given truncation order.

We emphasize that these problems are absent using effective action 
techniques. The fact that all equations of motion are obtained from
the same approximation of the effective action puts stringent 
conditions on their form. More precisely, a self-consistently
complete approximation has the property that the order of 
differentiation of, say, $\Gamma[D,V]$ with respect to the
propagator $D$ or the vertex $V$ does not affect the equations of
motion. Consider for instance: 
\beq\db
\frac{\delta \Gamma[D,{\rr V=V(D)}]}{\delta D}
\,=\, \frac{\delta \Gamma}{\delta D}\Big|_{V}
\,{\rr +\, \frac{\delta \Gamma}{\delta V}\Big|_{D}
\, \frac{\delta V}{\delta D} \, .}
\eeq
If $V = V(D)$ is the result of the stationary condition
$\delta \Gamma/\delta V = 0$ then the above 
corresponds to the correct stationarity condition for
the propagator for fixed $V$: $\delta \Gamma/\delta D = 0$. 
In contrast, with some ansatz $V=f(D)$ that does not
fulfill the stationarity condition of the 
effective action, the equation of motion for the
propagator would receive additional corrections
$\sim \delta V/\delta D$. In particular, it would be
inconsistent to use the equation of motion for the
propagator $\delta \Gamma/\delta D = 0$
(cf.~e.g.~Fig.~\ref{fig:ymmerge3} which corresponds 
to the SD equation result) but not the equation $\delta \Gamma/\delta V = 0$
for the vertex (cf.~Fig.~\ref{fig:ymver3}).

In turn, one can conclude that a wide class of employed truncations 
of exact SD equations cannot be obtained from the $n$PI effective 
action: this concerns those approximations which use the
exact SD equation for the propagator but make an ansatz
for the vertices that differs from the one displayed in 
Fig.~\ref{fig:ymver3}. The differences are, however, typically
higher order in the perturbative coupling expansion and there
may be many cases, in particular in vacuum or thermal equilibrium,
where some ansatz for the vertices is a very efficient way to  
proceed. Out of equilibrium however, as mentioned above, 
the ``conserving'' property of the effective action approximations
can have important consequences, since the 
effective loss of initial conditions and the presence of basic constants 
of motion such as energy conservation is crucial.

\subsubsection{Nonequilibrium evolution equations}
\label{sec:nonequilibrium}

Up to $\mathcal{O}(g^6)$ corrections in the self-consistently
complete expansion of the effective action,
the four-vertex parametrizing the diagrams of 
Figs.~\ref{fig:ymmerge3}--\ref{fig:ymver3} corresponds to the classical vertex.
At this order of approximation there is, therefore, no
distinction between the coupling expansion of the 3PI and 4PI 
effective action. To discuss the relevant differences between
the 2PI and 3PI expansions for time evolution problems, we will
use the language of QED for simplicity, where no four-vertex
appears. However, the evolution equations of this section can be
straightforwardly transcribed to the nonabelian case by 
taking into account in addition to the equation for the gauge--fermion 
three-vertex those for the gauge--ghost and gauge three-vertex
(cf.~Fig.~\ref{fig:ymver3}).        
In the following the effective action is a functional of the gauge
field propagator $D_{\mu\nu}(x,y)$, the fermion propagator 
$\Delta(x,y)$ and the gauge-fermion vertex 
$V_{3\, \mu}^{\rm (f)}(x,y;z)$, where we suppress Dirac
indices and we will write
$V_{3}^{\rm (f)} \equiv V$. According to 
Eqs.~(\ref{eq:exact4PI})---(\ref{eq:gamma2}) one has in this case
\beq\db
\Gamma_2[D,\Delta,V] 
\,=\, \Gamma_2^0[D,\Delta,V]
+ \Gamma_2^{\rm int}[D,\Delta,V] \, , 
\eeq
with
\beq\db
\Gamma_2^0 \,=\, - i g^2 \int_{x y z u} \tr\left[ \gamma_\mu
\Delta(x,y) V_{\nu}(y,z;u) \Delta(z,x) D^{\mu\nu}(x,u) \right] \, ,
\label{eq:gamma2qed}
\eeq
where the trace acts in Dirac space. For the given order of
approximation there are two 
distinct contributions to $\Gamma_2^{\rm int}$:
\bea\db 
\Gamma_2^{\rm int}&\db =&\db \Gamma_2^{(a)} + \Gamma_2^{(b)} 
+ \mathcal{O}\left(g^6\right)\, ,
\label{eq:gamma2qedint}\\
\db \Gamma_2^{(a)} &\db =&\db  \frac{i}{2} g^2 \int_{x y z u v w}
\tr\left[ V_{\mu}(x,y;z)
\Delta(y,u) V_{\nu}(u,v;w) \Delta(v,x) D^{\mu\nu}(z,w) \right] \, ,
\nonumber\\\db 
\Gamma_2^{(b)} &\db =&\db   
\frac{i}{4} g^4 \int_{x y z u v w x' y' z' u' v' w'}
\tr\left[ V_{\mu}(x,y;z)
\Delta(y,u) V_{\nu}(u,v;w) \Delta(v,x') 
 \right. \nonumber\\
&&\db  \left. V_{\rho}(x',y';z') \Delta(y',u') V_{\sigma}(u',v';w') 
\Delta(v',x) D^{\mu\rho}(z,z') D^{\nu\sigma}(w,w') \right] \, .
\nonumber
\eea
The equations of motions for the propagators and vertex 
are obtained from the stationarity conditions (\ref{eq:statprop})
and (\ref{eq:statver}) for the effective action. 
To convert (\ref{eq:SDforD}) for the photon propagator 
into an equation which is more suitable 
for initial value problems, we convolute with $D$ from the right 
and obtain for the considered case of vanishing ``background'' fields, 
e.g.~for covariant gauges (cf.~also the discussion in 
Sec.~\ref{sec:exactevoleq}):
\bea\db 
\left[{g^\mu}_\gamma \square - (1-\xi^{-1}) \partial^\mu 
\partial_{\gamma} \right]_x D^{\gamma\nu}(x,y)
- i \int_z {\Pi^\mu}_{\gamma}(x,z) D^{\gamma\nu}(z,y) 
\nonumber \\
\db  = i g^{\mu\nu} \delta_\mathcal{C}(x-y)  \, .
\label{eq:evolD}
\eea 
Similarly, the corresponding equation of (\ref{eq:SDforG}) yields 
the evolution equation for the fermion propagator given in 
Eq.~(\ref{eq:evolDel}). Using the above results the self-energies are
\bea\db 
\Sigma^{(f)}(x,y) &\db =&\db  - g^2 \int_{z'z''} D_{\mu\nu}(z',y)
V^{\mu}(x,z'';z') \Delta(z'',y) \gamma^{\nu} \, ,
\label{eq:selffermexact}\\\db 
\Pi^{\mu\nu}(x,y) &\db =&\db  g^2 \int_{z'z''} \tr\, \gamma^{\mu}
\Delta(x,z') V^{\nu}(z',z'';y) \Delta(z'',x)  \, .
\label{eq:selfexact}
\eea
Note that the form of the self-energies is exact for known three-vertex.
To see this within the current framework, we note that the self-energies can
be expressed in terms of $\Gamma_2^0$ only. The latter receives no further 
corrections at higher order in the expansion (cf.~Sec.~\ref{sec:calcSUN}), 
and thus the expression is exactly known: With
\bea\db 
\int_z \Sigma^{(f)}(x,z) \Delta(z,y)
= - i \int_z \left( \frac{\delta \Gamma_2^0}{\delta \Delta(z,x)} 
+ \frac{\delta \Gamma_2^{\rm int}}{\delta \Delta (z,x)} \right) \Delta(z,y) 
\, ,
\eea
and since $\Gamma_2^{\rm int}$ is only a functional of $V \Delta D^{1/2}$
(cf.~Sec.~\ref{sec:calcSUN}) one can use the identity
\bea\db 
\int_z \frac{\delta \Gamma_2^{\rm int}}{\delta \Delta(z,x)} \Delta(z,y)
&\db =&\db   \int_{z z'} V_{\mu}(x,z;z')  
\frac{\delta \Gamma_2^{\rm int}}{V_{\mu}(y,z;z')} \nonumber\\
&\db =&\db  - \int_{z z'} V_{\mu}(x,z;z')  
\frac{\delta \Gamma_2^0}{V_{\mu}(y,z;z')}
\label{eq:delverrew}
\eea
to express everything in terms of the known\footnote{This
can also be directly verified from (\ref{eq:gamma2qed}) 
to the given order of approximation.} $\Gamma_2^0$.
The last equality in (\ref{eq:delverrew}) uses that
$\delta (\Gamma_2^0+\Gamma_2^{\rm int})/\delta \Delta = 0$.  
A similar discussion can be done for the photon self-energy.
As a consequence,
all approximations are encoded in the equation for the vertex, 
which is obtained from (\ref{eq:gamma2qedint}) as 
\bea\db 
V^{\mu} (x,y;z) &\db =&\db  V_0^{\mu} (x,y;z)  
- g^2 \int_{v w x' y' u' w'}
V_{\nu} (x,v;w) \Delta(v,x')V^{\mu} (x',y';z)
 \nonumber\\
&&\db  \Delta(y',u') V_{\sigma} (u',y;w') D^{\sigma\nu}(w',w) 
+ \mathcal{O}\left(g^4\right)\, ,
\label{eq:vertexeom}
\eea
where
\beq\db 
V_0^{\mu} (x,y;z) = \gamma^{\mu} \delta(x-z) \delta(z-y) \, . 
\eeq
For the self-consistently complete two-loop approximation
the self-energies are given by  
\bea\db 
\Sigma^{(f)}(x,y) &\db =&\db  - g^2 D_{\mu\nu}(x,y)
\gamma^{\mu} \Delta(x,y) \gamma^{\nu} + \mathcal{O}\left(g^4\right) 
\, ,\\[0.2cm]
\db \Pi^{\mu\nu}(x,y) &\db =&\db  g^2\, \tr\, \gamma^{\mu}
\Delta(x,y) \gamma^{\nu} \Delta(y,x)  + \mathcal{O}\left(g^4\right)\, .
\label{eq:selfg2}
\eea
Following the discussion of Sec.~\ref{sec:specstat},
we decompose the two-point functions into spectral and 
statistical components using the identities (\ref{eq:decompgauge}) for
gauge fields and (\ref{eq:fermdec}) for fermions.
Then $\rho_D$ corresponds to the gauge field spectral function 
and $F_D$ is the statistical two-point 
function, while $\rho^{(f)}$ and $F^{(f)}$ are the 
corresponding fermion two-point functions.
The same decomposition can be done for the corresponding
self-energies:
\bea\db 
\Pi^{\mu\nu}(x,y) 
&\db =&\db  \Pi_{(F)}^{\mu\nu}(x,y) - \frac{i}{2} \Pi_{(\rho)}^{\mu\nu}(x,y) 
\, {\rm sign} (x^0-y^0) \, , \\
\db \Sigma^{(f)}(x,y) 
&\db =&\db  \Sigma^{(f)}_{F}(x,y) 
- \frac{i}{2} \Sigma^{(f)}_{\rho} (x,y)\, {\rm sign} (x^0-y^0) \, ,
\label{eq:sigmadec}
\eea
and similarly for the fermions as described by (\ref{eq:selffermdec}).
Since the above decomposition for the propagators and self-energies
makes the time-ordering explicit, we can evaluate 
the r.h.s.~of (\ref{eq:evolD}) along the time contour following
the discussion of Sec.~\ref{sec:noneqeveq}. One finds the evolution 
equations:
\bea\db 
\left[{g^\mu}_\gamma \square - (1-\xi^{-1}) \partial^\mu 
\partial_{\gamma} \right]_x \rho_D^{\gamma\nu}(x,y)
&\db =&\db  \int_{y^0}^{x^0}\! {\rm d}z\, 
\Pi_{(\rho)}^{\mu \gamma}(x,z) {\rho_{D,\gamma}}^{\nu}(z,y) \,\,\, ,
\label{eq:rho}\\[0.2cm]
\db \left[{g^\mu}_\gamma \square - (1-\xi^{-1}) \partial^\mu 
\partial_{\gamma} \right]_x F_D^{\gamma\nu}(x,y)
&\db =&\db  \int_{t_0}^{x^0}\! {\rm d}z\, 
\Pi_{(\rho)}^{\mu \gamma}(x,z) {F_{D,\gamma}}^{\nu}(z,y) 
\nonumber\\
&\db -&\db  \int_{t_0}^{y^0}\! {\rm d}z\, 
\Pi_{(F)}^{\mu \gamma}(x,z) {\rho_{D,\gamma}}^{\nu}(z,y) \,\,\, ,
\label{eq:F}
\eea
where we use the abbreviated notation $\int_{t_1}^{t_2}
{\rm d}z \equiv \int_{t_1}^{t_2} {\rm d}z^0 
\int_{-\infty}^{\infty} {\rm d}^d z$. Here the initial time
is denoted by $t_0$, which was taken without loss of
generality to be zero in the
respective equations (\ref{eq:exactrhoF}) for scalars. The equations of
motion for the fermion spectral and statistical correlators
are obtained in a similar way from (\ref{eq:evolDel}) as
described in Sec.~\ref{sec:noneqeveq} and are given in Eq.~(\ref{eq:Fexact}).

A similar discussion as for the two-point functions can also be done 
for the higher correlation functions. For the three-vertex 
we write
\beq\db 
V^{\mu} (x,y;z) =  V_0^{\mu} (x,y;z) + \bar{V}^{\mu} (x,y;z)\, .
\label{eq:separation}
\eeq
and the corresponding decomposition into spectral and
statistical components reads
\bea\db 
\lefteqn{ \bar{V}^{\mu} (x,y;z) = }
\nonumber\\
&&\db  U_{(F)}^{\mu}(x,y;z)\, {\rm sign} (y^0-x^0)\, {\rm sign} (z^0-x^0) 
- \frac{i}{2} U_{(\rho)}^{\mu}(x,y;z)\, {\rm sign} (y^0-z^0)
\nonumber\\
&\db \!+\!\!&\db  
V_{(F)}^{\mu}(x,y;z)\, {\rm sign} (x^0-z^0)\, {\rm sign} (y^0-z^0) 
- \frac{i}{2} V_{(\rho)}^{\mu}(x,y;z)\, {\rm sign} (x^0-y^0)
\nonumber\\
&\db \!+\!\!&\db  
W_{(F)}^{\mu}(x,y;z)\, {\rm sign} (z^0-y^0)\, {\rm sign} (x^0-y^0) 
- \frac{i}{2} W_{(\rho)}^{\mu}(x,y;z)\, {\rm sign} (z^0-x^0) .
\label{eq:vbar}
\eea
To discuss this in more detail we use the short-hand notation
\beq\db 
\Theta(x^0,y^0,z^0) \equiv \Theta(x^0-y^0)\Theta(y^0-z^0) \, .
\eeq
With the separation of Eq.~(\ref{eq:separation}), the  
time-ordered three-vertex can be written as (cf.~also the
corresponding discussion for two-point functions in Sec.~\ref{sec:specstat})
\bea\db 
\bar{V}^{\mu} (x,y;z) &\db =&\db  V_{(1)}^{\mu} (x,y;z) \Theta(x^0,y^0,z^0)
+ V_{(2)}^{\mu} (x,y;z) \Theta(y^0,z^0,x^0)
\nonumber\\
&\db +&\db V_{(3)}^{\mu} (x,y;z) \Theta(z^0,x^0,y^0)
+ V_{(4)}^{\mu} (x,y;z) \Theta(z^0,y^0,x^0) \qquad
\nonumber\\
&\db +&\db  V_{(5)}^{\mu} (x,y;z) \Theta(x^0,z^0,y^0)
+ V_{(6)}^{\mu} (x,y;z) \Theta(y^0,x^0,z^0) \, ,
\label{eq:vstart}
\eea
with `coefficients' $V_{(i)}^{\mu} (x,y;z)$, $i = 1,\ldots, 6$. 
These coefficients can be expressed in terms of three spectral vertex
functions $U_{(\rho)}^{\mu}(x,y;z)$, $V_{(\rho)}^{\mu}(x,y;z)$ and 
$W_{(\rho)}^{\mu}(x,y;z)$, as well as the corresponding statistical components 
$U_{(F)}^{\mu}(x,y;z)$, $V_{(F)}^{\mu}(x,y;z)$ and $W_{(F)}^{\mu}(x,y;z)$
that have been employed in Eq.~(\ref{eq:vbar}). One finds,
suppressing the space-time arguments:
\bea\db 
V_{(1)}^{\mu} &\db \equiv&\db  U_{(F)}^{\mu} + V_{(F)}^{\mu} - W_{(F)}^{\mu}
- \frac{i}{2} \Big( U_{(\rho)}^{\mu} + V_{(\rho)}^{\mu} 
- W_{(\rho)}^{\mu} \Big) \, ,
\nonumber\\\db 
V_{(2)}^{\mu} &\db \equiv&\db  U_{(F)}^{\mu} - V_{(F)}^{\mu} + W_{(F)}^{\mu}
- \frac{i}{2} \Big( U_{(\rho)}^{\mu} - V_{(\rho)}^{\mu} 
+ W_{(\rho)}^{\mu} \Big) \, ,
\nonumber\\\db 
V_{(3)}^{\mu} &\db \equiv&\db  - U_{(F)}^{\mu} + V_{(F)}^{\mu} + W_{(F)}^{\mu}
- \frac{i}{2} \Big( - U_{(\rho)}^{\mu} + V_{(\rho)}^{\mu} 
+ W_{(\rho)}^{\mu} \Big) \, ,
\nonumber\\\db 
V_{(4)}^{\mu} &\db \equiv&\db  U_{(F)}^{\mu} + V_{(F)}^{\mu} - W_{(F)}^{\mu}
+ \frac{i}{2} \Big( U_{(\rho)}^{\mu} + V_{(\rho)}^{\mu} 
- W_{(\rho)}^{\mu} \Big) \, ,
\\\db 
V_{(5)}^{\mu} &\db \equiv&\db  U_{(F)}^{\mu} - V_{(F)}^{\mu} + W_{(F)}^{\mu}
+ \frac{i}{2} \Big( U_{(\rho)}^{\mu} - V_{(\rho)}^{\mu} 
+ W_{(\rho)}^{\mu} \Big) \, ,
\nonumber\\\db 
V_{(6)}^{\mu} &\db \equiv&\db  - U_{(F)}^{\mu} + V_{(F)}^{\mu} + W_{(F)}^{\mu}
+ \frac{i}{2} \Big( - U_{(\rho)}^{\mu} + V_{(\rho)}^{\mu} 
+ W_{(\rho)}^{\mu} \Big) \, .
\nonumber
\eea
In terms of the coefficients $V_{(i)}^{\mu}$ these are given by:
\bea\db 
U_{(F)}^{\mu} &\db =&\db  \frac{1}{4} \Big(V_{(1)}^{\mu} 
+ V_{(2)}^{\mu} + V_{(4)}^{\mu} + V_{(5)}^{\mu}
\Big)\, , \quad 
U_{(\rho)}^{\mu} = \frac{i}{2} \Big(V_{(1)}^{\mu} 
+ V_{(2)}^{\mu} - V_{(4)}^{\mu} - V_{(5)}^{\mu}
\Big)\, , \nonumber \\\db 
V_{(F)}^{\mu} &\db =&\db  \frac{1}{4} \Big(V_{(1)}^{\mu} 
+ V_{(3)}^{\mu} + V_{(4)}^{\mu} + V_{(6)}^{\mu}
\Big)\, , \quad
V_{(\rho)}^{\mu} = \frac{i}{2} \Big(V_{(1)}^{\mu} 
+ V_{(3)}^{\mu} - V_{(4)}^{\mu} - V_{(6)}^{\mu}
\Big)\, , \quad \nonumber \\\db 
W_{(F)}^{\mu} &\db =&\db  \frac{1}{4} \Big(V_{(2)}^{\mu} 
+ V_{(3)}^{\mu} + V_{(5)}^{\mu} + V_{(6)}^{\mu}
\Big)\, , \quad
W_{(\rho)}^{\mu} = \frac{i}{2} \Big(V_{(2)}^{\mu} 
+ V_{(3)}^{\mu} - V_{(5)}^{\mu} - V_{(6)}^{\mu}
\Big) \nonumber \, .
\eea
Insertion shows the equivalence of (\ref{eq:vstart}) and
(\ref{eq:vbar}).

\subsection{Kinetic theory}
\label{sec:kinetic}

To make contact with frequent discussions in the literature, 
we will consider for the above gauge field and fermion
nonequilibrium equations a standard ``on-shell'' approximation  
which is typically employed 
to derive kinetic equations for effective particle number
densities. This part can be viewed as a continuation of 
Sec.~\ref{sec:detourboltzmann}, where scalar fields and the 
limitations of ``on-shell'' approximations have been discussed.

\subsubsection{``On-shell'' approximations}
\label{eq:onshelllimits}

The evolution equations (\ref{eq:rho})--(\ref{eq:Fexact}) to
order $g^2$ and higher contain ``off-shell'' and ``memory'' 
effects due to their time integrals on the r.h.s.~(cf.~also 
Sec.~\ref{eq:scatoffmem}). 
To simplify the description one conventionally considers 
a number of additional assumptions which finally lead
to effective kinetic or Boltzmann-type descriptions for ``on-shell''
particle number distributions. The derivation of kinetic equations
for the two-point functions $F^{\mu\nu}(x,y)$ and $\rho^{\mu\nu}(x,y)$ 
of Sec.~\ref{sec:nonequilibrium} can be based on 
(i)~the restriction that the initial condition for the time
evolution problem is specified in the remote past,
i.e.~$t_0 \to - \infty$, (ii) a derivative expansion in the center
variable $X = (x+y)/2$, and (iii) a ``quasiparticle'' picture.    
In contrast to the discussion for scalar fields in 
Sec.~\ref{sec:detourboltzmann}, within this approach one first sends
the initial time $t_0$ to the remote past in the equations 
(\ref{eq:rho})--(\ref{eq:Fexact}). This allows one to use 
standard derivative expansion techniques in a straightforward
way. The procedure of Sec.~\ref{sec:detourboltzmann} has the advantage
that one can discuss which contributions to the evolution
are lost in this limit. On the other hand, the advantage of the 
derivative expansion is that it may in principle be used to include 
higher order corrections. However, the complexity of a derivative 
expansion grows rapidly beyond the lowest order. 

For the sake of simplicity (not required), we consider the
Feynman gauge $\xi = 1$ in the following. We will also consider 
a chirally symmetric theory, i.e.~no vacuum fermion mass,
along with parity and $CP$ invariance. Therefore, the system is
charge neutral and, in particular, the most general fermion
two-point functions can be written in terms of vector components
only: $F^{(f)}(x,y) = \gamma_{\mu} F^{(f) \mu}(x,y)$,
$\rho^{(f)}(x,y) = \gamma_{\mu} \rho^{(f) \mu}(x,y)$,
with hermiticity properties $F^{(f) \mu}(x,y) = 
[F^{(f) \mu}(y,x)]^*$, $\rho^{(f) \mu}(x,y)
= - [\rho^{(f) \mu}(y,x)]^*$. For the gauge fields the
respective properties of the statistical and spectral
correlators read $F_D^{\mu\nu}(x,y) = 
[F_D^{\nu\mu}(y,x)]^*$, $\rho_D^{\mu\nu}(x,y)
= - [\rho_D^{\nu\mu}(y,x)]^*$. 

In order to Fourier transform with respect to the relative coordinate
$s^{\mu} = x^{\mu} - y^{\mu}$, we write
\bea\db
\tilde{F}_D^{\mu\nu} \left( X, k \right) &\db =&\db 
 \int {\rm d}^4{s} \; e^{i k s} 
F_D^{\mu\nu}
\left( X + \frac{s}{2}, X - \frac{s}{2} \right) \;, 
\\\db
\tilde{\varrho}_D^{\mu\nu}
\left( X, k \right) &\db =&\db - i \int {\rm d}^4{s} \; e^{i k s} 
\rho_D^{\mu\nu} \left( X + \frac{s}{2}, X - \frac{s}{2} \right) \;, 
\label{eq:wigner}
\eea
and equivalently for the fermion statistical and spectral
function, $\tilde{F}^{(f)} (X,k)$
and $\tilde{\varrho}^{(f)} (X,k)$.
Here we have introduced a factor $-i$ in the definition of the 
spectral function transform for convenience.
For the Fourier transformed quantities we note the 
following hermiticity properties, for the
gauge fields:
$[\tilde{F}_D^{\mu\nu}(X,k)]^* = 
\tilde{F}_D^{\nu\mu}(X,k) \,\, , \,\,
[\tilde{\varrho}_D^{\mu\nu}(X,k)]^* = 
\tilde{\varrho}_D^{\nu\mu}(X,k)$,
and for the vector components of the fermion fields:
$[\tilde{F}^{(f) \mu}(X,k)]^* = 
\tilde{F}^{(f) \mu}(X,k) \,\, , \,\,
[\tilde{\varrho}^{(f) \mu}(X,k)]^* = 
\tilde{\varrho}^{(f) \mu}(X,k)$.
After sending 
$t_0 \to - \infty$ the derivative expansion can be
efficiently applied to the exact Eqs.~(\ref{eq:rho})---(\ref{eq:Fexact}). 
Here one considers the difference of (\ref{eq:rho}) and the
one with interchanged coordinates $x$ and $y$, and equivalently for
the other equations. We use 
\bea\db
\int\! {\rm d}^4{s} \; e^{i k s} \int\! {\rm d}^4{z}
f(x,z) g(z,y) &\db \!=\!&\db \tilde{f}(X,k) \tilde{g}(X,k) + \ldots 
\label{eq:grad1} \,\, \label{eq:grad2}\\
\db \int\! {\rm d}^4{s} \; e^{i k s} \int\! {\rm d}^4{z}\int\! {\rm d}^4{z'}
f(x,z) g(z,z') h(z',y) &\db \!=\!&\db \tilde{f}(X,k) \tilde{g}(X,k) 
\tilde{h}(X,k) + \ldots 
\nonumber 
\eea 
where the dots indicate derivative terms, which will be
neglected. E.g.~the first derivative corrections
to (\ref{eq:grad1}) can be written as a Poisson bracket, 
which is in particular important if ``finite-width'' effects of the spectral
function are taken into account. However, a typical quasiparticle picture 
which employs a free-field or ``zero-width'' form 
of the spectral function is consistent with
neglecting derivative terms in the scattering part.
We also note that the 
quasiparticle/free-field form of the two-point functions implies 
\bea\db
F_D^{\mu\nu}(X,k) \to - g^{\mu\nu} F_D(X,k)
\quad , \quad \rho_D^{\mu\nu}(X,k) \to - g^{\mu\nu} \rho_D(X,k) \, .
\label{eq:simpleindex}
\eea  
At this point the only use of the above replacement is that all 
Lorentz contractions can be done. This doesn't affect the derivative
expansion but keeps the notation simple. Similar to 
Eq.~(\ref{eq:wigner}), we define the Lorentz contracted self-energies:
\bea\db 
- 4 \tilde{\Pi}_{(F)} (X,k) &\db \equiv&\db  \int {\rm d}^4{s} \; e^{i k s} 
\Pi_{(F) \mu}^{\mu} \left( X + \frac{s}{2}, X - \frac{s}{2} \right) \, ,
\\\db
-4 \tilde{\Pi}_{(\varrho)} (X,k) &\db \equiv&\db  
-i \int {\rm d}^4{s} \; e^{i k s} 
\Pi_{(\rho)\mu}^{\mu} \left( X + \frac{s}{2}, X - \frac{s}{2} \right) \, . 
\eea
Without further assumptions, i.e.~using the above notation and 
applying the approximation (\ref{eq:grad1}) and 
(\ref{eq:simpleindex}) to the exact evolution equations
one has\footnote{The relation
to a more conventional form of the equations can be seen by writing:
\bea
\lefteqn{\left(
\tilde{\Pi}_{(\varrho)} \tilde{F}_D  
- \tilde{\Pi}_{(F)} \tilde{\varrho}_D \right) (X,k) = }
\nonumber \\
&&
\left( \left[ \tilde{\Pi}_{(F)} + \frac{1}{2} \tilde{\Pi}_{(\varrho)} \right]
\left[\tilde{F}_D - \frac{1}{2} \tilde{\varrho}_D \right] -
\left[ \tilde{\Pi}_{(F)} - \frac{1}{2} \tilde{\Pi}_{(\varrho)} \right]
\left[\tilde{F}_D + \frac{1}{2} \tilde{\varrho}_D \right] 
\right) (X,k) \, .\nonumber
\eea
The difference of the
two terms on the r.h.s.~can be directly interpreted as the
difference of a so-called ``loss'' and a ``gain'' term in
a Boltzmann-type description.
} 
\bea\db
2\, k^{\mu} \frac{\partial}{\partial X^{\mu}} \tilde{F}_D(X,k)
&\db =&\db  \tilde{\Pi}_{(\varrho)} (X,k)\, \tilde{F}_D (X,k)  
- \tilde{\Pi}_{(F)} (X,k)\, 
\tilde{\varrho}_D (X,k) \, , \label{eq:F2} \qquad \\
\db 2\, k^{\mu} \frac{\partial}{\partial X^{\mu}} 
\tilde{\varrho}_D (X,k) &\db =&\db 0 \, . \label{eq:rho2}
\eea 
One observes that the equations (\ref{eq:F2})
and (\ref{eq:rho2}) have a 
structure reminiscent of that for the exact equations for vanishing
``background'' fields, (\ref{eq:rho}) and (\ref{eq:F}),
evaluated at equal times $x^0 = y^0$. 
However, one should keep 
in mind that (\ref{eq:F2}) and (\ref{eq:rho2}) are, in 
particular, only valid for initial
conditions specified in the remote past and neglecting gradients
in the collision part. 

From (\ref{eq:rho2}) one observes that in this approximation
the spectral function receives no contribution from scattering
described by the r.h.s.~of the exact equation (\ref{eq:rho}).
As a consequence, the spectral function obeys the
free-field equations of motion. In particular, 
$\rho_D^{\mu\nu}(x,y)$ 
have to fulfill the equal-time commutation relations 
$[\rho_D^{\mu\nu}(x,y)]_{x^0=y^0} = 0$ and
$[\partial_{x^0} \rho_D^{\mu\nu}(x,y)]_{x^0=y^0} = - g^{\mu\nu}
\delta(\bx -\by)$ in Feynman gauge. The Wigner transformed
free-field solution solving (\ref{eq:rho2}) then reads   
$\tilde{\varrho}_D (X,k) = \tilde{\varrho}_D (k) = 
 2 \pi\, {\rm sign}(k^0)\, \delta(k^2)$.
A very similar discussion can be done as well 
for the evolution equations (\ref{eq:Fexact}) for fermions, which
is massless due to chiral symmetry as stated above.
Again, in lowest order in the derivative expansion the fermion spectral 
function obeys the free-field equations of motion and one has
$\tilde{\varrho}^{(f)} (X,k) =
\tilde{\varrho}^{(f)} (k) = 2 \pi k\slash\, {\rm sign}(k^0)\, 
\delta(k^2)$.

Assuming a ``generalized
fluctuation-dissipation relation'' or so-called
``Kadanoff-Baym ansatz'':
\bea\db
\tilde{F}_D (X,k) &\db =&\db \left[ \frac{1}{2} + n_D(X,k)  \right]
\tilde{\varrho}_D(X,k) \, , \nonumber\\\db
\tilde{F}^{(f)} (X,k) &\db =&\db \left[ \frac{1}{2} - n^{(f)}(X,k) \right]
\tilde{\varrho}^{(f)} (X,k) \, ,
\label{eq:genflucdiss}
\eea
one may extract the kinetic equations for the effective
photon and fermion particle numbers $n_D$ and $n^{(f)}$, respectively. 
Considering spatially homogeneous, isotropic systems for 
simplicity, we define the on-shell quasiparticle numbers 
($t \equiv X^0$)
\beq\db
n_D(t,\bk) \equiv n_D(t,k)|_{k^0 = \bk} \quad ,
\quad n^{(f)}(t,\bk) \equiv n^{(f)}(t,k)|_{k^0 = \bk}
\eeq
and look for the evolution equation for 
$n_D(t,\bk)$. Here it is useful to note the symmetry properties
\bea\db
\tilde{F}_D(t,-k) &\db =&\db 
\tilde{F}_D(t,k) \,\, , \,\,
\tilde{\varrho}_D(t,-k) = 
- \tilde{\varrho}_D(t,k) \, ,
\nonumber\\\db
\tilde{F}^{(f)}(t,-k)^{\mu} &\db =&\db 
- \tilde{F}^{(f)}(t,k)^{\mu} \,\, , \,\,
\tilde{\varrho}^{(f)}(t,-k)^{\mu} = 
 \tilde{\varrho}^{(f)}(t,k)^{\mu} \, .
\eea
Applied to the quasiparticle ansatz (\ref{eq:genflucdiss}) these
imply
\bea\db
n_D(t,-k) = - \left[ n_D(t,k) + 1 \right] \,\,\, , \,\,\,
n^{(f)}(t,-k) = - \left[ n^{(f)}(t,k)-1 \right] \, .
\eea
This is employed to rewrite terms with negative values of $k^0$.
To order $g^2$ the self-energies read (cf.~Eq.~(\ref{eq:selfg2})):
\bea\db
\tilde{\Pi}_{(F)} (X,k) &\db =&\db 2 g^2 \int \frac{{\rm d}^4 p}{(2\pi)^4}
\Big[\tilde{F}^{(f)}(X,k+p)^{\mu}\tilde{F}^{(f)}(X,p)_{\mu} 
\nonumber\\
&&\db - \frac{1}{4} \tilde{\varrho}^{(f)}(X,k+p)^{\mu}
\tilde{\varrho}^{(f)}(X,p)_{\mu} \Big] \, ,
\nonumber\\[0.1cm]
\db \tilde{\Pi}_{(\varrho)} (X,k) &\db =&\db 
2 g^2 \int \frac{{\rm d}^4 p}{(2\pi)^4}
\Big[\tilde{F}^{(f)}(X,k+p)^{\mu}\tilde{\varrho}^{(f)}(X,p)_{\mu} 
\nonumber\\
&&\db - \tilde{\varrho}^{(f)}(X,k+p)^{\mu}
\tilde{F}^{(f)}(X,p)_{\mu} \Big] \, .
\label{eq:derivself}
\eea 
From the equations (\ref{eq:F2}) and (\ref{eq:genflucdiss})
one finds at this order: ($\bq \equiv \bk - \bp$)  
\bea\db
\lefteqn{
\partial_t n_D(t,\bk) = g^2 k^2 \int \frac{{\rm d}^3 p}{(2\pi)^3}\,
\frac{1}{2 \bk 2 \bp 2 \bq} \Bigg\{ }
\nn
&&\db \Big( n^{(f)}(t,\bp)\, n^{(f)}(t,\bq) \left[n_D(t,\bk) + 1\right]
\nn
&&\db - \left[n^{(f)}(t,\bp)-1\right] \left[n^{(f)}(t,\bq)-1\right] 
n_D(t,\bk) \Big)
\db 2 \pi \delta(\bk - \bp - \bq)
\nonumber \\[0.1cm]
&\db +&\db 2 \Big( \left[n^{(f)}(t,\bp)-1\right] n^{(f)}(t,\bq) 
\left[n_D(t,\bk) + 1\right]
\nn
&&\db - n^{(f)}(t,\bp) \left[n^{(f)}(t,\bq)-1\right] n_D(t,\bk) \Big)
\db2 \pi \delta(\bk + \bp - \bq)
\nonumber \\[0.1cm]
&\db +&\db \Big( \left[n^{(f)}(t,\bp)-1\right] \left[n^{(f)}(t,\bq)-1\right] 
\left[n_D(t,\bk) + 1\right]
\nn
&&\db - n^{(f)}(t,\bp)\, n^{(f)}(t,\bq)\, n_D(t,\bk) \Big)
\db 2 \pi \delta(\bk + \bp + \bq)
\Bigg\} \, .
\label{eq:nevol}
\eea
The r.h.s.~shows the standard ``gain term'' minus ``loss term'' structure.
E.g.~for the case $k^2 > 0$, $k^0 >0$ the interpretation is given by the
elementary processes $e \bar{e} \to \gamma$,
$e \to e \gamma$, $\bar{e} \to \bar{e} \gamma$ and ``$0$'' 
$\to e \bar{e} \gamma$ from which only the first one is not kinematically
forbidden. From (\ref{eq:nevol}) one also recovers the fact 
that the ``on-shell'' evolution with $k^2 = 0$ vanishes identically 
at this order. A nonvanishing result is obtained if one takes into
account ``off-shell'' corrections for a fermion line in the loop of the 
self-energy (\ref{eq:derivself}). As a consequence the first 
nonzero contribution to the self-energy starts at ${\mathcal O}(g^4)$.

\begin{figure}[t]
\centerline{
\epsfig{file=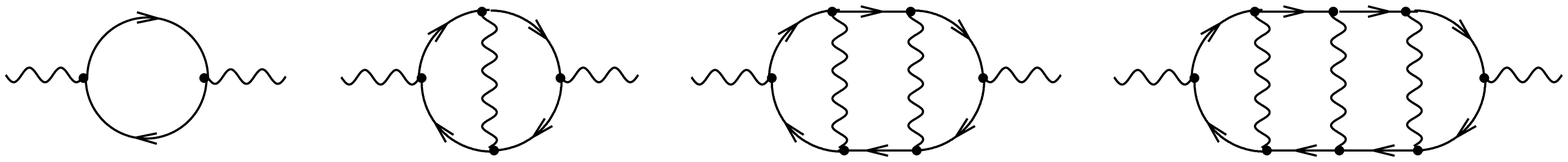,width=12.cm}
}
\vspace*{-0.5cm}
\caption{Infinite series of self-energy contributions with
dressed propagator lines and classical vertices.}
\label{fig:ymselfinf}
\end{figure}
Since the lowest order contribution to the kinetic equation
is of ${\mathcal O}(g^4)$, the 3PI effective action provides 
a self-consistently complete starting point for its description. 
At this order the self-energies and vertex are given by 
Eqs.~(\ref{eq:selffermexact}), (\ref{eq:selfexact}) and (\ref{eq:vertexeom}). 
Starting from the three-vertex (\ref{eq:vertexeom}) consider for a moment 
the vertex resummation for the photon leg only, i.e.~approximate the
fermion-photon vertex by the classical vertex. As a consequence,
one obtains:  
\bea\db
V^{\mu}(x,y;z) &\db \simeq &\db \gamma^{\mu} 
\delta(x - z)\delta(z - y) \label{eq:detver}\\
&&\db - g^2 \int_{x' y'}
\gamma^{\nu} \Delta (x,z)
V^{\mu}(x',y';z) \Delta (y',y)
\gamma^{\sigma} D_{\sigma \nu} (y,x) \, . 
\nonumber
\eea
Using this expression for the photon self-energy~(\ref{eq:selfexact}), by
iteration one observes that this resums all 
the ladder diagrams shown in Fig.~\ref{fig:ymselfinf}. 
Here propagator lines correspond to self-energy resummed
propagators whereas all vertices are given by the classical ones. 
In the context of kinetic equations,
relevant for sufficiently homogeneous systems, the dominance of this sub-class 
of diagrams has been discussed in detail in the weak coupling limit
in the literature (cf.~the bibliography at the
end of this section). One may decompose the contributions
to the kinetic equation into $2 \leftrightarrow 2$ particle processes,
such as $e \bar{e} \to \gamma \gamma$ annihilation in the context of QED,
and inelastic ``$1 \leftrightarrow 2$'' processes, such as the 
nearly collinear bremsstrahlung process. For the description of
``$1 \leftrightarrow 2$'' processes, 
once Fourier transformed with respect to the 
relative coordinates, the gauge field propagator in (\ref{eq:detver}) is
required for space-like momenta. Furthermore, as seen 
from (\ref{eq:nevol}), the proper inclusion of nonzero 
contributions from $2 \leftrightarrow 2$ processes requires to go beyond
the naive on-shell limit. 
In the context of the evolution equations (\ref{eq:rho}) and
(\ref{eq:F}) this can be achieved by employing the following identities:   
\bea\db 
F_D^{\mu\nu}(x,y) &\db =&\db \lim_{t_0 \to -\infty}
\int_{t_0}^{x^0} {\rm d}z \int_{t_0}^{y^0} {\rm d}z'  
\left[ \rho_D(x,z) \Pi_{(F)}(z,z') \rho_D(z',y) \right]^{\mu\nu}
\,\, \nonumber \\
&\db =&\db - \int_{-\infty}^{\infty} {\rm d}z {\rm d}z' 
\left[ D_R(x,z) \Pi_{(F)}(z,z') D_A(z',y)\right]^{\mu\nu} \,\, 
\nonumber \\
\db \rho_D^{\mu\nu}(x,y) &\db =&\db \lim_{t_0 \to -\infty}
\int_{t_0}^{x^0} {\rm d}z \int_{t_0}^{y^0} {\rm d}z'  
\left[ \rho_D(x,z) \Pi_{(\rho)}(z,z') \rho_D(z',y) \right]^{\mu\nu}
\nonumber \\
&\db =&\db - \int_{-\infty}^{\infty} {\rm d}z {\rm d}z' 
\left[ D_R(x,z) \Pi_{(\rho)}(z,z') D_A(z',y)\right]^{\mu\nu} \, ,
\label{eq:Frhoident}
\eea
written in terms of the retarded and advanced propagators, 
$D_R(x,y)^{\mu\nu} = \Theta (x^0-y^0) \rho_D(x,y)^{\mu\nu}$
and $D_A(x,y)^{\mu\nu} = - \Theta (y^0-x^0) \rho_D(x,y)^{\mu\nu}$,
in order to have an unbounded time integration. The above
identity follows from a straightforward application of the
exact evolution equations and using the anti-symmetry property of
the photon spectral function, $\rho_D^{\mu\nu}(x,y)|_{x^0=y^0} = 0$. 
We emphasize that the
identity does not hold for an initial value problem
where the initial time $t_0$ is finite. Similarly, one
finds from (\ref{eq:Fexact}) 
for the fermion two-point functions using 
$\gamma^0 \rho^{(f)}(x,y)|_{x^0=y^0} = i \delta(\bx - \by)$:
\bea\db
F^{(f)}(x,y) 
&\db =&\db - \int_{-\infty}^{\infty} {\rm d}z {\rm d}z' 
\Delta_R(x,z) \Sigma^{(f)}_{(F)}(z,z') \Delta_A(z',y) \, , 
\nonumber \label{eq:fermidF}\\
\db \rho^{(f)} (x,y) 
&\db =&\db - \int_{-\infty}^{\infty} {\rm d}z {\rm d}z' 
\Delta_R(x,z) \Sigma^{(f)}_{(\rho)}(z,z') \Delta_A(z',y) \,\, ,
\label{eq:fermidrho}
\eea
with $\Delta_R(x,y) = \Theta (x^0-y^0) \rho^{(f)}(x,y)$ and
$\Delta_A(x,y) = - \Theta (y^0-x^0) \rho^{(f)}(x,y)$.
Neglecting all derivative
terms, i.e.~using (\ref{eq:grad2}), and the above notation
these give:\footnote{As for the spectral function $\varrho (X,k)$
in Eq.~(\ref{eq:wigner}), the Fourier transform of the retarded
and advanced propagators includes a factor
of $-i$.}  
\bea\db
\tilde{F}_D(X,k) &\db \simeq&\db 
 \tilde{D}_R(X,k) \tilde{\Pi}_{(F)}(X,k) 
\tilde{D}_A(X,k) \, ,
\nonumber\\
\db \tilde{\varrho}_D (X,k) &\db \simeq&\db 
 \tilde{D}_R(X,k) \tilde{\Pi}_{(\varrho)}(X,k) 
\tilde{D}_A(X,k) \, ,
\label{eq:psident}
\eea
and equivalently for the fermion two-point functions.
Applied to one fermion line in the one-loop contribution of 
Fig.~\ref{fig:ymselfinf}, it is straightforward to recover the 
standard Boltzmann equation for $2 \leftrightarrow 2$ processes, using
the ${\mathcal O}(g^2)$ fermion self-energies:
\bea\db
\tilde{\Sigma}_{(F)} (X,k)^{\mu} &\db =& \db
- 2 g^2 \int \frac{{\rm d}^4 p}{(2\pi)^4}
\Big[\tilde{F}_D(X,p) \tilde{F}^{(f)}(X,k-p)^{\mu} 
\nonumber\\
&&\db + \frac{1}{4} \tilde{\varrho}_D(X,p)
\tilde{\varrho}^{(f)}(X,k-p)^{\mu} \Big] \, ,
\nonumber\\[0.1cm]
\db \tilde{\Sigma}_{(\varrho)} (X,k)^{\mu} 
&\db =&\db  - 2 g^2 \int \frac{{\rm d}^4 p}{(2\pi)^4}
\Big[\tilde{F}_D(X,p) \tilde{\varrho}^{(f)}(X,k-p)^{\mu} 
\nonumber\\
&&\db + \tilde{\varrho}_D(X,p)
\tilde{F}^{(f)}(X,k-p)^{\mu} \Big] \, .
\label{eq:derivselfferm}
\eea 
For the Boltzmann equation
$\Delta_R$ and $\Delta_A$ are taken to enter the scattering matrix element,
which is evaluated in (e.g.~``hard thermal loops'' resummed) 
equilibrium, whereas all other lines are taken to be ``on-shell''. 
The contributions from the 
$1 \leftrightarrow 2$ processes may be obtained with the help
of (\ref{eq:psident}) with the ${\mathcal O}(g^2)$ photon 
self-energies~(\ref{eq:derivself}).
Of course, simply adding the contributions from $2 \leftrightarrow 2$ 
processes and $1 \leftrightarrow 2$ processes entails the problem of
double counting since a diagram enters twice.    
This occurs whenever the internal line in a $2 \leftrightarrow 2$
process is kinematically allowed to go on-shell and has to be
suppressed.

\subsubsection{Discussion}
\label{sec:limitations}

In view of the generalized fluctuation-dissipation relation
(\ref{eq:genflucdiss}) employed in the above ``derivation'',
one could be tempted to say that for consistency 
an equivalent relation should be valid for the self-energies as well:
\bea\db
\tilde{\Pi}_{(F)} (X,k) &\db =&\db \left[ \frac{1}{2} + n_D(X,k)  \right]
\tilde{\Pi}_{(\varrho)}(X,k) \, .
\label{eq:flucdissself}
\eea
Such a relation is indeed valid in thermal equilibrium, where all 
dependence on the center coordinate $X$ is lost. 
Furthermore, the above relation can be shown to be a consequence
of (\ref{eq:genflucdiss}) using the identities (\ref{eq:Frhoident}) 
in a lowest-order derivative expansion: Together with Eq.~(\ref{eq:psident})
the above relation for the self-energies is a direct consequence of
the ansatz (\ref{eq:genflucdiss}). However, clearly
this is too strong a constraint since the evolution equation 
(\ref{eq:F2}) would become trivial in this case: Eq.~(\ref{eq:genflucdiss})
and (\ref{eq:flucdissself}) lead to a vanishing
r.h.s.~of the evolution equation for $\tilde{F}_D (X,k)$ and 
there would be no evolution.

The above argument is just a manifestation of the well-known
fact that the kinetic equation is not a self-consistent approximation
to the dynamics. The discussion of 
Sec.~\ref{eq:onshelllimits} takes 
into account the effect of scattering for the dynamics of effective 
occupation numbers, while keeping the spectrum   
free-field theory like. In contrast, the same scattering does 
induce a finite width for the spectral function in the 
self-consistent $n$PI approximation discussed in Sec.~\ref{sec:nonequilibrium} 
because of a nonvanishing imaginary part of the self-energy
(cf.~also the discussion and explicit solution of a similar 
Yukawa model in Sec.~\ref{sec:prethermal} and the discussion 
in Sec.~\ref{eq:scatoffmem}). 

Though particle number is not well-defined in an 
interacting relativistic quantum field theory in the
absence of conserved charges,
the concept of time-evolving effective particle numbers in an
interacting theory is useful in the
presence of a clear separation of scales.
Much progress has been achieved in the quantitative 
understanding of kinetic descriptions in the vicinity of thermal 
equilibrium for gauge theories at high temperature, which is well 
documented in the literature and for further reading we refer
to the bibliography below.

For gauge theories the employed on-shell limit circumvents problems of
gauge invariance or subtle aspects of renormalization. We emphasize
that renormalization for linear symmetries as realized in QED can 
be treated along similar lines as discussed in Sec.~\ref{sec:renormalize}. 
The generalization to nonabelian gauge theories is, however, technically more 
involved and needs to be further investigated.

A derivative expansion is typically not valid 
at early times where the time evolution can
exhibit a strong dependence on $X$,
and the homogeneity requirement underlying kinetic descriptions
may only be fulfilled at sufficiently late times. This has
been extensively discussed in the context of 
scalar and fermionic theories in Sec.~\ref{sec:neqevolution}. 
Homogeneity is certainly realized at late times sufficiently close to 
the thermal limit, since for thermal equilibrium the correlators do 
strictly not depend on $X$. 
Of course, by construction kinetic equations 
are not meant to discuss the detailed early-time
behavior since the initial time $t_0$
is sent to the remote past. For practical purposes, in this context 
one typically specifies the initial condition for the effective 
particle number distribution at some finite
time and approximates the evolution by the 
equations with $t_0 \to - \infty$. 
The role of finite-time effects has been controversially discussed
in the recent literature in the context of photon production
in relativistic plasmas at high temperature. Here a solution
of the proper initial-time equations as discussed
in Sec.~\ref{sec:nonequilibrium} seems mandatory.

\section{Acknowledgements}

I thank Gert Aarts, Daria Ahrensmeier, Rudolf Baier, 
Szabolcs Bors\'anyi, J{\"u}rgen Cox, Markus M.~M{\"u}ller,
Urko Reinosa, Julien Serreau and Christof Wetterich 
for very fruitful collaborations on this topic.

\section{Bibliographical notes}

I apologize for the omission of many interesting contributions
to this wide topic in the annotated list below, which concentrates on
relativistic quantum field theory applications related
to the content of the text presented above.   
\bi
\item A rather recent short review with a more comprehensive
{\rr list of references} can be found in:
{\em Progress in nonequilibrium quantum field theory}, 
J.~Berges and J.~Serreau, in Strong and Electroweak Matter 2002,
ed. M.G.~Schmidt (World Scientific, 2003), http://arXiv:hep-ph/0302210.
\item General discussions on {\rr $n$PI effective actions} include:  
{\em Effective Action For Composite Operators},
J.~M.~Cornwall, R.~Jackiw and E.~Tomboulis,
Phys.\ Rev.\ D {\bf 10} (1974) 2428.
{\em Higher effective actions for bose systems},
H.~Kleinert, Fortschritte der Physik {\bf 30} (1982) 187.
{\em Functional Methods in Quantum Field Theory and Statistical
Physics}, A.N.~Vasiliev, Gordon and Breach Science Pub.~(1998).
{\em $n$PI effective action techniques for gauge theories},
J.~Berges, Phys.\ Rev.\ D in print, http://arXiv:hep-ph/0401172.
{\em Gauge-fixing dependence of Phi-derivable approximations},
A.~Arrizabalaga and J.~Smit, Phys.\ Rev.\ D {\bf 66} (2002) 065014.
\item The nonperturbative 
{\rr 2PI $1/N$ expansion} is derived in: 
{\em Controlled nonperturbative dynamics of quantum fields out of
equilibrium}, J.~Berges, Nucl.\ Phys.\ {\bf A699} (2002) 847;
{\em Far-from-equilibrium dynamics with broken symmetries
from the 1/N expansion of the 2PI effective action}, G.~Aarts,
D.~Ahrensmeier, R.\ Baier,
J.~Berges and J.~Serreau, Phys.\ Rev.\ {\bf D66} (2002) 045008.
The latter discusses also the relation to the similar approximation scheme of  
{\em Resumming the large-N approximation for time evolving quantum systems,}
B.~Mihaila, F.~Cooper and J.~F.~Dawson, Phys.\ Rev.\ D {\bf 63} (2001) 096003
\item {\rr Far-from-equilibrium quantum fields and thermalisation} 
are discussed in:
{\em Thermalization of Quantum Fields from Time-Reversal Invariant 
Evolution Equations}, J.~Berges and J.~Cox, Phys.\ Lett.\ {\bf B517} (2001) 
369-374. 
{\em Nonequilibrium time evolution of the spectral function in 
quantum field theory}, G.~Aarts and J.~Berges,
Phys.\ Rev.\ D {\bf 64} (2001) 105010.
{\em Controlled nonperturbative dynamics of quantum fields out of
equilibrium}, J.~Berges, Nucl.\ Phys.\ {\bf A699} (2002) 847.
{\em Quantum dynamics of phase transitions in broken symmetry 
$\phi^4$ field theory},
F.~Cooper, J.~F.~Dawson and B.~Mihaila, Phys.~Rev.~{\bf D67}
(2003) 056003. 
{\em Thermalization of fermionic quantum fields},
J.~Berges, Sz.~Bors{\'a}nyi and J.~Serreau, Nucl.\ Phys.\ {\bf B660}
(2003) 52.
{\em Bose-Einstein condensation without chemical potential},
D.~J.~Bedingham, Phys.\ Rev.\ D {\bf 68} (2003) 105007.
{\em Quantum dynamics and thermalization for out-of-equilibrium phi**4-theory},
S.~Juchem, W.~Cassing and C.~Greiner,
Phys.\ Rev.\ D {\bf 69} (2004) 025006.
\item The phenomenon of {\rr prethermalization} is discussed in:
{\em Prethermalization}, J.~Berges, Sz.~Bors{\'a}nyi and C.~Wetterich, 
Phys.\ Rev.\ Lett.\ in print, http://arXiv:hep-ph/0403234 
\item {\rr Far-from-equilibrium dynamics of macroscopic fields with large
fluctuations} using the 2PI $1/N$ expansion are discussed in: 
{\em Parametric resonance in quantum field theory},
J.~Berges and J.~Serreau,
Phys.\ Rev.\ Lett.~{\bf 91} (2003) 111601. 
The leading-order large-$N$ description has been given in:  
{\em Analytic and numerical study of preheating dynamics},
D.~Boyanovsky, H.~J.~de Vega, R.~Holman and J.~F.~Salgado,
Phys.\ Rev.\ D {\bf 54} (1996) 7570.
\item A detailed discussion about {\rr Gaussian initial density 
matrices and dynamics} can be found in:
{\em Nonequilibrium dynamics of symmetry breaking in lambda Phi**4 field},
F.~Cooper, S.~Habib, Y.~Kluger and E.~Mottola,
Phys.\ Rev.\ D {\bf 55} (1997) 6471.
Gaussian dynamics for inhomogeneous fields are discussed in:
{\em Particle production and effective thermalization in inhomogeneous mean  
field theory}, G.~Aarts and J.~Smit, Phys.\ Rev.\ D {\bf 61} (2000) 025002.
{\em Staying thermal with Hartree ensemble approximations},
M.~Salle, J.~Smit and J.~C.~Vink,
Nucl.\ Phys.\ B {\bf 625} (2002) 495. 
{\em Dynamical behavior of spatially inhomogeneous relativistic lambda  
phi**4 quantum field theory in the Hartree approximation},
L.~M.~Bettencourt, K.~Pao and J.~G.~Sanderson,
Phys.\ Rev.\ D {\bf 65} (2002) 025015.
For a leading-order study including fermions see:
{\em Nonequilibrium dynamics of fermions in a spatially homogeneous scalar  
background field}, J.~Baacke, K.~Heitmann and C.~P\"atzold,
Phys.\ Rev.\ D {\bf 58} (1998) 125013.
\item The {\rr renormalization} of 2PI approximation 
schemes is discussed in:
{\em Renormalization in self-consistent approximations schemes at 
finite  temperature. I: Theory}, H.~van Hees and J.~Knoll,
Phys.\ Rev.\ D {\bf 65} (2002) 025010. 
{\em Renormalization of self-consistent approximation schemes II:  
Applications to the sunset diagram}, H.~Van Hees and J.~Knoll,
Phys.\ Rev.\ D {\bf 65} (2002) 105005.
{\em Renormalization in self-consistent approximation schemes at finite
temperature III: Global symmetries}, H.~van Hees and J.~Knoll,
Phys.\ Rev.\ D {\bf 66} (2002) 025028.
{\em Renormalizability of Phi-derivable approximations in scalar phi**4
theory}, J.~P.~Blaizot, E.~Iancu and U.~Reinosa,
Phys.\ Lett.\ B {\bf 568} (2003) 160.
{\em Renormalization of $\Phi$-derivable approximation
schemes in scalar field theories}, J.-P.~Blaizot,
E.~Iancu and U.~Reinosa, Nucl.\ Phys.\ A {\bf 736} (2004) 149.
{\em Renormalizing the Schwinger-Dyson equations in the auxiliary field
formulation of lambda phi**4 field theory},
F.~Cooper, B.~Mihaila and J.~F.~Dawson, http://arXiv:hep-ph/0407119.
{\em Renormalized thermodynamics from the 2PI effective action},
J.~Berges, Sz.~Borsanyi, U.~Reinosa and J.~Serreau, 
http://arXiv:hep-ph/0409123.
\item  Classical aspects of nonequilibrium quantum fields and 
{\rr precision tests} of the 2PI $1/N$ expansion are examined in: 
{\em Classical aspects of quantum fields far from equilibrium}, 
G.~Aarts and J.~Berges, Phys.\ Rev.\ Lett.\ {\bf 88}
(2002) 0416039. 
{\em Nonequilibrium quantum fields and the classical field theory limit},
J.~Berges, Nucl.\ Phys.\ A {\bf 702} (2002) 351.
{\em Tachyonic preheating using 2PI - 1/N dynamics and the classical
approximation}, A.~Arrizabalaga, J.~Smit and A.~Tranberg,
http://arXiv:hep-ph/0409177.
\item {\rr Classical statistical field theory studies} related to
approximation schemes in QFT can be found in:
{\em Exact and truncated dynamics in nonequilibrium field theory},
G.~Aarts, G.~F.~Bonini and C.~Wetterich,
Phys.\ Rev.\ D {\bf 63} (2001) 025012.
{\em On thermalization in classical scalar field theory},
G.~Aarts, G.~F.~Bonini, C.~Wetterich,
Nucl.\ Phys.\  {\bf B587} (2000) 403. 
{\em Time evolution of correlation functions and thermalization},
G.~F.~Bonini and C.~Wetterich,
Phys.\ Rev.\ D {\bf 60} (1999) 105026.
{\em Classical limit of time-dependent quantum field theory: A  
Schwinger-Dyson approach}, F.~Cooper, A.~Khare and H.~Rose,
Phys.\ Lett.\ B {\bf 515} (2001) 463.
K.~Blagoev, F.~Cooper, J.~Dawson and B.~Mihaila,
Phys.\ Rev.\ D {\bf 64} (2001) 125003.
For diagrammatics in classical field theory see:
{\em Finiteness of hot classical scalar field theory and the plasmon
damping rate}, G.~Aarts and J.~Smit,
Phys.\ Lett.\ B {\bf 393} (1997) 395.
{\em Classical statistical mechanics and Landau damping},
W.~Buchm\"uller and A.~Jakov\'ac,
Phys.\ Lett.\ B {\bf 407} (1997) 39.
{\em Classical approximation for time-dependent quantum field
theory:  Diagrammatic analysis for hot scalar fields},
G.~Aarts and J.~Smit,
Nucl.\ Phys.\ {\bf B511} (1998) 451.
\item The relation to {\rr kinetic equations} is discussed in:
{\em Nonequilibrium Quantum Fields: Closed Time Path Effective Action, 
Wigner Function And Boltzmann Equation}, E.~Calzetta and B.~L.~Hu,
Phys.\ Rev.\ D {\bf 37} (1988) 2878; 
{\em Exact Conservation Laws of the Gradient Expanded Kadanoff-Baym Equations},
J.~Knoll, Y.~B.~Ivanov and D.~N.~Voskresensky, Annals 
Phys.\  {\bf 293} (2001) 126. 
{\em Nonequilibrium quantum fields with large fluctuations},
J.~Berges and M.~M.~Muller, in
Progress in Nonequilibrium Green's Functions II, 
Eds.~M.~Bonitz and D.~Semkat, World Scientific (2003)
[http://arXiv:hep-ph/0209026].
Cf.~also 
{\em Stochastic dynamics of correlations in quantum field theory: From
Schwinger-Dyson to Boltzmann-Langevin equation}, E.~Calzetta and B.~L.~Hu,
Phys.\ Rev.\ D {\bf 61} (2000) 025012.
\item For a discussion of {\rr transport coefficients} employing 2PI effective
actions see:
{\em Hydrodynamic transport functions from quantum kinetic field theory},
E.~A.~Calzetta, B.~L.~Hu and S.~A.~Ramsey,
Phys.\ Rev.\ D {\bf 61} (2000) 125013.
{\em Transport coefficients from the 2PI effective action},
G.~Aarts and J.~M.~Martinez Resco,
Phys.\ Rev.\ D {\bf 68} (2003) 085009.
{\em Shear viscosity in the O(N) model},
G.~Aarts and J.~M.~Martinez Resco,
JHEP {\bf 0402} (2004) 061.
{\em Transport coefficients at leading order: Kinetic theory versus diagrams},
G.~D.~Moore, in Strong and Electroweak Matter 2002,
ed. M.G.~Schmidt (World Scientific, 2003), http://arXiv:hep-ph/0211281.
Cf.~also {\em Transport coefficients in high temperature gauge theories II: 
Beyond leading log}, P.~Arnold, G.~D.~Moore and L.~G.~Yaffe,
JHEP {\bf 0305} (2003) 051.
\item For an application to {\rr photon production} rates in
a non-equilibrium medium see: 
{\em Out-of-equilibrium electromagnetic radiation}, J.~Serreau,
JHEP {\bf 0405} (2004) 078.
\ei

\paragraph{\rr Figures}

Figs.~\ref{Figpeakphi}--\ref{FigLogFrho}: 
J.~Berges, Nucl.\ Phys.\ {\bf A699} (2002) 847. 
Fig.~\ref{fig:lateuni} left: 
J.~Berges and J.~Cox, Phys.\ Lett.\ {\bf B517} (2001) 
369-374. 
Fig.~\ref{fig:lateuni} right: G.~Aarts and J.~Berges,
Phys.\ Rev.\ D {\bf 64} (2001) 105010.
Figs.~\ref{fig:join_fn}--\ref{fig:Tkinevolrat}:
J.~Berges, Sz.~Bors{\'a}nyi and C.~Wetterich, 
Phys.\ Rev.\ Lett.\ in print, http://arXiv:hep-ph/0403234.
Figs.~\ref{fig:energy}--\ref{fig:field}: 
J.~Berges and J.~Serreau,
Phys.\ Rev.\ Lett.~{\bf 91} (2003) 111601. 
Figs.~\ref{fig:classearly}--\ref{fixedpointclassical}:
G.~Aarts and J.~Berges, Phys.\ Rev.\ Lett.\ {\bf 88}
(2002) 0416039.

\end{document}